\documentstyle[11pt,epsf]{article}

\newcommand{\be}{\begin{equation}}
\newcommand{\ee}{\end{equation}}
\newcommand{\bea}{\begin{eqnarray}}
\newcommand{\eea}{\end{eqnarray}}

\renewcommand{\theequation}{\arabic{section}.\arabic{equation}}

\begin{document}

\begin{titlepage}

\begin{flushright}
hep-th/9909061\\
to appear in {\em Physics Reports}
\end{flushright}

\vspace{.5in}

\begin{center}
\huge
{\bf Superstring Cosmology}

\vspace{.5in}

\large{
James E. Lidsey$^1$, David Wands$^2$ and E. J. Copeland$^3$}

\normalsize
\vspace{.2in}

{\em $^1$\ Astronomy Unit, School of Mathematical 
Sciences,  \\ 
Queen Mary \& Westfield, Mile End Road, London, E1 4NS, U.K.\\
j.e.lidsey@qmw.ac.uk}

\vspace{.2in}

{\em $^2$\ School of Computer Science and Mathematics,
University of Portsmouth,\\
Portsmouth, PO1 2EG, United Kingdom\\
david.wands@port.ac.uk}

\vspace{.2in}

{\em $^3$\ Centre for Theoretical Physics, University of Sussex,\\
Falmer, Brighton, BN1 9QJ, United Kingdom\\
e.j.copeland@sussex.ac.uk}

\vspace{.5in}

\begin{abstract}
Aspects of superstring cosmology are reviewed with an emphasis on the
cosmological implications of duality symmetries in the theory. The
string effective actions are summarized and toroidal compactification
to four dimensions reviewed.  Global symmetries that arise in the
compactification are discussed and the duality relationships between
the string effective actions are then highlighted.
Higher--dimensional Kasner cosmologies are presented and interpreted
in both string and Einstein frames, and then given in dimensionally
reduced forms.  String cosmologies containing both non--trivial
Neveu--Schwarz/Neveu--Schwarz and Ramond--Ramond fields are derived by
employing the global symmetries of the effective actions.  Anisotropic
and inhomogeneous cosmologies in four--dimensions are also developed.
The review concludes with a detailed analysis of the pre--big bang
inflationary scenario. The generation of primordial spectra of
cosmological perturbations in such a scenario is discussed. 
Possible future directions offered in the 
Ho\v{r}ava--Witten theory are outlined.

\end{abstract}

\end{center}

\end{titlepage}

\tableofcontents

\newpage

\section{Introduction}

\label{Section1}

\setcounter{equation}{0}

\def\theequation{\thesection.\arabic{equation}}

Superstring theory represents the most promising candidate for a
unified theory of the fundamental interactions, including gravity 
\cite{GreSchWit87,Polchinski98}. One
of the strongest constraints on the theory is that it should be
consistent with the standard model of the very early universe.  The
cosmological implications of string theory are currently receiving
considerable attention. This interest has been inspired in part by the
recent advances that have been made towards a non--perturbative
formulation of the theory.  There are five anomaly--free,
supersymmetric perturbative string theories known as the type I, type
IIA, type IIB, ${\rm SO}(32)$ heterotic and ${\rm E}_8 \times {\rm
E}_8$ heterotic theories.  There is now evidence that these theories
are related by a set of dualities and may in fact represent different
manifestations of a more fundamental quantum theory, often termed
M--theory \cite{Witten95}.  
Supersymmetry implies that the quantization of the string
is only consistent if spacetime is ten--dimensional.  On the other
hand, $M$--theory, defined originally in terms of the strongly coupled
limit of the type IIA superstring, is an eleven--dimensional
theory. Eleven--dimensional supergravity fits into the picture as the
low--energy limit of this new theory \cite{Witten95,Townsend95}.
The goal of superstring cosmology is to examine the dynamical
evolution in these theories and re-examine cosmological questions in
the light of our new understanding of string theory.

String theory contains a number of massless degrees of freedom that
are separated from the massive states by an energy gap of order
$\sqrt{\hbar c/\alpha'}$, where $\alpha'$ is the inverse string tension
and is usually taken to be close to the Planck scale (see below).  
In the `sigma--model' approach one considers only the massless modes
of the string, and conformal invariance is imposed by insisting that
the $\beta$--functions for the fields vanish. These constraints may
then be interpreted as field equations that are derived by varying an
effective spacetime action~\cite{FraTse85a,FraTse85b,CalFriMar85,Sen85,Lovelace86}.  To
lowest--order in the perturbation theory and in the low--energy limit,
the massless sectors of the superstring theories are determined by the
corresponding supergravity actions.

The standard approach in string cosmology is to analyse
time--dependent solutions to the lowest--order string equations of
motion.  This approach applies on scales below the string scale but
above those energies where the string symmetries are broken. It is
valid if the fields evolve sufficiently slowly that their
higher--derivative terms can be neglected. Solutions derived in this
context may be viewed as perturbative approximations to exact
solutions of the full theory and it is anticipated that they should
exhibit at least some of the features of these more general solutions.

A definitive prediction of string theory is the existence of a scalar
field, $\varphi$.  This is referred to as the dilaton and it couples
directly to matter.  There are two further massless excitations that
are common to all five perturbative string theories. These are the
tensor field, $g_{\mu \nu}$, known as the graviton, and a rank two
anti-symmetric tensor field, $B_{\mu \nu}$.
The appearance of the dilaton in the string spectrum was first
discussed by Scherk and Schwarz \cite{SchSch74a,SchSch74b}.
Its vacuum expectation value determines the strengths
of both the gauge and gravitational couplings.  
The inverse string tension $\alpha'$ defines the characteristic string length
scale:
\begin{equation}
l_s \equiv \sqrt{\hbar c\alpha'} \ ,
\end{equation}
but the effective Planck length is dependent upon both $\alpha'$ and
the value of the dilaton:
\begin{equation}
\label{deflPlD}
l_{\rm Pl}^{(D)} \equiv e^{\varphi/(D-2)}\, \sqrt{\hbar c\alpha'} \ ,
\end{equation}
in a $D$--dimensional spacetime (see \cite{Veneziano98} for a review). 
Henceforth we set $\hbar=c=1$ but retain units of length (or
equivalently mass$^{-1}$).
The gauge coupling strength is given by~\cite{Veneziano98}
\begin{equation}
\alpha_{\rm gauge} \sim g_s^2 \equiv e^\varphi = \left( {l_{\rm Pl} \over l_s}
\right)^{D-2} \ ,
\end{equation}
and thus we enter the weak coupling regime of string theory for
$e^\varphi \ll 1$.  In such a regime, one may treat the dilaton as a
massless particle in perturbative string theory.  The cosmological
consequences of the dilaton field in this regime are profound and its
dynamical effects lead to a radical departure from the standard
picture of early universe cosmology based on Einstein gravity with
a fixed Planck length.

A central paradigm of modern theoretical models of the early universe
is cosmological inflation, where the universe undergoes an epoch of
accelerated expansion in its most distant past. (An extended bibliography 
can be found in the recent reviews   
\cite{LidLyt93,LidLidKol97,LytRio99,KamKos99}). If sufficient
inflation occurs, many of the problems of the hot big bang model, such
as the horizon and flatness problems, can in principle be
resolved~\cite{Guth81}. Inflation also provides a causal mechanism for
generating a primordial spectrum of density
inhomogeneities~\cite{LidLyt93}. This is essential for producing the
observed large--scale structure in the universe as well as the
temperature anisotropies of the cosmic microwave background radiation.
A crucial question that must be addressed in string cosmology,
therefore, is whether the theory admits realistic inflationary
solutions. 
It is known that supergravity corrections make it difficult to obtain
sufficiently flat potentials to drive conventional slow-roll
inflation~\cite{LytRio99}. 
In recent years an inflationary model known as the pre--big bang
scenario has been developed~\cite{GasVen93a} employing the concepts of
string duality within a cosmological setting. In this model, the
accelerated expansion of the universe is driven by the kinetic energy
of the dilaton field. This differs significantly from the standard
chaotic inflation picture, where the expansion is driven by potential
energy. Although there presently remain unresolved problems with
this scenario, it does have a number of important astrophysical
consequences that in principle could be detectable within the next few
years. This opens up the tantalizing prospect of directly constraining
string theory via current and forthcoming cosmological observations.

In this review, we discuss the nature of cosmological solutions that
are derived from the string effective actions, with particular
emphasis on the cosmological implications of duality symmetries.
Prior to the developments that led to the discovery of string
dualities, the majority of studies in string cosmology focused on the
${\rm E}_8 \times {\rm E}_8$ heterotic theory, since this is
considered to be the theory most relevant to particle physics
phenomenology. (For a review see, e.g., Ref. \cite{Tseytlin92} and
references therein). Given such a change in perspective, however, the
type II theories and eleven--dimensional supergravity may also be
relevant to cosmology. One of the purposes of this review is to focus
on some of the cosmological aspects of the type II theories.

This review is intended for a wide audience including 
particle physicists, cosmologists and relativists. 
It is anticipated, therefore, 
that the typical reader will be more expert in some areas of the review 
than in others. With this in mind, we have divided the review into three parts 
of roughly equal length. The first part (Sections 2--5) reviews 
the subject of string dualities at the level of the supergravity theories. 
The second part (Sections 6--8) develops and studies 
different classes of string cosmologies in a variety 
of settings and the final part (Sections 9--10)
reviews the pre--big bang inflationary scenario.

More specifically, we begin in Section \ref{Section2} with a summary of the 
different fields that arise in 
ten-- and eleven--dimensional supergravity theories. 
The toroidal compactifications of the different sectors of the theories 
are then considered in Section \ref{Section3}. Section 
\ref{Section4} proceeds to discuss 
the non--compact, global symmetries of the dimensionally reduced 
actions that are relevant to later Sections in the review and Section 
\ref{Section5} provides an introductory overview of the web of dualities 
that link the perturbative superstring theories. 

In Section \ref{Section6} we review the higher--dimensional Kasner
solutions based on a toroidal spacetime that are the fundamental
building blocks of many studies in string cosmology. We present the
different interpretations that are possible according to whether one
works in the higher--dimensional, or dimensionally reduced theory, and
according to whether the solutions are presented in terms of the
string length or the Planck scale [whose definition given in
Eq.~(\ref{deflPlD}) is dependent up on the number of dimensions].
Section \ref{Section7} contains a detailed analysis of
four-dimensional string cosmologies containing non--trivial NS--NS
fields. We consider the class of spatially isotropic and homogeneous
Friedmann--Robertson--Walker (FRW) universes, together with the
spatially anisotropic Bianchi models and the inhomogeneous
Einstein--Rosen space--times. Section \ref{Section8} extends the
analysis to include the Ramond--Ramond sector of the type IIB theory.

A review of the pre--big bang inflationary scenario is presented 
in Sections \ref{Section9} and \ref{Section10}. 
Section \ref{Section9} addresses the main unresolved issues
in this scenario, including the question of fine--tuning in 
the initial conditions and the problem of exiting from the 
inflationary phase. Section \ref{Section10} 
discusses the generation of primordial perturbation spectra 
from quantum vacuum fluctuations in the massless fields of the string
effective actions. It is shown how the spectra 
are intimately related to the duality symmetries 
that arise in string theory.
The formalism developed therefore provides a link 
between these duality symmetries and the observed large--scale universe. 

We conclude in Section \ref{Section11} with a 
discussion on the cosmological solutions that 
arise in the Ho\v{r}ava--Witten interpretation 
of the ${\rm E}_8\times {\rm E}_8$ heterotic superstring 
\cite{HorWit96a,HorWit96b}. 
Appendices summarise some of the mathematical concepts used in the
review, including conformal transformations, the modular group of the
torus and the Bianchi classification of homogeneous spacetimes.
Unless otherwise stated, units are chosen such that $\hbar=c=1$.  
Our sign--conventions are those of Wald~\cite{Wald84}, denoted $(+++)$
in Misner, Thorne and Wheeler~\cite{MisThoWhe73}.

\newpage
\section*{PART I}

\addcontentsline{toc}{section}{PART I}

\section{M--theory and Superstring Effective Actions}
\label{actionSection}
\label{Section2}

\setcounter{equation}{0}

\def\theequation{\thesection.\arabic{equation}}

In this Section we present a brief overview of the bosonic sectors of
the effective supergravity actions of the five superstring theories
together with that of $D=11$ supergravity. More detailed introductory
reviews to perturbative string theories are provided in
\cite{GreSchWit87,Polchinski98,LusThe89,Polchinski94}, where a full
bibliography can also be found. Recent extended reviews on
supergravity theories in higher dimensions can be found in
\cite{West98,Tanii98}.

The world--sheet action for a free, closed superstring with string
tension $T=1/(2\pi \alpha')$, is 
\begin{equation}
\label{worldsheet}
S= \frac{1}{4\pi \alpha'} \int d^2\sigma \eta_{\mu\nu} \left[ 
\eta^{\alpha\beta} \partial_{\alpha} X^{\mu} \partial_{\beta} 
X^{\nu} +i \bar{\psi}^{\mu} \gamma^{\alpha} \partial_{\alpha} 
\psi^{\nu} \right]
\end{equation}
The coordinates on the world--sheet are $\sigma^{\alpha} =(\tau ,
\sigma )$ $(\alpha = 0,1)$ and the gauge has been chosen such that the
world--sheet has a flat metric.  The coordinates $X^{\mu} (\tau
,\sigma )$ $(\mu =0, 1, \dots , D-1 )$ are the coordinates of the
string world--sheet propagating in $D$--dimensional Minkowski
spacetime with (flat) metric $\eta_{\mu\nu}$. The spinors on the
world--sheet are denoted by $\psi^{\mu}=\psi^{\mu} (\tau , \sigma )$.
The matrices $\gamma^{\alpha}$ are $2\times 2$ matrices satisfying the
algebra $\{ \gamma_{\alpha} , \gamma_{\beta} \}_+
=2\eta_{\alpha\beta}$ and can be taken to be real.

The world--sheet action (\ref{worldsheet}) is supersymmetric
in the sense that it is invariant under the infinitesimal transformation: 
\begin{equation}
\label{susy}
\delta X^{\mu}  = i \bar{\epsilon} \psi^{\mu} , 
\qquad \delta \psi^{\mu} = \gamma^{\alpha} 
\partial_{\alpha} X^{\mu} \epsilon
\end{equation}
where $\epsilon$ represents a constant anti-commuting 
spinor. This  world--sheet supersymmetry relates the spacetime coordinates 
$X^{\mu}(\tau, \sigma )$ to the fermionic variables 
$\psi^{\mu}(\tau , \sigma )$. 

A suitable choice of basis for 
the $\gamma$--matrices is: 
\begin{equation}
\gamma_0 =\left( \begin{array}{cc}
0 & 1 \\
-1 & 0 \end{array} \right) , \quad 
\gamma_1 = \left( \begin{array}{cc}  
0 & 1 \\
1 & 0 \end{array} \right)
\end{equation}
Defining the matrix $\gamma_2\equiv \gamma_0 
\gamma_1$ then implies that 
the spinor $\psi$ may be separated into two chiral components:
\begin{equation}
\label{chiral}
\psi = \left( \begin{array}{c} 
\psi_R \\
\psi_L \end{array} \right)
\end{equation}
where 
\begin{equation}
\psi = \frac{1}{2} (1+\gamma_2)\psi_R +\frac{1}{2}
(1-\gamma_2)\psi_L
\end{equation}

The equation of motion for the fermionic 
degrees of freedom is the two--dimensional 
Dirac equation:
\begin{equation}
\label{dirac}
\gamma^{\alpha}\partial_{\alpha} \psi^{\mu} =0
\end{equation}
The importance of the basis (\ref{chiral}) is that it allows Eq. 
(\ref{dirac}) to become separated into two parts: 
\begin{eqnarray}
\left( \frac{\partial}{\partial \sigma} +\frac{\partial}{\partial\tau} 
\right) \psi^{\mu}_R =0 \\
\left( \frac{\partial}{\partial \sigma} -\frac{\partial}{\partial \tau} 
\right) \psi^{\mu}_L =0
\end{eqnarray}
This implies that left and right moving fermions 
have negative and positive chirality, respectively. 
Moreover, since they are decoupled, they can be 
treated separately. 

Boundary conditions must be imposed on solutions to 
these equations. For closed strings both the left and 
right moving fermions are either periodic or anti 
periodic, i.e., $\psi^{\mu}_{L,R} (\sigma =2\pi ) = 
\pm \psi^{\mu}_{L,R} (\sigma =0)$. The sector corresponding to  
periodic boundary conditions is referred to as the Ramond (R) 
sector \cite{Ramond71}, 
whereas the sector for anti periodic boundary conditions
is known as the Neveu--Schwarz (NS) sector \cite{NevSch71a,NevSch71b}. 
It follows that for both left and right moving fermions there are four
possible sectors in total:  
NS--NS, RR, NS--R and R--NS. The spacetime 
bosonic excitations arise from  the NS--NS and RR sectors, 
whereas the fermions arise from the NS--R and 
R--NS sectors. 

In the zero--slope limit, $\alpha' \rightarrow 0$, all massive modes
in the superstring spectrum decouple, and only the massless sector
remains.  The effective action for the massless excitations of the
superstring can be derived by rewriting the world--sheet action
(\ref{worldsheet}) for a curved spacetime background and imposing the
constraint that quantum corrections do not break conformal invariance. 
This implies the vanishing of the $\beta$--functions and
these constraints can in turn be interpreted as field equations
derived from an effective
action~\cite{CalFriMar85,FraTse85a,FraTse85b,Sen85,Lovelace86}.

\paragraph{Type IIA superstring.}
The effective bosonic action of the type IIA superstring 
is $N=2$, $D=10$, non--chiral supergravity and is given by 
\begin{eqnarray}
\label{TYPEIIA}
S_{\rm IIA} = \frac{1}{16\pi \alpha^{\prime4}} \left\{ 
\int d^{10}x \sqrt{|g_{10}|} 
\left[ e^{-\Phi}  \left( R_{10} +
\left( \nabla \Phi \right)^2  -\frac{1}{12} H_3^2 \right) \right. \right.
\nonumber \\
\left. \left. -\frac{1}{4} F^2_2 -\frac{1}{48} ({F_4}')^2 \right] 
+\frac{1}{2} \int B_2 \wedge F_4 \wedge F_4 \right\}
\end{eqnarray}
where $R_{10}$ is the Ricci scalar curvature of the spacetime with metric
$g_{MN}$ and $g_{10} \equiv {\rm det}g_{MN}$.  Strings sweep out
geodesic surfaces with respect to the metric $g_{MN}$. The corresponding 
action in the conformally related Einstein frame was presented 
in Refs. \cite{HuqNam85,GiaPer84,CamWes84}. The
dilaton field, $\Phi$, determines the value of the string coupling
parameter, $g_s^2 =e^{\Phi}$. It is interesting to note that the
dilaton--graviton sector of this action may be interpreted as a
ten--dimensional Brans--Dicke theory \cite{BraDic61}, 
where the coupling between the
dilaton and graviton is specified by the Brans--Dicke parameter
$\omega =-1$.  The antisymmetric tensor field strengths are defined by
$H_3=dB_2$, $F_2 =dA_1$, $F_4 =dA_3$ and $F'_4 =F_4 +A_1 \wedge H_3$,
where in general $X_{p}$ denotes an antisymmetric $p$--form potential
and $d$ is the exterior derivative. The last term in
Eq. (\ref{TYPEIIA}) is a Chern--Simons term and is a necessary
consequence of supersymmetry. For the backgrounds we consider in this
review, however, this term can be neglected and we do not consider
it further.
Action (\ref{TYPEIIA}) represents the zeroth--order expansion in both
the string coupling and the inverse string tension. The NS--NS sector
of the action contains the graviton, the antisymmetric 2--form
potential and the dilaton field. The RR sector contains antisymmetric
$p$--form potentials, where $p$ is odd. The NS--NS sector couples
directly to the dilaton, but the RR fields do not
\cite{Witten95,Polchinski96}.

There are two gravitini in the spectra of both the type IIA and type
IIB theories. They therefore have $N=2$ supersymmetry \cite{GreSch82}. What
distinguishes the two theories is that the gravitini have {\em
opposite} chirality in the type IIA theory, and this theory is
therefore non--chiral. Conversely, the gravitini have the same
chirality in the type IIB theory and this theory is chiral.

\paragraph{Type IIB superstring.}
The bosonic massless excitations arising in the 
NS--NS sector of the type IIB superstring 
are the dilaton, $\Phi$, the metric,  $g_{MN}$, 
and the antisymmetric, 2--form potential, denoted 
here by $B^{(1)}_{MN}$. 
The RR sector contains a scalar axion field, 
$\chi$, a 2--form potential,  $B^{(2)}_{MN}$, 
and a 4--form potential, $D_{MNPQ}$.
The field equations correspond to 
those of $N=2$, $D=10$ chiral supergravity \cite{Schwarz83,GreSch83,SchWes83,HowWes84}. 
The field equation for the 4--form implies that its 
5--form field strength should be self--dual. This latter 
constraint can not 
be derived from a covariant ten--dimensional action \cite{MarSch82}. 
However, one may drop this self--duality constraint 
by introducing new degrees of freedom at the level of the action. 
The constraint can then be imposed as a consistent truncation 
of the field equations derived from this more general action
\cite{BerBooOrt96}. The appropriate bosonic action is given by 
\cite{BerHulOrt95,BerBooOrt96}
\begin{eqnarray}
\label{TYPEIIB}
S_{\rm IIB}=\frac{1}{16\pi \alpha^{\prime4}} \left\{ 
\int d^{10}x \sqrt{|g_{10}|} \left[ e^{-\Phi} \left( R_{10}+\left( 
\nabla \Phi \right)^2 -\frac{1}{12} (H_3^{(1)})^2 
\right) \right. \right. \nonumber \\
\left. \left. -\frac{1}{2} \left( \nabla \chi \right)^2 -
\frac{1}{12}(H_3^{(2)}+ \chi H_3^{(1)})^2 -\frac{1}{240} ({F_5})^2  \right] 
+\int A_4 \wedge H_3^{(2)} \wedge H_3^{(1)} \right\}
\end{eqnarray}
where the RR field strengths are defined by 
$H_3^{(2)} =dB_2^{(2)}$ and 
$F_5 =dA_4 +B_2^{(2)} \wedge 
H_3^{(1)}$. 
The NS--NS sector for the type IIB theory has the same form as that 
of the type IIA action. Once again the RR fields 
do not couple directly to the dilaton field, 
but in contrast to the 
type IIA theory, the kinetic terms correspond to those 
for $p$--form potentials, where $p$ is even.

\paragraph{Type I superstring.}
The theory that admits open string states is 
the type I theory. The boundary conditions for an open 
string halve the number of supersymmetries to $N=1$ \cite{GreSchWit87}. 
Open strings can carry charges at their ends. However, the theory 
is only free from anomalies and  quantum mechanically consistent if the 
gauge group is uniquely chosen to be ${\rm SO}(32)$ \cite{GreSch84}. 
The bosonic sector of the effective action for the massless 
excitations of the type I superstring is $N=1$, $D=10$ 
supergravity coupled to ${\rm SO}(32)$ super Yang--Mills 
theory \cite{GreSch82,GreSch84,GreSch85a,GreSch85b}. 
The action for this supergravity theory was originally derived in Refs. 
\cite{Chamseddine81a,BerRooWit82,BerRooWit83,ChaMan83}. 
In the string frame it takes the form \cite{Witten95}:
\begin{equation}
\label{TYPEI}
S_{\rm I} =\frac{1}{16\pi \alpha^{\prime4}} \int d^{10}x
\sqrt{|g_{10}|}\left[ e^{-\Phi}  
\left( R_{10} +\left( \nabla \Phi \right)^2 \right) -\frac{1}{12}
H^2_3 -\frac{1}{4} e^{-\Phi /2} F^2_2 \right]
\end{equation}
where $F^2_2$ is the Yang--Mills field strength taking values in the 
gauge group $G= {\rm SO(32)}$ and $H_3 =dB_2$ is the field strength of a 
2--form potential, $B_2$. We note that this field strength 
is not coupled to the dilaton field in this frame.

\paragraph{Heterotic superstring.}
The origin of the two heterotic theories derives from the 
fact that the left-- and right--moving modes of a closed string 
can be considered independently \cite{GreSchWit87}. 
Thus, supersymmetry need only 
be imposed in the right--moving sector. This reduces the supersymmetry 
to $N=1$. Quantization of the left--moving sector then requires 
the gauge groups to be either ${\rm SO}(32)$ 
or ${\rm E}_8\times {\rm E}_8$, depending on the 
fermionic boundary conditions that are imposed \cite{GroHarMar85a}. 
The effective action of the heterotic superstring is 
\begin{equation}
\label{HET}
S_{\rm H}=\frac{1}{16\pi \alpha^{\prime4}} \int d^{10} x
\sqrt{|g_{10}|} e^{-\Phi}  
\left[ R_{10}+\left( \nabla \Phi \right)^2 -\frac{1}{12} H_3^2 
-\frac{1}{4} F_2^2 \right]
\end{equation}
where $F_2^2$ is the field strength corresponding to 
the gauge groups ${\rm SO}(32)$ or 
${\rm E}_8\times {\rm E}_8$ \cite{GroHarMar85a,GroHarMar85b,GroHarMar86}. 
The heterotic and type I theories have the same particle content. 
Their effective actions differ, however, because 
all bosonic degrees of freedom couple 
directly to the dilaton field in the heterotic theory, 
whereas the 2--form potential is a RR degree of freedom in the type 
I theory. 

\paragraph{Eleven--dimensional supergravity.}
We conclude this Section with a discussion on a 
further supergravity theory that is closely 
related to the ten--dimensional theories discussed above. 
This is $N=1$, $D=11$ supergravity 
\cite{CreJulSch78,CreJul78,CreJul79}. It is now widely 
believed that this theory represents the low--energy limit of M--theory
\cite{Witten95,Townsend95}. 
The simplest way to appreciate why the $D=11$ theory is special 
is to consider the field content of 
$D=4$ supergravity theories by counting the 
number of on--shell massless degrees of freedom. (For a review, see, 
e.g., Ref. \cite{Freund86}). 
To construct the representations for an extended supergravity 
theory of type $N$, 
one begins with the state of maximum helicity
$\lambda_{\rm max}$. Each of the $N$ 
supersymmetry generators (charges) then acts successively to lower the 
helicity of the physical 
states by one--half unit, so that the lowest helicity 
in the super-multiplet is $\lambda_{\rm min} =\lambda_{\rm 
max} -N/2$ \cite{Strathdee87}. This implies that 
there must exist particles of helicity $|\lambda| \ge N/4$ 
in such an extended theory. Requiring that 
a theory in four dimensions contains no fields  
with helicity $|\lambda | \ge 5/2$ limits the maximum number of 
allowed supersymmetries to be $N=8$. In this case, 
the state of maximum helicity, $\lambda_{\rm max} =2$, 
represents the graviton (vierbein). 

The particle content of the $N=8$, $D=4$ 
supergravity can be identified with the massless degrees of 
freedom arising in certain higher--dimensional theories. 
The number of degrees of freedom of the bosonic and 
fermionic fields that arise in such theories in $D$ dimensions 
is summarized in Tables (\ref{tabledegree}) and (\ref{tableform}) 
\cite{Nahm78,DufNilPop86}. In four dimensions each of the spinors 
associated with the supersymmetry generators has four degrees 
of freedom. Thus, we have eight, four--component spinors, 
yielding a total of 32 degrees of freedom. 
In general, a supergravity theory is said to be maximal 
if it has $32$ conserved supercharges in any dimension.
Table (\ref{tabledegree}) implies that the number of spinor 
degrees of freedom increases with the 
dimensionality of the spacetime and the highest 
dimension consistent with 32 degrees of freedom is $D=11$.
Since a necessary condition for two theories to be related is that 
they should have the same number of bosonic and fermionic degrees 
of freedom, an identification between 
$N=8$, $D=4$ supergravity and a higher--dimensional theory is only possible if 
$D\le 11$, at least for a spacetime with 
signature $(1,D-1)$ \cite{Nahm78}. Remarkably, supersymmetry leads us to an 
{\em upper} limit on 
the dimensionality of spacetime. 

\begin{table}
\begin{center}
\begin{tabular}{||c|c||}
\hline \hline
Field & Degrees of Freedom   \\
\hline \hline
$D$--Bein           & $D(D-3)/2$ \\
\hline
Gravitino        & $2^{\alpha}(D-3)$ \\
\hline
Vector           & $(D-2)$ \\
\hline
Spinor           & $2^{\alpha}$ \\
\hline
Scalar           &  $1$ \\ 
\hline \hline
\end{tabular}
\end{center}
\caption[shortname]{The number of degrees of freedom for bosonic and 
fermionic fields in $D$ dimensions. 
The spinors in this table correspond to Dirac spinors. 
In this case, $\alpha =D/2$ 
if $D$ is even and $\alpha = (D-1)/2$ if the 
dimensionality of spacetime is odd.
The relationship between Dirac and 
Majorana spinors is discussed in Refs. \cite{Tanii98,DufNilPop86}. 
Majorana spinors have half the number of degrees of freedom as Dirac spinors.}
\label{tabledegree}
\end{table}


\begin{table}
\begin{center}
\begin{tabular}{||c|c||}
\hline \hline
Potential & Degrees of Freedom \\
\hline \hline
$A_{MNP}$      &    $(D-4)(D-3)(D-2)/6$ \\
\hline
$A_{MN}$       &    $(D-3)(D-2)/2$     \\
\hline
$A_M$          &    $(D-2)$            \\
\hline
$A$            &    $1$ \\
\hline \hline
\end{tabular}
\end{center}
\caption[shortname]{The number of degrees of freedom for antisymmetric 
tensor gauge fields in $D$ dimensions. In general, 
a gauge field potential with $n>2$ indices has $(D-n-1)(D-n) \ldots
(D-2)/n!$ degrees of freedom.} 
\label{tableform}
\end{table}

The unique theory in eleven dimensions 
is $N=1$ supergravity and the number of graviton degrees of freedom
in this case is $(11\times 8)/2 = 44$. 
The corresponding gravitino can be represented as a 
Majorana spinor in eleven dimensions and has  
$(2^5 \times 8)/2 =128$ degrees of freedom.
Supersymmetry therefore requires the introduction of a 
further 84 bosonic degrees of freedom. Table 
(\ref{tableform}) immediately suggests a plausible candidate 
in the form of 
a three--index, antisymmetric tensor gauge field. 
Indeed, it can be shown that such a field 
is the {\em only} field that could be introduced if supersymmetry is to 
be maintained \cite{Nahm78}. Introducing 84 spin--0 fields,  
for example, would violate supersymmetry. 

The graviton and three--form potential 
constitute the entire bosonic sector of $N=1$, $D=11$ supergravity. 
The action is given by \cite{CreJulSch78,CreJul78,CreJul79}
\begin{equation}
\label{Maction}
S_{\rm M}= \frac{1}{16\pi G_{11}} 
\left( \int d^{11}x \sqrt{|g_{11}|} \left[ R_{11} 
-\frac{1}{48}F_4^2 \right] +\frac{1}{6} \int A_3 \wedge F_4 \wedge F_4 
\right)
\end{equation}
where $F_4 = dA_3$ is the four--form field strength of the 
three--index gauge potential $A_3$. The eleven--dimensional 
Newton constant, $G_{11}$, is related to the Planck length 
by $G_{11} =l_{\rm Pl}^{(9)}$ and is the only parameter in the theory 
\cite{CasFreGia83}. The Chern--Simons term arises as a direct 
consequence of the supersymmetry \cite{CreJulSch78}. 

It can be shown that the bosonic degrees of freedom 
in this theory can be consistently identified with 
those of $N=8$, $D=4$ supergravity by 
compactifying the former theory on a 7--torus, $T^7$. Indeed, 
Cremmer and Julia first derived the Lagrangian 
for the $N=8$, $D=4$ theory by performing such a compactification 
\cite{CreJul78,CreJul79}.  
Moreover, compactification of $N=1$, $D=11$ supergravity 
on a circle, $S^1$, results in the type IIA supergravity theory
\cite{CamWes84,HuqNam85,GiaPer84}. 
This correspondence proves central when interpreting the strongly 
coupled limit of the type IIA superstring in terms of an 
eleven--dimensional theory \cite{Witten95,Townsend95}.

In conclusion, there exist five supersymmetric string theories that
each have a consistent weak coupling expansion in perturbative theory.
The type I theory and the two heterotic theories have
(ten--dimensional) $N=1$ supersymmetry. The type II theories have
$N=2$ supersymmetry.  Heterotic and type II superstrings are oriented,
unbreakable and closed. Open strings are only possible in the type I
theory where strings are unoriented and breakable.  There is no
freedom for introducing a super Yang--Mills gauge group in the type II
theories and the only gauge group that can be consistently introduced
in the type I theory is ${\rm SO}(32)$. The heterotic theories admit
both ${\rm SO}(32)$ and ${\rm E}_8\times {\rm E}_8$.  Supersymmetry
implies that quantum consistency is only possible in perturbative
string theory if spacetime is ten--dimensional. On the other hand, the
upper limit on the dimensionality of spacetime implied by
supersymmetry is $D=11$.  Indeed, requiring gravity to be unique in
four dimensions restricts the number of supersymmetries present in the
theory. If particles are to have helicity $|\lambda| \le 2$, the
number of supersymmetries can not exceed $N=8$ and this implies $D \le
11$.

The main features of the superstring effective actions 
discussed in this Section that are of primary importance to 
cosmology are that they are higher--dimensional, scalar--tensor 
theories of gravity  
that contain antisymmetric tensor fields of varying degree. 
In order to discuss four--dimensional 
cosmological solutions in these theories, therefore, 
we must first consider their compactification down to four dimensions. 
This is the topic of the next Section. 

\section{Toroidal Compactification of the String Effective Actions}

\label{Section3}

\def\theequation{\thesection.\arabic{equation}}

\setcounter{equation}{0}

\subsection{Toroidal Compactification of the NS--NS Action}

\label{Section3.1}

We begin this Section by considering the toroidal compactification of
the NS--NS sector of the string effective action which contains the
dilaton field, the graviton and a 2--form potential and is common to
both the type II and heterotic theories.  (For reviews on
Kaluza--Klein gravity see, e.g.,
\cite{ColMarSqu89,AppChoFre87,DufNilPop86,BaiLov87,OveWes97}).

In Kaluza--Klein dimensional reduction, the universe is viewed as the
product space ${\cal{M}} ={\cal{J}}\times {\cal{K}}$, where the
$D$-dimensional space-time ${\cal{J}}(x^{\rho})$ has metric
$g_{\mu\nu}(x^{\rho})$ $(\mu , \nu = 0, 1, \ldots , D-1)$ and the
$d$-dimensional internal space is denoted by ${\cal{K}}(y^{a})$ with
metric $h_{ab}(x^\rho)$ $(a, b = D, \ldots , D+d -1 )$. This space
must be Ricci flat if the matter fields are independent of its
coordinates $y^a$. For the purposes of the
present discussion, it is sufficient to assume that ${\cal{K}}$ is a 
$d$--dimensional torus, $T^d$.
A $d$--torus is the subspace of the $d$--dimensional 
complex Euclidean space described by $z_j =\exp (2\pi i 
\theta_j)$, where $0\le \theta_j <1$ \cite{AlvMee99}. Topologically, it 
is the Cartesian product of 
$d$ circles, $T^d = S^1\times S^1 \times \dots \times S^1$, and 
is metrically flat. This manifold has the maximal number 
of covariantly constant spinors. (This is related to the fact that 
flat tori are the only manifolds with trivial holonomy). 
Each of these spinors is  
associated with a supersymmetry that 
is unbroken by the compactification. For compactification 
to four or more dimensions, other basic choices for the 
internal manifolds are those with ${\rm SU}(n)$ holonomy, corresponding 
to Calabi--Yau $n$--forms, and those with ${\rm Sp}(n)$ holonomy. 
(For introductory reviews, see, e.g., \cite{Vafa97,OogYin96}). 
We do not consider such compactifications in this work, but 
note that in many settings, our results apply to these 
more general cases when the only modulus field 
that is dynamically important represents the volume of the 
internal space. 

When the higher-dimensional metric is compactified on a circle, it
splits into a lower--dimensional metric tensor, a 1--form potential
(gauge field) and a 0--form potential (scalar field). A $p$--form, on
the other hand, splits into a $p$--form and a $(p-1)$--form
\cite{SchSch79,CreJulLu98, Cremmer82}.  For compactification on $T^d$,
therefore, the higher-dimensional graviton produces 1 graviton, $d$
vector fields and $d(d+1)/2$ spin--0 fields. The NS--NS 2--form
potential splits into a 2--form potential, $d$ 1--form potentials, and
$d(d-1)/2$ 0--form potentials. More generally, a $p$--form
compactified on $T^d$ produces a total of $d!/[p!(d-p)!]$ scalar
moduli fields from its internal components.

Maharana and Schwarz \cite{MahSch93} have derived 
the form of the $(D+d)$--dimensional NS--NS action compactified on $T^d$. 
In $(D+d)$ dimensions, the action is given by
\begin{eqnarray}
\label{hataction}
\hat{S}=\int d^{D+d} x \sqrt{| \hat{g}_{D+d} |}
e^{-\hat{\Phi}} \left[ \hat{R}_{D+d}(\hat{g}) +\hat{g}^{AB} 
\hat{\nabla}_A \hat{\Phi} \hat{\nabla}_B \hat{\Phi} \right. 
\nonumber \\
\left. -\frac{1}{12} \hat{H}_{M_1M_2M_3}\hat{H}_{M_1'M_2'M_3'}
\hat{g}^{M_1M_1'} \hat{g}^{M_2M_2'} \hat{g}^{M_3M_3'} \right]
\end{eqnarray}
where  $(A,B) =(0, 1, \dots , D+d-1)$ and 
a hat denotes quantities in the $(D+d)$--dimensional 
spacetime.  
The complete $(D+d)$--dimensional metric can be expressed 
in the form
\begin{equation}
\label{metric}
\hat{g}_{AB}  =
     \left( \begin{array}{cc}  g_{\mu\nu}+ {A_{\mu}}^{c} 
A_{\nu c} & A_{\mu b} \\ A_{\nu a} & h_{ab}  
\end{array} \right),   
\end{equation}
where $\hat{g}_{ab} =h_{ab} $ is the metric on ${\cal{K}}$,
$\hat{g}_{\mu\nu} =g_{\mu\nu} +h_{ab}{A^a}_{\mu}
{A^b}_{\nu}$ 
and $\hat{g}_{\mu a} =h_{ab} {A^b}_{\mu}$. The 
$d(d+1)/2$ degrees of freedom, $h_{ab}$,  
represent `moduli' fields. 
The determinant of $\hat{g}_{AB}$ is 
given by $\hat{g}_{D+d} = g_Dh$, 
where $g_D \equiv {\rm det} g_{\mu\nu}$ 
and $h\equiv {\rm det}h_{ab}$  are 
the determinants of the metrics on 
${\cal{J}}$ and ${\cal{K}}$, respectively. The 
inverse of the $(D+d)$--dimensional metric is 
\begin{equation}
\label{inversehat}
\hat{g}^{AB} =
\left( \begin{array}{cc} 
g^{\mu\nu} &  -A^{\mu b} \\
-A^{\nu a} & h^{ab} + A^{\rho a} {A_{\rho}}^b \end{array} \right)
\end{equation}

It can be shown by employing the relation 
\begin{equation}
\partial_{\mu} \ln h =h^{ab} \partial_{\mu} h_{ab}
\end{equation}
that the 
$(D+d)$--dimensional Ricci curvature scalar is related 
to the Ricci scalar of the manifold ${\cal{J}}$ 
by \cite{SchSch79,Argurio98}
\begin{equation}
\label{hatR}
\hat{R}_{D+d} (\hat{g}) = R_D(g) +\frac{1}{4} \nabla_{\mu} 
h^{ab} \nabla^{\mu} h_{ab} + 
\nabla_{\mu} (\ln \sqrt{h}) \nabla^{\mu} (\ln \sqrt{h} )
-\frac{2}{\sqrt{h}} \Box \sqrt{h} -\frac{1}{4} h_{ab} 
{F_{\mu\nu}}^aF^{\mu\nu b}
\end{equation}
where 
${F_{\mu\nu}}^a =\partial_{\mu} {A_{\nu}}^a -\partial_{\nu} 
{A_{\mu}}^a$ is the field strength of ${A_{\mu}}^a$. 

Substituting Eq. (\ref{hatR}) into 
Eq. (\ref{hataction}) implies that, 
modulo a boundary term, the dilaton--graviton sector 
of the dimensionally reduced action is given by 
\begin{equation}
\label{nohat}
S_g=\int d^D x \sqrt{|g_D|} e^{-\varphi} 
\left[ R_D +\left( \nabla \varphi \right)^2 +\frac{1}{4}
\nabla h^{ab} \nabla h_{ab} -
\frac{1}{4} h_{ab} 
{F_{\mu\nu}}^aF^{\mu\nu b} \right]
\end{equation}
where the effective $D$--dimensional string coupling
is parametrized in terms of the `shifted' 
dilaton  field by \cite{Veneziano91,Buscher87,Buscher88,SmiPol91,Tseytlin91}
\begin{equation}
\label{defvarphi}
\varphi \equiv \hat{\Phi} -\frac{1}{2} \ln {\rm det} h_{ab}
\end{equation}
It is important to note that 
the coupling parameter between the 
$D$--dimensional dilaton and graviton 
is still given by $\omega =-1$ after the dimensional reduction. 
This result is independent of the dimensionality 
of the spacetime and the number of compactified dimensions.
It often proves 
convenient to parametrize the internal metric
$h_{ab}$ in the form 
\begin{equation}
h_{ab} =h^{1/d} f_{ab} , \qquad {\rm det} f_{ab} =1
\end{equation}
Equation~(\ref{nohat}) then reduces further to
\begin{eqnarray}
\label{stillnohat}
S_g=\int d^D x \sqrt{|g_D|} e^{-\varphi} 
\left[ R_D + \left( \nabla \varphi \right)^2 -\frac{1}{d} 
\left( \nabla \ln \sqrt{h} \right)^2 \right. \nonumber \\
\left. +\frac{1}{4} 
\nabla f^{ab} \nabla f_{ab} -\frac{1}{4}
e^{(2\ln \sqrt{h})/d} f_{ab} {F_{\mu\nu}}^aF^{\mu\nu b}
\right]
\end{eqnarray}

Equation (\ref{stillnohat}) also represents the action for the
toroidal compactification of Einstein gravity, $\hat{S} =\int d^{D+d}x
\sqrt{|\hat{g}_{D+d}|}\hat{R}_{D+d}$, when the higher--dimensional
dilaton field as given in Eq.~(\ref{hataction}) 
is trivial, $\hat{\Phi}=0$.  In
this case, the action reduces further to
\begin{equation}
\label{slaction}
S_g=\int d^D x \sqrt{|g_D|} e^{-\varphi} 
\left[ R_D -\omega \left( \nabla \varphi \right)^2 
+\frac{1}{4} \nabla f^{ab} \nabla f_{ab} 
-\frac{1}{4} e^{-2\varphi/d} f_{ab} {F_{\mu\nu}}^aF^{\mu\nu b}
\right]
\end{equation}
where \cite{Freund82}
\begin{equation}
\varphi \equiv - \ln\sqrt{h} \, ,
 \quad {\rm and} \quad 
\omega \equiv  -1+ \frac{1}{d} \, .
\end{equation} 
The first two terms in this expression correspond to the action for
the gravitational sector of the Brans--Dicke theory of scalar-tensor
gravity\cite{BraDic61}, where the coupling $\omega$ between the
Brans--Dicke field, $e^{-\varphi}$, and the graviton is determined by the
number of internal dimensions. The coupling is bounded such that $-1 <
\omega \le 0$. It it is interesting to note that the lower bound
corresponds to the value that arises in the string effective action
and is formally saturated in the limit $d \rightarrow \infty$. 
{}From this viewpoint it is easy to see why the string value,
$\omega=-1$, is a fixed-point under further dimensional reduction by a
finite number of dimensions.

We now proceed to consider how 
the field strengths of the form fields behave under toroidal 
compactification. We begin by compactifying the 
$(n-1)$--form $\hat{B}_{M_1 \ldots M_{n-1}}$
with field strength 
\begin{equation}
\hat{H}_{M_1\ldots M_n} =n \hat{\partial}_{[M_1} 
\hat{B}_{M_2 \ldots M_n ]} \,,
\end{equation}
on a circle of radius $\hat{g}_{yy} =e^{2\gamma}$.  
The square brackets imply total antisymmetrization 
over all indices. 
For simplicity, we assume that the gauge field that also arises in the
compactification is trivial, i.e., $\hat{g}_{\mu y} =0$.  In general,
such a field would lead to Abelian Chern--Simons terms in the
dimensionally reduced action, but a detailed discussion of these terms
is beyond the scope of this review. (See, e.g.,
Refs. \cite{MahSch93,LuPop96} for further details).

The $(n-1)$--form potential 
separates into two components, a $(n-1)$--form 
$B^{(n-1)}_{\mu_1 \mu_2 \ldots \mu_{n-1}} 
= \hat{B}_{\mu_1 \mu_2 \ldots \mu_{n-1}}$ and a 
$(n-2)$--form 
$B^{(n-2)}_{\mu_1 \ldots \mu_{n-2}} = \hat{B}_{\mu_1 \ldots \mu_{n-2}y}$. 
These have field strengths 
\begin{equation}
H^{(n)}_{\mu_1 \mu_2 \ldots \mu_n} =n 
\partial_{[\mu_1} B^{(n-1)} _{\mu_2 \ldots \mu_n]} = 
\hat{H}_{\mu_1 \ldots \mu_n}
\end{equation}
and 
\begin{equation}
H^{(n-1)}_{\mu_1 \ldots \mu_{n-1}} =(n-1) \partial_{[\mu_1} 
B^{(n-2)}_{\mu_2 \ldots \mu_{n-1}]} = 
\hat{H}_{\mu_1 \ldots \mu_{n-1} y}
\end{equation}
respectively. 
This implies that a higher--dimensional action of the form 
\begin{equation}
\hat{S}_H = - 
\int d^{D+1} x \sqrt{|\hat{g}_{D+1}|}
 \left[ \frac{1}{2n!} \hat{H}_{M_1M_2 \ldots M_n}
\hat{H}_{M_1'M_2' \ldots M_n'}\hat{g}^{M_1M_1'} \hat{g}^{M_2 M_2'} 
\ldots \hat{g}^{M_nM_n'} \right]
\end{equation}
reduces to \cite{MahSin97}
\begin{equation}
S_H=-\int d^Dx\sqrt{|g_D|}e^{\gamma} \left[ 
\frac{1}{2n!} H^{(n)}_{\mu_1 \ldots \mu_n} H^{(n) 
\mu_1 \ldots \mu_n} +\frac{1}{2(n-1)!}  e^{-2\gamma} 
H^{(n-1)}_{\mu_1 \ldots \mu_{n-1}}H^{(n-1) \mu_1 
\ldots \mu_{n-1}} \right]
\end{equation}

It follows that 
the compactification of the NS--NS 3--form field strength in action 
(\ref{hataction}) on $T^d$ produces field strengths 
for $n$-form potentials, where $n=\{ 0,1,2 \}$. 
A specific calculation yields the dimensionally 
reduced action \cite{MahSch93}: 
\begin{equation}
\label{nohatH}
S_H=- \int d^D x \sqrt{|g_D|} e^{-\varphi} 
\left[ \frac{1}{12} H_{\mu\nu\lambda}H^{\mu\nu\lambda} +
\frac{1}{4} H_{\mu\nu a}H^{\mu\nu a} +
\frac{1}{4} H_{\mu ab}H^{\mu ab} +\frac{1}{12} 
H_{abc}H^{abc} 
\right]
\end{equation}
where, since we are assuming that $\hat{g}_{\mu a} =0$, it follows that 
$H_{\mu\nu\lambda} =3 \partial_{[\mu} B_{\nu \lambda ]}$, 
$B_{\mu\nu} =\hat{B}_{\mu\nu}$, $H_{\mu ab} =\partial_{\mu}
B_{ab}$ and $B_{ab}=\hat{B}_{ab}$. 
We remark that since all fields are assumed to be independent of the
internal coordinates, $y^a$, then $H_{abc} \equiv 0$.  We will further
assume for simplicity that $\hat{B}_{AB}$ has block diagonal form,
i.e., $\hat{B}_{\mu a}=0$, and hence the 2--form field strengths of
the 1--form potentials vanish, i.e., $H_{\mu\nu a}=0$.  

Thus, the dimensionally reduced $D$--dimensional NS--NS string
effective action on a $d$--torus, with vector fields frozen out, is
given by
\begin{eqnarray}
\label{NSNSactionreduced}
S=\int d^D x \sqrt{|g_D|} e^{-\varphi} \left[ 
R_D +\left( \nabla \varphi \right)^2 -\frac{1}{12} 
H_{\mu\nu\lambda} H^{\mu\nu\lambda} \right. \nonumber \\
\left. +\frac{1}{4}\nabla_{\mu}  h^{ab} \nabla^{\mu}h_{ab} 
-\frac{1}{4} \nabla_{\mu} B_{ab} \nabla^{\mu} 
B_{cd} h^{ac} 
h^{bd} \right]
\end{eqnarray}

As discussed in Appendix \ref{appendixA}, 
the dilaton field may be minimally coupled to 
the graviton by performing the conformal 
transformation 
\begin{equation}
\label{tildeg}
\tilde{g}_{\mu\nu} = \Omega^2 g_{\mu\nu} , \qquad \Omega^2 \equiv 
\exp \left[ - \frac{2}{D-2} \varphi \right]
\end{equation}
on the $D$--dimensional metric, together with the field redefinition
\begin{equation}
\label{phivarphi}
\tilde\varphi \equiv \sqrt{\frac{2}{D-2}} \varphi
\end{equation}
This transforms the action (\ref{NSNSactionreduced}) into 
an  `Einstein--Hilbert' form:
\begin{eqnarray}
\label{confeinstein}
S=\int d^D x\sqrt{|\tilde{g_D}|} \left[ \tilde{R}_D
-\frac{1}{2} \left( \tilde{\nabla} \tilde\varphi \right)^2 -\frac{1}{12} 
e^{-\sqrt{8/(D-2)}\tilde\varphi} \tilde{H}_{\mu\nu\lambda}
\tilde{H}^{\mu\nu\lambda} \right. \nonumber \\
\left. +\frac{1}{4} 
\tilde{\nabla}_{\mu} h_{ab} \tilde{\nabla}^{\mu} 
h^{ab} -\frac{1}{4} \tilde{\nabla}_{\mu}B_{ab}\tilde{\nabla}^{\mu} 
B_{cd} h^{ac}h^{bd} \right]
\end{eqnarray}

In four dimensions an alternative 
formulation of the dimensionally reduced action 
can be made by expressing 
the three--form field strength, $H_{\mu\nu\lambda}$,
in terms of its Poincar\'e dual \cite{ShaTriWil91,Schwarz92,Sen93a}. 
This formulation is  
important when discussing the different classes of string cosmologies
that arise from the effective action. 
It is also relevant when investigating the global 
symmetry properties of the actions. 
In view of this, we discuss Poincar\'e duality of form fields further 
in the following Section. 

\subsection{Antisymmetric Tensor Fields and Poincar\'e Duality}

\label{Section3.2}
\label{PoincareDual}

We have already seen that 
the effective actions in string theory contain one or more 
antisymmetric tensor field
strengths. In this Section we consider 
a single $n$-form field strength, $H$, derived from an
$(n-1)$-form potential:  
\begin{equation}
\label{Poincareaction}
H_{A_1A_2\ldots} \equiv n \partial_{[A_1}B_{A_2\ldots]} \ ,
\end{equation}
These fields typically appear in the action through a term
\begin{equation}
{\cal L}_H = - {e^{a\Phi} \over 2n!} H^2
 = - {e^{a\Phi} \over 2n!} g^{A_1B_1}g^{A_2B_2} \ldots
H_{A_1A_2\ldots} H_{B_1B_2\ldots} \ .
\end{equation}
where $\Phi$ denotes a linear combination of scalar fields and 
$a$ is a constant, whose numerical value is determined by 
the degree of the form field and the specific type 
of compactification that is assumed. 
The field equation derived from the action (\ref{Poincareaction}) is then
\begin{equation}
\label{firstHeom}
\nabla_{A_1} \left( e^{a\Phi} H^{A_1A_2\ldots} \right) = 0 \ .
\end{equation}
In addition there is a Bianchi identity (closure condition)
\begin{equation}
\label{BianchiH}
\partial_{[A_1}H_{A_2 A_3 \ldots]} \equiv  0 \ ,
\end{equation}
which follows from the definition of $H$ in terms of an 
antisymmetric  potential.

In order to describe the degrees of freedom associated with the form
field, $H$, it is sometimes more 
convenient to work with a dynamically equivalent 
field strength, $H'$. This 
correspondence is possible because an $n$-form
field in $D$ dimensions is dual to a $(D-n)$-form field (see, e.g., 
Ref. \cite{MisThoWhe73}). 
This duality is established by means of the 
covariantly-conserved, 
totally-antisymmetric volume form $\epsilon_{A_1A_2\ldots A_D}$. 
This has several
important properties (we follow the conventions of Wald \cite{Wald84}):
\begin{enumerate}
\label{properties}
\item
$\nabla_A \epsilon_{BC\ldots} = 0$.
\item
$\epsilon_{123\ldots D} = \sqrt{|g_D|}$.
\item
$\epsilon^{A_1A_2\ldots A_jA_{j+1}\ldots A_D}
 \epsilon_{A_1A_2\ldots A_jB_{j+1}\ldots B_D} = (-1)^s \, j! \, (D-j)!
 \, \delta^{[A_{j+1}\ldots}_{B_{j+1}\ldots} \delta^{A_D]}_{B_D}$
\end{enumerate}
where $s$ is determined by the signature of the metric, and is equal to $0$
for Euclidean space, and $1$ for Lorentzian spacetime. 

We define the $(D-n)$-form dual to $H$ as
\begin{equation}
\label{dualtoH}
^*H_{B_1B_2\ldots B_{D-n}} \equiv
 {1\over n!} \epsilon_{A_1A_2\ldots A_nB_1\ldots B_{D-n}} H^{A_1\ldots A_n}
 \ .
\end{equation}
Conversely, taking the dual of $^*H$, we recover
\begin{equation}
\label{HasDual}
H_{A_1A_2\ldots A_n} = {(-1)^s \over (D-n)!}
 \epsilon_{A_1A_2\ldots A_nB_1\ldots B_{D-n}}\, ^*H^{B_1\ldots B_{D-n}}
\ .
\end{equation}
Substituting equation (\ref{HasDual}) into the field equation
(\ref{firstHeom}), we obtain
\begin{equation}
\label{Heom}
\epsilon^{A_1A_2\ldots A_nB_1\ldots B_{(D-n)}} \nabla_{A_1} \left(
e^{a\Phi}\, ^*H_{B_1\ldots B_{(D-n)}} \right) = 0
\ ,
\end{equation}
which is automatically satisfied if
\begin{equation}
^*H_{B_1B_2\ldots B_{(D-n)}} = e^{-a\Phi} \bar{H}_{B_1\ldots B_{(D-n)}}
\ ,
\end{equation}
where $\bar{H}$ is itself an antisymmetric $(D-n)$-form field strength
derived from a potential
\begin{equation}
\bar{H}_{B_1B_2\ldots B_{(D-n)}} \equiv
(D-n) \partial_{[B_1} \bar{B}_{B_2\ldots B_{(D-n)}]}
\ .
\end{equation}
The field equation~(\ref{Heom}) is then satisfied due to the Bianchi
identity for the field strength $\bar{H}$: 
\begin{equation}
\partial_{[A}\bar{H}_{BC\ldots]} = 0
\ .
\end{equation}

However, we see that the original Bianchi identity for $H$, 
as given in equation
(\ref{BianchiH}), is only satisfied if we also require that 
\begin{equation}
\epsilon_{[A_1\ldots A_nB_1\ldots B_{(D-n)}}\nabla_{C]} e^{-a\Phi}
\bar{H}^{B_1\ldots B_{(D-n)}} = 0
\end{equation}
and this implies that $\bar{H}$ must obey the new field equation
\begin{equation}
\label{newHeom}
\nabla_{B_1} \left( e^{-a\Phi} \bar{H}^{B_1\ldots B_{(D-n)}} \right) =
0 \ .
\end{equation}
Thus, the role of the field equation and the Bianchi
identity is interchanged if we work in terms of the $(D-n)$-form field strength
$\bar{H}$ rather than the $n$-form field strength $H$.
The new field equation (\ref{newHeom}) for $\bar{H}$ can be derived
from the Lagrangian 
\begin{equation}
{\cal L}_{\bar{H}} = - {e^{-a\Phi} \over 2(D-n)!} \bar{H}^2 \ .
\end{equation}
It should be emphasized that in view of the relationship 
\begin{equation}
{\cal L}_{\bar{H}} = (-1)^s {\cal L}_{H} \,,
\end{equation}
this Lagrangian is {\em not} in general equal to the original
Lagrangian in a Lorentzian spacetime.

The energy-momentum tensor, defined as
\begin{equation}
\label{defTH}
^{(H)}T^{AB} = \frac{2}{\sqrt{-g}} \frac{\partial}{\partial g_{AB}}
 \left( \sqrt{-g} {\cal L}_H \right) \ ,
\end{equation}
is given in terms of $H$ by
\begin{equation}
^{(H)}T^{AB} = {e^{a\Phi} \over n!} \left( n H^A_{~~C_1C_2\ldots}
H^{BC_1C_2\ldots} - {1\over2} g^{AB} H^2 \right) \ .
\end{equation}
This is equivalent to the energy-momentum tensor derived from the Lagrangian
${\cal L}_{\bar{H}}$:
\begin{equation}
^{(\bar{H})}T^{AB} = 
{e^{-a\Phi} \over (D-n)!} \left( (D-n) \bar{H}^A_{~~C_1C_2\ldots}
\bar{H}^{BC_1C_2\ldots} - {1\over2} g^{AB} \bar{H}^2 \right)
 \ .
\end{equation}

Throughout this review we will make extensive use of the Poincar\'e dual of
field strengths appearing in the string effective actions. In
Section~\ref{Section6} we shall compare `elementary' higher-dimensional
cosmological solutions with `solitonic' solutions obtained in terms of
their Poincar\'e dual. The Poincar\'e dual is particularly useful in studying
four dimensional cosmologies where the NS-NS 3-form field strength is
dual to a 1-form related to the derivative of a pseudo-scalar axion
field. In the pre--big bang scenario discussed in
Section~\ref{Section10}, such axion fields have distinctive
perturbation spectra that could have important observational
consequences.

We will now employ the Poincar\'e duality discussed above to derive dual
effective actions in four dimensions firstly for the NS-NS sector, and
then go on to include the RR fields of the type II theories.

\subsection{Dual Effective NS--NS Action in Four Dimensions}

\label{Section3.3}

In four dimensions there exists 
a duality between the NS--NS three--form 
field strength and a one--form. This one--form may be 
interpreted as the gradient of a scalar degree of freedom. 
Applying this duality to 
the toroidally compactified NS--NS action 
(\ref{NSNSactionreduced}) implies that we may define 
a pseudo--scalar axion field, $\sigma$ \cite{ShaTriWil91,Sen93a}: 
\begin{equation}
\label{dualHstring}
H^{\mu\nu\lambda} = \epsilon^{\mu\nu\lambda\kappa} 
\, e^{\varphi} \, \nabla_{\kappa} \sigma
\end{equation}
The field equations 
may then be derived from the dual effective action
\begin{eqnarray}
\label{NSNS4d}
S_*=\int d^4 x \sqrt{|g|} e^{-\varphi} 
\left[ R +\left( \nabla \varphi \right)^2 
-\frac{1}{2} e^{2\varphi} \left( \nabla \sigma \right)^2 
\right. \nonumber \\
\left. +\frac{1}{4} \nabla_{\mu} h^{ab}\nabla^{\mu}h_{ab} 
-\frac{1}{4} \nabla_{\mu} B_{ab} 
\nabla^{\mu} B_{cd} h^{ac} h^{bd} \right]
\end{eqnarray}

It is important to emphasize that the action $S_*$ in
Eq.~(\ref{NSNS4d}) is {\em not} identical to the original action $S$ in
Eq.~(\ref{NSNSactionreduced}) because the roles of the Bianchi
identities and the field equations are interchanged by the Poincar\'e
duality as discussed in the preceding subsection. 
Nevertheless, the two descriptions are {\em dynamically
equivalent} as long as the field equations are satisfied. In any case,
either form of the action should only be viewed as an effective action
which reproduces the correct equations of motion.  As we shall see in
later Sections, the dual description of the action often provides the
most convenient framework for discussing the global symmetries of the
field equations.
In the corresponding Einstein frame, where the dilaton field is
minimally coupled to gravity, the action is given by
Eq.~(\ref{confeinstein}) and the appropriate duality transformation
in this case is
\begin{equation}
\label{dualityexpression}
\tilde{H}^{\mu\nu\lambda} =
\tilde{\epsilon}^{\mu\nu\lambda\kappa} \, e^{2\varphi} \,  
\tilde{\nabla}_{\kappa} \sigma
\end{equation}
Note from Eq.~(\ref{phivarphi}) that in four dimensions
$\tilde\varphi=\varphi$.  The field equations in the Einstein frame
may then be derived by extremizing the dual action
\begin{eqnarray}
\label{dualactionfirst}
S_*=\int d^4 x \sqrt{|\tilde{g}|} \left[ 
\tilde{R} -\frac{1}{2} \left( \tilde{\nabla} \varphi \right)^2 
- \frac{1}{2} e^{2\varphi} \left( \tilde{\nabla} \sigma \right)^2 
 \right. \nonumber \\
\left. +\frac{1}{4} \tilde{\nabla}_{\mu} h_{ab} \tilde{\nabla}^{\mu}
h^{ab} -\frac{1}{4} \tilde{\nabla}_{\mu} B_{ab} \tilde{\nabla}^{\mu} 
B_{cd} h^{ac}h^{bd} \right]
\end{eqnarray}
This action could also be derived directly from
Eq. (\ref{NSNSactionreduced}) by first applying the Poincar\'e duality
(\ref{dualHstring}) and then the conformal transformation
(\ref{tildeg}). (See also Eq.~\ref{CTH} in Appendix \ref{appendixA}).

The $d^2$ moduli fields arising from the internal 
degrees of freedom $h_{ab}$ and $B_{ab}$ behave collectively 
as a set of massless scalar fields. 
{}From the point of view of homogeneous, 
four--dimensional cosmologies, where all fields are 
uniform on the surfaces of homogeneity, 
the dynamics of the external spacetime can then be determined 
by considering the effects of a single 
modulus field, $\beta$. This is formally equivalent to 
considering the compactification of the $(4+d)$--dimensional
effective NS--NS action (\ref{hataction}) on an isotropic $d$--torus, where 
the components of the 
two--form potential on the internal space 
are assumed to be trivial. The radius, 
or `breathing mode' of the internal 
space,  is then 
parametrized by the modulus field, $\beta$, and determines 
the volume of the internal dimensions. 
In many settings, therefore, it is sufficient to assume that 
the $(4+d)$--dimensional metric is of the form  
\begin{equation}
\label{isotropictorus}
ds^2 =-dt^2 +g_{ij} dx^idx^j +
e^{\sqrt{2/d} \beta} \delta_{ab} dX^a dX^b
\end{equation}
where indices run from $(i,j)=(1, 2, 3)$ and $(a,b)=(4, \ldots , 
3+d)$ and $\delta_{ab}$ is the $d$--dimensional 
Kronecker delta. The modulus field $\beta$ is normalized
in such a way that it becomes minimally coupled to gravity 
in the Einstein frame. 

The effective dilaton in the four--dimensional spacetime is then given
by Eq.~(\ref{defvarphi}) as
\begin{equation}
\label{4Dvarphi}
\varphi \equiv \hat{\Phi} -\sqrt{\frac{d}{2}} \beta
\end{equation}
and substituting $h_{ab} =e^{\sqrt{2/d}\beta}
\delta_{ab}$ into the action (\ref{NSNSactionreduced})
yields
\begin{equation}
\label{reduced1}
S_*=\int d^4 x \sqrt{|g|} e^{-\varphi} \left[ R +
\left( \nabla \varphi \right)^2 -\frac{1}{2} 
\left( \nabla \beta \right)^2 -\frac{1}{2} 
e^{2\varphi} \left( \nabla\sigma \right)^2 
\right]
\end{equation}
The dual, four--dimensional  
action in the Einstein frame (\ref{dualactionfirst}) simplifies to
\begin{equation}
\label{reduced2}
S_*=\int d^4 x\sqrt{|\tilde{g}|} \left[ \tilde{R} 
-\frac{1}{2} \left( \tilde{\nabla} \varphi \right)^2 
-\frac{1}{2} \left( \tilde{\nabla} \beta \right)^2 
-\frac{1}{2} e^{2\varphi} \left( \tilde{\nabla} 
\sigma \right)^2 \right]
\end{equation}

The dimensionally reduced actions (\ref{reduced1}) and
(\ref{reduced2}) may be viewed as the prototype actions  for string cosmology
because they contain many of the key features common to more general
actions. Cosmological solutions to these actions have been extensively
discussed in the literature, both in the homogeneous and inhomogeneous
contexts (see Section \ref{Section7}). Many of these solutions play a
central role in the pre--big bang inflationary scenario reviewed in
Sections \ref{Section9} and \ref{Section10}.  
However, it is also important to determine
whether non--trivial RR fields can have significant effects on the
dynamics of the universe. In view of this, we derive in the next two
subsections truncated versions of type IIA and type IIB effective
actions. As in the above analysis we compactify the theories on an
isotropic six--torus when only the variations of the form fields on
the four--dimensional spacetime are included. This allows a direct
comparison to be made with the NS--NS actions (\ref{reduced1}) and
(\ref{reduced2}). A detailed comparison of the type IIB cosmologies is
presented in Section \ref{Section8}.

\subsection{Dual Effective Action for the Type IIA Theory in Four Dimensions}

\label{Section3.4}

The type IIA string effective action, 
in the absence of vector fields and the Chern--Simons term, is given 
by Eq. (\ref{TYPEIIA}): 
\begin{equation}
S=\int d^{10}x \sqrt{|g_{10}|} \left[ e^{-\Phi} 
\left( R_{10}+\left( \nabla \Phi \right)^2 -\frac{1}{12} 
H_{ABC}H^{ABC} \right) -\frac{1}{48} F_{ABCD}F^{ABCD} 
\right]
\end{equation}
We compactify on an isotropic six--torus corresponding to 
Eq. (\ref{isotropictorus}) and assume that the only 
non--trivial components of the form--fields are those on the 
external four--dimensional spacetime. We also normalize the action 
such that $16\pi \alpha^{\prime4} =1$. 

The dimensionally reduced action is given by 
\begin{equation}
S=\int d^4 x \sqrt{|g|} \left[ e^{-\varphi} 
\left( R + \left( \nabla \varphi \right)^2 -\frac{1}{2} 
\left( \nabla \beta \right)^2  -\frac{1}{12} H_{\mu\nu\lambda}
H^{\mu\nu\lambda} \right) -\frac{1}{48} e^{\sqrt{3}\beta} 
F_{\mu\nu\lambda\kappa}F^{\mu\nu\lambda\kappa} \right]
\end{equation}
in the string frame, where 
\begin{equation}
\label{shifteddilaton}
\varphi \equiv \Phi - \sqrt{3} \beta
\end{equation}
represents the effective four--dimensional dilaton 
field. The corresponding action 
in the Einstein frame, $\tilde{g}_{\mu\nu} 
=e^{-\varphi}g_{\mu\nu}$, takes the form: 
\begin{equation}
S=\int d^4 x \sqrt{|\tilde{g}|} \left[ \tilde{R} -
\frac{1}{2} \left( \tilde{\nabla} \varphi \right)^2 -\frac{1}{2} 
\left( \tilde{\nabla} \beta \right)^2 -\frac{1}{12}
e^{-2\varphi} \tilde{H}_{\mu\nu\lambda} \tilde{H}^{\mu\nu\lambda} 
-\frac{1}{48} e^{\sqrt{3}\beta -2\varphi} 
\tilde{F}_{\mu\nu\lambda\kappa}\tilde{F}^{\mu\nu\lambda\kappa} 
\right]
\end{equation}

In the Einstein frame the dual of the NS--NS  three--form field strength 
is given by Eq. (\ref{dualityexpression}). 
The field equation for 
the four--form is deduced directly from Eq. (\ref{firstHeom}): 
\begin{equation}
\label{Ffield}
\tilde{\nabla}_{\mu} \left( e^{\sqrt{3} \beta -2\varphi} 
\tilde{F}^{\mu\nu\kappa\lambda} \right) =0
\end{equation}
and the Bianchi identity for this field strength is 
\begin{equation}
\label{Fidentity}
\partial_{[\alpha}\tilde{F}_{\beta\gamma\delta\epsilon ]} \equiv 0
\end{equation}
The discussion of Section \ref{Section3.2} implies that 
Eqs. (\ref{Ffield}) and (\ref{Fidentity}) are 
solved by 
\begin{equation}
\label{Fsolution}
\tilde{F}^{\mu\nu\lambda\kappa} = Q 
e^{-\sqrt{3}\beta +2\varphi} \tilde{\epsilon}^{\mu\nu\lambda\kappa} 
\end{equation}
where $ Q$ is an arbitrary constant. 

It follows, therefore, that the truncated 
type IIA field equations in the Einstein frame 
can be derived from the dual action
\begin{equation}
\label{IIAactiond}
S_*=\int d^4x \sqrt{|\tilde{g}|} \left[ \tilde{R} 
-\frac{1}{2} \left( \tilde{\nabla} \varphi \right)^2 -\frac{1}{2} 
\left(  \tilde{\nabla} \beta \right)^2 -\frac{1}{2} 
e^{2\varphi} \left( \tilde{\nabla} \sigma \right)^2 -\frac{1}{2} 
Q^2 e^{2\varphi - \sqrt{3} \beta} \right]
\end{equation}
It is seen that including the 
excitations of the RR 3--form potential on the 
external spacetime introduces an effective, exponential 
interaction potential for the four--dimensional dilaton and modulus 
fields. 

\subsection{Dual Effective Action for the Type IIB Theory in Four Dimensions}

\label{Section3.5}

The low--energy limit of the type IIB 
superstring is $N=2$, $D=10$ chiral supergravity 
\cite{Schwarz84,HowWes84}. 
In what follows, we 
assume that the RR 4--form vanishes. The field equations for 
the remaining degrees of freedom can then be derived by extremizing 
the action~\cite{BerHulOrt95}: 
\begin{eqnarray}
\label{IIB}
S_{\rm IIB} =\int d^{10} x \sqrt{|g_{10}|} \left\{ e^{-\Phi} 
\left[ R_{10} +\left( \nabla \Phi \right)^2 
-\frac{1}{12} \left( H^{(1)} \right)^2  \right]
 \right. \nonumber \\ 
\left.
 -\frac{1}{2} \left( \nabla \chi \right)^2 -\frac{1}{12} 
\left(  H^{(1)} \chi +H^{(2)} \right)^2
 \right\} .
\end{eqnarray}

The toroidal compactification of the type IIB theory (\ref{IIB}) has
been discussed at different levels of 
complexity \cite{BerHulOrt95,Maharana97,Roy98,BerBooOrt96,AndDauFer97a}. 
Maharana \cite{Maharana97} and Roy 
\cite{Roy98} compactified down to $D$ dimensions 
and showed how the ${\rm SL}(2,R)$ symmetry 
of the ten--dimensional theory is
respected in lower dimensions (see Section \ref{Section8.1})
\cite{HulTow95,Schwarz95,BerBooOrt96}. 
In this Section we compactify the ten--dimensional action (\ref{IIB}) 
to four dimensions with the toroidal 
ansatz (\ref{isotropictorus}) \cite{CopLidWan98a,CopLidWan98b}. 
The reduced four--dimensional  
effective actions in the string and 
Einstein frames are then given by 
\begin{eqnarray}
\label{4daction}
S_4&=& \int d^4 x \sqrt{-g} \left\{ e^{-\varphi} \left[ R + 
\left( \nabla\varphi \right)^2
 - {1\over2} \left( \nabla \beta \right)^2 
-\frac{1}{12} (H^{(1)})^2 \right]
 \right. \nonumber \\ 
&& \left. \qquad 
-\frac{1}{2} e^{\sqrt{3}\beta} \left( \nabla \chi \right)^2 -\frac{1}{12} 
e^{\sqrt{3}\beta} (H^{(1)} \chi +H^{(2)}  )^2 \right\}   ,
\end{eqnarray}
and
\begin{eqnarray}
\label{confaction}
S_4=\int d^4 x \sqrt{-\tilde{g}} \left\{ \tilde{R}_4 -\frac{1}{2} 
\left( \tilde{\nabla} \varphi \right)^2 - {1\over2} \left( \tilde{\nabla} 
\beta \right)^2 -\frac{1}{12} e^{-2\varphi} ( \tilde{H}^{(1)} )^2
\right. \nonumber \\
\left. 
- \frac{1}{2} e^{\sqrt{3}\beta +\varphi} \left( 
\tilde{\nabla} \chi \right)^2 
-\frac{1}{12} e^{\sqrt{3}\beta -\varphi} ( \tilde{H}^{(1)} \chi + 
\tilde{H}^{(2)} )^2
\right\}   .
\end{eqnarray}
respectively, where the dilaton field, $\varphi$, is defined in Eq.  
(\ref{shifteddilaton}). 

For the remainder of this Section 
our discussion takes place within 
the context of the Einstein frame and we therefore 
drop tildes for notational simplicity. 
We now proceed to derive an action that is dual to
Eq. (\ref{confaction}) by exploiting the Poincar\'e duality that
exists between the form fields. The discussion is somewhat more
complicated than that presented in Section \ref{Section3.2}, however,
due to the non--linear couplings that exist between the fields in the
action (\ref{confaction}).

The field equations for the three--forms are given by
\begin{eqnarray}
\label{hfield2}
\nabla_{\mu} \left[ e^{\sqrt{3}\beta -\varphi} \left( \chi H^{(1) \, 
\mu\nu\lambda} +
H^{(2) \, \mu\nu\lambda} \right) \right] =0 \\
\label{hfield1}
\nabla_{\mu} \left[ e^{-2\varphi} H^{(1) \, \mu\nu\lambda} + \chi 
e^{\sqrt{3}\beta -\varphi} \left( \chi H^{(1) \, \mu\nu\lambda} +
H^{(2) \, \mu\nu\lambda}
\right) \right]  = 0 
\end{eqnarray}
and, since the three--form field strengths are dual to one--forms
in four dimensions, we may write
\begin{eqnarray}
\label{dualK}
H^{(1)}_{\mu\nu\lambda} \equiv \epsilon_{\mu\nu\lambda\kappa} K^{\kappa}  \\
\label{dualJ}
H^{(2)}_{\mu\nu\lambda} \equiv \epsilon_{\mu\nu\lambda\kappa} J^{\kappa}  
\end{eqnarray}
The field equations (\ref{hfield2}) and 
(\ref{hfield1}) then take the form 
\begin{eqnarray}
\label{Zeom}
\epsilon^{\mu\nu\lambda\kappa} \nabla_\mu 
 \left[ e^{\sqrt{3}\beta -\varphi} \left( \chi K_\kappa +
  J_\kappa \right) \right] =0 \\
\label{Yeom}
\epsilon^{\mu\nu\lambda\kappa} \nabla_\mu
 \left[ e^{-2\varphi} K_\kappa + \chi 
 e^{\sqrt{3}\beta 
-\varphi} \left( \chi K_\kappa + J_\kappa \right) \right]  = 0 
\end{eqnarray}
when written in terms of the dual one--forms. 

These dual forms may be written in terms
of the gradients of two pseudo--scalar `axion' fields.
Equation~(\ref{Zeom}) requires that
\begin{equation}
\label{Jeq}
e^{\sqrt{3}\beta -\varphi} \left( \chi K_\kappa + J_\kappa \right)
 = \nabla_\kappa \sigma_2   ,
\end{equation}
where $\sigma_2$ is any scalar function. Substituting this into
Eq.~(\ref{Yeom}) implies that 
\begin{equation}
\label{Keq}
e^{-2\varphi} K_\kappa + \chi \nabla_\kappa\sigma_2
 = \nabla_\kappa \sigma_1  ,
\end{equation}
where $\sigma_1$ 
is a second arbitrary scalar 
function. These together imply that the field equations 
(\ref{hfield2}) and (\ref{hfield1}) are automatically
satisfied by \cite{CopLidWan98b}
\begin{eqnarray}
\label{defH1}
H^{(1) \, \mu\nu\lambda} &=& \epsilon^{\mu\nu\lambda\kappa}
 e^{2\varphi} \left( \nabla_\kappa\sigma_1 - \chi\nabla_\kappa\sigma_2
 \right) \\
\label{defH2}
H^{(2) \, \mu\nu\lambda} &=& \epsilon^{\mu\nu\lambda\kappa}
  \left[ e^{\varphi-\sqrt{3}\beta}\nabla_\kappa\sigma_2 - \chi e^{2\varphi}
 \left( \nabla_\kappa\sigma_1 - \chi\nabla_\kappa\sigma_2 \right)
 \right]  .
\end{eqnarray}

It should be emphasised that the definitions of both scalar fields
$\sigma_i$ are arbitrary up to a redefinition $\sigma_i \rightarrow
\sigma_i +f_i$, where $f_i$ represents an arbitrary scalar function.
In this sense, therefore, there is no unique definition of the
pseudo--scalar axion fields. However, when $H^{(2)}_{\mu\nu\lambda} =
\chi=0$, we see that our definition of $\sigma_1$ coincides with the
usual definition presented in Eq. (\ref{dualityexpression}) 
for the axion field that is dual to the NS--NS three--form
field strength. 

Although the field
equations~(\ref{hfield2}) and~(\ref{hfield1})
for the three--forms are now automatically satisfied by 
the dual ansatz, we must also impose the Bianchi identities 
\begin{equation}
\label{bianchiidentity}
\nabla_{[\mu} H^{(i)}_{\nu\lambda\kappa]} \equiv 0 
\end{equation}
that arise because the three--form field strengths are defined 
in terms of the  
gradients of two--form potentials. These correspond to the 
constraint equations
\begin{eqnarray}
\label{fieldsigma1}
\nabla_{\rho} \left[ e^{2\varphi} \left( 
\nabla^{\rho} \sigma_1 -\chi \nabla^{\rho} \sigma_2 \right) 
\right] =0 \\
\label{fieldsigma2}
\nabla_{\rho} \left[ e^{-\sqrt{3}\beta +\varphi} \nabla^{\rho} \sigma_2
-\chi e^{2\varphi} \left( \nabla^{\rho} \sigma_1 -\chi \nabla^{\rho} 
 \sigma_2 \right) \right] =0
\end{eqnarray}
on the fields $\sigma_i$. They are interpreted in the dual ansatz as field 
equations that can be derived from
the effective action \cite{CopLidWan98a,CopLidWan98b}
\begin{eqnarray}
\label{solitonicaction}
S_{4*}&=&\int d^4 x \sqrt{-g} \left[ R -\frac{1}{2} 
\left( \nabla \varphi \right)^2 - {1\over2} \left( 
\nabla \beta \right)^2  -\frac{1}{2} e^{\sqrt{3}\beta +\varphi}
\left( \nabla \chi \right)^2 
 \right. \nonumber \\ 
&& \left. \qquad
-\frac{1}{2} e^{-\sqrt{3}\beta +\varphi}
\left( \nabla \sigma_2 \right)^2  -\frac{1}{2} 
e^{2\varphi} \left( \nabla \sigma_1 -\chi \nabla \sigma_2 \right)^2 
\right]   .
\end{eqnarray}
We discuss the cosmological implications of this truncated type IIB
effective action in Sections \ref{Section8} and \ref{Section10}.

The equations of motion for 
the five scalar fields are given by Eqs.~(\ref{fieldsigma1})
and~(\ref{fieldsigma2}), together with 
\begin{eqnarray}
\label{fieldphi}
\Box \varphi = {1\over2} e^{\sqrt{3}\beta +\varphi} (\nabla\chi)^2
 + {1\over2} e^{-\sqrt{3}\beta +\varphi} (\nabla\sigma_2)^2
 + e^{2\varphi} \left( \nabla\sigma_1 - \chi\nabla\sigma_2 \right)^2
\\
\label{fieldy}
\Box \beta  =  {\sqrt{3}\over2} e^{\sqrt{3}\beta +\varphi}(\nabla\chi)^2
 -  {\sqrt{3}\over2} e^{-\sqrt{3}\beta +\varphi}(\nabla\sigma_2)^2
\\
\label{fieldchi}
\nabla^\mu ( e^{\sqrt{3} \beta +\varphi}\nabla_\mu\chi )
  =  - e^{2\varphi} \nabla^\mu\sigma_2 \left( \nabla_\mu\sigma_1 -
 \chi\nabla_\mu\sigma_2 \right)   .
\end{eqnarray}

The one--forms $K$ and $J$ defined in
Eqs.~(\ref{dualK}) and~(\ref{dualJ}) can be written in terms of the
pseudo--scalar axion fields using Eqs.~(\ref{Jeq}) and~(\ref{Keq}). This 
yields 
\begin{eqnarray}
\label{Kcurrent}
K_\mu = e^{2\varphi} ( \nabla_\mu\sigma_1 - \chi \nabla_\mu\sigma_2 )
\\
\label{Jcurrent}
J_\mu = e^{-\sqrt{3} \beta +\varphi} \nabla_\mu\sigma_2 - \chi K_\mu
\end{eqnarray}
and the Bianchi identities~(\ref{fieldsigma1}) and~(\ref{fieldsigma2})
simply correspond to the requirement that these currents are conserved:
\begin{equation}
\nabla^\mu K_\mu =0 \ , \qquad \nabla^\mu J_\mu =0 .
\end{equation}
In terms of the original three--form field strengths, these currents are
topologically conserved due to the Bianchi identities, but in the dual
formulation they are Noether currents conserved due to the symmetry of
the action~(\ref{solitonicaction}).

The conserved currents allow us to integrate out the kinetic terms
for the pseudo--scalar axion fields $\sigma_i$. 
This reduces the field equations~(\ref{fieldphi}--\ref{fieldchi}) to
\begin{eqnarray}
\label{redfieldphi}
\Box \varphi = {1\over2} e^{\sqrt{3} \beta +\varphi} (\nabla\chi)^2
 - {\partial V \over \partial\varphi}
\\
\label{redfieldy}
\Box \beta  =  {\sqrt{3}\over2} e^{\sqrt{3} \beta +\varphi}(\nabla\chi)^2
 - {\partial V \over \partial \beta}
\\
\label{redfieldchi}
\nabla^\mu ( e^{\sqrt{3} \beta +\varphi}\nabla_\mu\chi )
  = - {\partial V \over \partial\chi} \ ,
\end{eqnarray}
where the effective interaction potential, $V$, for the fields
$\varphi$, $\beta$ and $\chi$ is given by
\begin{equation}
\label{effV}
V = - \frac{1}{2} g^{\mu\nu}
 \left[ e^{\sqrt{3}\beta  -\varphi} \left( J_\mu + \chi K_\mu \right) 
    \left( J_\nu + \chi K_\nu \right) 
  + e^{-2\varphi} K_\mu K_\nu \right] .
\end{equation}

Note that the field equation for the field $\chi$,
Eq.~(\ref{fieldchi}), can also be written as
\begin{equation}
\nabla^\mu ( e^{\sqrt{3} \beta +\varphi}\nabla_\mu\chi )
  =  - K_\mu \nabla^\mu\sigma_2   .
\end{equation}
Since $\nabla^\mu K_\mu=0$, we deduce that
\begin{equation}
\label{Lcurrent}
L_\mu = e^{\sqrt{3} \beta +\varphi}\nabla_\mu\chi + \sigma_2 K_\mu   ,
\end{equation}
where $\nabla^\mu L_\mu=0$. Thus, $L_\mu$ is the third conserved
current for the form--fields, independent of $K_\mu$ and $J_\mu$.
This does not allow us to simplify the
equations~(\ref{redfieldphi}--{\ref{redfieldchi}) any further as we
simply swap our ignorance of $\nabla\chi$ for our ignorance of
$\sigma_2$.  However, it is indicative of a further symmetry of the
dual action. We discuss this symmetry when we present cosmological
solutions from this action in Section \ref{Section8}. 

This concludes our discussion of the four--dimensional string
effective actions that are of particular relevance to cosmology. 
In the following Section we proceed to investigate the global
symmetries that are exhibited by these actions.  In general,
supergravity theories containing scalar fields exhibit non--compact,
global symmetries.  The duality symmetries of superstring theory are
discrete subgroups of these global symmetry groups. Considerable
insight into the nature of duality can therefore be gained by studying
the symmetries that arise in the compactified supergravity theories.
In particular, the action (\ref{NSNS4d}) exhibits a global ${\rm
SL}(2,R)$ symmetry that acts non--linearly on the dilaton and axion
fields.  Furthermore, the NS--NS action (\ref{hataction}) exhibits a
global ${\rm O}(d,d)$ symmetry when compactified on a $d$--torus.

\section{Global Symmetries of the Toroidally Compactified NS--NS Action}

\setcounter{equation}{0}

\def\theequation{\thesection.\arabic{equation}}

\label{Section4}

\subsection{Non--Linear Sigma--Models}

\label{Section4.1}

A global symmetry of a supergravity theory is generally associated  
with a non--compact Lie group, $G$. The scalar fields 
$\{ \phi^i \}$ in the theory parametrize the coset $G/H$, 
where $H$ represents the maximal compact subgroup of $G$. 
(This condition is required in order to avoid ghosts \cite{CreJul79}). 
This space is referred to as the {\em target space}, $\bar{\Phi}$, 
and is a non--compact, Riemannian symmetric space with a  
metric $\gamma_{ij} (\phi^k)$. The corresponding line element 
may be written as
\begin{equation}
\label{tsm}
ds^2_{\rm target} = \gamma_{ij}(\phi^k) d\phi^i d\phi^j
\end{equation}
and the scalar fields 
$\{ \phi^i \}$ may therefore be viewed as coordinates on $\bar{\Phi}$.
This implies that the number of scalar fields is given by the dimensionality 
of the coset space: 
\begin{equation}
\label{numbercoset}
N(\phi^i) = {\rm dim} \, {\rm G} -{\rm dim} \, {\rm H} , \quad i=1, 2, 
\ldots, N
\end{equation}
In many settings the target space is also referred to as the `moduli space' 
and the scalar fields are called `moduli' fields. 
The method of describing scalar fields with coset spaces was described in 
Ref. \cite{CalColWes69}. 

The scalar fields couple to gravity in the form 
of a non--linear sigma--model. 
A summary of such models can be found in \cite{West98,BreMaiGib88}.
In the simplest case, we may consider $D$--dimensional Einstein 
gravity coupled to $N$ scalar fields 
$\{ \phi^i \}$. The action of such a system is  
\begin{equation}
\label{sigmamodel}
S=\int d^D x \sqrt{|g_D|} \left[ R_D -
g^{\mu\nu}(x)\partial_{\mu} \phi^i (x) \partial_{\nu} 
\phi^j (x)  \gamma_{ij}(\phi(x)) \right]
\end{equation}
where $g_{\mu\nu} (x)$ is the spacetime metric. 
These fields  
are functions of the spacetime coordinates $x^{\alpha}$
and solutions to the scalar field equations define a map from the spacetime 
manifold with metric $g_{\mu\nu}$ to the target space manifold with metric 
$\gamma_{ij}$. 

We are interested in models where the target space 
metric may be parametrized in terms of a suitable matrix 
representation of a group element ${\cal{P}} \in G$. In other words, 
we write the line element of the target space as 
\begin{equation}
\label{tsm1}
ds^2_{\rm target} = -\frac{1}{4} 
{\rm Tr} \left( d {\cal{P}} d  {\cal{P}}^{-1} \right)
\end{equation}
and the action (\ref{sigmamodel}) as
\begin{equation}
\label{sigmamodel1}
S=\int d^D x \sqrt{|g_D|} \left[ R_D +\frac{1}{4} 
{\rm Tr} (\nabla {\cal{P}} \nabla {\cal{P}}^{-1} )
\right]
\end{equation}
In practice, it is often convenient to introduce the 
`square root' of ${\cal{P}}$: 
\begin{equation}
\label{squareroot}
{\cal{P}} \equiv {\cal{V}}^{\rm T} {\cal{V}} , 
\qquad {\cal{V}} \in G
\end{equation}
where a superscript `T' denotes the transpose. The matrix ${\cal{V}}$ 
plays a role analogous to that of the vierbein in four--dimensional 
general relativity. The target space metric may then be written as
\begin{equation}
\label{tsm2}
ds^2_{\rm target} = \frac{1}{2} {\rm Tr} \left[ 
(d{\cal{V}} {\cal{V}}^{-1}) (d {\cal{V}} {\cal{V}}^{-1} )
+ (d{\cal{V}} {\cal{V}}^{-1}) 
(d{\cal{V}} {\cal{V}}^{-1} )^{\rm T} \right]
\end{equation}

Eq. (\ref{sigmamodel1}) is invariant under 
the {\em global} $G$ symmetry transformations 
\begin{equation}
\label{symtra}
g_{\mu\nu} \rightarrow g_{\mu\nu} , \qquad {\cal{V}} \rightarrow {\cal{V}} 
U^{\rm T} , \qquad 
{\cal{P}} \rightarrow U {\cal{P}} U^{\rm T}
\end{equation}
where $U \in G$ is a constant matrix.  
The action (\ref{sigmamodel1}) is also invariant under 
an independent local $H$ symmetry  
transformation that depends on the target space coordinates 
$\{ \phi^i \} $.
This local $H$ symmetry may be 
employed to choose a gauge where the scalar fields belonging to 
the $H$ subgroup are consistently set to zero. This is 
why the number of independent scalar fields in the model is given by 
Eq. (\ref{numbercoset}) rather than by the dimensionality 
of ${\rm G}$. 

A general feature of non--linear sigma--models of this type 
is that the gravitational sector transforms as a singlet, 
i.e., the spacetime metric is {\rm invariant} under the 
symmetry transformations (\ref{symtra}). 
This proves important in Sections \ref{Section7} and \ref{Section8} 
when we employ some of the non--compact global symmetries of the 
string effective actions to generate new string cosmologies 
from previously known solutions. 

A global symmetry of this type was first discovered within the context of 
$N=4$, $D=4$ supergravity by performing an appropriate 
Poincar\'e duality transformation \cite{CreSchFer78,CreSch78}. The symmetry 
group in this theory is $G={\rm SL}(2,R)$ and $H={\rm U}(1)$. The 
${\rm SL}(2, R)/{\rm U}(1)$ coset is an important model 
and proves central to our subsequent discussions. 
We therefore consider this model in some detail 
in the next subsection. 

\subsection{The SL(2,R)/U(1) Coset}

\label{Section4.2}

The group ${\rm SL}(2,R)$ is the group of all real $2 \times 2$ matrices 
with unit determinant. A matrix $N$ is an element of 
this group if it satisfies the invariance condition \cite{Oraifeartaigh86}
\begin{equation}
\label{Jinvariance}
N^{\rm T} J N =J
\end{equation}
where 
\begin{equation}
\label{J}
J= \left( \begin{array}{cc} 
0 & 1 \\
-1 & 0
\end{array} \right) , \qquad J^2 = - {\rm I}_2
\end{equation}
is the ${\rm SL}(2,R)$ metric and ${\rm I}_2$ 
is the $2 \times 2$ identity matrix. This implies that the inverse of 
$N$ is given linearly by 
\begin{equation}
\label{inverseofN}
N^{-1} =-JN^{\rm T} J
\end{equation}

The action for the ${\rm SL}(2,R)/{\rm U}(1)$ non--linear sigma--model 
coupled to Einstein gravity in $D$ dimensions is  
\begin{equation}
\label{sl2raction}
S=\int d^D x \sqrt{|g_D|} \left[ R_D -\frac{1}{2} \left( \nabla 
\phi \right)^2 -\frac{1}{2} e^{2\phi} \left( \nabla \chi 
\right)^2 \right]
\end{equation}
and the 
target space metric is given by $ds^2_{\rm target} 
=\frac{1}{2}
d\phi^2 +\frac{1}{2} 
e^{2\phi}d\chi^2$. For $D=4$, this is precisely 
the axion--dilaton--graviton sector of the 
string effective action (\ref{dualactionfirst}) when 
formulated in terms of the Einstein frame metric (\ref{tildeg}). 
To establish that this action  is indeed invariant 
under a global ${\rm SL}(2,R)$ transformation on the scalar 
fields $\{ \phi ,\chi \}$, it is convenient to introduce the 
triangular ${\rm SL}(2,R)$ matrix
\begin{equation}
{\cal{V}} = \left( \begin{array}{cc}
e^{\phi /2} & \chi e^{\phi /2}  \\
0 & e^{-\phi /2}  
\end{array} \right) 
\end{equation}
It follows immediately that 
\begin{equation} 
\label{vvminus1}
d {\cal{V}} {\cal{V}}^{-1} = \left( \begin{array}{cc}
\frac{1}{2}d\phi & e^{\phi} d \chi \\
0 & -\frac{1}{2} d\phi 
\end{array} \right) 
\end{equation}

If we now define the symmetric matrix 
\begin{equation}
\label{M}
{\cal{M}} \equiv 
{\cal{V}}^{\rm T} {\cal{V}}  = 
\left( \begin{array}{cc} 
e^{\phi} & \chi e^{\phi} \\
\chi e^{\phi} & e^{-\phi} +\chi^2 e^{\phi} 
\end{array} \right) , 
\end{equation}
substitution of Eq. (\ref{vvminus1}) into Eqs. (\ref{tsm2}) 
and (\ref{tsm1})
implies that action (\ref{sl2raction})
may be written in the form
\begin{equation}
\label{sl2rsymaction}
S=\int d^D x \sqrt{|g_D|} \left[ R_D +\frac{1}{4} 
{\rm Tr} \left( \nabla {\cal{M}} \nabla {\cal{M}}^{-1} 
\right) \right]
\end{equation}

Thus, Eq. (\ref{sl2rsymaction}) is invariant under global 
${\rm SL}(2,R)$ transformations of the form (\ref{symtra}), where 
$U$ is defined as the constant ${\rm SL}(2,R)$ matrix 
\begin{equation}
\label{U}
U \equiv \left( \begin{array}{cc}
d & c \\
b & a 
\end{array} \right) , \qquad ad-bc =1 
\end{equation}
We emphasize that
this ${\rm SL}(2, R)$ transformation  acts non--linearly on the scalar fields 
but leaves the spacetime metric $g_{\mu\nu}$ invariant. 
Specifically, the two scalar fields transform to
\begin{eqnarray}
\label{sphi}
e^{\phi} \rightarrow c^2 e^{-\phi} +(d +c \chi )^2e^{\phi} \\
\label{schi}
\chi e^{\phi} \rightarrow ace^{-\phi} +
(b+a \chi )(d +c \chi ) e^{\phi}
\end{eqnarray}

This ${\rm SL}(2,R)$ transformation is global,  
in the sense that $U$ is independent of the scalar fields, but
there also exists a local transformation that leaves the action invariant. 
It is given by 
\begin{equation}
\label{localsl}
{\cal{V}} \rightarrow {\cal{O}} {\cal{V}} U^{\rm T}
\end{equation}
where ${\cal{O}} ={\cal{O}} (\phi^i)$ is field-dependent. 
Substitution of Eq. (\ref{localsl}) 
into Eq. (\ref{tsm2}) implies that the action 
(\ref{sl2rsymaction}) is invariant under Eq. (\ref{localsl}) 
if ${\cal{O}}^{\rm T} {\cal{O}} ={\rm I}_2$. Thus, 
${\cal{O}}$ must be  an element of the group ${\rm SO}(2)$. This 
is the maximal compact subgroup of ${\rm SL}(2, R)$ and is isomorphic 
to ${\rm U}(1)$. Thus, we may conclude that the scalar 
fields $\phi$ and $\chi$ in action (\ref{sl2raction}) parametrize the 
${\rm SL}(2,R)/{\rm U}(1)$ coset. 

There exists the isomorphism ${\rm SL}(2,R) = {\rm SU}(1,1)$ and the 
${\rm SU}(1,1)/{\rm U}(1)$ formulation of this coset is 
discussed in Refs. \cite{Roo85,Roo86,BerKohSez85,RooWag85}. 
The global ${\rm SL}(2, R)$ symmetry summarized  above 
may also be described in terms of the complex parameter: 
\begin{equation}
\label{tau}
\tau \equiv \chi +ie^{-\phi}
\end{equation}
The ${\rm SL}(2,R)$ transformation corresponding to 
Eq. (\ref{symtra})   
is then equivalent to the scalar field $\tau$ undergoing a fractional linear 
transformation
\begin{equation}
\label{tautrans}
\tau \rightarrow \frac{a \tau +b}{c \tau +d} , \qquad ad-bc =1
\end{equation}
The discrete subgroup ${\rm SL}(2,Z)$ 
represents the modular group of the 2--torus and, in this context, the 
field $\tau$ is the modular parameter of the torus. This 
fundamental connection
with the 2--torus is discussed further in Appendix \ref{appendixB}.

This concludes our discussion on the ${\rm SL}(2,R)/{\rm U}(1)$ 
coset. The dilaton and axion fields 
in the string effective action (\ref{dualactionfirst}) 
parametrize such a coset and this action 
is therefore invariant under a global ${\rm SL}(2,R)$ transformation.
This transformation leaves the Einstein frame metric, $\tilde{g}_{\mu\nu}$,
and the moduli fields, $h_{ab}$ and $B_{ab}$, invariant. The 
dilaton and axion transform according to Eqs. (\ref{sphi}) 
and (\ref{schi}). 
In the following subsection, we discuss a further symmetry of 
the action (\ref{dualactionfirst}). 

\subsection{The O(d,d)/O(d)$\times$O(d) Coset}

\label{Section4.3}

The $d^2$ moduli fields that arise in the toroidal compactification 
of the NS--NS string effective action parametrize the 
${\rm O}(d,d) / {\rm O}(d) \times {\rm O}(d)$ coset. 
The group ${\rm O}(d,d)$ is the non--compact, 
pseudo--orthogonal group in $2d$ dimensions (see, e.g., 
\cite{GivPorRab94,Oraifeartaigh86}). Its representation is given by 
\begin{equation}
\label{UOdd}
U= \left( \begin{array}{cc}
W & X \\
Y & Z \end{array} \right) \in {\rm O}(d,d)
\end{equation}
where $\{W,X, Y, Z \}$ are $d\times d$ 
matrices such that $U$ preserves the bilinear form $\eta$: 
\begin{equation}
\label{onn}
U^{\rm T} \eta U =\eta , \qquad 
\eta \equiv \left( \begin{array}{cc} 
0 & {\rm I}_d \\
{\rm I}_d & 0 \end{array} \right)
\end{equation}
and  ${\rm I}_d$ is the $d \times d$ identity matrix. 
This implies that 
\begin{equation}
W^{\rm T}Y + Y^{\rm T} W =0 , \quad X^{\rm T}Z+Z^{\rm T}X =0 , \quad 
W^{\rm T}Z+Y^{\rm T}X ={\rm I}_d
\end{equation}
Since $\eta^2 ={\rm I}_{2d}$, the inverse of $U$ is given 
linearly: 
\begin{equation}
\label{inverseonn}
U^{-1} =\eta U^{\rm T} \eta
\end{equation}
The maximal compact subgroup of ${\rm O}(d,d)$ 
is ${\rm O}(d) \times {\rm O}(d)$. This is generated by elements 
of the form 
\begin{equation}
U=\frac{1}{2} 
\left( \begin{array}{cc}
\sigma_1 +\sigma_2 & \sigma_1 -\sigma_2 \\
\sigma_1 -\sigma_2 & \sigma_1 + \sigma_2 \end{array} 
\right) 
\end{equation}
where $\sigma_k \sigma_k^{\rm T}  = {\rm I}_d$ $(k=1,2)$, 
i.e., $\sigma_1,\sigma_2 \in {\rm O}(d)$. 
The dimension of ${\rm O}(d) \times {\rm O}(d)$ 
is $d^2-d$ and the dimension 
of the coset space ${\rm O}(d,d)/{\rm O}(d)\times {\rm O}(d)$
is $d^2$. 

The parametrization 
of the ${\rm O}(d,d)/{\rm O}(d) \times {\rm O}(d)$ coset 
is determined by 
introducing the upper triangular $2d\times 2d$ matrix \cite{CreJulLu98}
\begin{equation}
\label{Vonn}
{\cal{V}} =\left( \begin{array}{cc}
S & R \\
0 & (S^{-1})^{\rm T} \end{array} \right)  
\end{equation}
where the condition $RS^{\rm T} =-SR^{\rm T}$ must 
be imposed for Eq. (\ref{onn}) to be valid. 
The symmetric matrix ${\cal{M}} ={\cal{V}}^{\rm T}{\cal{V}}$
is then given by 
\begin{equation}
\label{Monn}
{\cal{M}} = \left( \begin{array}{cc} 
S^{\rm T}S & S^{\rm T}R \\
R^{\rm T}S & [(S^{\rm T}S)^{-1} +R^{\rm T}R] \end{array} \right)
\longrightarrow \left( \begin{array}{cc}
(S^{\rm T}S)_{AB} & {(S^{\rm T}R)_A}^B \\
{(R^{\rm T}S)^A}_B & ((S^{\rm T}S)^{-1})^{AB} +(R^{\rm T}R)^{AB} 
\end{array} \right)
\end{equation}

We define $h^{AB}  = (S^{\rm T}S)^{AB}$ and $B^{AB} =-
(S^{-1}R)^{AB}$, where $h^{AB}=h^{BA}$ and $B^{AB}=-B^{BA}$. 
This implies that Eq. (\ref{Monn}) may be written in the form
\begin{equation}
\label{Mexpression}
{\cal{M}} =\left( \begin{array}{cc} 
h^{-1} & -h^{-1}B \\
Bh^{-1} & h - Bh^{-1}B \end{array} \right)
\end{equation}
Substituting Eq. (\ref{Vonn}) into Eq. (\ref{tsm2}) and noting from Eq. 
(\ref{inverseonn}) that the inverse of ${\cal{M}}$ is 
given by ${\cal{M}}^{-1}=\eta {\cal{M}} \eta$ then 
allows us to write 
the line element of the ${\rm O}(d,d)/{\rm O}(d)\times {\rm O}(d)$
coset space as 
\begin{equation}
\label{lineonn}
ds^2_{\rm target}
= - \frac{1}{8} {\rm Tr} ( \eta d{\cal{M}} \eta d{\cal{M}} ) 
= -\frac{1}{4} {\rm Tr} ( dh dh^{-1}  +h^{-1} dB h^{-1} dB )
\end{equation}

We may now relate this to  the toroidally compactified 
string effective action (\ref{NSNSactionreduced}) \cite{MahSch93}. 
This action may be written in matrix notation as
\begin{eqnarray}
\label{435}
S=\int d^D x \sqrt{|g_D|} e^{-\varphi} \left[ 
R_D +\left( \nabla \varphi \right)^2 -\frac{1}{12} 
H_{\mu\nu\lambda} H^{\mu\nu\lambda} 
\right. \nonumber \\
\left. +\frac{1}{4} {\rm Tr} \left[ \partial_{\mu}h^{-1} 
\partial^{\mu}h + h^{-1} \partial_{\mu}B h^{-1} \partial^{\mu} B 
\right] \right]
\end{eqnarray}
where in this expression
$h$ represents the internal metric on the $d$--torus. 
Comparison with Eq. (\ref{lineonn}) immediately implies that 
Eq. (\ref{435})  may be expressed in the form
\begin{equation}
\label{oddsymaction}
S=\int d^D x \sqrt{|g_D|} e^{-\varphi} \left[ 
R_D +\left( \nabla \varphi \right)^2 -\frac{1}{12} 
H_{\mu\nu\lambda}H^{\mu\nu\lambda} +\frac{1}{8} 
{\rm Tr} \left[ \nabla_{\mu} {\cal{M}} \nabla^{\mu} 
{\cal{M}}^{-1} \right] \right]
\end{equation}
As shown by Maharana and Schwarz \cite{MahSch93}, 
it then follows that the action (\ref{oddsymaction}) is symmetric 
under the global ${\rm O}(d,d)$ transformation 
\begin{equation} 
\label{NSNSOdd}
{\cal{M}} \rightarrow U {\cal{M}} U^{\rm T} , \qquad  
g_{\mu\nu} \rightarrow g_{\mu\nu} , \qquad \varphi \rightarrow 
\varphi , \qquad H_{\mu\nu\lambda} \rightarrow H_{\mu\nu\lambda}
\end{equation}
where $U$ satisfies Eq. (\ref{onn}) and ${\cal{M}}$ is given by 
Eq. (\ref{Mexpression}).  Thus, the moduli fields 
parametrize the ${\rm O}(d,d) /{\rm O}(d) \times {\rm O}(d)$ coset. 

The $D$--dimensional string coupling and spacetime metric 
transform as singlets under Eq. (\ref{NSNSOdd}), as does the 
3--form field strength $H_{\mu\nu\lambda}$. In general, 
such a symmetry arises when there exist $d$ abelian 
isometries in the model. (When $D+d=10$ 
and $d=(7,8)$, however, the symmetry groups are enlarged. See, e.g., Refs. 
\cite{MarSch83,Chamseddine81b,Bakas94,Maharana95,Kehagias95}
for details).  The case 
${\rm O}(2,2)$ is relevant when considering certain
classes of inhomogeneous string cosmologies, 
as is discussed further in Section \ref{Section7.3}. Moreover,  
the ${\rm O}(3,3)$ group is employed in Section \ref{Section7.2} 
to generate `elementary' Bianchi type I cosmological solutions 
\cite{MeiVen91}.

\subsection{Manifestly Invariant Field Equations}

\label{Section4.4}

We conclude this Section by deriving the 
scalar field equations for the 
${\rm SL}(2,R)/{\rm U}(1)$ and ${\rm O}(d,d)/{\rm O}(d) 
\times {\rm O}(d)$ non--linear sigma--models. These equations can be 
expressed in a manifestly symmetric form. This is important because the
evolution equations for perturbations around an arbitrary classical 
solution can also be written in a symmetric way, at 
least for the ${\rm SL}(2, R)/{\rm U}(1)$ model. This turns out 
to be an extremely powerful tool when considering inhomogeneous 
perturbations 
about homogeneous cosmological solutions derived from this 
model (see Section \ref{Section10}). 

We obtain the classical equations of motion for the 
${\rm SL}(2,R)/{\rm U}(1)$ model from the action given in 
Eq. (\ref{sl2rsymaction}), 
by considering first-order variations of the
matrix ${\cal M}$ defined in Eq. (\ref{M}). 
We should proceed with caution since this 
matrix is subject to several constraints. 
The variation must be performed in a way that is 
consistent with the properties of the group ${\rm SL}(2, R)$. 
We therefore consider
an infinitesimal ${\rm SL}(2,R)$ 
transformation ${\cal M}\to U{\cal M}U^{\rm T}$,  where
$U=1+\epsilon$,  and keep terms only up to first--order in $\epsilon$.
This yields the required perturbation:
\begin{equation}
\label{deltaM}
\delta{\cal M} = \epsilon{\cal M} + {\cal M}\epsilon^{\rm T} \ ,
\end{equation}
where Eq. (\ref{Jinvariance}) implies that $\epsilon$ must satisfy 
the constraint
\begin{equation}
\label{epsilonJsym}
\epsilon^T=J\epsilon J \ .
\end{equation}
This in turn implies that $\epsilon$ is traceless, ${\rm Tr} \epsilon 
=0$. 

The matter sector 
of the ${\rm SL}(2,R)$ invariant Lagrangian given in Eq. (\ref{sl2rsymaction})
may be written as 
\begin{equation}
\label{calL}
{\cal L} = - {1\over4} {\rm Tr} [J\nabla^\mu{\cal M}J\nabla_\mu{\cal
M}] \ ,
\end{equation}
by employing Eq.~(\ref{inverseofN}). 
Hence its first-order perturbation can be written as
\begin{equation}
\delta{\cal L} = - {1\over2} {\rm Tr} [J\nabla^\mu{\cal
M}J\nabla_\mu(\epsilon{\cal M} + {\cal M}\epsilon^T)] 
\end{equation}
{}From Eq.~(\ref{epsilonJsym}) and the identity
\begin{equation}
\nabla_\mu{\cal
M}J{\cal M}J = - {\cal M}J\nabla_\mu{\cal M}J
\end{equation}
one then obtains
\begin{equation}
\label{obtain}
\delta{\cal L} = {\rm Tr}[\nabla^\mu({\cal M}J\nabla_\mu{\cal M}J)\epsilon]
 - \nabla^\mu \left({\rm Tr} [ {\cal M}J\nabla_\mu{\cal M}J\epsilon ] 
\right) \ .
\end{equation}

The total divergence in Eq.~(\ref{obtain}) vanishes if we only allow
variations, $\epsilon$, that vanish on the boundary.  Classical
solutions that extremize the action (\ref{sl2rsymaction}) must
therefore satisfy the field equation
\begin{equation}
\label{SL2eom}
\nabla^\mu ({\cal M}J\nabla_\mu{\cal M} J) = 0 \ .
\end{equation}
Thus, the ${\rm SL}(2,R)$ symmetry leads to the classically conserved current
\begin{equation}
\label{Kmu}
K_\mu \equiv {\cal M}J\nabla_\mu{\cal M}J \ .
\end{equation}
We remark that the cyclical 
property of the trace and Eq. (\ref{inverseofN}) 
together imply that 
Tr$[K_\mu]=0$. 

We can obtain an effective action for first--order perturbations about
any classical, ${\rm SL}(2,R)$ background 
solution by considering second-order
perturbations of the Lagrange density in Eq.~(\ref{calL}).
We assume that the second-order perturbations are of the form given in
Eq. (\ref{deltaM}) and, by employing the background equation of
motion~(\ref{SL2eom}), we obtain the second-order effective action for
$\epsilon$: 
\begin{equation}
\delta^2{\cal L} = - {\rm Tr} [ (\nabla\epsilon)^2
 + ({\cal M}J\nabla\epsilon)^2
 - K^\mu (\epsilon\nabla_\mu\epsilon - \nabla_\mu\epsilon\epsilon) ] \ .
\end{equation}

By considering a variation of $\epsilon$,  we 
are now able to derive the
Euler-Lagrange equation for perturbations about a given background solution:
\begin{equation}
\nabla^\mu ( \nabla_\mu\epsilon + {\cal M}J\nabla_\mu\epsilon{\cal M}J
+ \epsilon K_\mu - K_\mu \epsilon ) = 0 \ .
\end{equation}
Thus, we have derived a conserved current for the perturbations: 
\begin{equation}
\label{perturbedcurrent}
\delta K_\mu \equiv 
\nabla_\mu\epsilon + {\cal M}J\nabla_\mu\epsilon{\cal M}J +
\epsilon K_\mu - K_\mu \epsilon 
\end{equation}
This current can also be derived directly by perturbing
Eq.~(\ref{Kmu}) which provides a useful consistency check.

The perturbed current in Eq.~(\ref{perturbedcurrent}) is necessarily 
traceless. Nevertheless, it is possible to 
construct two non--trivial perturbations that remain 
invariant under a global ${\rm SL}(2,R)$ transformation 
of the background solution. They are given by
\begin{eqnarray}
\label{vectoru}
u_\mu &\equiv& 2 {\rm Tr} [ K_\mu \epsilon ] \
 =  \ {\rm Tr} [ J\nabla_\mu{\cal M}J\delta{\cal M} ] \ , \\
\label{vectorv}
v_\mu &\equiv& 2 {\rm Tr} [ - {\cal M} J K_\mu \epsilon ] \ 
 = \ {\rm Tr} [ - J{\cal M}J\nabla_\mu{\cal M}J\delta{\cal M} ] \ .
\end{eqnarray}
respectively, where we have employed Eq. (\ref{deltaM}).
These spacetime vectors and their derivatives are the only
non-trivial ${\rm SL}(2,R)$ invariant first-order perturbations that can be
constructed.  
Their divergences yield ${\rm SL}(2,R)$ invariant scalars
\begin{eqnarray}
\label{perturbedU}
U &\equiv& \nabla^\mu u_\mu
\ =\ {\rm Tr} [ K_\mu \delta K^\mu ] \ ,\\
\label{perturbedV}
V &\equiv& \nabla^\mu v_\mu
\ =\ {\rm Tr} [ - {\cal M} J K_\mu \delta K^\mu ] 
\end{eqnarray}

The energy-momentum tensor derived from the ${\rm SL}(2,R)$ invariant
Lagrange density in Eq.~(\ref{calL}) can be written as
\begin{equation}
T^{\mu\nu}
 = - {1\over2} \left(
   g^{\mu\lambda}g^{\nu\kappa} - {1\over2}g^{\mu\nu}g^{\lambda\kappa} \right)
   {\rm Tr} [ K_\lambda K_\kappa ] \ .
\end{equation}
Thus, the scalar $U$ defined in Eq.~(\ref{perturbedU}) can be
identified with the perturbation of the trace of the energy-momentum
tensor, and hence must lead to a perturbation of the spacetime
curvature. By contrast, $V$ may remain non-zero even when the
energy-momentum tensor is unperturbed and thus describes an
isocurvature perturbation of the fields.  The time components of the
vectors (\ref{vectoru}) and (\ref{vectorv}) are directly related to
the variables that describe cosmological curvature and entropy
perturbations in pre--big bang cosmologies (see Eqs. (\ref{9.90}) and
(\ref{9.91}) in Section \ref{Section10}). These manifestly ${\rm
SL}(2,R)$ invariant quantities will allow us to fully generalize the
cosmological perturbation spectra calculated in vacuum models to
include the effects of the pseudo--scalar axion field in Section
\ref{Section10}.

The derivation of the ${\rm O}(d,d)$ invariant equations of motion was
presented in Ref.~\cite{MeiVen91} and is very similar to the above
derivation for the ${\rm SL}(2,R)$ model. This is due to the similar
properties of the two groups.  The only significant difference in the
derivation is that the ${\rm O}(d,d)$ metric $\eta$, as given in
Eq. (\ref{onn}), is symmetric rather than antisymmetric. This implies
that the variation, $\epsilon$, must satisfy
$\epsilon^T=-\eta\epsilon\eta$.  By defining the traceless matrix
\begin{equation}
L_\mu\equiv{\cal M}\eta\nabla_\mu{\cal M}\eta\ , 
\end{equation}
we may write the Euler-Lagrange equations as $\nabla^\mu L_\mu=0$
\cite{MeiVen91}.  The effective action for perturbations about an ${\rm
O}(d,d)$ invariant background solution is derived in an exactly
analogous fashion to that discussed above and we recover the perturbed
equation of motion $\nabla^\mu\delta L_\mu=0$, where
\begin{equation}
\delta L_\mu = \nabla_\mu\epsilon + {\cal M}\eta\nabla_\mu\epsilon{\cal
M}\eta + \epsilon L_\mu - L_\mu \epsilon \ .
\end{equation}

We can construct an ${\rm O}(d,d)$ invariant spacetime vector perturbation
\begin{equation}
w_\mu \equiv 2Tr[\eta\nabla_\mu{\cal M}\eta\delta{\cal M}] \ ,
\end{equation}
analogous to $u_\mu$ in Eq.~(\ref{vectoru}), whose divergence,
$W\equiv\nabla^\mu w_\mu$, describes the ${\rm O}(d,d)$ invariant
curvature perturbation. But the symmetry of the ${\rm O}(d,d)$ metric,
$\eta$ in Eq.~(\ref{onn}), means that the vector analogous to $v_\mu$
vanishes identically, and we are unable to construct an ${\rm O}(d,d)$
invariant isocurvature perturbation.

\section{S--, T-- and U--Dualities}

\setcounter{equation}{0}

\def\theequation{\thesection.\arabic{equation}}

\label{Section5}
\label{targetsection}

In Section \ref{Section2} we wrote down the bosonic sectors of the 
effective actions for the five perturbative superstring theories, 
together with that of $N=1$, $D=11$ supergravity. Whilst one 
could proceed to develop cosmological 
models from these apparently different 
theories by directly solving the field equations, 
it is constructive to first discuss the deep relationships 
that exist between them. These relationships manifest themselves 
in the duality symmetries of the theories. A number of authors  
have recently surveyed the dramatic 
progress that has been made in this subject \cite{AlvAlvLoz95,Duff98,Tanii98,Li98,Kiritsis97,HaaKorLus99,West98,Sen98,Schwarz98a,Schwarz98b,Schwarz96,Duff96,Vafa97,Sen94,Gibbons98,AlvMee99,ObePio98,Argurio98,GivPorRab94,Polchinski98,Sezgin98}. 
Our primary aim here is to 
convey the key points of string duality to those with an interest 
in its cosmological implications. With this in mind, 
we follow an heuristic approach and 
restrict our discussion to the level of the effective 
supergravity actions. 
Whilst such an approach has obvious limitations, it represents a first step 
towards studying string dualities and  
proves important in locating 
theories that are dual to one another, or are self--dual. 

Three fundamental types of duality arise in string theory. 
These are referred to as S--, T-- and U--duality, 
respectively. Two 
string theories, ${\cal{A}}$ and ${\cal{B}}$, are said  to be T--dual 
to one another if the compactification 
of theory ${\cal{A}}$ on a 
space of `small' volume is equivalent to the compactification of 
theory ${\cal{B}}$ on a space of `large' volume, and vice--versa. 
If the two theories 
${\cal{A}}$ and ${\cal{B}}$ are equivalent under S--duality, 
then the strong coupling limit of one theory is mapped 
onto the weak coupling limit of the other, and vice--versa.
Finally, ${\cal{A}}$ and ${\cal{B}}$ 
are related by U--duality if the compactification 
of ${\cal{A}}$ on a space of large (small) volume
is equivalent to ${\cal{B}}$ at strong (weak) coupling. 
In some cases, a given theory is mapped 
onto itself by a duality transformation and the duality 
then represents an exact symmetry of the theory. 
An example is the S--duality of the type IIB superstring in ten dimensions
\cite{HulTow95}. 

The best understood of the three dualities is T--duality 
\cite{Buscher87,Buscher88,Nepomechie84,Teitelboim86}.
(For reviews on T--duality, see Refs. 
\cite{GivPorRab94,AlvAlvLoz95,AlvMee99,HaaKorLus99,Polchinski96}). 
This is a perturbative symmetry in the 
sense that the transformation takes the weak coupling limit 
of one theory onto the weak coupling limit of another theory. 
It can therefore be tested at the level of perturbation theory 
and shown to be exact order by order  
\cite{RocVer92,GivPorRab94}. By contrast, 
S-- and U--dualities are non--perturbative symmetries in 
the string coupling. Qualitatively, one may view U--duality as the 
unification of S-- and T--duality, i.e., $U = S\times T$.
The first suggestion that string 
theory exhibits a strong/weak coupling S--duality was made within the 
context of the heterotic theory compactified on $T^6$ 
\cite{FonIbaLus90,Rey91} and further supporting evidence was 
subsequently provided \cite{Sen93a,Sen93b,Sen93c,Sen93d,Sen94,SchSen93,SchSen94}. 

In general, equivalent string backgrounds are related by a group 
of duality transformations that corresponds to 
a discrete subgroup of a non--compact, global symmetry 
of the effective supergravity actions. 
These latter symmetries are broken to discrete subgroups 
by quantum effects. For example, 
the type IIB supergravity theory in ten dimensions 
is invariant under a global ${\rm SL}(2,R)$ symmetry 
\cite{GreSch83,SchWes83,Schwarz83,HowWes84,Hull95}. The 
discrete subgroup ${\rm SL}(2,Z)$ is then the conjectured S--duality of 
the type IIB superstring \cite{HulTow95}. 
Likewise, the discrete subgroup ${\rm O}(d,d;Z)$ of the group
${\rm O}(d,d)$ is the T--duality 
group of the toroidally compactified string
\cite{Narain86,NarSarWit87,ShaWil89,GivRabVen89,MeiVen91,GivMalRab89,GivMalRab90}. 

In this paper we employ the term `duality' to 
refer to the continuous global symmetries of the effective
actions. Much of the subsequent discussion is based around the
${\rm SL}(2,R)$ and ${\rm O}(d,d)$ symmetries summarized in 
Section \ref{Section4}. 
These are relevant to the compactified NS--NS sectors of 
the type II and heterotic effective actions. The former is 
related to S--duality, since in many settings, the scalar field $\phi$ 
that arises in the non--linear sigma--model (\ref{sl2raction}) 
can be identified with the dilaton and, hence, the string coupling.  
The group ${\rm O}(d,d)$, on the other hand, is 
relevant to T--duality. 

\subsection{Target Space and Scale Factor Duality}

\label{Section5.1}

The simplest manifestation of T--duality is found by compactifying 
a string on a circle of radius $r$, where the compactification is  associated
with a given coordinate $X (\tau ,\sigma)$ 
\cite{KikYam84,SakSen86,AlvOso89}. (We suppress 
spacetime indices here for notational simplicity). The compactification 
of $X$ is equivalent to the identification $X \approx X +2\pi mr$, where 
$m$ is an arbitrary integer. 
The string world--sheet action (\ref{worldsheet}) 
implies that the bosonic coordinates $X(\tau , \sigma )$ 
satisfy the $(1+1)$--dimensional wave equation.  
Thus, in general they can be expanded in terms of 
left-- and right--moving modes, $X(\tau , \sigma ) =
X_L(\tau +\sigma) +X_R (\tau -\sigma)$. 
In particular, the zero modes of the oscillator expansion 
can be expressed as 
\begin{eqnarray}
X_R = \sqrt{\frac{\alpha'}{2}} p_R (\tau -\sigma ) 
\nonumber \\
X_L =\sqrt{\frac{\alpha'}{2}} p_L (\tau + \sigma )
\end{eqnarray}
where $p_{L,R}$ represent the centre of mass momenta. 
The total conjugate momentum is $P=(p_R+p_L)/\sqrt{2\alpha'}$
and the zero mode of $X$ can therefore be expressed as 
\begin{equation}
\label{tmom}
X= \alpha' P \tau + \sqrt{\frac{\alpha'}{2}}
\left( p_L -p_R \right) \sigma
\end{equation} 

If $X$ is not compact, then 
the second term in Eq. (\ref{tmom}) vanishes, because for a closed string 
the periodicity of $X$ with respect to $\sigma$, i.e., 
$X(\sigma =2\pi) =  X(\sigma )$,  implies that 
$X$ can not depend linearly on $\sigma$, so we must have $p_L =p_R$.
If $X$ is a compact dimension, however, this constraint 
need not apply, because the  string can wrap around this 
direction an arbitrary number of times. Wrapping the string around the circle 
$m$ times stretches it by an amount $2\pi mr$ and this increases
its energy by a discrete amount. 

Furthermore, if $X$ is compact, the momentum 
along this direction must be quantized in units of $1/r$, 
i.e., $P=n /r$, where $n$ is an arbitrary integer. 
It follows that the quantities $p_{L,R}$ can be 
of the form
\begin{eqnarray}
\label{LRP}
p_R =\frac{1}{\sqrt{2}} \left( \frac{\sqrt{\alpha'}}{r} n 
-\frac{r}{\sqrt{\alpha'}} m \right) \nonumber \\
p_L =\frac{1}{\sqrt{2}} \left( \frac{\sqrt{\alpha'}}{r}n 
+ \frac{r}{\sqrt{\alpha'}} m \right)
\end{eqnarray}
where the arbitrary integer $m$ may now be interpreted 
as the winding number. 

The Hamiltonian is given by 
$H=(p^2_R+p^2_L)/2$. It can be seen from 
Eq. (\ref{LRP}) that this is invariant under the transformation
\begin{equation}
\label{Rduality}
\frac{r}{\sqrt{\alpha'}} \rightarrow \frac{\sqrt{\alpha'}}{r} , 
\qquad m \leftrightarrow n
\end{equation}
The duality transformation (\ref{Rduality}) interchanges the winding 
modes with  the momenta and simultaneously 
inverts the radius of the circle. It is 
referred to as {\em target space duality}. 

It can be shown 
that this symmetry is a symmetry of the interacting string 
if the vacuum expectation value of the (higher--dimensional) dilaton 
also transforms as \cite{Buscher87,GinVaf87,GivPorRab94,RocVer92} 
\begin{equation}
\label{Rphi}
\Phi \rightarrow \Phi -2\ln r
\end{equation}
This implies that target 
space duality is an {\em exact} symmetry of string perturbation theory
because it is valid order by order in the string coupling. 
In the above example, it implies 
that the compactification of a string on a small circle of radius 
$r$ is {\em equivalent} to the compactification of the string on a large 
circle of radius $1/r$. The symmetry group in this 
example is ${\rm Z}_2$. Target space duality also implies the 
existence of a minimal length scale, $l_s =\sqrt{\alpha'}$, 
because a string on scales smaller than this can be considered 
in terms of a string on scales larger than $l_s$. 
The transformation (\ref{Rphi}) leaves invariant the 
lower--dimensional effective coupling, in the same way that 
the lower--dimensional dilaton of the toroidally compactified NS--NS 
string action (\ref{oddsymaction}) is invariant under 
the global ${\rm O}(d,d)$ symmetry transformation (\ref{NSNSOdd}). 

Target space duality is closely related to a 
scale factor duality of the string effective action 
(\ref{NSNSactionreduced}). This latter symmetry is a subgroup of the global
${\rm O}(d,d)$ symmetry (\ref{NSNSOdd}) discussed in 
Section \ref{Section4.3}. It can be better understood by encoding the 
$d^2$ data of the ${\rm O}(d,d)/{\rm O}(d)\times {\rm O}(d)$ 
coset in terms of the background parameter matrix
\begin{equation}
\label{Emat}
E_{AB} = h_{AB} +B_{AB} , \qquad A,B = 1, 2, \ldots , d
\end{equation}
The number of independent 
degrees of freedom in Eq. (\ref{Emat}) is then composed 
of $d(d+1)/2$ from the (symmetric) internal 
metric, $h_{AB}$, and $d(d-1)/2$ from the (antisymmetric) matrix
$B_{AB}$. The ${\rm O}(d,d)$ 
transformation (\ref{NSNSOdd}) on ${\cal{M}}$ is then equivalent 
to \cite{GivRabVen89,ShaWil89,GivMalRab89,GivMalRab90}
\begin{equation}
\label{Etrans}
E \rightarrow (WE+X)(YE+Z)^{-1}
\end{equation}
where $\{ W, X, Y , Z \}$ are defined in Eq. (\ref{UOdd}). 

Scale factor duality corresponds to the transformation generated 
by the ${\rm O}(d,d;Z)$ element 
\begin{equation}
U=\eta =\left( \begin{array}{cc} 
0 & {\rm I}_d \\
{\rm I}_d & 0 \end{array} \right)
\end{equation}
It can be seen directly from Eq. (\ref{Etrans}) 
that this inverts the parameter matrix $E\rightarrow E^{-1}$ 
\cite{GivRabVen89,ShaWil89}. 
In terms of $h$ and $B$ it results in the transformation
\begin{eqnarray}
\label{BGtrans}
h\rightarrow (h-BhB)^{-1} \\
B \rightarrow (B-hB^{-1}h)^{-1}
\end{eqnarray}
such that the combination $h^{-1}B 
\rightarrow -Bh^{-1}$ is invariant. We see that 
in the absence of a 2--form potential, $B=0$, 
Eq. (\ref{BGtrans}) inverts the metric on the internal 
space, $h \rightarrow h^{-1}$. In this sense, therefore, 
scale factor duality is a T--duality. This symmetry 
has numerous cosmological implications and is central to 
the pre--big bang inflationary scenario \cite{Veneziano91} 
(see Section \ref{Section9}).

\subsection{Relationships between the String Effective Actions}

\label{Section5.2}

We are now in a position to explore some of the duality 
relationships between the five superstring theories. 
We find  hints for these relationships at the level
of the effective actions by counting the relevant powers of 
$e^{\Phi}$ that arise in each of the terms. Our 
discussion follows the format presented by Witten \cite{Witten95}. 
We begin by considering 
$N=1$, $D=11$ supergravity \cite{CreJul79}. 
We may write the action for such a theory in the schematic form
\begin{equation}
\label{Msc}
S \approx \int d^{11}x \sqrt{|g_{11}|} \left[ R_{11} - |dC_3|^2 -
\ldots \right] 
\end{equation}
where dots denote the extra terms that arise in Eq. (\ref{Maction})
and, in what follows,  $dX_n$ 
denotes the field strength of an $n$--form potential, $X_n$.  
Compactification of this theory on a circle $S^1$
of radius $r_{11} = e^{\gamma}$ 
with the {\em ansatz} $ds^2_{11} =g_{\mu\nu} dx^{\mu}dx^{\nu}
+e^{2\gamma}(dx_{11}-A_{\mu}dx^{\mu})^2$ then leads to 
a ten--dimensional action containing the terms
\cite{Witten95,GiaPer84,CamWes84,HuqNam85} 
\begin{equation}
\label{IIAterms}
S \approx \int d^{10}x\sqrt{|g_{10}|} \left[ e^{\gamma} 
\left( R_{10} + |\nabla \gamma |^2 - |dC_3|^2 \right) -
e^{3\gamma} |dA_1|^2 -e^{-\gamma}|dB_2|^2 \right] -
\ldots
\end{equation}
Performing the conformal transformation (see Appendix~\ref{confsec})
\begin{equation} 
\label{Mconformal}
\tilde{g}_{\mu\nu} =e^{\gamma} g_{\mu\nu} 
\end{equation}
rewrites the action (\ref{IIAterms}) in the form 
\begin{equation}
\label{IIAc}
S\approx \int d^{10} x \sqrt{|\tilde{g}_{10}|} \left[ e^{-3\gamma}
\left( \tilde{R}_{10} +|\tilde\nabla \gamma |^2 -|dB_2|^2 \right) 
-|dA_1|^2 -|dC_3|^2 \right] -\ldots
\end{equation}

Comparison between the terms that arise in 
the type IIA string effective action 
(\ref{TYPEIIA}) and the dimensionally reduced action 
(\ref{IIAc}) indicates that the terms 
match \cite{CamWes84,HuqNam85,GiaPer84}, in the 
sense that the powers of the dilaton field are the same term by term, if we 
identify the ten--dimensional string coupling with the 
radius of the eleventh dimension \cite{Witten95}:
\begin{equation}
\label{identifystring}
r^3_{11} = g_s^2
\end{equation}
where $g_s^2 =e^{\Phi}$. 
This is the first piece of evidence indicating that the strong coupling 
limit of the type IIA superstring is related to an eleven--dimensional 
theory. Indeed, one definition 
of M--theory is that it is the strong coupling limit of the type IIA 
superstring \cite{Witten95,Townsend95}. 
It has $N=1$, $D=11$ supergravity as its low energy limit. 

We may also consider the strong coupling limits of the other 
string theories. {}From Eq. (\ref{HET}), the action for the 
${\rm SO}(32)$ heterotic string is given schematically by 
\begin{equation}
\label{hetscale}
S \approx \int d^{10}x \sqrt{|g_{10}|} e^{-\Phi} 
\left[ R_{10} +|\nabla \Phi |^2 - |dB_2|^2 -F_2^2 \right]
\end{equation}
where $F_2^2$ denotes the Yang--Mills field strength. 
The dilaton-gravity sector of the action remains
invariant under a conformal transformation
$\tilde{g}_{\mu\nu}=e^{-\Phi /2} g_{\mu\nu}$ together with the field
redefinition $\tilde{\Phi} =-\Phi$ (see Appendix~\ref{confsec}) and
the full rescaled action has the form
\begin{equation}
\label{rescaleI}
S \approx \int d^{10}x \sqrt{|\tilde{g}_{10}|} 
\left[ e^{-\tilde{\Phi}} (\tilde{R}_{10} +| \tilde\nabla \tilde{\Phi} |^2 )
-e^{-\tilde{\Phi} /2} \tilde{F}_2^2 -| d\tilde{B}_2 |^2 \right]
\end{equation}
The terms scale with $e^{\tilde\Phi}$ in the same way as 
in the type I effective action (\ref{TYPEI}). Moreover, 
since the dilaton has effectively gone from $\Phi$ to $-\Phi$, 
the string couplings in the two theories are related by 
$g_s \leftrightarrow g^{-1}_s$. This suggests that 
the strong coupling limit of the ${\rm SO}(32)$ string 
is dual to the weak coupling limit of the type I string, 
and vice--versa \cite{Hull95,Witten95,PolWit96}. 

We now consider the strong coupling 
limit of the type IIB string. For the moment let us ignore the RR axion 
field in Eq. (\ref{TYPEIIB}). (A more complete analysis
of the global symmetries of the action is made in the following 
subsection). The remaining  
NS--NS and RR fields scale with the dilaton as
\begin{equation}
\label{scaleIIB}
S \approx \int d^{10}x \sqrt{|G|} \left[ e^{-\Phi} 
\left( R+|\nabla \Phi |^2 -| dB_2^{(1)} |^2 \right) 
-|dB_2^{(2)}|^2 -|dD_4|^2 \right] -\ldots
\end{equation}
With the rescalings $G= e^{\Phi /2} g$ and $\tilde{\Phi} =-\Phi$, 
this action transforms to
\begin{equation}
\label{transIIB}
S \approx \int d^{10}x \sqrt{|g|} \left[ 
e^{-\tilde{\Phi}} \left( R +|\nabla \tilde{\Phi} |^2 -
| dB_2^{(2)} |^2  \right) -
| dB_2^{(1)} |^2  -|dD_4|^2 \right] -\ldots
\end{equation}
Thus, the two actions (\ref{scaleIIB}) and (\ref{transIIB}) 
are equivalent up to an interchange of the NS--NS and RR 
3--form field strengths. This suggests  that the 
IIB superstring is self--dual, in the 
sense that the strong coupling limit of this 
theory can be interchanged with its weakly coupled limit \cite{HulTow95}. 

It is also interesting to consider the relationships 
between the different theories below ten dimensions. 
In particular, it is known that there is only one 
$N=2$ supergravity theory when $D\le 9$ (see, e.g., \cite{West98}) and this 
implies that the type IIA and type IIB theories should 
be related. Furthermore, both type II theories have the same
number of massless degrees of freedom 
when compactified on a circle. Indeed, it can be shown 
that at the level of the effective 
actions, a type IIA string background 
containing one abelian isometry can be mapped onto 
a type IIB string background containing one abelian isometry 
\cite{CamWes84,BerHulOrt95}. 
The RR sectors of the theories are interchanged by such a transformation 
and the radii of the 
compactified dimension in the two theories are related by the duality 
$r_{10}^{\rm (IIA)} = 1/r_{10}^{\rm (IIB)}$ (in string units). 
In this sense, the type IIA string compactified on a circle of radius 
$r_{10}$ is equivalent to the type IIB string compactified 
on a circle of radius $r^{-1}_{10}$. This is a perturbatively 
exact result and the type IIA and IIB theories are T--dual
\cite{Ginsparg87,DinHueSei89,DaiLeiPol89}.

A similar relationship between the two heterotic strings applies 
below ten dimensions. That is, 
the ${\em E}_8\times {\rm E}_8$ heterotic string compactified on a circle 
of radius $r$ is equivalent to the ${\rm SO}(32)$ heterotic string 
compactified on a circle of radius $1/r$ 
\cite{Narain86,NarSarWit87,Ginsparg87}. This is also a perturbatively 
exact result. Thus, below ten dimensions there are only three distinct 
theories -- type I , type II and heterotic. Furthermore, the 
${\rm SO}(32)$ heterotic string can be related to the type 
I string in ten dimensions \cite{Witten95,PolWit96}. 
Can the heterotic and type II strings be related in a similar way 
in lower dimensions? 

The heterotic string has $N=1$ supersymmetry 
in the ten--dimensional setting, whereas the type II strings have 
$N=2$ supersymmetries. A necessary condition 
for the heterotic and type II strings to be dual to each other
in lower dimensions 
is that the number of supersymmetries should coincide after 
compactification. The number of unbroken supersymmetries 
that survive compactification depends on the 
holonomy group of the internal space. 
The metrically flat $d$--torus, $T^d$, has trivial holonomy
and admits the maximal 
number $(d)$ of covariantly constant spinors \cite{Vafa97}.  

The simplest compact, Ricci--flat manifold after the torus 
is the K3 surface. This is a four--dimensional space 
with holonomy group ${\rm SU}(2)$. It was first employed as a compactifying 
space within the context of $D=11$ supergravity \cite{DufNilPop83}
and has played a central role  in string duality. (For a detailed 
review of its properties, see \cite{Aspinwall96}). 
Since the K3 manifold  has an  ${\rm SU}(2)$ holonomy group, it 
admits two covariantly constant spinors and therefore 
preserves half of the original supersymmetry \cite{HawPop78}. 
Indeed, the K3 manifold is the only four--dimensional space, 
together with $T^4$, that admits covariantly constant spinors 
\cite{Vafa97}. Consequently,  compactifying a $N=2$ theory on K3 and a 
$N=1$ theory on $T^4$ results in two theories with 
the same number of supersymmetries. 

The metric for K3 exists but is unknown \cite{Yau78}. However, 
an approximate description of the K3 manifold is
to view it as an orbifold of $T^4$, i.e., 
${\rm K3} \approx T^4/{\rm Z}_2$ \cite{Page78,GibPop79}. The construction 
of this space is described in  Ref. \cite{LuPopSte98}. 
It involves identifying the coordinates 
$y^i \approx y^i +2\pi$ in $\Re^4$ $(i=1,2,3,4)$ and then making 
the further identifications $y^i \approx -y^i$. 

We now proceed to qualitatively consider 
the compactification of the heterotic string 
on $T^4$ and the compactification of the type IIA string on K3 by 
keeping the internal dimensions fixed \cite{Witten95}. The heterotic 
action then has  the form 
\begin{equation}
\label{het6}
S_{\rm H} \approx \int d^6x \sqrt{|g_H|} 
e^{-\Phi_H} \left[ R_H +|\nabla \Phi_H |^2 -| dB_{2H} |^2 -
|dA_{1H}|^2 \right]
\end{equation}
in six dimensions and the type IIA string effective action scales as 
\begin{equation}
\label{IIA6approx}
S_{\rm IIA} \approx \int d^6 x \sqrt{|g_A|} 
\left[ e^{-\Phi_A} \left( R_A +|\nabla \Phi_A |^2 -| dB_{2A} |^2 \right)
- | dA_{1A}|^2 \right]
\end{equation}
after compactification. 

Performing the conformal transformation
\begin{equation}
\label{confHA}
\tilde{g}_A = e^{-\Phi_A} g_A \,,
\end{equation}
on the type IIA action leads to an effective theory  
where all the terms scale as $e^{\Phi_A}$, with the exception of the 
$B_{2A}$ term. In this case, the scaling goes as $e^{-\Phi_A}$. However,  
a 3--form is Poincar\'e dual to another  3--form in six dimensions. 
The field equation for the 
2--form potential $B_{2A}$ in the new frame 
is $d ^*(e^{-\Phi_A} dB_{2A})=0$ and this is trivially  
solved by the {\em ansatz}  $^*(d \bar{B}_{2A}) = 
e^{-\Phi_A} dB_{2A}$. The field equations derived from the type IIA 
action (\ref{IIA6approx}) are therefore formally equivalent to 
those derived from 
\begin{equation}
\label{IIA6}
S_{\rm IIA} \approx \int d^6 x \sqrt{|\tilde{g}_A|} e^{\Phi_A} \left[ 
\tilde{R}_A + |\tilde\nabla \Phi_A|^2 - |d \bar{B}_2 |^2 -
|dA_1|^2 \right]
\end{equation}

It can now be seen that the two effective actions (\ref{het6}) and
(\ref{IIA6}) can be transformed into one another by a suitable
identification of the fields \cite{Seiberg88}.  Since $\Phi_H
=-\Phi_A$, the strong coupling limit of one theory is transformed to
the weak coupling limit of the other, $g_{s(H)} =g_{s(A)}^{-1}$. This
is the basis for string--string duality in six dimensions
\cite{Witten95,DufLu91,DufKhu94,DufLu94,DufMin95,HulTow95,DufKhuLu95,Sen95,HarStr95}. It
can be shown that string/string duality in six dimensions follows as a
consequence of heterotic/type I duality in ten dimensions
\cite{Vafa97}.

To summarize, there is evidence even at the level of the effective
actions to suggest that the five perturbative string theories,
together with $N=1$, $D=11$ supergravity, are fundamentally related by
duality symmetries. For consistency, therefore, all theories should be
considered equally. This provides strong motivation for studying the
cosmology of the type II theories in particular since these 
contain non--trivial RR sectors. In the following subsections, we
consider the S-- and U--dualities of the type IIB theory in more
detail. This provides the basis for studying the cosmological
consequences of this theory in Sections \ref{Section8} and
\ref{Section10}.

\subsection{SL(2,R) and S--Duality of the Type IIB Superstring}

\label{Section5.3}

The strong/weak coupling ${\rm SL}(2,Z)$
S--duality of the type IIB theory \cite{HulTow95}
can be understood in terms of the 
compactification of $D=11$ supergravity on a 2--torus, 
$T^2 =S^1\times S^1$ \cite{Schwarz95,Aspinwall95}. 
The relationship between the 2--torus and the modular group ${\rm SL}(2,Z)$ 
is summarized in Appendix \ref{appendixB}. 
For simplicity, we assume that the torus is rectangular. 
Compactification of $D=11$ supergravity on a circle of radius $r_1$ 
yields the type IIA theory \cite{GiaPer84,CamWes84,HuqNam85}, 
where the string coupling 
is determined by $g_A = r_1^{3/2}$ \cite{Witten95}. 
The ten--dimensional 
spacetimes in the M--theory and type IIA pictures are related by 
Eq. (\ref{Mconformal}). Thus, compactification 
of the type IIA theory on a circle of radius 
$r_A =g_A^{1/3} r_2$ is equivalent 
to compactification of $D=11$ supergravity on a 2--torus  
with radii $r_1$ and $r_2$. The T--duality $r_B =1/r_A$ then 
transforms the $D=9$ type IIA theory into the type IIB theory 
\cite{CamWes84,BerHulOrt95,Ginsparg87,DinHueSei89,DaiLeiPol89}. 

The effective ten--dimensional coupling of the type IIB theory is determined 
from the condition that the coupling in nine dimensions be invariant 
under this T--duality, i.e., $g_B^2/r_B = g_A^2 /r_A$. 
The type IIB coupling can then be written as 
$g_B = r_1/r_2$. Now, the area of the M--theory 
2--torus is $A_M =r_1r_2$ and this can be written as 
$A_M \propto  r_B^{-1}$. Thus, $r_B \rightarrow 
\infty$ as $A_M \rightarrow 0$. In this 
sense, therefore, the type IIB theory may be viewed as the compactification 
of M--theory on a 2--torus in the limit where the area of the torus vanishes
\cite{Schwarz95,Aspinwall95}. 
Moreover, invariance under the simultaneous interchange 
of the two cycles of the torus, $r_1 \leftrightarrow 
r_2$, in the M--theory 
picture  then implies $g_B \rightarrow g_B^{-1}$. This 
is the S--duality of the type IIB theory \cite{HulTow95}.

The above discussion provides a geometrical 
interpretation of the S--duality of the type IIB theory 
in terms of a 2--torus. 
This has led to the suggestion that 
the origin of the type IIB superstring may be found in a 
twelve--dimensional theory, labelled F--theory, that 
has been compactified on $T^2$ \cite{Vafa96,Vafa97,MorVaf96a,MorVaf96b,Sen96}. 

We now establish that the 
type IIB supergravity theory exhibits a global 
${\rm SL}(2,R)$ symmetry \cite{SchWes83}. 
The symmetry becomes manifest in the ten--dimensional Einstein frame 
\cite{Hull95,Schwarz95}. 
Performing the conformal transformation 
\begin{equation}
\label{conften}
\hat{g}_{MN} = e^{-\Phi/4} g_{MN}
\end{equation}
implies that the action (\ref{IIB}) 
becomes 
\begin{eqnarray}
\label{IIBEinstein}
S_{\rm IIB} =\int d^{10} x \sqrt{-\hat{g}_{10}} \left\{   
 \hat{R}_{10} - {1\over8} \left( \hat\nabla \Phi \right)^2
 -\frac{1}{2} e^{\Phi} \left( \hat\nabla \chi \right)^2 
\right. \nonumber \\ \left. 
 -\frac{1}{12} e^{\Phi/2} ( \hat{H}^{(1)} )^2 
-\frac{1}{12} e^{-\Phi/2}
 \left( \hat{H}^{(1)} \chi + \hat{H}^{(2)} \right)^2 \right\}
\end{eqnarray}

It can be seen from Eq. (\ref{IIBEinstein}) that 
the scalar fields and 2--form potentials appear as pairs. Indeed, 
comparison with the non--linear sigma--model 
action (\ref{sl2raction}) implies that the 
the dilaton and RR axion fields parametrize the ${\rm SL}(2,R)/{\rm U}(1)$ 
coset. Action (\ref{IIBEinstein}) can be written in a manifestly 
${\rm SL}(2,R)$ invariant form by introducing the `metric' 
\begin{equation}
\label{matrixM}
M \equiv   \left( \begin{array}{cc} e^{\Phi/2} &  \chi e^{\Phi/2} \\ 
\chi e^{\Phi/2} & e^{-\Phi/2} + \chi^2 e^{\Phi/2} \end{array} \right) ,
\end{equation}
and the two--component vector
\begin{equation}
\label{vectorH}
\hat{H}  = \left( \begin{array}{c} 
\hat{H}^{(2)} \\ \hat{H}^{(1)} \end{array} \right) .
\end{equation}
Since the matrix (\ref{matrixM}) satisfies the constraint 
$M^{\rm T}JM=J$, where $J$ is the ${\rm SL}(2,R)$ metric defined 
by Eq. (\ref{J}), it follows that action (\ref{IIBEinstein}) 
may be written as \cite{Hull95}
\begin{equation}
\label{10DSLaction}
S_{\rm IIB} =\int d^{10} x \sqrt{-\hat{g}_{10}} \left\{  
 \hat{R}_{10}
 + {1\over4} {\rm Tr} \left( \hat\nabla M \hat\nabla M^{-1} \right)
 - \frac{1}{12} \hat{H}^T M \hat{H} \right\}  ,
\end{equation}

The action (\ref{10DSLaction}) is invariant under the global 
${\rm SL}(2,R)$ transformation \cite{HulTow95,Schwarz95,BerBooOrt96}:  
\begin{equation}
\label{schwarz}
\bar{M} = \Sigma M \Sigma^T, \qquad \bar{\hat{g}}_{MN} = \hat{g}_{MN} , 
\qquad \bar{\hat{H}} = \left( \Sigma^T \right)^{-1} \hat{H}  ,
\end{equation}
where 
\begin{equation}
\label{sigma}
\Sigma \equiv \left( \begin{array}{cc} D & C \\ 
B  & A \end{array} \right) , \qquad AD-BC =1 .
\end{equation}
The 2--form potentials therefore 
transform as a doublet under the ${\rm SL}(2,R)$ symmetry.
The invariance of the ten--dimensional Einstein metric $\hat{g}_{MN}$ implies 
that the string metric $g_{MN}$ transforms under Eq. (\ref{schwarz}) as
\begin{equation}
\bar{g}_{MN} e^{-\bar{\Phi} /4} = g_{MN} e^{-\Phi /4}  .
\end{equation}
The four--form transforms as a singlet and therefore remains frozen in 
this analysis. 

For vanishing RR scalar field $\chi$ the particular transformation 
with $A=D=0$ and $C=-B=1$ yields 
\begin{equation}
\label{strongweakexact}
\bar{g}_B = \frac{1}{g_B} , \quad 
\bar{B}^{(1)}_{MN} =  - B^{(2)}_{MN} , \quad \bar{B}^{(2)}_{MN} 
= B^{(1)}_{MN}
\end{equation}
where $g^2_B =e^{\Phi}$ is the string coupling. This is 
the S--duality discussed above. It is intrinsically 
a non--perturbative symmetry because it exchanges the strongly coupled 
regime with the weakly coupled regime. It is important to note that
the S--duality ${\rm SL}(2,Z)$ relates different regimes 
of the {\em same} theory. 
Moreover, the NS--NS 2--form potential is interchanged with the 
RR 2--form potential under Eq. (\ref{strongweakexact}). 
This implies that both fields should be treated on the same level. 
This has important cosmological consequences, as will be seen 
in Sections \ref{Section8} and \ref{Section10}.

\subsection{U--Duality of the Type II Superstring}

\label{UdualitySection}

Thus far, we have seen that the NS--NS sector of the type II string 
effective action compactified on $T^{d-1}$ 
exhibits a global ${\rm SO}(d-1,d-1)$ 
symmetry \cite{MahSch93}, whereas the type IIB theory is invariant under 
a global ${\rm SL}(2,R)$ symmetry in ten dimensions \cite{SchWes83}. 
This latter symmetry interchanges NS--NS and RR fields. However, the 
different characteristics of the NS--NS and RR sectors arise 
directly from the perturbative formulation of string theory and  
the two sectors become unified within the context 
of eleven--dimensional supergravity \cite{CreJul79}. Moreover, 
since this latter theory exhibits a global 
${\rm SL}(d,R)$ symmetry when compactified on $T^d$,  
the toroidally compactified type II 
theory should exhibit a continuous symmetry group $G_d$  
that is generated by the non--commuting subgroups 
${\rm SO}(d-1,d-1)$ and ${\rm SL}(d,R)$. 

The group $G_d$ corresponds to a non--compact form of the exceptional group
${\rm E}_d$, denoted by ${\rm E}_{d(d)}$ \cite{Cremmer81,Julia81}. 
The existence of the ${\rm E}_{d(d)}$ symmetries in maximal 
supergravity theories was originally conjectured 
in \cite{CreJul78,CreJul79,Julia81,Cremmer81}. 
They were then found by direct construction \cite{SalSez89}. 
It was only recently, however, 
that the symmetries for all dimensionalities 
were obtained directly from the  
toroidal compactification of the eleven--dimensional 
theory \cite{CreJulLu98}. This unified previous 
works, where the symmetries 
for the particular cases of $D=9$ \cite{BerHulOrt95}, $D=4$ 
\cite{CreJul78,CreJul79} and 
$D=3$ \cite{Mizoguchi98} 
had been found directly by compactification.

The exceptional groups arise when performing the toroidal 
compactification down to $D$ dimensions and 
Poincar\'e dualising all field strengths with degree greater 
than $D/2$ \cite{CreJulLu98}. The identity 
of $G_d$ is deduced by establishing that the scalar fields in the model
parametrise the $G_d/H_d$ coset, where $H_d$ is the maximal 
compact subgroup of $G_d$. Once the identity of $H_d$ and 
the number of scalar fields 
present has been determined, the group $G_d$ can be found. 
The symmetry is a symmetry of the field 
equations when the dimensionality of spacetime is even and 
is also a symmetry of the action in odd dimensions \cite{CreJulLu98}.

Another important property is that the symmetry group $G_d$ 
does not act on the uncompactified $D$--dimensional 
spacetime \cite{HulTow95}. This implies that it survives further dimensional 
reduction and appears as a subgroup of the larger 
symmetry group $G_{d'}$, where $d' > d$. The origin of the 
exceptional groups can then be understood in terms of the 
${\rm SO}(d-1,d-1)$ T--duality and the ${\rm SL}(2,R)$ 
S--duality of the type IIB theory. The discrete subgroups of
these exceptional groups are the 
conjectured U--duality groups of the type II superstring
\cite{HulTow95}. The structure of 
the U--duality groups of M--theory are 
discussed in Refs. \cite{ObePio98,CreJulLu98,AndDauFer97a,AndDauFer97b,AndDauFer97c}. 

We now highlight some examples. In eight dimensions,
there is a ${\rm SL}(3,R)$ subgroup arising 
from M--theory on $T^3$ and a ${\rm SO}(2,2)$ 
subgroup arising from the T--duality. The group 
${\rm SO}(2,2)$ is isomorphic to ${\rm SL}(2,R) 
\times {\rm SL}(2,R)$ \cite{Oraifeartaigh86}. The origin of one of the 
${\rm SL}(2,R)$ subgroups can be traced to the 
toroidal compactification of the eleven--dimensional
theory. Thus, the global symmetry is ${\rm SL}(3,R) 
\times {\rm SL}(2,R)$ and this is isomorphic to ${\rm E}_{3(3)}$. 
In seven dimensions, the toroidal group 
is ${\rm SL}(4,R)$ and the T--duality group is 
${\rm SO}(3,3)$. The latter is isomorphic 
to ${\rm SL}(4,R)$ \cite{Oraifeartaigh86} and this does not 
commute with the other ${\rm SL}(4,R)$ subgroup. 
Together they form ${\rm E}_{4(4)} = {\rm SL}(5,R)$ \cite{CreJulLu98}. 
In six dimensions, we have the non--commuting subgroups 
${\rm SL}(5,R)$ and ${\rm SO}(4,4)$ and the smallest 
group containing both is ${\rm E}_{5(5)} ={\rm SO}(5,5)$ \cite{Vafa97}. 

As far as the scalar manifolds are concerned, no dualization of the 
higher degree form--fields is necessary to establish the coset structure 
of the axionic scalar lagrangians for $D\ge 6$. The 
situation is more complicated, however, below six dimensions.
In these cases, the exceptional groups emerge as the symmetry 
only after the $(D-2)$--form gauge potentials have been 
Poincar\'e dualized to axionic scalars \cite{CreJulLu98}. The 
dualization changes the coset structure to that 
of ${\rm E}_{(11-D)(11-D)}$.

One of the key points of U--duality is that it acts on the dilaton, 
the moduli of the metric and the axionic scalars 
that originate from the form fields. 
Thus, the ${\rm E}_{d(d)}$ symmetry unifies into 
a single quantity the 
dilaton, the RR fields and the scalar degrees of freedom 
arising from the metric compactification. It is in this sense that U--duality
maps a strongly coupled theory compactified on 
a small dimension onto a weakly coupled theory compactified on a 
large dimension, and vice--versa. Indeed, U--duality implies that {\em all} 
the coupling constants in the theory are equally important. 
It is inherently a non--perturbative symmetry of string theory.

In Section \ref{Section8}, we consider some of the cosmological implications 
of U--duality within the context of the type IIB theory.

\newpage
\section*{PART II}

\addcontentsline{toc}{section}{PART II}

\section{Higher Dimensional Kasner Cosmology}
\label{kasnersection}
\label{Section6}

\setcounter{equation}{0}

\def\theequation{\thesection.\arabic{equation}}

The perturbative formulation of string theory requires spacetime to be
ten--dimensional.  On the other hand, recent developments indicate
that the correct description of the universe may correspond to
M--theory, with a low--energy limit given by eleven--dimensional
supergravity.  In either case, a consistent cosmology must be able to
account for a universe where three of the spatial dimensions are large
relative to the hidden ones. This then raises the question of relating
apparently different cosmological solutions in different frames.

The purpose of the present Section is to survey a variety homogeneous
and spatially flat cosmological solutions in various dimensions and
frames. We begin by considering vacuum Einstein gravity solutions for
a $(D+d+1)$-dimensional torus. These solutions represent the
higher--dimensional generalization of the Kasner solution found in
four--dimensional vacuum Bianchi type I models. 
Kasner--like solutions were studied in a string 
context by Gibbons and Townsend \cite{GibTow87}. 
For $D+d+1=11$, they
represent vacuum cosmological solutions in the low energy limit of
M--theory~\cite{KalKogOli98,BanFisMot99}.  Interpreting one of the spatial
dimensions as an internal degree of freedom then leads to a
$(D+d)$-dimensional solution of the low-energy limit of string theory
which we refer to as a `dilaton--vacuum' cosmology. We then
reinterpret these solutions as $D$-dimensional cosmologies, where the
$d$ dimensions form a compact internal space. These dimensions play
the role of moduli fields and the solutions are referred to as
`dilaton--moduli--vacuum' solutions in the dimensionally reduced
theory. Attention has recently focussed upon the ability to
reinterpret singular solutions obtained from the dimensional reduced
action by ``oxidising'' back up to a non-singular solution in higher
dimensions~\cite{GibHorTow95,LarWil97,FeiVaz99}. At each stage of
dimensional reduction the dilaton-gravity solution can be re-written
via a conformal rescaling of the metric as a solution in Einstein
gravity with minimally coupled fields. It is also possible that the
singular nature of cosmological solutions in one frame may appear to
be non-singular in a conformally related
frame~\cite{ShaTriWil91,KalOli97,Quiros99}.

Different dilaton--moduli--vacuum solutions are related by the
symmetries of the low-energy action and provide simple illustrations
of the symmetries of the theories. The importance of these solutions
is that they represent the critical points in the phase-space of a
wide class of string cosmologies containing additional degrees of
freedom~\cite{BilColLid99a,BilColLid99b}. 
As an example, in Section~\ref{kasnerwithforms}, we present
homogeneous solutions containing form fields.

\subsection{$(D+d+1)$-dimensional Einstein-vacuum solutions}
\label{Section6.1}

Here we present $(D+d+1)$-dimensional cosmological
solutions of vacuum Einstein gravity. The gravitational action
is given by
\begin{equation}
\label{EinsteinDd1}
S = {1\over16\pi G_{D+d+1}} \int_{M_{D+d+1}} d^{D+d+1}x
  \, \sqrt{-\bar{g}_{D+d+1}} \, \bar{R}_{D+d+1}
\ ,
\end{equation}
where $\bar{R}_{D+d+1}$ is the Ricci scalar curvature of the 
$(D+d+1)$--dimensional manifold with metric $\bar{g}_{AB}$ and 
$G_{D+d+1}$ is the $(D+d+1)$--dimensional Newtonian constant.

We assume the topology of the universe is 
that of a rectilinear torus with $D+d$ independent scale factors,
$e^{\bar\alpha_I}$. The line element is  therefore of the form 
\begin{equation}
d\bar{s}_{D+d+1}^2 =
 -d\bar{t}^2 + \sum_{I=1}^{D+d} e^{2\bar{\alpha}_I} dx_I^2 \ ,
\end{equation}
and the Ricci scalar curvature is given by 
\begin{equation}
\label{Dd1Ricci}
\bar{R}_{D+d+1} = \sum_I 2\ddot{\bar\alpha}_I + \dot{\bar\alpha}_I^2
 + \dot{\bar\alpha}_I \sum_J\dot{\bar\alpha}_J \ .
\end{equation}
We also give the alternative curvature invariant
\begin{equation}
\label{Dd1R2}
\bar{R}_{ABCD}\bar{R}^{ABCD} 
 = 4 \sum_I \left(\ddot{\bar\alpha}_I +\dot{\bar\alpha}_I^2\right) 
   + 2 \sum_{J\neq I} \left(\dot{\bar\alpha}_I\dot{\bar\alpha}_J\right)^2 \
 .
\end{equation}
The action (\ref{EinsteinDd1}) then reduces to
\begin{equation}
S = {1\over16\pi G_{D+d+1}} \int d\bar{t} \int d^{D+d}x \left( \prod_K
 e^{\bar\alpha_K} 
 \right)
 \sum_I \left[ 2\ddot{\bar\alpha}_I + \dot{\bar\alpha}_I^2
 + \dot{\bar\alpha}_I \sum_J\dot{\bar\alpha}_J \right]
\end{equation}
which can be integrated by parts to yield an effective Lagrangian
\begin{equation}
\label{D+dLagrangian}
L = \left( \prod_K e^{\bar\alpha_K} \right) 
 \sum_I \left[ \dot{\bar\alpha}_I^2
 - \dot{\bar\alpha}_I\sum_J\dot{\bar\alpha}_J \right]
\end{equation}

The equations of motion that follow from varying this Lagrangian with
respect to each $\bar\alpha_I$ have the form
\begin{equation}
\label{fielddavid}
{d \over dt} \left[ \left( \prod_K e^{\bar\alpha_K} \right) \left(
2\dot{\bar\alpha}_I -2\sum_J\dot{\bar\alpha}_J \right) \right] = L
\ ,
\end{equation}
subject to the Hamiltonian constraint
\begin{equation}
\label{Dd1constraint}
L = 0 \ .
\end{equation}
Equations (\ref{fielddavid}) and (\ref{Dd1constraint}) 
are readily integrated to yield the power--law solutions
\begin{equation}
\label{Einsteinpi}
\bar\alpha_I = \alpha_{I0} + \bar{p}_I \ln \bar{t} \ ,
\end{equation}
where $\alpha_{I0}$ are arbitrary constants 
and the constants of integration, $\bar{p}_I$, obey the Kasner constraints
\begin{equation}
\label{Kasner}
\sum_I \bar{p}_I = 1 \ , \qquad \sum_I \bar{p}_I^2 = 1 \ .
\end{equation}
If the indices, $\bar{p}_I$, are viewed as coordinates on $\Re^{D+d}$, 
the constraints in Eq.~(\ref{Kasner}) imply that the spatially 
flat and homogeneous solutions to vacuum Einstein gravity
must lie on the intersection of the unit sphere with a fixed plane.

The Ricci scalar given in equation~(\ref{Dd1Ricci}) vanishes at all
times along this classical solution, $\bar{R}_{D+d+1}=0$, as must be
the case for any vacuum solution of Einstein's equations.
Nonetheless these solutions may have a curvature singularity due to the
divergence of other curvature invariants.
$\bar{R}_{ABCD}\bar{R}^{ABCD}$ is given from Eqs.~(\ref{Dd1R2}),
(\ref{Einsteinpi}) and~(\ref{Kasner}) as
\begin{equation}
\bar{R}_{ABCD}\bar{R}^{ABCD} 
 = 2 \left[ 3 + \sum_I \left( \bar{p}_I - 4 \right) \bar{p}_I^3 \right]
 \bar{t}^{-4} \ ,
\end{equation}
which is singular as $\bar{t}\to0$ except the particular case
$\bar{p}_I=1$ for one dimension $I$ and $\bar{p}_{J\neq
I}=0$~\cite{FeiVaz99}. This particular case corresponds to $11-D$
Minkowski spacetime in Rindler coordinates so that $t=0$ is just a
coordinate singularity. However we shall see that this non-singular
solution in eleven dimensions can give rise to apparently singular
dimensionally reduced cosmologies~\cite{FeiVaz99}.

It is sometimes illuminating to rewrite Eqs. (\ref{fielddavid}) in
terms of the overall expansion and shear of the spatial hypersurfaces, 
$\bar{t}={\rm constant}$, given by 
\begin{equation}
\bar\theta \equiv \sum_I \dot{\bar\alpha}_I \ ,
\end{equation}
and 
\begin{equation}
\bar\sigma^2 
\equiv 
\sum_{I,J} \left( \dot{\bar\alpha}_I - \dot{\bar\alpha}_J \right)^2 
\ ,
\end{equation}
respectively. 
The $D+d-1$ independent degrees of freedom associated with the shear 
of the spatial sections can be defined as
\begin{equation}
\dot{\bar\sigma}_I \equiv \sqrt{{I \over I+1}} \left( \dot{\bar\alpha}_{I+1} -
{1\over I} \sum_{J\leq I} \dot{\bar\alpha}_J \right) \ ,
\end{equation}
and the Lagrangian (\ref{D+dLagrangian}) can then be rewritten as 
\begin{equation}
L = \left( \prod_K e^{\bar\alpha_K} \right) 
 \left[ - \left( {D+d-1 \over D+d} \right) \bar{\theta}^2
 + {1\over 2} \sum_{I<D+d} \dot{\bar\sigma}_I^2 \right] \ .
\end{equation}
The constraint
equation~(\ref{Dd1constraint}) now takes the form 
\begin{equation}
{D+d-1 \over D+d} \bar\theta^2
 = {1\over2} \sum_{I<D+d} \dot{\bar\sigma}_I^2 \ .
\end{equation}
where the $D+d-1$ shear degrees of freedom evolve as (damped) free
fields, $\bar\sigma_I\propto\ln\bar{t}$, and drive the expansion,
$\bar\theta=1/\bar{t}$.

\subsection{$(D+d)$-dimensional dilaton-vacuum solutions}

We now compare the Kasner--type solutions discussed in the 
previous subsection with the related solutions that arise in 
string cosmologies. We begin with the action (\ref{EinsteinDd1}) 
but first compactify the theory on a circle and treat the scale factor,
$\bar\alpha_{D+d}$, as an internal degree of freedom 
propagating on the
$(D+d)$-dimensional manifold $M_{D+d}$. The 
reduced gravitational
action can be written by extracting the relevant 
information from Eq. (\ref{slaction}) as
\begin{equation}
\label{nokinetic}
S = {1\over16\pi G_{D+d}} \int_{M_{D+d}} d^{D+d}x \, 
  \sqrt{-\bar{g}_{D+d}} \, e^{\bar\alpha_{D+d}} \, \bar{R}_{D+d}
\ ,
\end{equation}
where $G_{D+d}\equiv G_{D+d+1}/\int dx_{D+d}$.  This appears to be a
rather unusual scalar-tensor theory of gravity because there is no
explicit kinetic term for the dilaton field, i.e., the effective
Brans--Dicke parameter takes the value $\bar\omega=0$. However, as
discussed in Appendix~\ref{confsec}, the effective Brans--Dicke
parameter changes under a conformal rescaling of the metric.
When other matter
fields are absent, this conformal rescaling can be arbitrary. 
In particular, for the class of conformally--related frames
defined by 
\begin{equation}
\label{alphaCT}
g_{AB} = e^{2\lambda\bar\alpha_{D+d}} \bar{g}_{AB} \ ,
\end{equation}
where $\lambda$ is an arbitrary constant, the 
$(D+d)$-dimensional action (\ref{nokinetic}) 
takes the standard form of the vacuum Brans--Dicke theory:
\begin{equation}
\label{SD+dstring}
S = {1\over16\pi G_{D+d}} \int_{M_{D+d}} d^{D+d}x
  \sqrt{-g_{D+d}} e^{-\phi} \left[ R_{D+d} - \omega (\nabla\phi)^2 \right] 
\ ,
\end{equation}
where
\begin{equation}
\phi = [(D+d-2)\lambda-1]\bar\alpha_{D+d}
\end{equation}
and
\begin{equation} 
\label{omegaD+d}
\omega = {(D+d-1)\lambda[2-(D+d-2)\lambda] \over [1-(D+d-2)\lambda]^2} \ ,
\end{equation}

In this frame, the rescaled scale-factors for $I<D+d$ are  
\begin{equation}
\alpha_I = \bar\alpha_I +\lambda\bar\alpha_{D+d} \ .
\end{equation}
and, for the higher--dimensional solutions presented in
Eq.~(\ref{Einsteinpi}), they can be written as the dilaton--vacuum 
solutions
\begin{equation}
\label{stringpi}
\label{rolling}
\alpha_I = \alpha_{I0} + p_I \ln t \ , \qquad 
\phi = \phi_0 + K \ln t \ ,
\end{equation}
where the cosmic time in the rescaled frame becomes
\begin{equation}
\label{barttot}
t \propto \bar{t}^{(1+\lambda \bar{p}_{D+d})} \ ,
\end{equation}
and hence
\begin{equation}
\label{pidefinition}
p_I = {\bar{p}_I + \lambda\bar{p}_{D+d} \over 1 + \lambda\bar{p}_{D+d}} \ ,
\end{equation}
and
\begin{equation}
\label{Kdefinition}
K = {[(D+d-2)\lambda-1]\bar{p}_{D+d} \over 1 + \lambda\bar{p}_{D+d}} \ .
\end{equation}
The rescaling of the cosmic time coordinate in Eq.~(\ref{barttot})
leads to the singular limit, $\bar{t}\to0$, being mapped onto a
non-singular past or future infinity, $|t|\to\infty$, (or vice versa) when
$\lambda\bar{p}_{D+d}\leq-1$. A necessary condition for this is thus
$|\lambda|\geq1$.

We are interested here in the special case of string dilaton--gravity,
where $\omega=-1$ and we refer to the rescaled metric $g_{AB}$ as the
string frame metric. The relevant value of $\lambda$ is given from
Eq.~(\ref{omegaD+d}) as
\begin{equation}
\label{lambdaval}
\lambda = 
{1 \over \pm \sqrt{D+d-1} - 1} \ .
\end{equation}
Note that there are two possible choices for $\lambda$ corresponding
to two dilaton-gravity theories related by an S-duality transformation,
with the dilaton
\begin{equation}
\phi = \pm \sqrt{D+d-1} \bar\alpha_{D+d} \ .
\end{equation}
In the case of the eleven dimensional low energy M-theory action
presented in Eq.~(\ref{Maction}), one obtains the $D+d=10$ type IIA
string theory effective action given in Eq.~(\ref{TYPEIIA}) when
$\lambda=1/2$ and $\phi=3\bar\alpha_{10}$.

The solution (\ref{rolling}) corresponds to 
the `rolling radii' solution first considered by Mueller 
within the context of the bosonic string, where $D+d=26$
\cite{Mueller90}. 
When $\omega=-1$ the constants of integration (\ref{pidefinition}) 
and (\ref{Kdefinition}) obey the generalised Kasner
constraints 
\begin{equation}
\label{stringKasner}
\sum_{I=1}^{D+d-1} p_I = 1+K \ , \qquad \sum_{I=1}^{D+d-1} p_I^2 = 1 \ .
\end{equation}
These dilaton-vacuum indices in the string frame, $p_I$, lie on the
intersection of the unit sphere with a plane whose distance from the
origin depends on the kinetic energy of 
the dilaton field, as parametrized by 
the constant of integration, $K$. Hence, this constant is bounded:
\begin{equation}
\label{kasnerbound}
-1-\sqrt{D+d-1} \leq K \leq -1+\sqrt{D+d-1} \ .
\end{equation}
We recover the general relativistic Kasner vacuum solutions
in $D+d$ dimensions as a particular solution when $\bar{p}_{D+d}=0$
and $K=0$.

Substituting equations (\ref{stringpi}) and (\ref{stringKasner}) into
the expression  for the $(D+d)$--dimensional Ricci scalar,
\begin{equation}
\label{DRicci}
R_{D+d} = \sum_{I=1}^{D+d-1} 2\ddot\alpha_I + \dot\alpha_I^2
 + \dot\alpha_I \sum_{J=1}^{D+d-1} \dot\alpha_J \ ,
\end{equation}
yields
\begin{equation}
R_{D+d} = {K^2 \over t^2} \ . 
\end{equation}
This implies that there is now a singularity in the
$(D+d)$-dimensional scalar curvature as $t\to0$ whenever $K\neq0$.
This is true even when $\bar{p}_{D+d}=1$ and the $(D+d+1)$-dimensional
Einstein-vacuum solution was flat Minkowski spacetime in Rindler
coordinates and there was no curvature singularity.  In this case the
$(D+d)$-dimensional cosmology is isotropic. As noted in
Ref.~\cite{FeiVaz99} is remarkable that regularity in the higher
dimensional spacetime requires isotropy in the lower dimensional
spacetime.

On the other hand, note from Eq.~(\ref{barttot}) that singular
behaviour when $t\to0$ in the $(D+d)$-dimensional dilaton-gravity
cosmology may correspond to non-singular behaviour where
$\bar{t}\to\pm\infty$ in the $(D+d+1)$-dimensional Einstein gravity if
$|\lambda|>1$ which, from Eq.~(\ref{lambdaval}), requires $D+d<5$.
This is another way in which a curvature singularity in the
dimensionally reduced dilaton-gravity solutions may be mapped onto
non-singular evolution in the higher-dimensional Einstein-vacuum
theory. 

We now consider the corresponding solutions for the 
$(D+d)$-dimensional Einstein metric. {}From Section~\ref{CTdilgravaction},
this frame is related to the string frame by a further
conformal transformation:
\begin{equation}
\label{D+dEinstein}
\hat{g}_{AB}=e^{-2\phi/(D+d-2)}g_{AB}\ ,
\end{equation}
To arrive at this metric, one may equivalently choose
$\lambda=1/(D-2)$ in equation~(\ref{alphaCT}).
We emphasize that this metric therefore differs by a conformal factor
from the $(D+d)$-dimensional part of the original
$(D+d+1)$-dimensional Einstein metric $\bar{g}_{AB}$.
The action (\ref{SD+dstring}) then becomes
\begin{equation}
\label{SD+dEinstein}
S = {1\over16\pi G_{D+d}} \int_{M_{D+d}} d^{D+d}x \sqrt{-\hat{g}}
\left[ \hat{R} - {1\over D+d-2} 
(\hat\nabla\phi)^2 \right]
\end{equation}

The solutions for the rescaled scale factors and dilaton 
field can be written as
\begin{equation}
\hat{\alpha}_i = \alpha_i - \frac{\phi}{D+d-2}
 = \hat{\alpha}_{i0} + \hat{p}_i \ln \hat{t} 
\end{equation}
and 
\begin{equation}
\phi = \hat{\phi}_0 + \hat{K} \ln \hat{t} \ ,
\end{equation}
respectively, where 
\begin{equation}
\label{D+dEindices1}
\hat{p}_i = \frac{(D+d-2)p_i - K}{D+d-2-K} 
\end{equation}
and
\begin{equation}
\label{D+dEindices*}
\hat{K} = \frac{(D+d-2)K}{D+d-2-K} \ ,
\end{equation}
and the proper time for comoving observers in the $(D+d)$-dimensional
Einstein frame becomes
\begin{equation}
\hat{t} \propto t^{(D+d-2-K)/(D+d-2)} \ .
\end{equation}
Since the quantity  $D+d-2-K$ is positive definite for $D+d>2$, 
$\hat{t}\to0$ as $t\to0$. The Ricci scalar is given by 
$\hat{R}=\hat{K}^2/[(D+d-2)\hat{t}^2]$, so there is again a curvature
singularity whenever $\hat{t}\to0$ for $\hat{K}\neq0$.

Finally, 
the generalised Kasner constraints in the $(D+d)$-dimensional Einstein
frame become
\begin{equation}
\sum_{i=1}^{D+d-1} \hat{p}_i = 1 \ , \qquad 
\sum_{i=1}^{D+d-1} \hat{p}_i^2 = 1 - \frac{\hat{K}^2}{D+d-2}
\end{equation}
where
\begin{equation}
- \frac{D+d-2}{\sqrt{D+d-1}} \leq \hat{K} \leq + \frac{D+d-2}{\sqrt{D+d-1}} 
\end{equation}
In contrast to the Kasner constraints (\ref{stringKasner}), 
the indices, $\hat{p}_i$, lie on the intersection of a fixed plane with
a sphere whose radius depends on $\hat{K}$.

\subsection{$D$-dimensional dilaton-moduli-vacuum solutions}

The dilaton--vacuum solutions presented in Eqs. (\ref{stringpi})
and (\ref{stringKasner}) can re-interpreted as dilaton-moduli-vacuum
solutions in a $D$-dimensional spacetime, where the line element is
given by 
\begin{equation}
\label{Dds}
ds^2 = -dt^2 + \sum_{i=1}^{D-1} e^{2\alpha_i} dx_i^2 \ ,
\end{equation}
The extra $d$ dimensions are now compactified in the form of a 
Ricci--flat internal space with a diagonal 
metric given by $ds^2_{\rm int} = h_{ab} dx^a dx^b$
$(a,b = D, \ldots , D+d)$, where $h_{aa} =e^{2\alpha_a}$. 
The dimensionally reduced
action then corresponds to a truncated form of Eq. (\ref{nohat}):
\begin{equation}
\label{SDstring}
S = \int_{M_D} d^Dx
  \sqrt{-g} e^{-\varphi} \left[ R_D + (\nabla\varphi)^2 
- \sum_{a=D}^{D+d-1} (\nabla\alpha_a)^2 \right] 
\ .
\end{equation}
There are no vector terms in Eq. (\ref{SDstring}) because we have assumed 
that the internal space is a 
rectilinear torus. The internal scale factors, $\alpha_a$, may 
be interpreted as moduli fields in this action. 

The effective Kasner constraints for the $D-1$ external dimensions
are
\begin{equation}
\label{DstringKasner}
\sum_{i=1}^{D-1} p_i = 1 + {\cal K} \ , \qquad 
\sum_{i=1}^{D-1} p_i^2 = 1 - B^2\ ,
\end{equation}
where
\begin{equation}
\label{calK}
{\cal K} \equiv  K - \sum_{a=D}^{D+d-1} p_a \ .
\end{equation}
and 
\begin{equation}
B^2 = \sum_{a=D}^{D+d-1}  p_a^2
\end{equation}
are constants satisfying the bounds
\begin{equation}
0 \leq B^2 \leq 1 
\end{equation}
and 
\begin{equation}
-1-\sqrt{(D-1)(1-B^2)} \leq {\cal K} \leq -1+\sqrt{(D-1)(1-B^2)} \ .
\end{equation}
The effective dilaton in $D$-dimensions then becomes
\begin{equation}
\label{varphi}
\varphi \equiv \phi - \sum_{a=D}^{D+d-1} \alpha_a 
 = \varphi_0 + {\cal K} \ln t \ ,
\end{equation}

When compared with the original dilaton-vacuum constraints given in
Eq.~(\ref{stringKasner}),  the dilaton-moduli-vacuum indices, $p_i$, lie
on the intersection of a sphere, whose radius now depends on $B^2$, and
a plane, whose position is determined by ${\cal K}$. Thus, 
in this interpretation, $B^2$ is an
additional constant of integration that parametrizes 
the evolution of the moduli fields, i.e., the internal space.

The $D$-dimensional Ricci scalar curvature is $R_D=({\cal
K}^2-B^2)/t^2$. 

It follows from the analysis of Section \ref{CTdilgravaction} that
the conformal transformation to the $D$-dimensional Einstein--frame metric
is given by 
\begin{equation}
\label{DEinstein}
\tilde{g}_{\mu\nu}=e^{-2\varphi/(D-2)}g_{\mu\nu}\ .
\end{equation}
and the reduced action in this frame becomes
\begin{equation}
\label{SDEinstein}
S = \int_M d^Dx \sqrt{-\tilde{g}} \left[ \tilde{R} - {1\over D-2}
(\tilde\nabla\varphi)^2 - \sum_{a=D}^{D+d-1} (\tilde\nabla\alpha_a)^2
\right]
 \ .
\end{equation}
The solutions for the rescaled scale factors and effective dilaton can
then be written as
\begin{equation}
\tilde{\alpha}_i = \alpha_i - \frac{\varphi}{D-2}
 = \tilde{\alpha}_{i0} + \tilde{p}_i \ln \tilde{t} \ , \qquad
\varphi = \tilde{\varphi}_0 + \tilde{K} \ln \tilde{t} \ ,
\end{equation}
where the proper time for comoving observers in the $D$-dimensional
Einstein frame is determined by 
\begin{equation}
\tilde{t} \propto t^{(D-2-{\cal K})/(D-2)} \ .
\end{equation}
and the constants $\tilde{p}_i$ and $\tilde{K}$ satisfy 
\begin{equation}
\tilde{p}_i = \frac{(D-2)p_i-{\cal K}}{D-2-{\cal K}} 
\end{equation}
and 
\begin{equation}
\tilde{{\cal K}} = \frac{(D-2){\cal K}}{D-2-{\cal K}} \ ,
\end{equation}
respectively. 
Note again that $D-2-{\cal K}$ is positive definite for $D>2$, so that
$\tilde{t}\to0$ whenever $t\to0$.

The generalised Kasner constraints in the $D$-dimensional Einstein
frame become
\begin{equation}
\sum_{i=1}^{D-1} \tilde{p}_i = 1 \ , \qquad 
\sum_{i=1}^{D-1} 
\tilde{p}_i^2 = 1 - \tilde{B}^2 - \frac{\tilde{{\cal K}}^2}{D-2}
\end{equation}
where
\begin{equation}
\tilde{B} = \frac{(D-2)B}{D-2-{\cal K}} \ .
\end{equation}
The indices, $\tilde{p}_i$, lie on the intersection of a fixed plane
with a sphere whose radius depends on $\tilde{{\cal
K}}^2/(D-2)+\tilde{B}^2$, which requires
\begin{equation}
0 \leq \frac{\tilde{{\cal K}}^2}{D-2} + \tilde{B}^2 \leq 1 \ .
\end{equation}
The Ricci scalar is $\tilde{R}=[\tilde{{\cal
K}}^2/(D-2)+\tilde{B}^2] / \tilde{t}^2$, so there is a curvature
singularity whenever $\tilde{t}\to0$ for $\tilde{{\cal K}}\neq0$ or
$\tilde{B}\neq0$.

Note that there is a conformal factor relating the $D$-dimensional
Einstein metric to the $D$-dimensional part of the $(D+d)$-dimensional
Einstein metric!
\begin{eqnarray}
\tilde{g}_{\mu\nu}&=&
\exp \left(
\frac{2(D-2)\sum_{a=D}^{D+d-1}\alpha_a-2d\varphi}{(D-2)(D+d-2)} \right) \,
\hat{g}_{\mu\nu}\ , \nonumber\\
&=& \exp \left(
\frac{2}{D-2} \sum_{a=D}^{D+d-1}\hat{\alpha}_a \right) \,
\hat{g}_{\mu\nu}\ .
\end{eqnarray}
The two different possible definitions for the Einstein frame are
related by a constant rescaling only if the volume of the
$d$-dimensional internal space remains constant in the
$(D+d)$-dimensional Einstein frame.


\subsection{Symmetries of cosmological vacuum solutions}
\label{Section6.4}
\label{vaccossym}

\paragraph{T-duality:}
The cosmological vacuum solutions presented above provide a simple
illustration of some of the symmetry properties exhibited by the
dimensionally reduced string effective actions. In particular, 
the compactified action (\ref{SDstring}) is manifestly invariant under
the transformation
\begin{equation}
\label{atominusa}
\alpha_a \to -\alpha_a\ ,
\end{equation}
for any or all of the $d$ moduli fields $\alpha_a$, while the
$D$-dimensional metric, $g_{\mu\nu}$, and $D$--dimensional 
effective dilaton,
$\varphi$, remain invariant. Note that the $D$-dimensional Einstein
metric, $\tilde{g}_{\mu\nu}$, given in Eq.~(\ref{DEinstein}), is also
invariant under this transformation.  
This symmetry is the simplest example of a T-duality
and corresponds to the scale factor 
duality discussed in Section \ref{targetsection}.
The symmetry is also present in the original $(D+d)$-dimensional
action given in Eq.~(\ref{SD+dstring}) if the geometry is toroidal. 

Note, however, that the original
$(D+d)$-dimensional dilaton is not left invariant by 
Eq. (\ref{atominusa}). 
Eq.~(\ref{varphi}) implies that 
it transforms as $\phi\to\phi-2\sum_a \alpha_a$.  Thus the
$(D+d)$-dimensional Einstein metric given in Eq.~(\ref{D+dEinstein})
is conformally rescaled to 
\begin{equation}
\hat{g}_{AB} \to e^{4\sum_a\alpha_a/(D+d-2)} \, \hat{g}_{AB} \ .
\end{equation}
under this symmetry transformation. 
In terms of the Kasner indices in the string frame, scale-factor
duality corresponds to a reflection $p_a\to-p_a$. The $D$-dimensional
constants of integration, $B$ and ${\cal K}$, remain invariant which
implies from Eq.~(\ref{calK}) that 
\begin{equation}
\label{KtoK}
K \to K-2\sum_a p_a \ .
\end{equation}

Interesting particular cases arise
for the isotropic $(D+d)$--dimensional solution. 
In this case, Eq.~(\ref{stringKasner}) implies that $p_i = p 
=\pm (D+d-1)^{-1/2}$ and one of the bounds in Eq.~(\ref{kasnerbound}) 
is therefore saturated. If we consider the positive root solution and 
take the dual under the transformation (\ref{atominusa}) for 
$d$ of the scale factors, one can verify from Eq.~(\ref{KtoK}) 
that the $(D+d)$--dimensional dilaton field 
is mapped onto a constant value by this symmetry transformation 
when the dimensions of the spacetime satisfy
the relationship~\cite{Pollock89,Pollock92}
\begin{equation}
\label{pollock}
(D-1-d)^2=D-1+d
\end{equation}
Solutions to Eq.~(\ref{pollock}) are $(D=4,d=1)$, $(D=4,d=6)$ and
$(D=11,d=15)$.  The second case implies that an anisotropic cosmology
with fixed dilaton field and with three spatial dimensions expanding
and six contracting is dual to the ten--dimensional isotropic
solution. We shall see in our discussion of the pre--big scenario in
Section~\ref{Section10} era that this particular case can lead to a
scale invariant spectrum of primordial axion perturbations.

\paragraph{S-duality:}
The $D$-dimensional compactified Einstein action (\ref{SDEinstein}) is
also manifestly invariant under the transformation
\begin{equation}
\varphi \to -\varphi \ ,
\end{equation}
while the metric, $\tilde{g}_{\mu\nu}$, and moduli fields, $\alpha_a$,
remain invariant. This is an example of an S--duality 
mapping the strongly coupled regime onto the weakly coupled, 
and vice--versa. 
This does not leave the string metric invariant and
Eq.~(\ref{DEinstein}) implies that 
\begin{equation}
g_{\mu\nu} \to e^{-4\varphi/(D-2)} \, g_{\mu\nu} \ .
\end{equation}
In terms of the Kasner indices we have that $\tilde{{\cal K}}\to
-\tilde{{\cal K}}$, while $\tilde{p_i}$ and $\tilde{B}$ remain fixed.
This leads to a non-trivial transformation in terms of the original
string frame indices:
\begin{equation}
p_i \to {(D-2-K)p_i-K \over D-2-2K} \, \quad
K \to {-(D-2)K \over D-2-2K} \ , \qquad
B \to {(D-2)B \over D-2-2K} \ .
\end{equation}

\paragraph{2/5-transformation:}
The 2/5 transformation is a symmetry of M-theory, whose low-energy
effective action contains the Einstein action presented in
equation~(\ref{EinsteinDd1}) with $D+d+1=11$. Thus the higher
dimensional cosmological solutions presented in
equation~(\ref{Einsteinpi}) correspond to low-energy solutions in
M-theory. When M-theory is compactified on a 3-torus it is invariant
under the 2/5 symmetry which exchanges the M2-brane with the M5-brane
wrapped around the 3-torus~\cite{Aharony96}.

M-theory can be related to type IIA string theory with $D+d=10$ by
compactifying the 11th dimension on a circle, as discussed in
Section~\ref{actionSection}, where the string dilaton
$\phi=3\bar\alpha_{10}$. These compactified string theory solutions
also possess T-duality (scale-factor duality). In particular, the
dilaton-vacuum solutions compactified on a 2-torus (i.e. $D=8$ and
$d=2$) are invariant under the scale-factor duality
$\alpha_8\to-\alpha_8$ and $\alpha_9\to-\alpha_9$. Combined with a
permutation of $\alpha_8$ and $\alpha_9$ this corresponds to the 2/5
M-theory transformation.

The 2/5 symmetry yields a transformation of the Kasner indices
\begin{equation}
p_8 \to -p_9 \ , \qquad 
p_9 \to -p_8 \ ,
\end{equation}
while $B$ and ${\cal K}$ remain invariant in the $D=8$ dimensional
dilaton-moduli-vacuum solution. This implies, from Eq.~(\ref{KtoK}),
\begin{equation}
K\to K - 2p_8 -2p_9 \ ,
\end{equation}
in the $D+d=10$ dimensional dilaton-vacuum solutions. 

In terms of the original $D+d+1=11$ dimensional Einstein vacuum
solutions we therefore have
\begin{equation}
\bar\alpha_I \to \bar\alpha_I + {1\over3} \bar\sigma \ \ \forall\ I<8
\quad , \quad
\bar\alpha_I \to \bar\alpha_I - {2\over3} \bar\sigma \ \ \forall\
I\geq8 \ ,
\end{equation}
where $\bar\sigma=\bar\alpha_8+\bar\alpha_9+\bar\alpha_{10}$.  
The transformation is performed at fixed time $t$ in the dimensionally
reduced dilaton-gravity theory, rather than at fixed time $\bar{t}$ in
the eleven dimensional Einstein frame, and this leads to the
transformation of the Kasner indices
\footnote
{As a result our transformation for the Kasner indices differs from
that presented in Ref.~\cite{BanFisMot99} but agrees with
Ref.~\cite{FeiVaz99}.} 
\begin{equation}
\bar{p}_I \to {3\bar{p}_I + \bar{s} \over 3+\bar{s}}
\ \ \forall\ I<8
 \quad , \quad
\bar{p}_I \to {3\bar{p}_I - 2\bar{s} \over 3+\bar{s}}
\ \ \forall\ I\geq8
\end{equation}
where $\bar{s}=\bar{p}_8+\bar{p}_9+\bar{p}_{10}$.


\subsection{Cosmological solutions with form-fields}

\label{kasnerwithforms}
\label{Section6.5}

We now extend our review of higher-dimensional cosmological solutions
to consider the dynamical role of the form-fields that are also
present in low-energy effective
actions~\cite{Tseytlin92,GolPer94,CopLahWan94,Kaloper97,LuMukPop96,LuMukPop97,LuMahMuk98,BroGreIva97,LukOvrWal97a,LukOvrWal97b,PopSch97,BehFor94a,Forste98}.
In particular, we consider an additional contribution to the
$D$-dimensional effective action Eq.~(\ref{SDstring}) given by
\begin{equation}
\label{Haction}
\Delta S= - {1\over2n!} \int d^Dx \, \sqrt{-g} \, e^\Phi \, H^2
\end{equation}
where $H$ represents an $n$--form field strength derived from an $(n-1)$-form
potential, i.e.,  $H_{\mu\nu\ldots}\equiv n\partial_{[\mu}B_{\nu\ldots]}$.
Here $\Phi$ represents some scalar 
function of the dilaton and moduli fields.
The field equation for $H$ derived from the
action (\ref{Haction}) is given by 
\begin{equation}
\label{Hfieldeom}
\nabla^\mu \left( e^\Phi H_{\mu\nu\ldots} \right) = 0 \ .
\end{equation}
When written in terms of the $D$-dimensional Einstein metric, 
this additional term in the action takes the form
\begin{equation}
\label{HEinsteinAction}
\Delta S= - {1\over2n!} \int d^Dx \, \sqrt{-\tilde{g}} \, e^{\tilde\Phi} \,
\tilde{H}^2 \ ,
\end{equation}
where we have employed  Eqs.~(\ref{DEinstein}), (\ref{CTrootg})
and~(\ref{CTH}) to obtain
\begin{equation}
\tilde\Phi = \Phi + {(D+2n)\varphi \over D-2}  \ .
\end{equation}

In general, a tensor field strength is not compatible with a maximally
symmetric spacetime. Indeed, there are only two values of $n$ in $D$
dimensions that are compatible with the assumption of spatial isotropy
and homogeneity.  These are given by $n=D$ and $n=D-1$, respectively.
The connection between these cosmological solutions and solitonic
$p$-brane solutions has been discussed previously
\cite{LuMukPop96,LarWil97,BroGreIva97,LukOvrWal97b,PopSch97,BehFor94a,Forste98}.
The $n=D$ case is referred to as the {\em elementary} ansatz for the
field strength and the $n=D-1$ solution as the {\em solitonic}
ansatz. We consider both cases separately in what follows and will
discuss the brane interpretation for the case $D=4$ in more detail in
Section~\ref{branesubsubsection}.

\subsubsection{Elementary Solutions}

In the elementary (or electric) ansatz, 
the antisymmetric tensor potential $B$ is assumed to be spatially homogeneous,
i.e., it is assumed to be a function only of time, 
$B =B (t)$. Since the field strength, $H$,
determines the form of the energy--momentum tensor and is derived from an 
exterior derivative of $B$, this 
assumption is only compatible with a homogeneous and
isotropic spacetime if $D=n$.  The $n$-form field strength $H$ is
then proportional to the totally antisymmetric, covariantly conserved
$n$-form $\epsilon$. 
Thus, the field equation~(\ref{Hfieldeom}) and Bianchi identity, 
$\partial_{[\mu}H_{\nu\rho\ldots]}=0$, are
automatically satisfied by
\begin{equation}
H_{\mu\nu\ldots} = Q e^{-\Phi} \epsilon_{\mu\nu\ldots} \ ,
\end{equation}
where $Q$ is an arbitrary constant.
This yields $H^2=- n!Q^2 e^{-2\Phi}$ and leads to an effective
exponential potential for the field $\Phi$.

We place our present discussion 
within the context of the $D$-dimensional Einstein frame, where
$\tilde{H}_{\mu\nu\ldots} = Q e^{-\tilde\Phi} \tilde\epsilon_{\mu\nu\ldots}$.
The field strength leads to an energy-momentum tensor of the form  [see
equation~(\ref{defTH})] 
\begin{equation}
^{(H)}\tilde{T}_\mu^{~\nu} = - {Q^2 e^{-\tilde\Phi} \over 2} 
\delta_\mu^{~\nu} 
\end{equation}
and for $\tilde\Phi=0$ [i.e., $\Phi=-(D+2n)\varphi/(D-2)$],
this is equivalent to a cosmological constant. More generally, 
although the dilaton and 
moduli fields are minimally coupled with respect to the
metric in the Einstein frame, the antisymmetric tensor field leads
to an effective potential for $\tilde\Phi$:
\begin{equation}
V(\tilde\Phi) = {Q^2\over4} e^{-\tilde\Phi} \ .
\end{equation}
Cosmological solutions containing scalar fields with an exponential
potential energy have been widely discussed in the cosmological
literature~\cite{LucMat85,Halliwell87,BurBar88,Wetterich88,CopLidWan98d}.

The effective Lagrangian for the homogeneous fields which follows from
Eqs.~(\ref{SDEinstein}) and~(\ref{HEinsteinAction}) is therefore given by 
\begin{equation}
\label{lagform}
L = \left( \prod_k e^{\tilde\alpha_k} \right) 
 \left[  \sum_i \dot{\tilde\alpha}_i^2
 - \dot{\tilde\alpha}_i\sum_j\dot{\tilde\alpha}_j 
 + {1\over D-2} \dot\varphi^2 + \sum_a \dot\alpha_a^2
 - {Q^2 \over 2} e^{-\tilde\Phi} \right] 
\end{equation}
after integration over the spatial variables, 
where a dot denotes differentiation with respect to the proper time
in the $D$-dimensional Einstein frame. 

In solving the field equations, it is helpful to introduce a 
shifted time variable in the $D$-dimensional Einstein frame:
\begin{equation}
\label{shiftedtime}
d\tilde\tau = {d\tilde{t} \over \prod_k e^{\tilde\alpha_k}}
\end{equation}
This then re-scales the Lagrangian (\ref{lagform}): 
\begin{equation}
\tilde{L} = \sum_i \tilde\alpha_i^{\prime2}
 - \tilde\alpha_i'\sum_j\tilde\alpha_j' 
 + {1\over D-2} \varphi^{\prime2} + \sum_a \alpha_a^{\prime2}
 - {Q^2 \over 2} \exp\left(-\tilde\Phi+2\sum_k \tilde\alpha_k \right) 
\ .
\end{equation}
where a prime denotes differentiation with respect to $\tilde\tau$.
The equations of motion become
\begin{eqnarray}
\tilde\alpha_i'' &=& \sum_j \tilde\alpha_j''
 - {Q^2\over2} \exp \left(
-\tilde\Phi+2\sum_j\tilde\alpha_j \right) \\
\varphi'' &=& {(D-2)Q^2 \over 4} \exp \left(
-\tilde\Phi+2\sum_j\tilde\alpha_j \right) {\partial\tilde\Phi \over
\partial\varphi} \\
\alpha_a'' &=& {Q^2 \over 4} \exp \left(
-\tilde\Phi+2\sum_j\tilde\alpha_j \right) {\partial\tilde\Phi \over
\partial\alpha_a} \ ,
\end{eqnarray}
together  with the constraint equation: 
\begin{equation}
\sum_i \tilde\alpha_i^{\prime2}
 - \tilde\alpha_i'\sum_j\tilde\alpha_j' 
 + {1\over D-2} \varphi^{\prime2} + \sum_a \alpha_a^{\prime2}
 = {Q^2 \over 2} \exp\left(-\tilde\Phi+2\sum_k \tilde\alpha_k \right) 
 \ .
\end{equation}

For a linear function of the form 
\begin{equation}
\label{linphi}
\tilde\Phi = \lambda_\varphi \varphi + \sum_a \lambda_a \alpha_a \ ,
\end{equation}
we can define
\begin{equation}
X = \tilde\Phi -2 \sum_j \tilde\alpha_j
 = \lambda_\varphi \varphi + \sum_a \lambda_a \alpha_a -2 \sum_j
 \tilde\alpha_j 
\end{equation}
so that
\begin{equation}
\label{Liouville}
X'' = C Q^2 e^{-X}
\end{equation}
where
\begin{equation}
C = {D-2\over4} \lambda_\varphi^2 + {1\over4}\sum_a \lambda_a^2
 - {D-1 \over D-2} \ .
\end{equation}
The first integral of Eq. (\ref{Liouville}) yields
\begin{equation}
{1\over2} X'^2 + C Q^2 e^{-X} = E \ .
\end{equation}
which is the energy of a 1-D system with potential energy
$CQ^2e^{-X}$.  The character of the solution depends on the sign of $C$
and $E$. Integrating again we obtain
\begin{equation}
\label{eX}
e^X =
\left\{
\begin{array}{cc}
\left|{CQ^2\over E}\right| \cosh^2 \left( \sqrt{{|E|\over2}}\tilde\tau
\right) & C>0,\ E>0 \\
\exp\left( \sqrt{2E} \tilde\tau \right) & C=0,\ E>0 \\
\left|{CQ^2\over E}\right| \sinh^2 \left( \sqrt{{|E|\over2}}\tilde\tau
\right) & C<0,\ E>0 \\
{|CQ^2|\over 2} \tilde\tau & C<0,\ E=0 \\
\left|{CQ^2\over E}\right| \sin^2 \left( \sqrt{{|E|\over2}}\tilde\tau
\right) & C<0,\ E<0 \\
\end{array}
\right. \ .
\end{equation}

\begin{itemize}
\item
For $C>0$ we require $E>0$ and there is a minimum allowed value of
$X_{\rm min}=\ln(CQ^2/E)$. Asymptotically, at early or late times we have
$X\to+\infty$ and we recover the vacuum solutions
$X\propto\sqrt{2E}|\tilde\tau|$. The form field dynamically
interpolates between two vacuum solutions whose Kasner indices are
related by a reflection in the line $X=X_{\rm min}$. 
\item
For $C<0$ the potential is unbounded from below and there is a runaway
solution where $X\to-\infty$ and the general solution approaches the
$E=0$ solution with $X \propto \ln\tau$. If $E<0$ there is a maximum
value $X_{\rm max}=\ln(CQ^2/E)$, but for $E\geq0$ $X$ is unbounded
from above as well as from below.
\end{itemize}

\subsubsection{Solitonic solutions}
\label{solitonic}

An $n$-form field strength is dual to a $(D-n)$-form field
strength in $D$ dimensions. This implies 
that it is also possible to have a homogeneous and
isotropic spacetime when $D=n+1$ if the field strength $H$ is dual to
the gradient of a homogeneous scalar field (see
Section~\ref{PoincareDual}). This choice leads to 
the class of solitonic (or magnetic) cosmological form field 
solutions. 

The field equation~(\ref{Hfieldeom}) is
satisfied by
\begin{equation}
H_{\mu\nu\ldots} =
 e^{-\Phi} \epsilon_{\lambda\mu\nu\ldots} \nabla^\lambda \sigma \ ,
\end{equation}
where $\epsilon_{\lambda\mu\nu\ldots}$ is the covariantly-conserved
volume-form in the $D$-dimensional spacetime.  In particular, the
effective energy-momentum tensor for the form field is the same as
that for a massless scalar field, $\sigma$, coupled to $\Phi$, with
an energy--momentum tensor given by 
\begin{equation}
^{(H)}T_\mu^\nu = e^{-\Phi} \left[ \nabla_\mu\sigma \nabla^\nu\sigma
 - {1\over2} g_\mu^\nu (\nabla\sigma)^2 \right] \ .
\end{equation}
Thus,  a homogeneous field, $\sigma(t)$, 
behaves as a stiff fluid in the
$(n+1)$-dimensional spacetime with an isotropic pressure equal to the
energy density $\rho_{(H)}=e^{-\Phi} \dot\sigma^2/2$.

The Bianchi identity, $\partial_{[\mu}H_{\nu\rho\ldots]}=0$, requires
\begin{equation}
\nabla_\mu ( e^{-\Phi} \nabla^\mu \sigma ) = 0 \ .
\end{equation}
and for a homogeneous scalar field in a $D$-dimensional toroidal spacetime
(with a line element given in equation~(\ref{Dds})), this reduces to 
the constraint
\begin{equation}
\dot\sigma = P e^{\Phi- \sum_i\alpha_i}\ .
\end{equation}
where $P$ is an 
arbitrary constant. Hence, we obtain $H^2=P^2n!e^{-2\sum_i\alpha_i}$.

In the $D$-dimensional Einstein frame we have
$\tilde{H}^2=n! P^2 e^{-2\sum_i\tilde\alpha_i}$ and the
effective Lagrangian for the other homogeneous fields becomes
\begin{equation}
\tilde{L} = \sum_i \tilde\alpha_i^{\prime2}
 - \tilde\alpha_i'\sum_j\tilde\alpha_j' 
 + {1\over D-2} \varphi^{\prime2} + \sum_a \alpha_a^{\prime2}
 - {P^2 \over 2} e^{\tilde\Phi}
\end{equation}
where a prime denotes differentiation 
with respect to the shifted time variable
$\tilde\tau$, defined in equation~(\ref{shiftedtime}).
The equations of motion become
\begin{eqnarray}
\tilde\alpha_i'' &=& \sum_j \tilde\alpha_j''\\
\varphi'' &=&
 - {(D-2)P^2 \over 4} e^{\tilde\Phi} {\partial\tilde\Phi \over \partial\varphi} \\
\alpha_a'' &=&
 - {P^2 \over 4} e^{\tilde\Phi} {\partial\tilde\Phi \over \partial\alpha_a} \ ,
\end{eqnarray}
and the Friedmann constraint equation is given by 
\begin{equation}
\sum_i \tilde\alpha_i^{\prime2}
 - \tilde\alpha_i'\sum_j\tilde\alpha_j' 
 + {1\over D-2} \varphi^{\prime2} + \sum_a \alpha_a^{\prime2}
 = {P^2 \over 2} e^{\tilde\Phi} 
 \ .
\end{equation}

For a linear function $\tilde\Phi$, as given in
equation~(\ref{linphi}), we have the simple equation of motion:
\begin{equation}
\tilde\Phi'' = \bar{C} P^2 e^{\tilde\Phi}
\end{equation}
with the first integral
\begin{equation}
{1\over2} \tilde\Phi^{\prime2} + \bar{C} P^2 e^{\tilde\Phi} = \bar{E}
\end{equation}
where
\begin{equation}
\bar{C} = {D-2 \over4}\lambda_\varphi^2 + {1\over4} \lambda_a^2
\end{equation}

The solution for $\tilde\Phi$ is analogous to the elementary solution
presented in equation~(\ref{eX}) for $-X$ in terms of the 
constant $C$. The crucial difference, however, is that
$\bar{C}$ is positive definite in the solitonic case, and hence $\bar{E}$
must be as well. Consequently, the only physically relevant 
solution is 
\begin{equation}
\label{barCequation}
e^{-\tilde\Phi} =
{\bar{C}P^2\over \bar{E}}
 \cosh^2 \left( \sqrt{{\bar{E}\over2}}\tilde\tau \right)
\ .
\end{equation}
We conclude from Eq. (\ref{barCequation}), therefore, that the solitonic 
form field enforces an {\em upper} bound on the coupling
function $\tilde\Phi\leq \ln(\bar{E}/\bar{C}P^2)$. This leads to a
solution that interpolates between two vacuum solutions 
that are related by an
S-duality transformation, $\tilde\Phi\to-\tilde\Phi$. Specifically, we have 
\begin{equation}
\tilde\Phi \to
\left\{
\begin{array}{c}
\sqrt{2{\bar{E}}} \, \tilde\tau \qquad {\rm as}\ \tau\to-\infty \\
- \sqrt{2{\bar{E}}} \, \tilde\tau \qquad {\rm as}\ \tau\to+\infty
\end{array}
\right.
\ .
\end{equation}
This generic feature proves central in our discussions 
of cosmological solutions including 3-form field strengths in
Sections~\ref{Section7} and~\ref{Section8}.

\section{Four--dimensional NS--NS Cosmology}

\label{Section7}

\setcounter{equation}{0}

\def\theequation{\thesection.\arabic{equation}}

In this Section we consider in more detail the cosmological solutions
to the four-dimensional NS--NS effective action (\ref{reduced1}) or,
equivalently, (\ref{reduced2}).  
The spatially flat homogeneous models correspond to particular cases
of the higher-dimensional toroidal Kasner solutions presented in
Section~\ref{Section6}. Here we extend the discussion to include
spatial curvature and inhomogeneity. 

Recall that the NS--NS sector is common to both the heterotic and type
II string theories and is comprised of the dilaton, graviton and
antisymmetric two--form potential. In this Section 
we also include the effects of the breathing mode of the internal
space, $\beta$, in the actions (\ref{reduced1})
and (\ref{reduced2}). This degree of freedom represents a modulus field.

The field equations derived by varying the action (\ref{reduced1}) 
with respect to the massless excitations $g_{\mu\nu}$, 
$B_{\mu\nu}$, $\beta$ and $\varphi$, respectively, are given by 
\begin{eqnarray}
\label{vary1}
R_{\mu\nu} -\frac{1}{2} Rg_{\mu\nu} =
\frac{1}{12} \left( 3H_{\mu\lambda\kappa}{H_{\nu}}^{\lambda\kappa}
-\frac{1}{2}H^2g_{\mu\nu} \right) +
\frac{1}{2} \left( {g_{\mu}}^{\lambda}
{g_{\nu}}^{\kappa} -\frac{1}{2} g_{\mu\nu} g^{\lambda\kappa} 
\right)\nabla_{\lambda}\beta \nabla_{\kappa} \beta \nonumber \\
-\frac{1}{2} g_{\mu\nu} \left( \nabla \varphi \right)^2 +\left( 
g_{\mu\nu}g^{\lambda\kappa} - {g_{\mu}}^{\lambda}{g_{\nu}}^{\kappa} 
\right) \nabla_{\lambda}\nabla_{\kappa} \varphi =0 \\
\label{vary2}
\nabla_{\mu} \left( e^{-\varphi} H^{\mu\nu\lambda} \right) =0 \\
\label{vary3}
\nabla_{\mu} \left( e^{-\varphi} \nabla^{\mu} \beta \right) =0 \\
\label{vary4}
2\Box \varphi =-R +\left( \nabla \varphi \right)^2 +
\frac{1}{2} \left( \nabla \beta 
\right)^2 +\frac{1}{12} H^2 
\end{eqnarray}
where $H^2 \equiv H_{\mu\nu\lambda}H^{\mu\nu\lambda}$. 
We remark that a consistent solution to these field equations 
is for the internal space to be static, $\nabla_{\mu} \beta =0$. 

The corresponding 
field equations derived from the dual action (\ref{reduced2}) take the form
\begin{eqnarray}
\label{varytilde1}
\tilde{R}_{\mu\nu} -\frac{1}{2}\tilde{R} \tilde{g}_{\mu\nu} =
\frac{1}{2} \left( \tilde{\nabla}_{\mu} \varphi \tilde{\nabla}_{\nu} 
\varphi -\frac{1}{2} \tilde{g}_{\mu\nu} \left( \tilde{\nabla}
\varphi \right)^2 \right) +\frac{1}{2} \left( 
\tilde{\nabla}_{\mu} \beta \tilde{\nabla}_{\nu} \beta -\frac{1}{2} 
\tilde{g}_{\mu\nu} \left( \tilde{\nabla} \beta \right)^2 \right) 
\nonumber \\
+ \frac{1}{2} e^{2\varphi} \left( \tilde{\nabla}_{\mu}\sigma 
\tilde{\nabla}_{\nu} \sigma -\frac{1}{2} \tilde{g}_{\mu\nu} 
\left( \tilde{\nabla} \sigma \right)^2  \right) \\
\label{varytilde2}
\tilde{\Box} \sigma +2\tilde{\nabla}^{\mu}\varphi \tilde{\nabla}_{\mu} 
\sigma =0   \\
\label{varytilde3}
\tilde{\Box} \beta =0 \\
\label{varytilde4}
\tilde{\Box} \varphi -e^{2\varphi} \left( \tilde{\nabla} \sigma 
\right)^2 =0
\end{eqnarray}
in the Einstein frame (\ref{tildeg}). 

\subsection{Spatially Homogeneous and Isotropic Cosmologies}

\label{Section7.1}

All homogeneous and isotropic external four--dimensional spacetimes can
be described by the Friedmann-Robertson-Walker (FRW) metric. We 
write the general  line element in the string frame as 
\begin{equation}
\label{FRW}
ds_4^2 = a^2(\eta) \left\{ -d\eta^2 + d\Omega_\kappa^2 \right\}
 \, ,
\end{equation}
where $a(\eta)$ is the scale factor of the universe, $\eta$ 
represents conformal time and
$d\Omega_\kappa^2$ is the line element on a 3-space with constant
curvature $\kappa$:
\begin{equation}
\label{constantline}
d\Omega^2_{\kappa} 
=d\psi^2 +\left( \frac{\sin \sqrt{\kappa}\psi}{\sqrt{\kappa}}
\right)^2 \left( d\theta^2 +\sin^2  \theta d\varphi^2 \right)
\end{equation}
To be compatible with a homogeneous and isotropic metric,
all fields, including 
the pseudo--scalar axion field, must be spatially homogeneous.

\subsubsection{Dilaton-moduli-vacuum solutions}

The models with vanishing form fields, but time-dependent
dilaton and moduli fields, are known as 
{\em dilaton-moduli-vacuum} solutions. 
In the Einstein--frame, these solutions may be interpreted 
as FRW cosmologies for a stiff perfect fluid, 
where the speed of sound equals the speed of light. 
The dilaton and moduli fields behave collectively as a massless, 
minimally coupled scalar field. 
The scale factor in the Einstein frame is given
by~\cite{MimWan95a,CopLahWan94} 
\begin{equation}
\label{einsteinscalefactor}
\tilde{a} =\tilde{a}_* \sqrt{\frac{\tau}{1+\kappa \tau^2}} 
\end{equation}
where $\tilde{a} \equiv e^{-\varphi /2}a$, $\tilde{a}_*$ is 
a constant and 
we have defined a new time variable: 
\begin{equation}
\label{wanmimtime}
\tau \equiv \left\{
\begin{array}{ll}
\kappa^{-1/2} |\tan(\kappa^{1/2}\eta)| & {\rm for}\ \kappa>0 \\
|\eta|  & {\rm for}\ \kappa=0 \\
|\kappa|^{-1/2} |\tanh(|\kappa|^{1/2}\eta)| & {\rm for}\ \kappa<0
\end{array}
\right.
\ .
\end{equation}
The time coordinate $\tau$ diverges at both early and late times in
models which have $\kappa\geq0$, but $\tau\to|\kappa|^{-1/2}$ in
negatively curved models.
There is a curvature singularity at $\eta =0$ with $\tilde{a} =0$ 
and the model expands away from it 
for $\eta >0$ or collapses towards it for $\eta < 0$.
The expanding, closed models recollapse at 
$\eta = \pm \pi /2$ and there are no bouncing solutions in 
this frame. 

The corresponding string frame scale factor and 
dilaton and modulus fields are given by~\cite{CopLahWan94}
\begin{eqnarray}
\label{dila}
a & = & a_* \sqrt{ {\tau^{1+\sqrt{3}\cos\xi_*} \over 1+\kappa\tau^2} } \, , 
\\  
\label{dilphi}
e^\varphi & = & e^{\varphi_*} \tau^{\sqrt{3}\cos\xi_*}
 \, ,\\
\label{dilbeta}
e^\beta & = & e^{\beta_*} \tau^{\sqrt{3}\sin\xi_*} \, 
\end{eqnarray}
These dilaton-moduli-vacuum solutions follow straight line
trajectories in the field space spanned by the dilaton, $\varphi$, and
modulus, $\beta$. Such an interpretation of these solutions is
important when discussing the more general cosmologies containing
non--trivial NS--NS and RR form fields in the type IIB string
cosmology (see Section \ref{Section8.2}).  The integration constant
$\xi_*$ determines the rate of change of the effective dilaton
relative to the volume of the internal dimensions.  Figures
\ref{figflatfrw1}--\ref{figopenfrw3} show the dilaton-vacuum solutions
in flat, closed and open FRW models when stable compactification has
occurred, so that the volume of the internal space is fixed, we have
$\xi_*\ {\rm mod}\ \pi=0$.

\begin{figure}
\begin{center}
\leavevmode\epsfysize=5.5cm \epsfbox{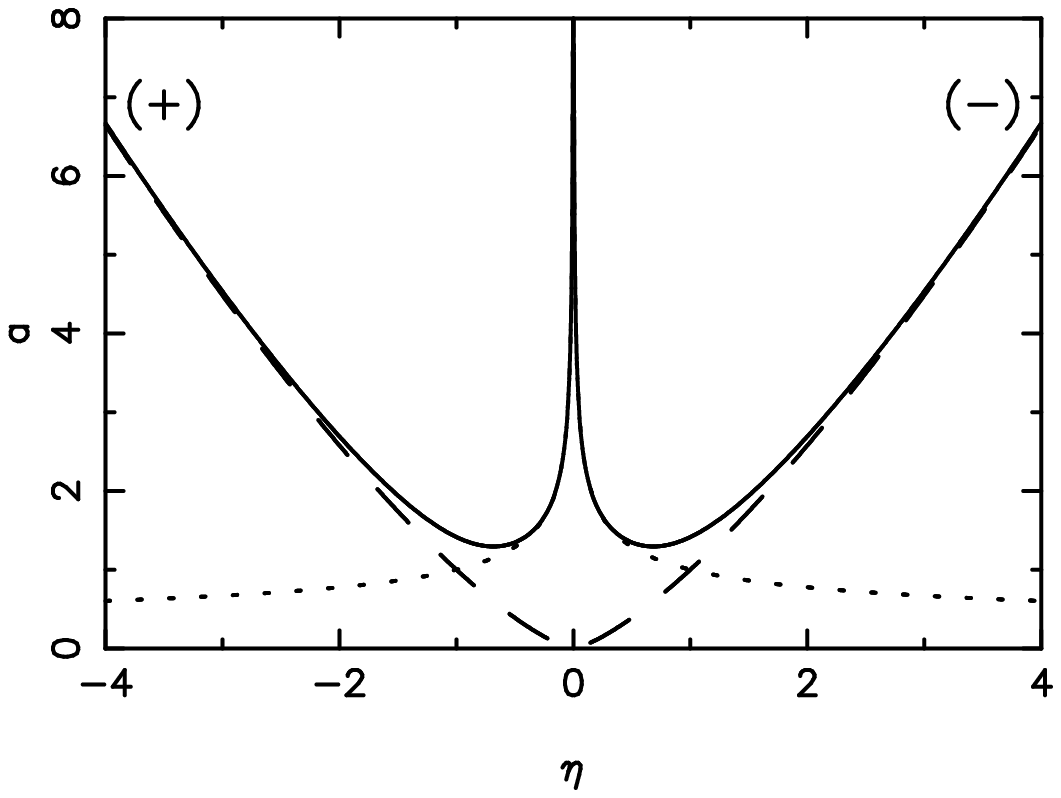}\\ 
\end{center}
\caption[Scale factor in flat FRW]
{String frame scale factor, $a$, as a function of conformal time, $\eta$, for
flat $\kappa=0$ FRW cosmology in dilaton-vacuum solution in
Eq.~(\ref{dila}) 
with $\xi_*=0$ (dashed-line), $\xi_*=\pi$ (dotted
line) and dilaton-axion solution in Eq.~(\ref{axia}) 
with $r=\sqrt{3}$
(solid line). The $(+)$ and $(-)$ branches are defined in 
Section~\ref{Section9.1}.
}
\label{figflatfrw1}
\end{figure}

\begin{figure}
\begin{center}
\leavevmode\epsfysize=5.5cm \epsfbox{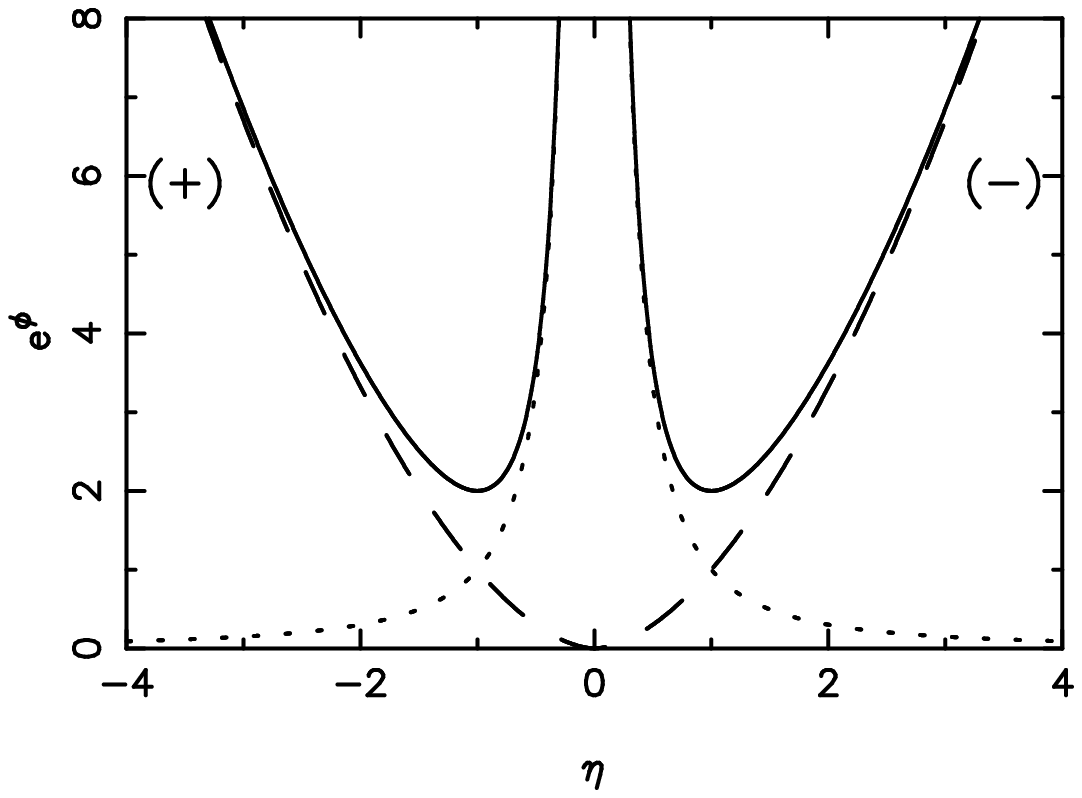}\\ 
\end{center}
\caption[Dilaton in flat dilaton-vacuum and dilaton-axion solutions]
{Dilaton, $e^\varphi$, as a function of conformal time, $\eta$, for
flat $\kappa=0$ FRW cosmology in dilaton-vacuum solution in
Eq.~(\ref{dilphi}) with $\xi_*=0$ (dashed-line), $\xi_*=\pi$ (dotted
line) and dilaton-axion solution in Eq.~(\ref{axiphi}) with $r=\sqrt{3}$
(solid line). The $(+)$ and $(-)$ branches are defined in
Section~\ref{Section9.1}.} 
\label{figflatfrw2}
\end{figure}

\begin{figure}
\begin{center}
\leavevmode\epsfysize=5.5cm \epsfbox{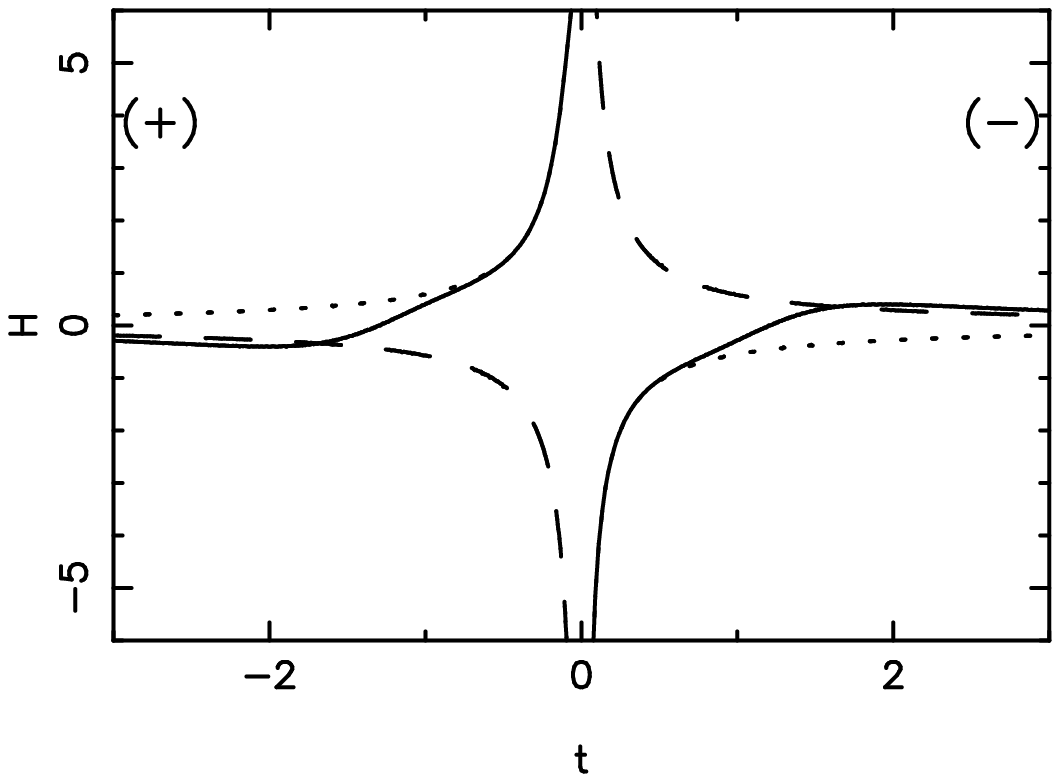}\\ 
\end{center}
\caption[Hubble rate in flat FRW]
{Hubble rate in the string frame, $H$, as a function of proper cosmic
time, $t$, for flat $\kappa=0$ FRW cosmology in dilaton-vacuum
solution in Eq.~(\ref{dila}) with $\xi_*=0$ (dashed-line), $\xi_*=\pi$
(dotted line) and dilaton-axion solution in Eq.~(\ref{axia}) with
$r=\sqrt{3}$ (solid line). The $(+)$ and $(-)$ branches are defined in
Section~\ref{Section9.1}.  }
\label{figflatfrw3}
\end{figure}

\begin{figure}
\begin{center}
\leavevmode\epsfysize=5.5cm \epsfbox{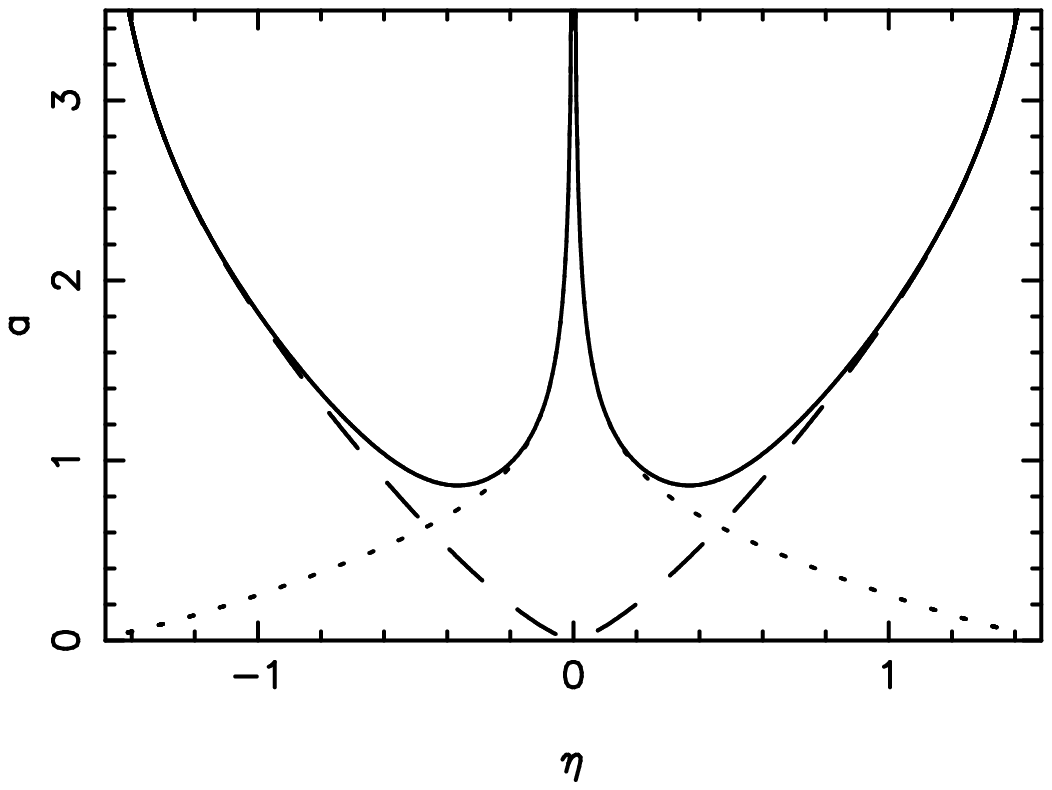}\\ 
\end{center}
\caption[Scale factor in closed FRW]
{String frame scale factor, $a$, as a function of conformal time, $\eta$, for
closed $\kappa=+1$ FRW cosmology in dilaton-vacuum solution in
Eq.~(\ref{dila}) 
with $\xi_*=0$ (dashed-line), $\xi_*=\pi$ (dotted
line) and dilaton-axion solution in Eq.~(\ref{axia}) 
with $r=\sqrt{3}$ (solid line).
}
\label{figclosedfrw1}
\end{figure}

\begin{figure}
\begin{center}
\leavevmode\epsfysize=5.5cm \epsfbox{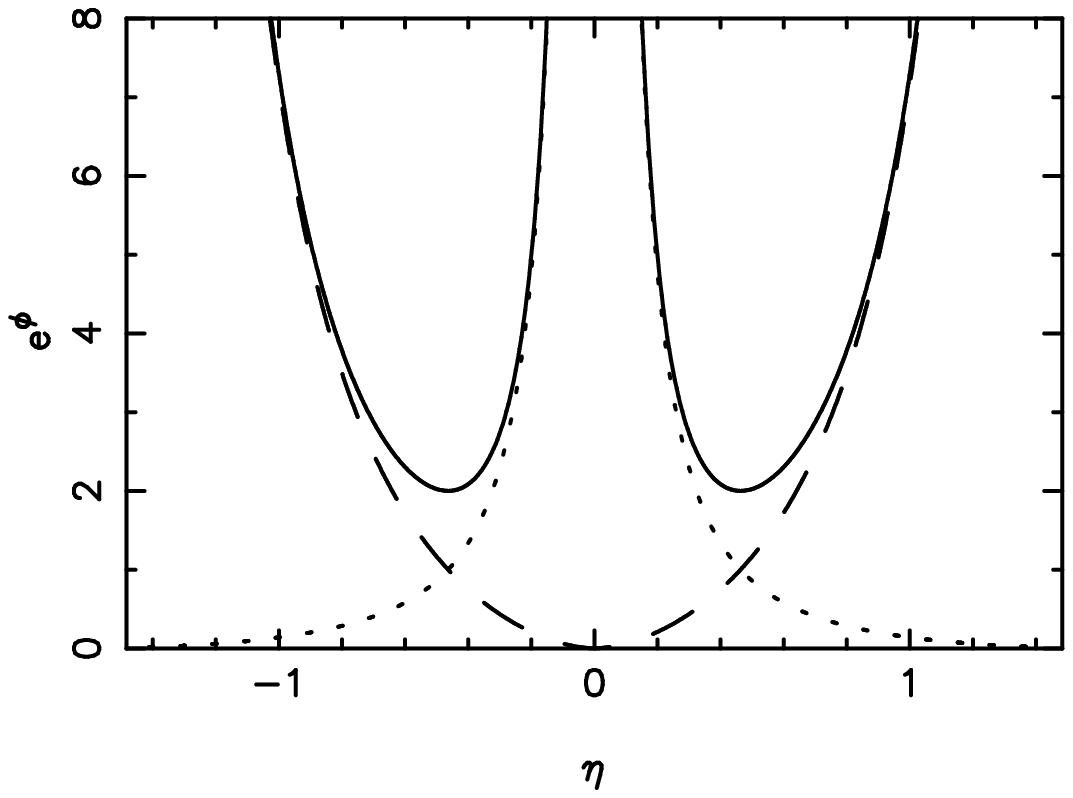}\\ 
\end{center}
\caption[Dilaton in closed dilaton-vacuum and dilaton-axion solutions]
{Dilaton, $e^\varphi$, as a function of conformal time, $\eta$, for
closed $\kappa=+1$ FRW cosmology in dilaton-vacuum solution in
Eq.~(\ref{dilphi}) with $\xi_*=0$ (dashed-line), $\xi_*=\pi$ (dotted
line) and dilaton-axion solution in Eq.~(\ref{axiphi}) with $r=\sqrt{3}$
(solid line).} 
\label{figclosedfrw2}
\end{figure}

\begin{figure}
\begin{center}
\leavevmode\epsfysize=5.5cm \epsfbox{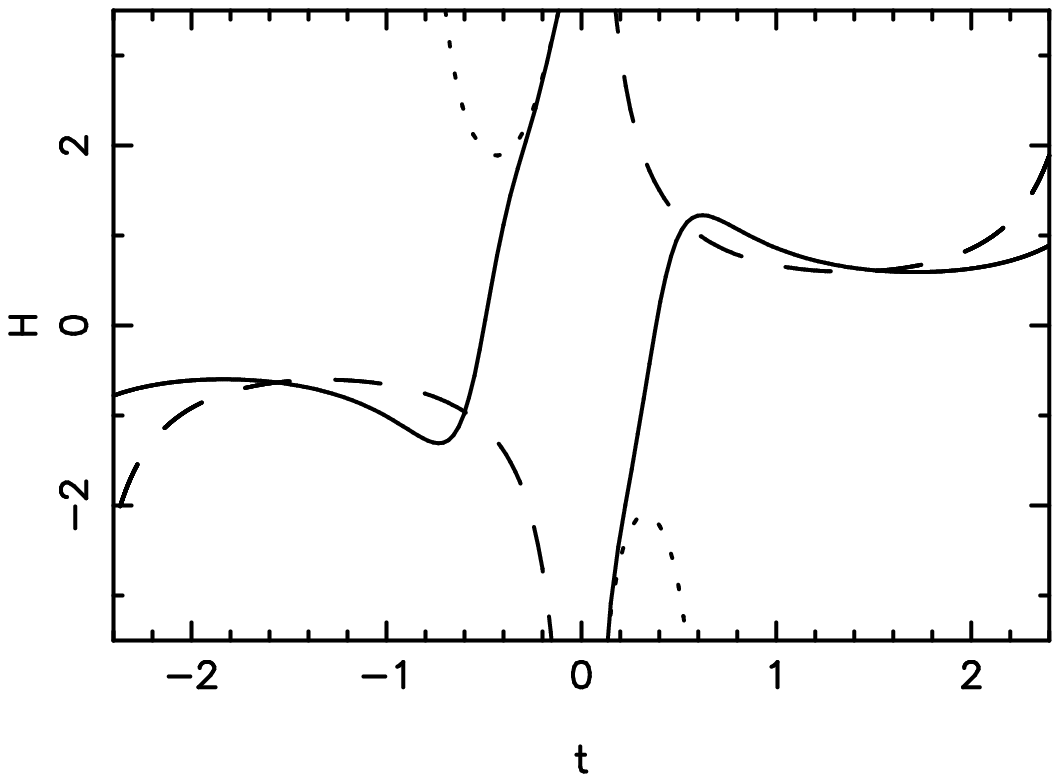}\\ 
\end{center}
\caption[Hubble rate in closed FRW]
{Hubble rate in the string frame, $H$, as a function of proper cosmic
time, $t$, for closed $\kappa=+1$ FRW cosmology in dilaton-vacuum
solution in Eq.~(\ref{dila}) with $\xi_*=0$ (dashed-line), $\xi_*=\pi$
(dotted line) and dilaton-axion solution in Eq.~(\ref{axia}) with
$r=\sqrt{3}$ (solid line).}
\label{figclosedfrw3}
\end{figure}

Note that in an expanding open FRW model the (negative) spatial
curvature comes to dominate the expansion. The dilaton approaches a
fixed value and the solution approaches flat Minkowski vacuum in Milne
coordinates at early or late times as $|\eta|\to\infty$.

\begin{figure}
\begin{center}
\leavevmode\epsfysize=5.5cm \epsfbox{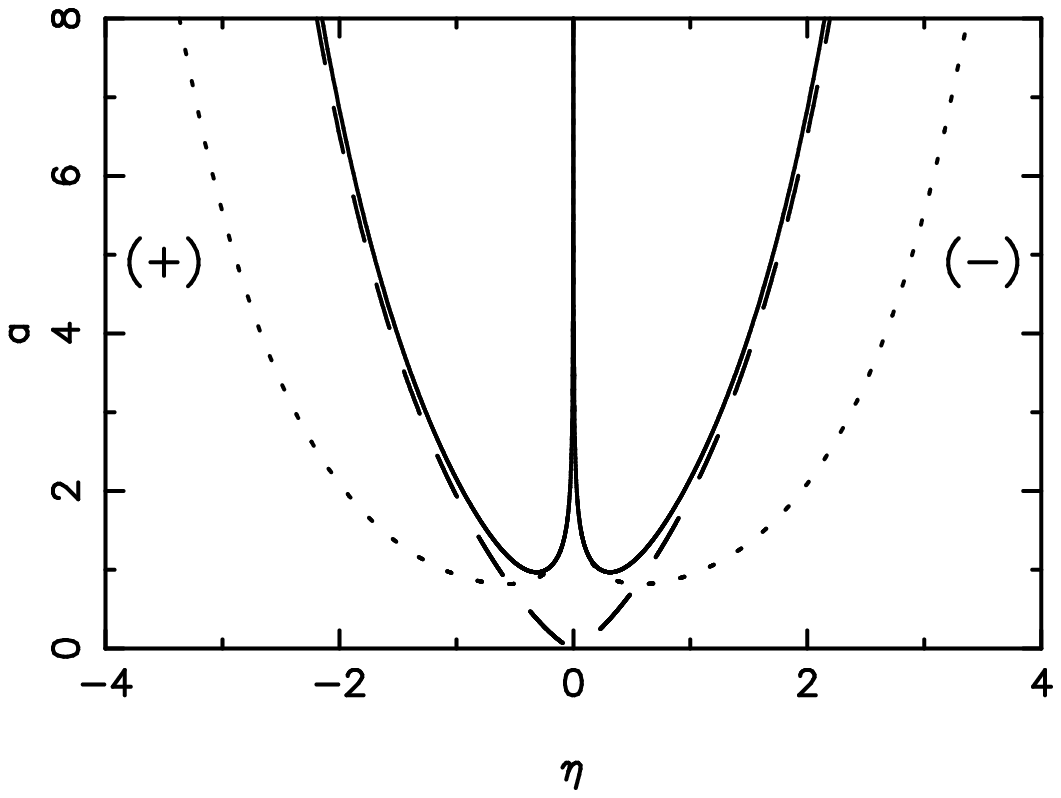}\\ 
\end{center}
\caption[Scale factor in open FRW]
{String frame scale factor, $a$, as a function of conformal time, $\eta$, for
open $\kappa=-1$ FRW cosmology in dilaton-vacuum solution in
Eq.~(\ref{dila}) 
with $\xi_*=0$ (dashed-line), $\xi_*=\pi$ (dotted
line) and dilaton-axion solution in Eq.~(\ref{axia}) 
with $r=\sqrt{3}$
(solid line). The $(+)$ and $(-)$ branches are defined in
Section~\ref{Section9.1}.}
\label{figopenfrw1}
\end{figure}

\begin{figure}
\begin{center}
\leavevmode\epsfysize=5.5cm \epsfbox{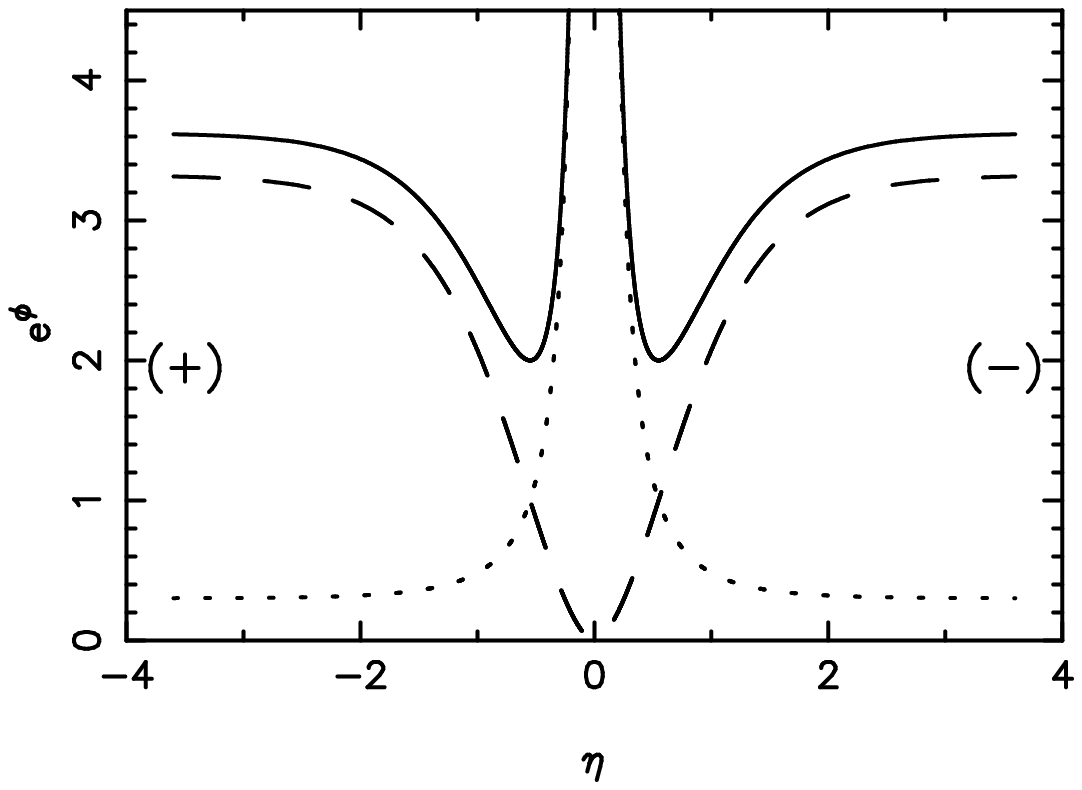}\\ 
\end{center}
\caption[Dilaton in open dilaton-vacuum and dilaton-axion solutions]
{Dilaton, $e^\varphi$, as a function of conformal time, $\eta$, for
open $\kappa=-1$ FRW cosmology in dilaton-vacuum solution in
Eq.~(\ref{dilphi}) with $\xi_*=0$ (dashed-line), $\xi_*=\pi$ (dotted
line) and dilaton-axion solution in Eq.~(\ref{axiphi}) with $r=\sqrt{3}$
(solid line). The $(+)$ and $(-)$ branches are defined in
Section~\ref{Section9.1}.} 
\label{figopenfrw2}
\end{figure}

\begin{figure}
\begin{center}
\leavevmode\epsfysize=5.5cm \epsfbox{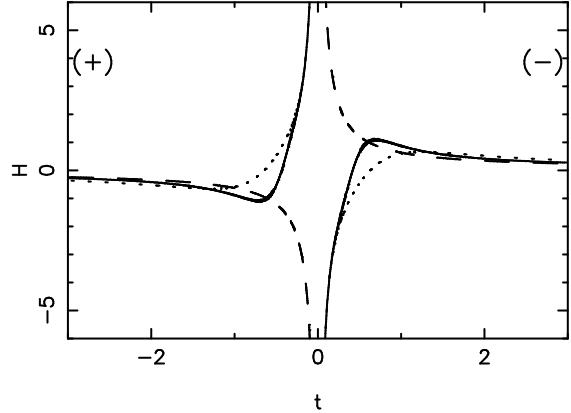}\\ 
\end{center}
\caption[Hubble rate in open FRW]
{Hubble rate in the string frame, $H$, as a function of proper cosmic
time, $t$, for open $\kappa=-1$ FRW cosmology in dilaton-vacuum
solution in Eq.~(\ref{dila}) with $\xi_*=0$ (dashed-line), $\xi_*=\pi$
(dotted line) and dilaton-axion solution in Eq.~(\ref{axia}) with
$r=\sqrt{3}$ (solid line). The $(+)$ and $(-)$ branches are defined in
Section~\ref{Section9.1}.  }
\label{figopenfrw3}
\end{figure}

We will discuss the spatially flat ($\kappa=0$) FRW models in the
context of the proposed pre--big bang scenario in
Section~\ref{Section9}.

\subsubsection{Dilaton-moduli-axion solutions}

The cosmologies containing a non--trivial 
axion field are referred to as 
{\em dilaton--moduli--axion} cosmologies. These  
may now be generated 
immediately due to the global ${\rm SL}(2,R)$ symmetry 
of the dual action (\ref{reduced2}), as discussed in Section \ref{Section4.2}. 
This symmetry exists because the dilaton and axion fields 
parametrize the ${\rm SL}(2,R)/{\rm U}(1)$ coset. 
In effect, we 
apply the ${\rm SL}(2,R)$ transformation summarized in 
Eqs. (\ref{sphi}) and (\ref{schi}) to the dilaton--moduli--vacuum 
solutions presented above. This transformation 
generates a solution with a dynamical axion field from one 
where this field is trivial. The dilaton and axion fields transform under 
a general ${\rm SL}(2,R)$ transformation as   
\begin{eqnarray}
\label{sphimodel}
e^{\bar{\varphi}} =c^2e^{-\varphi} +d^2 e^{\varphi} \\
\label{ssigmamodel}
\bar{\sigma} e^{\bar{\varphi}} =ac e^{-\varphi} 
+ bd e^{\varphi}
\end{eqnarray}
where the constants $\{ a,b,c,d \}$ satisfy $ad-bc=1$. The Einstein frame 
metric acts as a singlet under this transformation and the 
original stiff perfect fluid form of the scale factor 
(\ref{einsteinscalefactor}) therefore 
holds in this more general case. The modulus field, $\beta$, is also 
invariant. The string frame metric, 
however, is not invariant and transforms to
\begin{equation}
\label{smetricmodel}
d\bar{s}^2_{\rm string} = e^{\bar{\varphi} -\varphi} 
ds^2_{\rm string}
\end{equation}

Applying the ${\rm SL}(2,R)$ transformation to the 
dilaton--moduli--vacuum solution 
(\ref{dilphi}--\ref{dilbeta}) therefore yields~\cite{CopLahWan94}:
\begin{eqnarray}
\label{axiphi}
e^\varphi & = & {e^{\varphi_*} \over 2} \left\{
\left({\tau\over\tau_*}\right)^{-r} +
\left({\tau\over\tau_*}\right)^{r} 
\right\} \, ,\\
\label{axia}
a^2 & = & \frac{a_*^2}{2(1+\kappa \tau^2)} 
\left\{ \left({\tau\over\tau_*}\right)^{1-r}
 + \left({\tau\over\tau_*}\right)^{1+r} \right\} \, ,\\
\label{axibeta}
e^\beta & = & e^{\beta_*} \tau^{s} \, ,\\
\label{axisigma}
\sigma & = & \sigma_*
 \pm e^{-\varphi_*} \left\{ (\tau/\tau_*)^{-r} -
 (\tau/\tau_*)^{r} 
\over (\tau/\tau_*)^{-r} +
 (\tau/\tau_*)^{r} \right\} \, ,
\end{eqnarray}
where the exponents are related via
\begin{equation}
r^2+s^2=3 \ ,
\end{equation}
and without loss of generality we may take $r\geq0$.

In all cases, the dynamics of the axion field places a {\em lower}
bound on the value of the dilaton field, $\varphi\geq\varphi_*$. In
doing so, the axion smoothly interpolates between two
dilaton--moduli--vacuum solutions, where its dynamical influence
asymptotically becomes negligible.  The time--dependent axion
solutions are plotted in Figures \ref{figflatfrw1}--\ref{figNSNSsigma}
for open, closed and flat models when the modulus field is trivial
$(s=0)$. When the internal space is static, it is seen that the string
frame scale factors for all three models each exhibit a bounce.
However we still have a curvature singularity in the
Einstein frame as $\tau\to0$. In the string frame the
four--dimensional metric is non-singular as $\tau\to0$ in the
particular case $r=1$~\cite{BehFor94a}, but the dilaton and moduli
still diverge in this case leading to the breakdown of our effective
action~\cite{EasMaeWan96,KalMadOli96}.

\begin{figure}
\begin{center}
\leavevmode\epsfysize=5.5cm \epsfbox{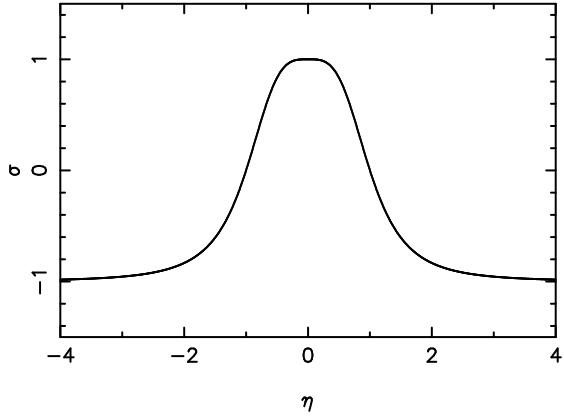}\\ 
\end{center}
\caption[Axion in flat dilaton-axion solutions]
{Axion, $\sigma$, as a function of conformal time, $\eta$, for flat
$\kappa=0$ FRW cosmology in dilaton-axion solution in Eq.~(\ref{axisigma})
with $r=\sqrt{3}$ (solid line).}
\label{figNSNSsigma}
\end{figure}

The spatially flat solutions reduce to the power law,
dilaton--moduli--vacuum solution given in
Eqs.~(\ref{dilphi}--\ref{dila}) at early and late times.  When
$\eta\to\pm\infty$ the solution approaches the vacuum solution with
$\sqrt{3}\cos \xi_* =+r$, while as $\eta\to0$ the solution approaches
the $\sqrt{3}\cos \xi_* =-r$ solution. Thus, the axion solution
interpolates between two vacuum solutions related by an S-duality
transformation $\varphi\to-\varphi$.  When the internal space is
static the scale factor in the string frame is of the form $a
\propto t^{1/\sqrt{3}}$ as $\eta\to\pm\infty$, while as $\eta\to0$ the
solution becomes $a \propto t^{-1/\sqrt{3}}$. These two vacuum
solutions are thus related by a scale factor duality that inverts the
spatial volume of the universe. We shall see in Section \ref{Section10}
that this asymptotic approach to dilaton--moduli--vacuum solutions at
early and late times leads to a particularly simple form for the
semi-classical perturbation spectra that is independent of the
intermediate evolution.

The solutions in the spatially curved models approach the flat
solutions only near the curvature singularity, $\eta =0$. At late
times, $\tau \rightarrow |\kappa |^{-1/2}$ in the open $(\kappa <0)$
models and the dilaton becomes frozen at a fixed value as the
curvature begins to dominate the evolution. Indeed, the asymptotic
form of the solution in this limit corresponds to the Milne form of
flat space. This metric plays a central role in establishing the generic 
initial conditions for the pre--big bang scenario, as we discuss in more 
detail in Section~\ref{Section9}. The closed
models also become curvature dominated at late times, but in this case
the scale factor diverges in a finite proper time, because $\tau
\rightarrow \infty$ as $\eta \rightarrow \pm \pi/2 $.

\subsubsection{Brane interpretation}

\label{branesubsubsection}

The overall dynamical effect of the axion field is negligible except near
$\tau\approx \tau_*$, when it leads to a bounce in the dilaton field.
Within the context of M--theory cosmology, it follows 
from Eq. (\ref{identifystring}) that the radius of the eleventh 
dimension is related to the dilaton by
$r_{11} \propto e^{\varphi /3}$ when the modulus 
field is fixed. This bound on the dilaton may therefore be reinterpreted 
as a lower bound on the size of the eleventh dimension.
In models where the modulus field is fixed, a bounce is also inevitable 
in the string frame scale factor.
This feature may be understood more fully by drawing on 
the analogy between these cosmological solutions and 
certain solitonic $p$--brane solutions \cite{LukOvrWal97a,LukOvrWal97b,LukOvrWal97b,PopSch97,BehFor94a,BehFor94b,LuMahMuk98,LuMukPop96,LarWil97,IvaMel98,BroGreIva97,GreIvaMel98}. 
(For a review of $p$--brane solitons,  see, e.g., 
Ref. \cite{DufKhuLu95}). In particular, cosmological solutions 
have been derived directly from black $p$--branes by 
exchanging the time and radial coordinates of the brane 
in a suitable regime \cite{BehFor94a,PopSch97,LarWil97,LuMukPop96}. This is 
analogous to the reinterpretation of the interior solution 
of the Schwarzschild black hole in terms of a cosmological model 
\cite{OppSny39}. This is possible since the time coordinate becomes spacelike 
in such a region and the radial coordinate becomes timelike. In this 
sense, therefore, the interior of a black $p$--brane may be viewed as 
a cosmological solution. 

It is more convenient in this discussion to employ the language of 
differential forms. We proceed by expressing the FRW line element (\ref{FRW}) 
in terms of an orthonormal tetrad:
\begin{equation}
ds^2 =\eta_{\alpha\beta} \lambda^{\alpha}\lambda^{\beta}
\end{equation}
where the one--forms $\lambda^{\alpha}$ are defined by
\begin{eqnarray}
\lambda^0 \equiv dt \nonumber \\
\lambda^1 \equiv a d\psi \nonumber \\
\lambda^2 \equiv a \left( \frac{\sin \sqrt{\kappa} \psi}{\sqrt{\kappa}} 
\right) d\theta \nonumber \\
\lambda^3 \equiv a \left( \frac{\sin \sqrt{\kappa} \psi}{\sqrt{\kappa}} 
\right) \sin \theta d \varphi
\end{eqnarray}
and $\eta_{\alpha\beta} \equiv {\rm diag} [-1, 1, 1, 1]$. 

The Poincar\'e dual of the three--form field 
strength is a one--form, defined in Eq. (\ref{dualtoH}). 
This may be expressed as 
\begin{equation}
^*H = \frac{1}{6} H^{\mu\nu\rho} \epsilon_{\mu\nu\rho\kappa} 
\lambda^{\kappa}
\end{equation}
Modulo an exponential function of the dilaton field, this is 
proportional to the gradient of the pseudo--scalar axion field, $\sigma$. 
Defining the one--forms $\omega^b \equiv a^{-1} \lambda^b$, 
where $b=(\psi , \theta , \varphi )$, 
the functions
\begin{eqnarray}
U \equiv \frac{1}{6} H^{abc}\epsilon_{abc0} \nonumber \\
V_b \equiv \frac{1}{6} H^{\mu\nu\rho}\epsilon_{\mu\nu\rho b} a
\end{eqnarray}
imply that 
\begin{equation}
^*H =U dt+V_b \omega^b
\end{equation}

For consistency 
with the FRW metric, we had to assume that the dilaton and axion fields 
were dependent only on time, $t$. The vector field $V_b$ 
is related to the spatial gradient of the axion and must therefore vanish. 
This implies that the time components 
of the field strength must necessarily vanish, $H^{0ab}=0$. 
As a result, a 
necessary condition on the two--form potential in a FRW background 
is that it must be independent of cosmic time. Thus, 
all NS--NS FRW string cosmologies are solitonic, in the sense discussed in 
Section \ref{Section6.5}. 

The remaining non--zero components of the form field can 
now be written as
\begin{equation}
\label{FRWHform}
H = H_{abc} \lambda^a \wedge \lambda^b \wedge \lambda^c 
\end{equation}
and this simplifies to 
\begin{equation}
H = 6 H_{\psi \theta \varphi} a^3  
\left( \frac{\sin \sqrt{\kappa} \psi}{\sqrt{\kappa}} \right)^2 \sin \theta 
d\psi \wedge d \theta \wedge d \varphi
\end{equation}
However, the volume element of a three--space of constant 
curvature is given by 
\begin{equation}
\epsilon \equiv \left( \frac{\sin \sqrt{\kappa} \psi}{\sqrt{\kappa}} 
\right)^2 \sin \theta  d\psi \wedge d \theta \wedge d \varphi 
\end{equation}
Consequently, the Bianchi identity, $dH \equiv 0$, 
implies that the field 
strength of the two--form potential must be proportional to the 
volume form of the three--space, i.e., 
$H=P \epsilon$, where $P = 6 H_{\psi \theta \varphi} a^3$
is a constant. This is precisely the type of 
behaviour exhibited 
by the NS--NS field strength in the solitonic 5--brane solution 
\cite{GibMae88,HorStr91,BehFor94a}. The constant $P$ represents 
the magnetic charge of the 
5--brane and in a cosmological context is related to the conjugate 
momentum of the pseudo--scalar axion field. 
Hence, the axion field may be interpreted in some 
sense as a membrane 
wrapped around the three--space of constant curvature. 
As the universe 
collapses, the membrane resists becoming squashed into 
a singular point and causes the universe to bounce into an expansionary 
phase. 

This concludes our analysis on the nature of the FRW cosmologies. 
We now proceed to extend our discussion to include models that are 
spatially anisotropic. 

\subsection{Spatially Homogeneous Cosmology}

\label{Section7.2}

In the FRW string models the spatial gradient of the pseudo--scalar 
axion field is constrained to be zero due to 
spatial isotropy. This condition need not necessarily  be 
imposed in some of the spatially homogeneous and anisotropic 
cosmologies, however. In this class of models, 
the components of the three--form field 
strength must depend only on time to ensure that the 
energy momentum tensor is homogeneous. 

In this Section we follow closely the notation of 
Barrow and Kunze \cite{BarKun97} and place our discussion within 
the context of the Einstein frame metric. We assume that the dilaton 
field is constant on the surfaces of homogeneity, $\varphi =\varphi (t)$.
Thus, the conformal transformation (\ref{ste}) that maps a given solution 
from the string frame to the Einstein frame, and vice--versa, does 
not change the geometry of the spacetime, i.e., a given Bianchi 
model is mapped onto the same Bianchi type.  

The four--dimensional homogeneous spacetimes are classified into 
two groups. These are the Bianchi metrics \cite{Bianchi98,RyaShe75,Wald84} 
and the Kantowski--Sachs 
models \cite{KomChe64,KanSac66}. 
The former admit a three--dimensional Lie group of 
isometries that acts simply transitively on the spatial 
hypersurfaces \cite{Wald84,MacCallum79}. 
The important features of these spacetimes 
are summarized in Appendix \ref{appendixC}. The Kantowski--Sachs 
universes are different to the Bianchi models, 
because the group of isometries 
does not act simply transitively on the homogeneous surfaces. 

When discussing Bianchi string cosmologies, it is convenient to 
employ the orthonormal--frame formalism \cite{HugJac69,RyaShe75}.
We also express all field strengths in terms of differential forms, because 
this allows a more general analysis to be made. 
We choose an orthonormal tetrad
\begin{equation}
\label{gbm}
ds^2 = \eta_{\alpha\beta}\lambda^{\alpha}\lambda^{\beta}
\end{equation}
for the spacetime metric, where $\eta_{\alpha\beta} ={\rm 
diag}[-1,1,1,1]$. The one--forms $\lambda^{\alpha}$ 
are given by \cite{HugJac69,RyaShe75}
\begin{equation}
\label{orthonorms}
\lambda^0 =dt , \qquad \lambda^a =e^{\alpha (t)} 
{b^a}_b(t) \omega^b
\end{equation}
where $\omega^b$ are defined in Eq. (\ref{omegadefinition}), 
$\alpha$ is defined in Eq. (\ref{3metricbianchi}) and 
the symmetric $3\times 3$ matrices ${b^a}_b$ are functions 
only of time, $t$. Comparison with Eq. (\ref{bianchimetric}) 
implies that the three--metric is given by
\begin{equation}
h_{ab}(t) =e^{2\alpha} \delta_{cd}{b^c}_a{b^d}_b
\end{equation}

The field equation for the three--form field strength is 
\begin{equation}
\label{formfieldequation}
d(^*H) -2(d \varphi ) \wedge (^*H) =0
\end{equation}
where $^*H$ is the Poincar\'e dual of $H$, defined 
in Eq. (\ref{dualtoH}). Eq. (\ref{formfieldequation}) 
is solved by
\begin{equation}
\label{sigmabianchi}
d\sigma = e^{-2\varphi} {^*}H
\end{equation}
where $\sigma$ represents the pseudo--scalar 
axion field. Eq. (\ref{sigmabianchi}) is equivalent to 
Eq. (\ref{dualityexpression}). 

We may express the one--form $d\sigma$ as 
\begin{equation}
d\sigma =\frac{1}{6} e^{-2\varphi} H^{\mu\nu\lambda} (t)
\epsilon_{\mu\nu\lambda\kappa}\lambda^{\kappa}
\end{equation}
or equivalently, as 
\begin{equation}
d\sigma =U(t) dt + V_b(t) \omega^b
\end{equation}
where 
\begin{eqnarray}
U(t) \equiv \frac{1}{6} e^{-2\varphi} H^{abc}\epsilon_{abc0} \nonumber \\
V_b(t) =\frac{1}{6} e^{\alpha -2\varphi} H^{\mu\nu\lambda}
\epsilon_{\mu\nu\lambda a}{b^a}_b
\end{eqnarray}
The field equation (\ref{formfieldequation}) reduces 
to the Bianchi identity $dd \sigma =0$ and may 
be directly integrated to yield the simultaneous constraints \cite{BarKun97}
\begin{eqnarray}
\label{firstint}
V_a = K_a \\
\label{secondint}
V_a {C^a}_{bc} =0
\end{eqnarray}
where $K_a$ is an arbitrary, constant 
spatial three--vector. 

This implies that 
the spatial dependence of the axion field is  
at most {\em linear} in the spatial variables. This is still consistent 
with the assumption of spatial homogeneity, because 
only the first derivatives 
of the axion field arise in the energy momentum tensor, implying 
that this tensor is a function of cosmic time, $t$, only. 
However, the second constraint (\ref{secondint}) restricts 
the number of non--zero components of $K_a$ for each Bianchi type. 
For the spatially flat, Bianchi I model, $K_a$ may be arbitrary 
since ${C^a}_{bc} =0$ for this type. On the other hand, 
at least one or more of its components 
must vanish for the other Bianchi types. 

Indeed, Eq. (\ref{secondint}) implies that 
\begin{equation}
\epsilon^{bcd}V_a {C^a}_{bc} =0
\end{equation}
and substituting in Eq. (\ref{Mab}) implies that \cite{BarKun97}
\begin{equation}
\label{gradientconstraint}
V_a \left( 2M^{ad} + A_c\epsilon^{cad} \right) =0
\end{equation}
where we have employed the properties summarized in (\ref{properties}). 
It is interesting to consider the Bianchi type IX 
cosmology, since this represents the 
anisotropic generalization of the 
positively curved FRW model. For the Bianchi class A, where $A_b =0$, 
Eq. (\ref{gradientconstraint}) implies that $V_aM^{ad} =0$. Remarkably, 
this equation is identical in structure 
to the Jacobi identity (\ref{jacobi1}). It implies that if 
$V_a \ne 0$, the rank of $M^{ab}$ can not exceed two. However, 
${\rm rank}[M^{ab}] =3$ for the Bianchi types VIII and IX. 
Thus, all three components of the spatial gradient of the pseudo--scalar axion 
field must vanish for these Bianchi types, i.e., the only 
possible Bianchi type VIII and IX string cosmologies  with 
NS--NS fields are solitonic, where the axion field is constant 
on the surfaces of homogeneity \cite{Batakis95,BarKun97}. 
In the type IX model, the Lie group of isometries is ${\rm SO}(3)$,
corresponding to the three--dimensional rotation group. 
If one of the components of $K_a$ were to be 
non-trivial, one direction would necessarily be 
selected over the others and this would break the rotational 
invariance of the model.

Bianchi string cosmologies have been classified 
by Batakis \cite{Batakis95} and by Barrow and Kunze 
\cite{BarKun97} under different assumptions. Both 
assume that the dilaton is a function only of cosmic time, $t$, but
Batakis restricts the analysis to metrics  that are diagonal 
and of the form given by  Eq. (\ref{gbm}), 
where $\lambda^b =a_b(t) \omega^b$ (no sum implied).
Barrow and Kunze place no restriction on the 
form of the metric and work within the orthonormal 
frame given by Eqs. (\ref{gbm}) and (\ref{orthonorms}). 
Batakis' classification is characterized by the orientation
with respect to the surfaces of homogeneity
of the dual, $^*H$. 
Barrow and Kunze, on the other hand,  
assume throughout that the two--form potential is spatially 
homogeneous, $B_{\mu\nu}=B_{\mu\nu}(t)$. 

In effect, Barrow and Kunze consider the class of 
elementary, anisotropic string cosmologies. They determine the most general 
form of the field strength $H$ consistent with their assumptions 
for all Bianchi models \cite{BarKun97}. It is found that 
the strongest restrictions on the form of $H$ arise in models with 
the most general geometries. The three--form can have at most 
three non--trivial components. The most general 
solutions arise for the Bianchi types III and ${\rm VI}_h$, 
where the group parameter takes the specific values $h=0,-
1/2, -2$. These solutions contain a total of eight independent constants. 
This is the necessary number of independent arbitrary 
constants that an anisotropic NS--NS string cosmology 
must exhibit if the solution is to be representative of the 
most general inhomogeneous solution to the field 
equations (\ref{vary1})--(\ref{vary4}). 

There are three classes of diagonal Bianchi string cosmologies in 
the Batakis classification. These are denoted by 
${\cal{X}}(\uparrow )$, 
${\cal{X}}(\rightarrow )$ and ${\cal{X}} (\nearrow )$,
respectively, where ${\cal{X}}$ refers to the specific 
Bianchi type and the arrow denotes 
the orientation of $^*H$ with respect to 
the surfaces of homogeneity, $\Sigma_t$ \cite{Batakis95}. In this picture, 
these surfaces should be viewed as horizontal. 
Specific solutions have been presented for all classes 
in Refs. \cite{Batakis95,Batakis95a,BatKeh95,BatKeh95a}. Solutions in the 
${\cal{X}}(\rightarrow )$ class exist only for Bianchi types 
I, II, III, ${\rm VI}_0$ and ${\rm VII}_0$. 
Non--trivial cosmologies are only possible in the 
${\cal{X}}(\nearrow )$ class if the Bianchi type is 
III,  V, or ${\rm VI}_h$.

The solitonic (magnetic) 
cosmologies, where the axion field is purely time--dependent, correspond to
the ${\cal{X}}(\uparrow )$ class, since in this case 
the congruences of $^*H$ are orthogonal to $\Sigma_t$. 
These solutions may be interpreted in the Einstein frame as orthogonal stiff 
perfect fluid cosmologies. In general, orthogonal 
Bianchi models in Einstein gravity are characterised by the condition that 
the fluid velocity vector is 
orthogonal to the group orbits (surfaces of homogeneity) 
\cite{KinEll73}. 
This is important because it implies that the problem of finding 
the class of orthogonal Bianchi string cosmologies  
can be reduced to finding the full set of 
orthogonal stiff perfect fluid 
solutions of general relativity  \cite{KraSteHer80}. 
The corresponding string cosmologies 
can then be generated from these seed solutions by employing the global 
${\rm SL}(2,R)$ symmetry of the NS--NS action (\ref{reduced2}) in 
precisely the same manner as outlined in Section \ref{Section7.1}. 

Suppose, for example, that the spacetime metric is known for 
a given Bianchi model containing an orthogonal stiff perfect fluid. 
When the axion field is trivial  $(\sigma =0)$, the dilaton 
field is massless and minimally coupled to gravity. Its 
gradient is then related to the four--velocity $u_{\mu}$ of the fluid 
by \cite{TauTab73}
\begin{equation}
\label{fluid}
u_{\mu} =\frac{\varphi_{,\mu}}{\sqrt{-\varphi_{,\nu}
\varphi^{,\nu}}}
\end{equation}
where a comma denotes ordinary differentiation.  
The equation of motion (\ref{varytilde4}) of the dilaton then reduces to 
\begin{equation}
\label{boxorthogonal}
\Box \varphi = \frac{1}{\sqrt{-g}} \left( \sqrt{-g} g^{\mu\nu}
\varphi_{,\mu} \right)_{,\nu} =0 . 
\end{equation}

Since the shear matrix $\beta_{ab}$ defined in 
Eq. (\ref{betametric}) is traceless, Eq. (\ref{boxorthogonal}) 
simplifies to 
\begin{equation}
\label{simplebox}
\ddot{\varphi} +3\dot{\alpha}\dot{\varphi} =0
\end{equation}
after substitution of Eq. (\ref{3metricbianchi}). 
Defining a new time coordinate \cite{BatKeh95a}
\begin{equation}
\label{neworthogtime}
\tau \equiv \int^t dt' e^{-3\alpha (t')}
\end{equation}
simplifies Eq. (\ref{simplebox}) still further: 
\begin{equation}
\label{moresimplebox}
\frac{d^2 \varphi}{d\tau^2} =0
\end{equation}
and Eq. (\ref{moresimplebox}) admits the general solution
\begin{equation}
\label{simplesolution}
\varphi =\varphi_0 +\varphi_1 \tau
\end{equation}
where $\varphi_{0,1}$ are arbitrary constants. 

After the  ${\rm SL}(2,R)$ transformation has been implemented, 
the string cosmologies are given 
by Eqs. (\ref{sphimodel})--(\ref{smetricmodel}), 
where the dilaton is determined by Eq. (\ref{simplesolution}). 
The definition (\ref{neworthogtime}) implies that 
$\tau$ is a monotonically increasing function of $t$, since 
$dt>0$ if and only if $d\tau >0$. Thus, the 
early and late time behaviour of (ever--expanding) orthogonal 
string cosmologies 
may be investigated by considering the limits $\tau \rightarrow 0$ 
and $\tau \rightarrow +\infty$, respectively. This applies to all 
Bianchi models with negative spatial 
curvature, but not necessarily to the Bianchi type 
IX cosmology. 

The simplest homogeneous metric that admits an elementary solution 
is the spatially flat, Bianchi type I model. In this case, 
the non--trivial components of the two--form potential and 
the spacetime metric 
are all independent of the spatial coordinates. This is not generally true of 
the metric components for the other Bianchi types, as can be 
seen directly from Table \ref{bianchiforms} in 
Appendix \ref{appendixC}. We may therefore integrate 
over the spatial variables in action (\ref{reduced1}) 
to derive an effective one--dimensional action. Since the topology of 
a closed, Bianchi type I metric is that of a three--torus, 
$T^3=S^1 \times S^1 \times S^1$, the reduced action 
is invariant under global ${\rm O}(3,3)$ transformations, as discussed in 
Section \ref{Section4.3}. 

Meissner and Veneziano have implicitly found the general form of the
elementary Bianchi type I string cosmology by employing this symmetry
of the field equations \cite{MeiVen91}. Here we briefly summarize the
derivation of the solution. Without loss of generality, we may express
the two-form potential and metric in the form
\begin{eqnarray}
\label{elem1}
{\cal{B}} = \left( \begin{array}{cc} 
0 & 0\\
0 & B(t) \end{array} \right) \\
\label{elem2}
{\cal{G}} = \left( \begin{array}{cc}
-1 & 0 \\
0  & h(t) \end{array} \right)
\end{eqnarray}
where $B(t)$ and $h(t)$ are $3\times 3$ matrices. The 
matrix $h$ corresponds to the three--metric on the spatial hypersurfaces. 
The one--dimensional action derived from Eq. (\ref{reduced1}) 
when the modulus, $\beta$, is trivial is then given by
\begin{equation}
\label{elem3}
S=\int dt e^{-\psi} \left[ \dot{\psi}^2  +\frac{1}{8} 
{\rm Tr} \left( \dot{M} \eta \dot{M} \eta \right) \right]
\end{equation}
where the matrix $M$ is defined in Eq. (\ref{Mexpression}), 
the matrix $\eta$ is the ${\rm O}(3,3)$ metric defined in 
Eq. (\ref{onn}) and 
\begin{equation}
\label{elem4}
\psi \equiv \varphi -\ln \sqrt{{\rm det} h}
\end{equation}
Action (\ref{elem3}) is invariant under the global 
${\rm O}(3,3)$ transformation $\bar{M} =UMU^{\rm T}$, $\bar{\psi} 
=\psi$, where $U$ is a constant ${\rm O}(3,3)$ matrix satisfying 
Eq. (\ref{onn}). 

The field equations may now be 
written in a manifestly ${\rm O}(3,3)$ invariant 
form. Variation of the action with respect to the non--dynamical 
lapse function yields the Hamiltonian constraint 
equation
\begin{equation}
\label{elem5}
\dot{\psi}^2 +\frac{1}{8} {\rm Tr} \left( 
\dot{M} \eta \dot{M} \eta \right) =0
\end{equation}
and the field equation for $\psi$ takes the form
\begin{equation}
\label{elem6}
\dot{\psi}^2 -2\ddot{\psi} -\frac{1}{8} {\rm Tr} 
\left( \dot{M} \eta \dot{M} \eta \right) =0
\end{equation}
The field equation for the matrix $M$ can be derived by employing the 
general techniques summarized in Section \ref{Section4.4}. It is given by
\begin{equation}
\label{elem7}
\partial_t \left( M\eta \dot{M} \right) =\dot{\psi} 
\left( M\eta \dot{M} \right)
\end{equation}
Eq. (\ref{elem7}) admits the first integral: 
\begin{equation}
\label{elem8}
e^{-\psi} \left( M\eta \dot{M} \right) =A
\end{equation}
where $A$ is a constant $6\times 6$ matrix that 
satisfies the conditions
\begin{equation}
\label{elem9}
A^{\rm T}=-A, \qquad M\eta A +A \eta M =0
\end{equation}
Substituting Eq. (\ref{elem8}) into Eq. (\ref{elem5}) 
implies that 
\begin{equation}
\label{elem10}
\dot{\psi}^2 =\frac{1}{8}e^{2\psi}{\rm Tr} (A \eta )^2
\end{equation}
and this can be readily integrated to yield the time 
dependence of the field $\psi$: 
\begin{equation}
\label{elem11}
e^{\psi} = \frac{C}{t} , \qquad C \equiv \sqrt{\frac{8}{{\rm Tr} 
(A\eta)^2}} 
\end{equation}
modulo a constant of proportionality. 

The general solution to Eq. (\ref{elem8}) is found 
by defining a new time variable \cite{MeiVen91}
\begin{equation}
\label{elem12}
dT \equiv e^{\psi} dt
\end{equation}
Integration of Eq. (\ref{elem8}) then implies that 
\begin{equation}
\label{elem13}
M(t) =M_0 e^{-A\eta T}
\end{equation}
where $M_0$ is a constant matrix that can be set to unity 
without loss of generality. Finally, 
substituting Eq. (\ref{elem11}) into Eq. (\ref{elem12}) 
and integrating implies that 
\begin{equation}
M = \exp \left( -C A \eta \ln t \right)
\end{equation}

Before concluding this Section, we discuss the one remaining homogeneous 
cosmology --
the Kantowski--Sachs universe \cite{BarDab96}. The metric 
admits a four--dimensional group of motions. However, it   
is not part of the Bianchi classification because the 
$G_3$ of isometries does not act simply transitively on 
the spatial hypersurfaces $\Sigma_t$ \cite{MacCallum79,KanSac66}. Instead, 
$G_3$  acts multiply--transitively and this implies that 
it acts on two--dimensional surfaces of 
maximal symmetry. The curvature of these two--spaces is constant 
and can be positive, zero or negative. 
The latter two cases correspond to the LRS Bianchi 
types I and III, respectively. The positive--curvature case 
is the Kantowski--Sachs model. It is a closed universe with spatial 
topology $S^1 \times S^2$ and reduces to the closed FRW model 
in the isotropic limit. Such a model was first discussed
by Kompanyeets and Chernov \cite{KomChe64}.  
The line element may be expressed in the form \cite{KanSac66}
\begin{equation}
\label{kantoski}
ds^2 = -dt^2 +a_1^2  dr^2 +a_2^2  \left[ d \theta^2  +\sin^2 \theta 
d\psi^2 \right]
\end{equation}
where  $a_i =a_i(t)$ are the scale factors of the universe. 

Barrow and Dabrowski have found new Kantowski--Sachs solutions to the
NS--NS field equations (\ref{vary1})--(\ref{vary4}) for the two cases
where the pseudo--scalar axion field is time--dependent and
time--independent \cite{BarDab96}. We do not reproduce their results
here. As in the orthogonal Bianchi models, the axion field is
equivalent to a stiff perfect fluid when the former ansatz is invoked
and this solution was considered earlier by Mimoso and Wands
\cite{MimWan95b}. When the axion is spatially--dependent, the
two--form potential is severely restricted and consistency requires
that the only non--zero component of its field strength is $H_{0\theta
\psi}$ \cite{BarDab96}.  This implies that the two--form potential
must be aligned along the two spatial directions associated with the
two--sphere.

To summarize thus far, we have presented in Section \ref{Section7.1} 
the general class of 
dilaton--moduli--axion FRW string cosmologies. These  
models play a central role in our discussions of the 
pre--big bang scenario in Sections \ref{Section9} and \ref{Section10}. 
We have also surveyed in this subsection the 
classes of spatially anisotropic solutions that have been discussed 
in the literature 
to date. However, given that our understanding 
of the geometry of the universe near the string scale 
is uncertain, such an approach is necessarily incomplete.  
It is therefore important to consider the role played by 
spatial inhomogeneities 
in early universe string cosmology. In the remainder of this Section, 
we discuss techniques that prove valuable when studying 
inhomogeneous string cosmologies. 

\subsection{Inhomogeneous String Cosmology}

\label{Section7.3}

The one--loop string equations of motion (\ref{vary1})--(\ref{vary4}) 
for inhomogeneous backgrounds are very difficult to solve, but further 
progress can be made by considering models where  
homogeneity is broken along one spatial direction. Metrics that 
admit two commuting space-like Killing vectors 
satisfy this property and 
are generically referred to Einstein--Rosen, or $G_2$, spacetimes
\cite{EinRos37}. 

String models admitting an abelian group, $G_2$, of isometries 
have a number of important physical applications. The 
spatially homogeneous Bianchi types I--${\rm VII}_h$ and 
locally rotationally symmetric (LRS) 
types VIII and IX admit a $G_2$ group of isometries
\cite{Tomita78}. These models may be naturally 
generalized to inhomogeneous Einstein--Rosen cosmologies \cite{CarChaMal81}. 
Non--linear inhomogeneities in the dilaton and axion fields can be 
investigated and, in principle, this allows density perturbations 
in string--inspired inflationary models such as the pre--big bang 
scenario to be studied \cite{Veneziano97,BuoMeiUng98,MahOnoVen98}. The  
propagation and collision of gravitational waves in string backgrounds 
may also be analyzed in terms of $G_2$ 
space--times \cite{AdaHelZim82,AdaHelZim85,BicGri94,FeiGri94,BicGri96,Szekeres71}.
Furthermore, the collision of self--gravitating plane waves can be 
modelled as the time reversal  
of a $G_2$ cosmology in the vicinity of the big bang 
singularity \cite{FeiIba89,Griffiths91}. Finally, it has been conjectured 
that $G_2$ metrics represent a first approximation to 
the general solution of Einstein gravity in the vicinity 
of a curvature  singularity \cite{BelKha69,BelKha70,BelKha71,BelLifKha82}. 
The high curvature regime 
is precisely the regime where stringy deviations from 
general relativity are expected to be significant. These models may therefore 
provide insight into the generic behaviour of cosmologies at very early 
times. 

A study of Einstein--Rosen string cosmologies is therefore 
well motivated. For reviews of the general properties of these spacetimes, 
see, e.g., \cite{CarChaMal81,Verdaguer93,Krasinski97}. 
A number of $G_2$ string cosmologies were recently 
derived by employing a variety of methods \cite{FeiLazVaz97,BarKun97a,ClaFeiLid99a,ClaFeiLid99b}. 
Barrow and Kunze studied an inhomogeneous generalization of the  
Bianchi type I string cosmology \cite{BarKun97a} and Feinstein, Lazkoz and 
Vazquez--Mozo derived a closed, inhomogeneous model by applying duality 
transformations on the LRS Bianchi type IX cosmology \cite{FeiLazVaz97}. 
Clancy {\em et al.} have found inhomogeneous generalizations 
of the Bianchi type ${\rm VI}_h$ universe and have  
studied their asymptotic behaviour \cite{ClaFeiLid99b}. These latter 
models are interesting because the  Bianchi type ${\rm VI}_h$ universe 
has a non--zero measure in the space 
of homogeneous initial data and includes the 
Bianchi type III as a special case. Furthermore, as we discussed in 
Section \ref{Section7.2}, 
the most general, elementary, homogeneous string cosmologies are 
the Bianchi types III and ${\rm VI}_h$, where 
the group parameter $h=\{ 0, -1/2 , -2 \}$ \cite{BarKun97}. 

The field equations (\ref{vary1})--(\ref{vary4}) 
reduce to a system of coupled, partial differential 
equations in two variables when spatial homogeneity is broken along 
a single direction. Unfortunately, 
these equations are still very complicated. 
On the other hand, solutions can be found due to the 
non--compact global symmetries of the string effective action. 
When the metric admits two commuting space-like Killing vectors, 
there exists an infinite--dimensional symmetry on the space of solutions
that may be identified infinitesimally 
with the ${\rm O}(2,2)$ current algebra 
\cite{Bakas94,Maharana95,Kehagias95}. This  
symmetry reduces to the Geroch group, 
corresponding to the ${\rm SL}(2,R)$ current 
algebra,  when the dilaton and two--form potential 
are trivial \cite{Geroch72,Kinnersley77,KinChi77,KinChi78}. 
The global ${\rm SL}(2,R)$ S--duality of 
the action (\ref{reduced2}) is contained within this symmetry \cite{Bakas94}. 
Moreover, we saw in Section \ref{Section4.3} that 
there exists a global ${\rm O}(d,d)$ symmetry in the 
string equations of motion when the background admits $d$ abelian 
isometries \cite{MahSch93}. Since the inhomogeneous Einstein--Rosen  
models we are considering in this 
Section admit two commuting Killing vectors, 
the field equations are  invariant 
under global ${\rm O}(2,2)$ symmetry transformations. 
This T--duality  is also contained within the  ${\rm O}(2,2)$ current algebra
\cite{Bakas94}. 

Application of both the S-- and T--duality 
symmetries leads to new, inequivalent 
solutions. Consequently, an inhomogeneous 
$G_2$ string cosmology containing a non--trivial two--form 
potential may be generated once a dilaton--vacuum solution has 
been specified.  We now employ these  
non--compact, global symmetries of the string effective action in 
a variety of settings and outline 
a number of different approaches that can be taken in deriving 
cosmologies of this type. The principles are  
similar to those of previous Sections. Our purpose here 
is to highlight the important features of the methods rather 
than restrict ourselves to a detailed analysis of specific 
solutions. We therefore provide a framework for future study 
in this topic.  

We begin our discussion in the Einstein frame and 
assume throughout that the modulus field in Eqs. (\ref{reduced1}) and 
(\ref{reduced2}) is stable. The class of Einstein--Rosen 
metrics is defined by the block diagonal line element 
\cite{CarChaMal81,Verdaguer93}
\be 
\label{9}
ds^2 =h_{\alpha\beta} (x^{\epsilon}) dx^{\alpha}dx^{\beta} 
+ \gamma_{ab} (x^{\epsilon}) dx^a dx^b    ,
\ee
where $\{ x^{\alpha} =\xi ,x \}$, $\{ x^a =y,z \}$ and 
$\xi$ represents the time coordinate. The functions 
$h_{\alpha\beta}$ and $\gamma_{ab}$ represent the longitudinal and transverse 
components of the spacetime metric, respectively, and 
depend only on the variables 
$\xi$ and $x$. Spatial homogeneity is therefore broken along the $x$--direction
\cite{CarChaMal81,Verdaguer93}. The two commuting 
space-like Killing vectors are $\partial /\partial x^a$. The 
abelian group, $G_2$, of isometries acts orthogonally--transitively
and the variables $x^a$ span the surfaces of transitivity. The local behaviour 
of these models is determined by the gradient $B_{\mu}
\equiv \partial_{\mu} \sqrt{{\rm det} \gamma_{ab}}$. 
Cosmological solutions arise if $B_{\mu}$ is globally time-like or 
when $B_{\mu}B^{\mu}$ changes sign  \cite{Tomita78}. 
When $B_{\mu}$ is globally space-like or null, the 
solutions represent cylindrical or gravitational plane waves, 
respectively \cite{CarChaMal81,Griffiths91,Szekeres71}. 
Since the longitudinal part of the metric is conformally flat, 
Eq. (\ref{9}) can be written as 
\begin{equation}
\label{simplermetric}
ds^2 = e^f \left( -d \xi^2 + dx^2 \right) + 
\gamma_{ab} (\xi , x)dx^adx^b
\end{equation}
where $f=f(\xi , x)$. 

The $G_3$ group 
of isometries for all Bianchi types I--${\rm VII}_h$ and the 
LRS types VIII and IX admits an abelian subgroup $G_2$ \cite{Tomita78}. 
These homogeneous models are therefore particular cases of 
$G_2$ cosmologies. In these cases, a coordinate system 
exists that allows the spacetime metric to be written in the form 
\begin{equation}
ds^2 = e^{f(\xi )} \left( -d \xi^2 +dx^2 \right)
+\eta_{cd} (\xi ){e^c}_a (x) {e^d}_b (x) dx^a dx^b
\end{equation}
where $a,b,c,d =1,2$ and the functional 
forms of the  ${e^c}_a (x)$ are determined by the specific Bianchi types. 
Inhomogeneous generalizations of these Bianchi models 
may therefore be 
considered by introducing a dependence on $x$ in the functions 
$f$ and $\eta_{cd}$. 

If one of the Killing vectors is orthogonal to the surfaces of 
transitivity, the transverse part of the metric
may be diagonalized. Without loss of generality, 
the metric (\ref{simplermetric}) may then be expressed in the form
\begin{equation}
\label{diagG2}
ds^2 =e^f \left( -d\xi^2 +dx^2 \right) +\xi \left( e^p 
dy^2 +e^{-p} dz^2 \right)
\end{equation}
where $p=p(\xi , x)$. {}From Eq. (\ref{tildeg})
the corresponding string--frame metric is then 
given by
\begin{equation}
\label{seedmetric}
ds^2_{\rm string}
= e^{f+\varphi} \left( -d\xi^2 +dx^2 \right) +\xi e^{\varphi}
\left( e^pdy^2 +e^{-p}dz^2 \right)
\end{equation}
where $\varphi$ is the four--dimensional dilaton field. 

In the absence of the two--form potential (axion field), 
the Einstein--scalar field 
equations (\ref{varytilde1}) and (\ref{varytilde4}) are given by 
\cite{ChaMal79}
\begin{eqnarray}
\label{G2field1}
\dot{f} =-\frac{1}{2\xi} +\frac{\xi}{2} \left( \dot{p}^2 +{p'}^2 
+\dot{\varphi}^2 +{\varphi'}^2 \right) \\
\label{G2field2}
f' =\xi \left( \dot{p}p' +\dot{\varphi} \varphi' \right) \\
\label{G2field3}
\ddot{p}+\frac{1}{\xi} \dot{p} -p'' =0 \\
\label{G2field4}
\ddot{\varphi}+\frac{1}{\xi} \dot{\varphi} -\varphi''
=0
\end{eqnarray}
for the metric (\ref{diagG2}), 
where a dot and prime denote partial differentiation 
with respect to $\xi$ and $x$, respectively, and it has been 
assumed that the dilaton field 
is independent of the transverse coordinates, $x^a$. 
The advantage of employing the coordinate system in 
Eq. (\ref{diagG2}) is that the field equations for the transverse 
component of the metric, $p$, and the dilaton, $\varphi$, 
are {\em identical}. Indeed, the field equations 
(\ref{G2field1})--(\ref{G2field4}) are symmetric under the 
simultaneous interchange
\begin{equation}
\label{pphisymmetry}
p \leftrightarrow \varphi
\end{equation} 

Moreover, Eqs. (\ref{G2field3}) and (\ref{G2field4}) 
are equivalent to the cylindrically symmetric wave equation in flat space
and the  general solution to this equation is formally known. 
The Einstein scalar field equations (\ref{G2field1})--(\ref{G2field4}) 
can then be solved, at least up to quadratures, by 
integrating Eqs. (\ref{G2field1}) and (\ref{G2field2}).
Integration of Eq. (\ref{G2field1}) leads to an expression for 
$f$ in terms of an arbitrary function $f_1(z)$. In many cases, 
it turns out that Eq. (\ref{G2field2}) is then automatically 
solved if $f_1(z)$ is taken to be a constant. 
The problem of solving the Einstein--scalar field equations for a diagonal 
$G_2$ background is 
therefore reduced to integrating Eq. (\ref{G2field1}). The 
general vacuum solution $(\varphi =0)$ 
to Eqs. (\ref{G2field1})--(\ref{G2field3}) consistent with toroidal 
boundary conditions is known \cite{Gowdy71,Gowdy74,ChaMal79}. 

Once a dilaton--vacuum solution has been found, string cosmologies 
with a non--trivial two--form potential can be derived directly 
by employing the global 
symmetries of the model. As in the homogeneous 
models, one can treat either the two--form potential 
or the pseudo--scalar axion field 
as the fundamental field. For example, application of the ${\rm 
SL}(2,R)$ transformation (\ref{sphimodel})--({\ref{ssigmamodel}) 
generates an inhomogeneous string cosmology 
with a non--trivial axion field. In general, 
the axion field in this `S--dual' solution tends to 
a constant value in the strong and weak coupling limits of the seed solution, 
$\varphi \rightarrow \pm \infty$. It is 
therefore dynamically negligible in these regimes. 
As in the isotropic models considered in 
Section \ref{Section7.1},  this implies that 
there exists a {\em lower} bound on the value of the string coupling for all 
inhomogeneous solutions generated by this symmetry transformation.

In view of this, it is important 
to discuss methods for solving the Einstein--scalar 
field equations of motion (\ref{G2field1})--(\ref{G2field4}). 
A technique for generating solutions with a minimally coupled, 
massless scalar field from a vacuum, $G_2$ cosmology was developed 
by Barrow \cite{Barrow77} and generalized by 
Wainwright, Ince and Marshman \cite{WaiIncMar79}. We now briefly summarize 
the algorithm. Suppose that 
\begin{equation}
\label{wv}
ds^2 =e^{2k} \left( -dt^2 + dx^2 \right) +R \left[ h 
(dy + w dz)^2 +h^{-1} dz^2 \right]
\end{equation}
is a $G_2$ vacuum  solution to the Einstein field equations, 
where $\{ R,h,w,k \}$ are functions 
of $(t,x)$. Then 
$R$ satisfies the one--dimensional wave equation
\begin{equation}
\label{wave}
R_{tt} -R_{xx} =0
\end{equation}
where subscripts denote partial differentiation. 
The functions $\{k,h,w\}$ can then be expressed in terms of quadrature, 
as discussed above. 

It can then be shown that the metric  
\begin{equation}
\label{masslesssolution}
ds^2 =e^{2k +\Omega} \left( -dt^2 +dx^2 \right)
+R \left[ h 
(dy + w dz)^2 +h^{-1} dz^2 \right]
\end{equation}
is a solution to the Einstein field equations for a minimally 
coupled, massless  scalar field, $\varphi$, if the coupled, 
differential equations  
\begin{eqnarray}
\label{wave1}
\varphi_{tt} + \frac{R_t}{R}\varphi_t -\varphi_{xx} 
-\frac{R_x}{R} \varphi_x =0 \\
\label{wave2}
\Omega_t R_t + \Omega_x R_x = \frac{1}{2} R \left( 
\varphi^2_t +\varphi^2_x \right)  \\
\label{wave3}
\Omega_t R_x +\Omega_x R_t =R \varphi_x \varphi_t 
\end{eqnarray}
are simultaneously satisfied, subject to the conditions
\begin{eqnarray}
R^2_t - R^2_x \ne 0 \nonumber \\
\varphi^2_t -\varphi^2_x >0
\end{eqnarray}
The transverse 
space is left unaltered in this 
procedure and the longitudinal component 
of the metric acquires a conformal 
factor. Wainwright {\em et al.} originally employed this method 
to derive inhomogeneous generalizations of a number of Bianchi cosmologies
\cite{WaiIncMar79}, 
but the cosmological interpretation of these solutions was subsequently 
criticized by Carmeli, Charach and Malin \cite{CarChaMal81}. 

The global ${\rm O}(2,2)$  symmetry of the system 
becomes manifest in the string frame \cite{MahSch93}.
For the purposes of the present discussion, it 
proves convenient to denote the components of 
the metric in this frame by 
$G_{\mu\nu} =e^{\varphi}g_{\mu\nu}$, where $g_{\mu\nu}$ is given by 
Eq. (\ref{simplermetric}), and to 
express the line element in the form
\begin{equation}
\label{G2stringmet}
ds^2_{\rm string} =e^{f+\varphi} \left( -d\xi^2 
+dx^2 \right) +\Gamma_{ab} dx^a dx^b
\end{equation}
where $\Gamma_{ab} =e^{\varphi}
\gamma_{ab}$ is the metric on the surfaces 
of transitivity in the string frame. Its determinant, $\Gamma \equiv {\rm 
det}\Gamma_{ab}$, determines the volume of 
these two--surfaces. 

A subset of the ${\rm O}(2,2)$ transformations corresponds to 
Buscher's duality symmetry \cite{Buscher87}. 
For each 
Abelian isometry, $\bar{x}^a = x^a +c$ $(c={\rm constant})$, 
associated with a given coordinate $x^a$, the field equations are 
invariant under the discrete transformation
\begin{eqnarray}
\label{buscher}
\bar{G}_{aa} =\frac{1}{G_{aa}} , \qquad  \bar{G}_{a\mu} = 
\frac{B_{a\mu}}{G_{aa}} , \qquad \bar{G}_{\mu\nu} =
G_{\mu\nu} -\frac{G_{a\mu}G_{a\nu} + B_{\mu a} B_{a\nu}}{G_{aa}} 
\nonumber \\
\bar{B}_{a\mu} =\frac{G_{a\mu}}{G_{aa}} , \qquad \bar{B}_{\mu\nu} 
=B_{\mu\nu} -\frac{G_{a\mu}B_{a\nu} + G_{a\nu}B_{\mu a}}{G_{aa}}
\nonumber \\
\bar{\varphi} =\varphi -{\rm ln} G_{aa} 
\end{eqnarray}
where $(\mu , \nu \ne a)$. In a cosmological setting this transformation 
may be viewed as a scale factor duality, since the $G_{aa}$ 
component of the metric is directly inverted. The off--diagonal 
component of the transverse part of the metric 
is interchanged with a component of the two--form potential. 
An important feature of Eq. (\ref{buscher}) is that 
a non--trivial dilaton field and two--form potential can be 
generated directly from a non--diagonal, 
$G_2$ vacuum solution of general relativity. 

Feinstein, Lazkoz and V\'azquez--Mozo have recently outlined a 
method for deriving inhomogeneous string cosmologies that 
employs the Buscher transformation (\ref{buscher})
\cite{FeiLazVaz97}. Beginning 
with a diagonal, Ricci--flat, $G_2$ cosmology, 
off--diagonal terms in the metric are generated by employing 
inverse scattering techniques 
\cite{BelKha70,BelKha71,HasFeiMan90,HoeKinXan79,Kitchingham84}
or an Ehlers rotation \cite{Ehlers57}. 
The method of Wainwright {\em et al.} is then 
employed to introduce a massless,  
minimally coupled scalar field \cite{WaiIncMar79}. The scalar 
field and metric are identified with the dilaton and    
Einstein frame metric, respectively. 
Performing the inverse of the conformal transformation 
(\ref{ste}) leads to a solution in the 
string frame with a trivial two--form potential. 
Finally, a non--trivial 
two--form is generated by applying Buscher's transformation 
(\ref{buscher}) \cite{Buscher87}. This results in a diagonal metric. 

Feinstein {\em et al.} employ this algorithm 
to generate a closed, inhomogeneous 
string cosmology with $S^3$ topology from a LRS Bianchi type IX 
solution \cite{FeiLazVaz97,CarChaFei83}. However, 
this algorithm involves a number of non--trivial operations. 
Indeed, it is necessary to solve Eqs. (\ref{wave1})--(\ref{wave3}) 
in order to generate a dilaton field. 
A more straightforward approach is to apply an ${\rm O}(2,2)$ 
transformation directly to the seed cosmology. 

To understand how the ${\rm O}(2,2)$ symmetry may be applied in this way, 
it is instructive to view the inhomogeneous 
$G_2$ backgrounds as $(1+1)$--dimensional 
cosmologies with a two--dimensional internal 
space. This is valid when the surfaces of transitivity are 
closed and have a toroidal topology
$T^2=S^1 \times S^1$. Thus, we may start with the four--dimensional 
effective action (\ref{reduced1}) and compactify on $T^2$. 
The dimensionally 
reduced action is given by the $D=2$ form of 
Eq. (\ref{oddsymaction}) with 
a trivial field strength $H_{\mu\nu\lambda} =0$, since 
a three--form must vanish identically in two dimensions. 

Let us identify the relevant terms in Eq. (\ref{oddsymaction}). 
It is consistent to assume that the only non--trivial component 
of the two--form potential is $B_{yz}=B_{yz}(\xi ,x)$. 
The `moduli' matrix ${\cal{M}}$ is then given by Eq. 
(\ref{Mexpression}), where
\begin{equation}
B = \left( \begin{array}{cc}
0 & B_{yz} \\
-B_{yz} & 0 \end{array} \right)
\end{equation}
and 
\begin{equation}
G =\left( \begin{array}{cc} 
\Gamma_{yy} & \Gamma_{yz} \\
\Gamma_{yz} & \Gamma_{zz} 
\end{array} \right)
\end{equation}
For a metric of the form 
(\ref{G2stringmet}) the relevant ${\rm O}(2,2)$ transformation is therefore 
given by
\begin{eqnarray}
\label{G2odd1}
\bar{\cal{M}} = U{\cal{M}} U^{\rm T} \\
\label{G2odd2}
\bar{\varphi} =\varphi +\frac{1}{2} \ln \left( \bar{\Gamma}
/\Gamma \right) \\
\label{G2odd3}
\bar{f} = f+ \varphi - \bar{\varphi} 
\end{eqnarray}
where $U$ is an arbitrary, constant ${\rm O}(2,2)$ matrix. 
Eq. (\ref{G2odd2}) follows because the effective two--dimensional 
dilaton field is a singlet under the ${\rm O}(2,2)$ transformation. 
Eq. (\ref{G2odd3}) follows due to the invariance of the longitudinal string 
frame metric. 

As we saw in Section \ref{Section4.3}, the four real degrees of freedom 
$\{ B_{yz} , \Gamma_{ab} \}$ parametrize the 
${\rm O}(2,2)/{\rm O}(2) \times {\rm O}(2) $ coset 
\cite{MahSch93,GivPorRab94,MeiVen91a,Sen91,Sen92,HasSen92}. 
However, there exists the important isomorphism
\begin{equation}
{\rm O}(2,2) = {\rm SL}(2, R) \times {\rm SL}(2,R)
\end{equation}
and this implies that the four degrees of freedom may be arranged 
into two complex coordinates \cite{DijVerVer88,BaiLovSab94}: 
\begin{eqnarray}
\label{taurho}
\tau \equiv \tau_1 +i\tau_2 = \frac{\Gamma_{yz}}{\Gamma_{zz}}
+i \frac{\sqrt{\Gamma}}{\Gamma_{zz}} \nonumber \\
\rho \equiv \rho_1 +i\rho_2 = B_{yz} +i 
\sqrt{\Gamma}
\end{eqnarray}
Substituting Eq. (\ref{taurho}) into Eq. (\ref{Mexpression})
implies that 
\begin{equation}
\label{matrixM1}
{\cal{M}}=\frac{1}{\tau_2\rho_2} 
 \left( \begin{array}{cccc} 
1 & -\tau_1 & -\tau_1\rho_1 & -\rho_1 \\
-\tau_1 & |\tau |^2 & \rho_1 |\tau |^2 & \tau_1 \rho_1 \\
-\tau_1 \rho_1 & \rho_1 | \tau |^2 & | \tau |^2 | \rho |^2 & 
\tau_1 | \rho |^2 \\
-\rho_1 & \tau_1 \rho_1 & \tau_1 | \rho |^2 & |\rho |^2 
\end{array} \right)  
\end{equation}

The ${\rm SL}(2,R)$ subgroups of ${\rm O}(2,2)$ are 
made more apparent by defining the $2\times 2$ 
matrix
\begin{equation}
\label{SG2}
S \equiv \frac{1}{\tau_2} \left( \begin{array}{cc}
1 & -\tau_1 \\
-\tau_1 & |\tau |^2 \end{array} \right)  , \qquad {\rm det} S =1
\end{equation}
It then follows that Eq. (\ref{matrixM1}) 
may be written in the block form
\begin{equation}
\label{blockform}
{\cal{M}} = \frac{1}{\rho_2} \left( \begin{array}{cc} 
S & -\rho_1 SJ \\
\rho_1 JS & |\rho |^2 S^{-1} \end{array} \right)
\end{equation}
where $J$ is the ${\rm SL}(2,R)$ metric defined in Eq. (\ref{J}). 
We now define two constant ${\rm O}(2,2)$ matrices
\cite{DijVerVer88,BaiLovSab94}: 
\begin{eqnarray}
\label{omegarho}
\Omega_{\rho} \equiv \left( \begin{array}{cc}
d{\rm I}_2 & -c J \\
bJ & a{\rm I}_2 \end{array} \right) \\
\label{omegatau}
\Omega_{\tau} \equiv
\left( \begin{array}{cc} 
A^{\rm T} & 0 \\
0 & A^{-1} \end{array} \right) 
\end{eqnarray}
where 
\begin{equation}
A \equiv 
\left( \begin{array}{cc}
d' & -b' \\
-c' & a' \end{array} \right)
\end{equation}
and 
\begin{equation}
ad-bc =a'd'-b'c' =1
\end{equation}
An ${\rm O}(2,2)$ transformation of the form (\ref{G2odd1}) is then 
generated by 
\begin{equation}
\Omega = \left( \begin{array}{cccc} 
dd' & -dc' & -cc' & -cd' \\
-db' & da' & ca' & cb' \\
-bb' & ba' & aa' & ab' \\
-bd' & bc' & ac' & ad' \end{array} \right)
\end{equation}

The matrix (\ref{omegarho}) generates the ${\rm SL}(2,R)$
transformation
\begin{equation}
\label{rhotransformation}
\bar{\rho} =\frac{a\rho +b}{c\rho +d}  , \qquad \bar{\tau} =\tau
\end{equation}
that leaves the complex scalar field $\tau$ 
invariant. The matrix (\ref{omegatau}), on the other 
hand, generates the ${\rm SL}(2,R)$ transformation
\begin{equation}
\label{tautransformation}
\bar{\tau}=\frac{a' \tau +b'}{c' \tau +d'} , \qquad 
\bar{\rho} =\rho
\end{equation}
that leaves $\rho$ invariant. 

Eq. (\ref{tautransformation}) leaves the two--form 
potential and volume of the transverse space invariant.  
This ${\rm SL}(2,R)$ transformation is equivalent to a general 
rotation and rescaling of the coordinates of the transverse space. 
In general, it generates a non--diagonal component 
$\Gamma_{yz}$ from a non--diagonal metric, 
but does not lead to a genuinely new solution. 
In this sense, it is similar to 
an Ehlers rotation \cite{Ehlers57}. On the other hand, it may 
be employed in conjunction with the other 
${\rm SL}(2,R)$ transformation (\ref{rhotransformation}). 
In particular, it may be employed within the context of the algorithm of 
Feinstein {\em et al.} to generate 
an off--diagonal component in the metric \cite{FeiLazVaz97}. 

Eq. (\ref{rhotransformation}) 
generates a non--trivial dilaton and two--form potential 
directly from a vacuum 
$G_2$ solution of general relativity.  
We now derive a class of inhomogeneous string cosmologies with 
this transformation \cite{ClaFeiLid99b}. We take as our seed solution 
a dilaton--vacuum cosmology of the form given by Eq. (\ref{seedmetric})
containing a non--trivial dilaton field, but a trivial two--form 
potential. The volume of the transverse space is given by 
\begin{equation}
\label{seeddet}
\Gamma =
\xi^2 e^{2\varphi}
\end{equation}
Eqs. (\ref{G2odd1})--(\ref{G2odd3}) then imply that 
\begin{eqnarray}
\label{tdual1}
\bar{\Gamma} =\frac{\Gamma}{\left( d^2 +c^2 \Gamma \right)^2} \\
\label{tdual2}
\bar{B}_{yz} =\frac{ac\Gamma +bd}{c^2 \Gamma +d^2} \\
\label{tdual3}
e^{\bar{\varphi}}=\frac{e^{\varphi}}{d^2 +c^2\Gamma} 
\end{eqnarray}
and the dual metrics in the both the string and Einstein frames 
are given  respectively by
\begin{equation}
\label{tdualstring}
d\bar{s}^2_{\rm string} =e^{f+\varphi} \left( 
-d\xi^2 +dx^2 \right) + \frac{\xi e^{\varphi}}{d^2 +c^2 \xi^2 e^{2\varphi}}
\left( e^p dy^2+e^{-p}dz^2 \right)
\end{equation}
and
\begin{equation}
\label{tdualeinstein}
d\bar{s}^2_{\rm einstein}=e^f \left( 
d^2 +c^2 \xi^2e^{2\varphi}\right) \left( -d\xi^2 +dx^2 
\right) + \xi \left( e^p dy^2 +e^{-p} 
dz^2 \right)
\end{equation}
The solution generated from a vacuum seed cosmology 
is determined by specifying $\varphi =0$ in Eqs. 
(\ref{tdual1})--(\ref{tdualeinstein}). 

In general, the 
transverse metric in the Einstein frame remains invariant 
under the transformation (\ref{rhotransformation}). 
The determinant of the transverse space in the string 
frame seed solution is given by Eq. (\ref{seeddet}) 
and determines the asymptotic form of the dual solutions. 
In the limit $\Gamma \rightarrow 0$, 
corresponding to the transverse space having 
a vanishingly small volume, the dual metric
in both the string and Einstein frames tends to its original form 
if $d \ne 0$. Similar behaviour is exhibited 
by the dilaton field and the two--form potential  
approaches a constant value. Thus, 
the dual solution asymptotically tends to 
the original dilaton--vacuum solution in this limit. 
In the opposite limit $\Gamma \rightarrow \infty$, 
where the volume of the transverse space 
diverges, the two--form potential tends towards a different 
constant value and again becomes dynamically negligible. 
The metric and dilaton field 
approach the limiting solution generated by the particular  
${\rm SL}(2,R)$ transformation (\ref{rhotransformation}) 
where $d=0$. 

\subsection{Mirror Images}

\label{Section7.4}

There exists a further discrete ${\rm Z}_2$ symmetry of the
class of $G_2$ string cosmologies that we are considering \cite{Bakas94}. 
This symmetry interchanges the scalar fields 
with the transverse metric degrees of freedom. It becomes 
manifest in the Einstein  frame after dualization of the two-form potential. 
When the metric is given by the line element (\ref{9}), we 
may integrate over the transverse  
variables in Eq. (\ref{reduced2}) to derive 
an effective two--dimensional action of the form 
\begin{equation}
\label{2d}
S=\int d^2 x \sqrt{-{{h}}} e^{-{\gamma}} \left[ 
{R}_2 +\frac{1}{2} \left( {\nabla} 
{\gamma} \right)^2 +
\frac{1}{4} {\rm Tr} \left( {\nabla} L {\nabla} 
L^{-1} \right) +\frac{1}{4} 
{\rm Tr} \left( {\nabla} N {\nabla} N^{-1} \right) \right]
\end{equation}
where $R_2$ is the Ricci curvature scalar of the $(1+1)$--dimensional 
manifold with metric $h_{\alpha\beta}$, $h\equiv {\rm det} h_{\alpha\beta}$, 
$\gamma \equiv -( \ln {\rm det} \gamma_{ab} )/2$ and both 
$L$ and $N$ denote symmetric, $2\times 2$ matrices. 
The latter is parametrized as usual by the dilaton and axion fields: 
\begin{equation}
N \equiv \left( \begin{array}{cc}
e^{\varphi} & \sigma e^{\varphi} \\
\sigma e^{\varphi} & e^{-\varphi} + \sigma^2 e^{\varphi} \end{array} 
\right) 
\ee
and the former is defined in terms of the transverse two--metric, 
$\gamma_{ab}$, 
such that $   \gamma_{ab} \equiv  e^{-\gamma} L_{ab}$. It may be expressed 
in the form
\begin{equation}
\label{defineL}
L \equiv \left( \begin{array}{cc} 
e^p & \omega e^p \\
\omega e^p & e^{-p} +\omega^2 e^p  \end{array} \right) 
\end{equation}
Thus, the scalar functions $p$ and $\omega$ parametrize a second 
${\rm SL}(2,R)/{\rm U}(1)$ coset. 

This implies that 
the action (\ref{2d}) is invariant under the simultaneous interchange
\cite{Bakas94}:
\begin{equation}
\label{mirror}
\bar{L}_{ab} =N_{ab} , \qquad \bar{N}_{ab} = L_{ab}
\end{equation}
or equivalently: 
\begin{equation}
\label{mirror1}
\bar{p} = \varphi , \quad \bar{\omega} =\sigma , \quad 
\bar{\varphi} =p , \quad \bar{\sigma} =\omega 
\end{equation}
Thus, the transverse metric degrees of freedom are 
interchanged with 
the dilaton and axion fields and this leads to a new solution 
with a {\em different} space--time interpretation. For diagonal, 
dilaton--vacuum solutions, where $\omega =\sigma =0$, 
this transformation reduces to that given in Eq. (\ref{pphisymmetry}). 

It should be emphasized that Eq. (\ref{mirror})  is {\em not} 
equivalent to the ${\rm SL}(2, R)$ or ${\rm O}(2,2)$ 
transformations discussed above. In the former case,  
the Einstein frame metric is invariant, but this is not 
the case when Eq. (\ref{mirror}) is applied. Furthermore, 
the longitudinal component of the string frame metric is 
a singlet under global ${\rm O}(2,2)$ transformations, 
as implied by Eq. (\ref{G2odd3}), but Eq. (\ref{mirror}) leaves 
invariant the longitudinal component of the Einstein frame
metric. Equivalence between the two transformations 
would therefore require the dilaton field to be invariant 
under Eq. (\ref{mirror}). In effect, Eq. (\ref{mirror}) 
interchanges the dilaton and axion fields with the components 
of the Einstein frame metric on the surfaces of orthogonality. The 
axion field is interchanged with the off--diagonal component, $\omega$, 
in Eq. (\ref{defineL}) and the dilaton field with the function $p$. 

We now outline how 
Eq. (\ref{mirror}) may be applied together with the 
${\rm SL}(2,R)$ and ${\rm O}(2,2)$ symmetries  to
derive inequivalent $G_2$ string cosmologies from dilaton--vacuum solutions
\cite{Lidsey98}. 
In some sense, these new backgrounds represent 
the `mirror images' of the former. For example, 
one may begin with the diagonal, dilaton--vacuum solution of the 
generic form given by Eq. (\ref{seedmetric}). 
The axion field is related to the off--diagonal component of the 
transverse metric. Such a term may be generated 
from a diagonal metric by employing the variety of techniques cited above, 
including the ${\rm SL}(2, R)$ 
transformation (\ref{tautransformation}). The simplest 
method for introducing an off--diagonal term  in the metric 
is to perform an ${\rm SL}(2, R)$ transformation  in the two--space 
of the Killing vectors,  $\partial / \partial x^a$: 
\be
\label{Ehlers}
\bar{L}= \Theta L \Theta^T , \qquad 
\Theta \equiv \left( \begin{array}{cc}
D & C \\
B & A \end{array} \right)   ,
\ee
where $AD -BC =1$ and all other variables are invariant 
\cite{Ehlers57,Griffiths91}. The transverse metric (\ref{defineL}) 
then transforms to  
\bea
\label{Eh1}
e^{\bar{p}} = C^2 e^{-p} +D^2 e^{p} \\
\label{Eh2}
\bar{\omega} = \frac{AC e^{-p} + BD e^{p}}{C^2 e^{-p} +D^2 e^{p}}  .
\eea
Eq. (\ref{Ehlers}) does not commute with the discrete 
transformation (\ref{mirror}) and the two may be employed together 
to generate the axion field. 
Application of Eq. (\ref{mirror1}) 
then implies that the mirror image will correspond to a string 
cosmology with a non--trivial axion field and a diagonal 
transverse metric.  

The same procedure applies when determining the mirror images 
of {\em any} vacuum, $G_2$ background of the 
form given in Eq. (\ref{diagG2}). In these cases, the new 
spacetime will have no preferred direction in the transverse space 
because $\bar{p} =\bar{\omega} =0$. This will correspond to a  
LRS,  $G_2$ cosmology, where there exists a 
one--parameter isotropy group in addition to the $G_2$ 
abelian isometry group. 
Alternatively, one may 
begin with the dilaton--axion solutions generated 
from the global ${\rm SL}(2,R)$ transformations 
(\ref{tautrans}). The mirror solution will 
correspond to a non--diagonal, dilaton--vacuum solution. A new 
axion field may be then be excited by a further 
${\rm SL}(2,R)$ transformation (\ref{tautrans}). 

We now discuss 
an interesting application of this $Z_2$ transformation. 
Let us first consider the mirror image of the vacuum Bianchi type I cosmology. 
The general, spatially homogeneous vacuum solution  
with a simply transitive Lie group $G_3 = \Re^3$ is 
the type I Kasner metric \cite{Kasner25}: 
\be
\label{Kasnermetric} 
ds^2 =-dt^2 +t^{2p_1} dx^2 + t^{2p_2} dy^2 + t^{2p_3} dz^2    , 
\ee
where $\sum_{i=1}^3 p_i =\sum_{i=1}^3 p_i^2 =1$ (see Section 
\ref{Section6.1}). 
Applying Eqs. (\ref{Ehlers}) and Eq. (\ref{mirror1}) yields the LRS type I 
solitonic string cosmology: 
\bea
\label{mirrorI}
ds^2 = -dt^2 +t^{2p_1}dx^2 +t^{p_2+p_3} \left( dy^2 + 
dz^2 \right) \nonumber \\
\Phi = \ln \left[ C^2 t^{p_3 -p_2} +D^2 t^{p_2 -p_3} \right]
\nonumber \\
\sigma = \frac{ACt^{p_3-p_2} +BDt^{p_2-p_3}}{C^2t^{p_3-p_2} 
+D^2t^{p_2-p_3}}   
\eea
and, when $p_2 = (1 -\sqrt{3})/3$ and $p_3 =
(1+\sqrt{3})/3$, 
this reduces to the general, spatially flat FRW string cosmology
presented in Section \ref{Section7.1} \cite{CopLahWan94}. 

The mirror symmetry 
(\ref{mirror}) provides a remarkably simple way of deriving 
this and other important string cosmologies. 
For example, we may consider 
the mirror image of a Bianchi type V solution. The Bianchi type V model 
represents one of the simplest 
homogeneous, spatially anisotropic models. It may 
be viewed as the anisotropic generalization of the 
negatively curved FRW universe and reduces to it 
in the isotropic limit \cite{MacCallum79,RyaShe75}. The one--forms for 
this model are given in Table \ref{bianchiforms}. 

A vacuum solution  $(R_{\mu\nu}=0)$ was found 
by Joseph \cite{Joseph66}:
\begin{equation}
\label{joseph}
ds^2 =\sinh 2t \left( -dt^2 +dx^2  
+e^{-2x} \left[ \left( {\rm tanh} t \right)^{\sqrt{3}} 
dy^2 + \left( {\rm tanh} t \right)^{-\sqrt{3}} dz^2 
\right] \right)  
\end{equation}
and we may derive the mirror image of this vacuum universe by applying 
the transformations (\ref{Ehlers}) and (\ref{mirror}). It is given by 
\cite{Lidsey98}
\begin{eqnarray}
\label{FRW-1}
ds^2 = \sinh 2t \left[ -dt^2 +dx^2 +e^{-2x}\left( dy^2 +dz^2 \right) 
\right] \nonumber \\
\varphi = \ln \left[ C^2 \left( {\rm tanh} t \right)^{-\sqrt{3}} 
+ D^2 \left( {\rm tanh} t \right)^{\sqrt{3}} \right]  \nonumber \\
\sigma = \frac{AC \left( {\rm tanh} t \right)^{-\sqrt{3}} +BD 
\left( {\rm tanh} t \right)^{\sqrt{3}}}{C^2 \left( 
{\rm tanh} t \right)^{-\sqrt{3}} +D^2 \left( 
{\rm tanh} t \right)^{\sqrt{3}}}    .
\end{eqnarray}
It can be shown after an appropriate redefinition of the variables 
that Eq. (\ref{FRW-1}) corresponds precisely to 
the general form of the negatively curved FRW string 
cosmology that we discussed in Section \ref{Section7.1} 
\cite{CopLahWan94,Lidsey98}. 
This highlights the important point that 
the mirror image has a different spacetime interpretation to that 
of the seed solution. In particular, an anisotropic 
cosmology is mapped onto a spatially isotropic cosmology by the 
transformation.

\subsection{Tilted String Cosmology}

\label{Section7.5}

We conclude this Section by applying the techniques discussed above 
to derive a class of models that exhibit some interesting 
properties. We begin our discussion with the  
Bianchi type ${\rm VI}_h$ vacuum cosmology found by 
Ellis and MacCallum \cite{EllMac69}: 
\begin{equation}
\label{emtypeVI}
ds^2 = \sinh 2t \left[ A^b \left( -dt^2 +dx^2 \right)
+A e^{2(1+b)x} dy^2 +A^{-1} e^{2(1-b)x} dz^2 \right]   ,
\end{equation}
where 
\be
A \equiv \left( \sinh 2t \right)^b  \left( {\rm tanh} t 
\right)^{\sqrt{3+b^2}}
\ee
and $b^2 \equiv -1/h$. This reduces  to the 
Joseph type V solution (\ref{joseph}) when $b=0$ \cite{Joseph66}. 
The mirror image of the Ellis--MacCallum type ${\rm VI}_h$ 
cosmology is given immediately 
by 
\begin{equation}
\label{mirror6}
ds^2 =\sinh 2t \left[ A^b \left( -dt^2 +dx^2 \right) +e^{2x} \left( dy^2 
+dz^2 \right) \right]  ,
\end{equation}
where the dilaton and axion fields 
are given by the right hand sides of Eqs. (\ref{Eh1}) and 
(\ref{Eh2}), respectively, with  
\begin{equation}
\label{pVI}
e^p =\left( \sinh 2t \right)^b \left( {\rm tanh} t \right)^{\sqrt{3+b^2}} 
e^{2bx}   .
\end{equation}

A calculation of the structure 
constants of the isometry group of 
Eq. (\ref{mirror6}) implies that it is a Bianchi type V 
cosmology. As in the above examples, the geometry of the two  
spacetimes is different. Moreover, the metric exhibits a one--parameter 
isotropy group, since the transverse space is isotropic, and 
this implies that Eq. (\ref{mirror6}) corresponds to a LRS type V 
cosmology. However, the interesting feature 
of the solution (\ref{mirror6})--(\ref{pVI}) is that 
the dilaton field has a direct dependence on one 
of the spatial variables. Consequently, it is 
not constant on the surfaces of homogeneity, $\Sigma_t$, and 
the solution is therefore not contained within the class of models 
discussed in Section \ref{Section7.2}. 

We have discussed previously how dilaton--axion solutions 
may be interpreted  in the Einstein frame as stiff 
perfect fluid models. When the axion field is trivial, for example, 
spatial homogeneity implies that the energy--momentum 
tensor of the dilaton field can depend only on time. In particular, 
its energy density must be constant on $\Sigma_t$. However, this does 
not necessarily imply that the dilaton field itself should 
be constant on these surfaces. If it is, then 
the fluid flow is orthogonal to $\Sigma_t$. 
If, on the other hand, the dilaton is not constant on these surfaces and 
exhibits an appropriate spatial dependence, the fluid flow will not 
be orthogonal to 
$\Sigma_t$. In this case, the model is said to 
be {\em tilted} \cite{KinEll73}. A tilted model is spatially homogeneous 
as seen by an observer with  a world line orthogonal to 
$\Sigma_t$, but appears inhomogeneous to an observer comoving with the fluid
\cite{ColEll79}. 

The solution (\ref{mirror6}) may therefore be interpreted as a tilted 
LRS Bianchi type V
string cosmology. A more general tilted type V string cosmology 
that does not exhibit local rotational symmetry may be generated 
from the duality symmetries of Section \ref{Section7.3} \cite{ClaFeiLid99a}. 
A tilted type V stiff perfect fluid solution to Einstein 
gravity was found by Maartens 
and Nel \cite{MaaNel76}, and discussed further by Wainwright {\em et al.} 
\cite{WaiIncMar79}. The metric is given by 
\begin{equation}
\label{mnV}
ds^2_e =e^f \left( -dt^2 +dx^2 \right) +e^{-2x} \sinh 2t 
\left( e^p dy^2 +e^{-p} dz^2 \right)
\end{equation}
where the longitudinal and transverse components 
of the metric have the form 
\begin{eqnarray}
\label{fV}
f =\frac{1}{2} \left( m^2 +\alpha^2 +\beta^2 -1 \right) \ln \sinh 
2t +\alpha \beta \tanh t \\
\label{pV}
p = m \ln \tanh t
\end{eqnarray} 
respectively, and the constants $\{ \alpha , \beta , m \}$ 
satisfy the constraint
\begin{equation}
\beta^2 -\alpha^2 +m^2 =3
\end{equation}

The fluid four--vector is given by Eq. (\ref{fluid}), where the 
scalar quantity, $\varphi$, has the form 
\begin{equation}
\label{varphifluid}
\varphi = -2\alpha x +\alpha \ln \sinh  2t +\beta \ln \tanh t
\end{equation}
The metric (\ref{mnV}) and scalar (\ref{varphifluid})
may be interpreted in our context 
as the Einstein frame metric and dilaton field, respectively. 
The fluid flow is not orthogonal to the group orbits if $\alpha \ne 0$ and 
the cosmology is in general tilted. It is orthogonal 
when $\alpha =0$ and reduces to the Joseph vacuum 
solution (\ref{joseph}) when $\alpha =\beta =0$. 

We have written the solution 
(\ref{mnV})--(\ref{varphifluid}) in such a way that 
the $G_2$ group of abelian isometries is manifest. 
We may therefore generate a new solution 
from this seed by application of the global ${\rm O}(2,2)$ symmetry 
(\ref{G2odd1})--(\ref{G2odd3}). We consider the
${\rm SL}(2,R)$ transformation (\ref{rhotransformation}). 
The dual metrics in the string-- and Einstein 
frames are 
\begin{equation}
\label{stringdualV}
d\bar{s}^2_s =e^{f+\varphi} \left( -dt^2 +
dx^2 \right) +\frac{\Gamma^{1/2}}{d^2 +c^2\Gamma} 
\left( e^p dy^2 +e^{-p} dz^2 \right) 
\end{equation}
and 
\begin{equation}
\label{einsteindualV}
d\bar{s}^2_e =e^f \left( d^2 +c^2 \Gamma \right)
\left( -dt^2 +dx^2 \right) + \Gamma^{1/2} e^{-\varphi} 
\left( e^p dy^2 +e^{-p} dz^2 \right)    ,
\end{equation}
respectively, where 
\begin{equation}
\label{Gammaform}
\Gamma = e^{2\varphi -4x} \sinh^2 2t
\end{equation}
and $f$ and $\varphi$ are given by Eqs. (\ref{fV}) 
and (\ref{varphifluid}). The two--form 
potential and dilaton field are given by Eqs. (\ref{tdual2}) and 
(\ref{tdual3}). 

In general, the dual solution (\ref{einsteindualV}) 
does not preserve the one--forms of the Bianchi type V 
metric. Two of the one--forms are left invariant, but 
the third, given by $dx$, can not be multiplied 
by a function of $x$ if the spacetime is to remain 
homogeneous. Since the longitudinal 
component of the metric (\ref{einsteindualV}) is 
$x$--dependent, homogeneity is broken along this direction 
and the solution therefore represents an inhomogeneous $G_2$ cosmology. 
In the limit where $\Gamma \rightarrow 0$, 
however, the original homogeneous solution 
is recovered for $d\ne 0$.

On the other hand, the determinant $\Gamma$ is independent 
of the spatial coordinates when $\alpha =-1$. In this 
case, we have $\Gamma = {\rm tanh}^{2\beta} t$ and the 
Einstein--frame metric is of the Bianchi 
type V. The model is tilted because the dilaton has a linear dependence 
on the spatial variable $x$. This 
implies that the string--frame metric (\ref{stringdualV}) 
is inhomogeneous, because the longitudinal component of this 
metric acquires a spatial dependence from the dilaton field. 

The late--time behaviour of the string cosmology (\ref{stringdualV}) 
is particularly interesting. In the limit that $t \rightarrow 
+\infty$, $\Gamma \rightarrow 1$. This implies that 
the two--form potential asymptotically tends to a constant value 
and the dilaton tends to its original form, modulo 
an arbitrary constant shift in its value. 
We may specify $d^2+c^2\Gamma =1$ without loss of generality and 
the original seed solution (\ref{mnV}) is therefore recovered 
in this limit. 

It is instructive to define a new coordinate pair:
\begin{eqnarray}
\label{uV}
u \equiv \frac{1}{2} e^{2(t-x)} \nonumber \\
v \equiv \frac{1}{8} e^{2(t+x)} \nonumber 
\end{eqnarray}
The line element (\ref{mnV}) then tends to the Rosen 
form of a plane wave \cite{Griffiths91}: 
\begin{equation}
\label{rosenwave}
ds^2_e =-dudv +u \left( dx^2 +dy^2 \right)
\end{equation}
after a trivial rescaling of the transverse coordinates and 
the dilaton field becomes null: 
\begin{equation}
\label{dilatonwave}
\varphi =\varphi_0 -\ln u
\end{equation}
where $\varphi_0$ is a constant. 

The dilaton field (\ref{dilatonwave}) and metric 
(\ref{rosenwave}) are independent 
of the variable $v$ and the solution admits a covariantly constant, null 
Killing vector field, $l^{\mu}$, such that $l_{\mu}l^{\mu} =
\nabla_{\mu} l_{\nu} =0$. Since the gradient of the dilaton field 
is proportional to $l_{\mu}$, the one--loop 
$\beta$--function equations reduce to the single constraint 
$R_{uu} =\left( \partial_{\mu} \varphi \right)^2 /2$ when expressed in terms 
of the Einstein--frame variables. It can be verified by direct 
substitution that this constraint is identically satisfied. Thus, the 
asymptotic form of the tilted solution in the late--time limit is 
itself an exact solution to the one--loop $\beta$--function equations. 

Moreover, all higher--order terms in the $\sigma$--model perturbation 
theory are second--rank tensors constructed 
from powers of the Riemann tensor, the gradient of the dilaton 
field, the metric, the three--form field strength and their 
covariant derivatives \cite{GreSchWit87,Polchinski98}. 
It can be shown by applying a geometrical argument 
due to Horowitz and Steif \cite{HorSte90} that a plane--wave background of the 
form (\ref{dilatonwave}) and (\ref{rosenwave}) acquires no 
$\alpha'$--corrections 
because the Riemann tensor is orthogonal to  $l^{\mu}$ and $\nabla^{\mu}$
on all its indices and because $\nabla_{\mu} \varphi$ 
is proportional to  $l_{\mu}$. 
Consequently, the asymptotic solution (\ref{dilatonwave}) 
and (\ref{rosenwave}) represents an {\em exact} 
solution to the classical string equations of motion to 
{\em all} orders in the inverse string tension.

\section{Type IIB Cosmology with Ramond--Ramond Fields}
\label{SectionIIB}

\label{Section8}

\setcounter{equation}{0}

\def\theequation{\thesection.\arabic{equation}}

The symmetries of the type II string theories 
have a number of important applications in cosmology. In particular,  
they may be employed to investigate the effects of RR fields 
on the classical dynamics of the universe and the generation 
of primordial density perturbations. 
In Section \ref{Section5.3} 
we discussed how the equations of motion of $N=2$,
$D=10$ type IIB supergravity are symmetric under
global ${\rm SL}(2,R)$ transformations \cite{SchWes83,Hull95}. These 
relate the strong-- and weak--coupling
regimes of the theory and interchange the NS--NS and RR 
charges. 

Moreover, type IIA and IIB supergravity theories are equivalent after toroidal
compactification. Compactification of the low-energy
effective type IIB action on a six--torus results in $N=8$
supergravity, which may also be derived by toroidally compactifying
$N=1$, $D=11$ supergravity on a seven--torus \cite{CreJul78,CreJul79}. 
This theory contains seventy scalar and
pseudo--scalar fields that parametrize the ${\rm E}_{7(7)}/[{\rm
SU}(8)/{\rm Z}_2]$ coset and the field equations are invariant under
the global action of the group ${\rm E}_{7(7)}$ \cite{CreJul79}. The discrete
subgroup ${\rm E}_{7(7)}({\rm Z})$ is the conjectured U--duality of
the type II superstring and contains the T--duality group ${\rm
O}(6,6;{\rm Z})$ and S--duality group ${\rm SL}(2,Z)$ \cite{HulTow95}.

One of the main purposes of the present Section 
is to consider the cosmology of the four--dimensional 
action (\ref{4daction}). This action was derived 
by compactifying the ten--dimensional theory 
(\ref{IIB}) on an isotropic six torus under the 
assumption that all fields are independent of the 
internal coordinates. The vector fields arising from 
the metric components and the moduli 
originating from the compactification of the form fields
were neglected. In effect, we reduced each form field 
to a single degree of freedom. 

This action may be viewed as 
a truncated sector of the $D=4$, $N=8$ 
supergravity theory. The motivation for 
considering such an action is that it
maintains enough 
simplicity that exact solutions to the field 
equations can be found analytically 
and analyzed in detail, but is sufficiently general that 
all the non-trivial 
interactions between the NS--NS and RR form fields on the external 
four--dimensional spacetime are preserved. 

We draw together much of the 
material discussed in previous Sections. 
Our approach in establishing the generic features of 
cosmologies of the type IIB theory is to employ the global 
symmetries exhibited by the field equations. This 
represents an extension of the method employed in Section \ref{Section7.1}
to derive the general solitonic FRW string cosmologies 
with a non--trivial NS--NS axion field. The symmetries of the 
field equations derived from 
action (\ref{4daction}) become manifest in the Einstein 
frame after the three--form field strengths have been dualised 
to one--forms. 
The dual action is given by Eq. (\ref{solitonicaction}).
We show that the five scalar and pseudo--scalar `axion' fields
parametrize the ${\rm SL}(3,R)/{\rm SO}(3)$ coset and 
this action therefore exhibits a non--compact, global ${\rm SL}(3,R)$ 
symmetry. The global
${\rm SL}(2,R)$ symmetry of the $D=10$ supergravity action is preserved in
four dimensions as a subgroup of this ${\rm SL}(3,R)$ symmetry. A discrete 
${\rm Z}_2$ symmetry can also be identified as 
a subgroup of the full O(6,6;Z)
T--duality, which leads to a further ${\rm SL}(2,R)$ symmetry that may be
viewed as a `mirror' image of the original ${\rm SL}(2,R)$ symmetry.
The ${\rm SL}(3,R)$ symmetry corresponds to a subgroup of the 
(continuous) U--duality group ${\rm E}_{7(7)}$ 
of $N=8$ supergravity. Consequently, the 
concepts of U--duality may be applied in a cosmological 
setting. 

\subsection{Global Symmetries}

\label{Section8.1}

\subsubsection{SL(3,R) symmetry}

The effective action (\ref{solitonicaction}) can be expressed as 
a four--dimensional, non--linear sigma--model in the form given 
by Eq. (\ref{sigmamodel}). We write it here as 
\begin{equation}
\label{nonlinsigma}
S_{4*} = \int d^4 x \sqrt{-g} \left[ R - \gamma_{ij} (\phi ) 
\nabla \phi^i \nabla \phi^j \right]   ,
\end{equation}
where the scalar fields $\phi^i = (\varphi , \beta ,\chi ,
\sigma_1 ,\sigma_2 )$, $(i,j)=(1,2, \dots ,  5)$
may be viewed as coordinates on the target space with metric
\begin{equation}
\label{tsmetric}
ds^2_{\rm target} = 
\frac{1}{2} d\varphi^2 + \frac{1}{2} d\beta^2 +\frac{1}{2} 
e^{2\varphi} \left( d\sigma_1 -\chi d\sigma_2 \right)^2 +
\frac{1}{2} e^{\varphi} \left[ e^{\sqrt{3} \beta} d\chi^2 
+e^{-\sqrt{3} \beta} d\sigma_2^2 \right]  .
\end{equation}

Eq.~(\ref{tsmetric}) is formally identical to the target space
considered by Gal'tsov, Garcia and Kechkin within the context of
five--dimensional 
Kaluza--Klein theory admitting two commuting Killing vectors
\cite{GalGarKec95}}. Maison first showed that this target space represents the
${\rm SL}(3,R)/{\rm SO}(3)$ coset corresponding to a homogeneous symmetric
Riemannian space, where the group ${\rm SL}(3,R)$ acts transitively
\cite{Maison79}. It can be shown by employing the Gauss decomposition of
the general ${\rm SL}(3,R)$ matrix that the action (\ref{solitonicaction}) may
be written in the form~\cite{GalGarKec95}
\begin{equation}
\label{Uaction}
S_{4*} =\int d^4 x \sqrt{-g} \left[ 
R +\frac{1}{4} {\rm Tr} \left[ \nabla U \nabla U^{-1} 
\right] \right] ,
\end{equation}
where 
\begin{equation}
\label{S}
U \equiv  e^{\varphi+\beta /\sqrt{3}} 
\left( \begin{array}{ccc} 
1 & \chi & \sigma_1 -\chi \sigma_2  \\ 
\chi & \chi^2 +e^{-\varphi-\sqrt{3}\beta} & \chi ( \sigma_1 -\chi \sigma_2 )
- \sigma_2 e^{-\varphi-\sqrt{3}\beta} \\
\sigma_1 -\chi \sigma_2 & \chi ( \sigma_1 -\chi \sigma_2 )
- \sigma_2 e^{-\varphi-\sqrt{3}\beta} & ( \sigma_1 -\chi \sigma_2 )^2
 + \sigma_2^2 e^{-\varphi-\sqrt{3}\beta} + e^{-2\varphi}
\end{array} \right)
\end{equation}
is a symmetric ${\rm SL}(3,R)$ matrix.

We may conclude, therefore, that the dual action
(\ref{solitonicaction}) is invariant under global ${\rm SL}(3,R)$
transformations. These correspond to U--duality transformations 
because they relate the modulus field that 
arises from the toroidal compactification with the 
four-dimensional string coupling (dilaton field). 
We now consider the relevant ${\rm SL}(2,R)$ subgroups that
prove useful in generating 
cosmological solutions with non--trivial RR and NS--NS
fields.

\subsubsection{SL(2,R)$_\chi$ symmetry}

The effective 
four--dimensional action (\ref{solitonicaction}) still exhibits the global
${\rm SL}(2,R)$ symmetry of the full ten--dimensional action manifest in
Eq.~(\ref{IIB}) ~\cite{Maharana97,Roy98}. This becomes apparent by defining  
new scalar fields: 
\begin{eqnarray}
\label{ufield}
{1\over2}\Phi \equiv u \equiv \frac{1}{2} \varphi + {\sqrt{3}\over2} 
\beta \\
\label{vfield}
v \equiv \frac{\sqrt{3}}{2} \varphi - {1\over2} \beta  .
\end{eqnarray}
The action given in Eq.~(\ref{solitonicaction}) then takes the form
\begin{eqnarray}
\label{uaction}
S_{4*}&=&\int d^4 x \sqrt{-g} \left[ R - \frac{1}{2} \left( \nabla u \right)^2 
-\frac{1}{2} e^{2u} \left( \nabla \chi \right)^2 - 
\frac{1}{2} \left( \nabla v \right)^2
 \right. \nonumber \\ 
&& \left. \qquad
 - \frac{1}{2} 
e^{\sqrt{3} v} \left( e^{-u} (\nabla\sigma_2)^2 + e^{u} ( \chi\nabla\sigma_2 -
\nabla\sigma_1 )^2 
\right) \right]   .
\end{eqnarray}

The ${\rm SL}(3,R)$ matrix $U$ given in Eq.~(\ref{S}) can be written as
\begin{equation}
U = \left( 
\begin{array}{cc}
e^{v/\sqrt{3}}M & -e^{v/\sqrt{3}} M\sigma \\
-e^{v/\sqrt{3}} \sigma^T M & e^{-2v/\sqrt{3}} + e^{v/\sqrt{3}}
\sigma^TM\sigma 
\end{array}
\right) \ ,
\end{equation}
where the symmetric $2 \times 2$ matrix $M$ is given in
Eq.~(\ref{matrixM}), and we have defined the vector
\begin{equation}  
\label{sigmavector}
\sigma \equiv \left( \begin{array}{c} -\sigma_1 \\ 
  \sigma_2 \end{array} \right) \ .
\end{equation}
This implies that Eq. (\ref{Uaction}) may be written as
\begin{equation}
\label{dualSLchiaction}
S_{4*}=\int d^4 x \sqrt{-g} \left[ R + \frac{1}{4} {\rm Tr} 
\left[ \nabla M \nabla M^{-1} \right] -\frac{1}{2} \left( \nabla 
v \right)^2
 -\frac{1}{2} e^{\sqrt{3} v} \nabla\sigma^T M \nabla\sigma \right]  .
\end{equation}

The action remains invariant under the sub-group
\begin{equation}
\bar{U} = \tilde\Sigma_\chi U \tilde\Sigma_\chi \ ,
\end{equation}
where
\begin{equation}
\tilde\Sigma_\chi = \left(
\begin{array}{cc}
\Sigma & 0 \\
0 & 1
\end{array}
\right) 
\end{equation}
and 
\begin{equation}
\Sigma \equiv \left( \begin{array}{cc}
D & C \\
B & A \end{array} \right) , \qquad AD-BC=1 .
\end{equation}
This corresponds to the ${\rm SL}(2,R)$ transformation
\begin{equation}
\label{sl2rchi}
\bar{M} = \Sigma M \Sigma^T, \qquad \bar{g}_{\mu\nu} = 
g_{\mu\nu} , \qquad \bar{\sigma} = \left( \Sigma^T \right)^{-1} \sigma, 
\qquad \bar{v} =v .
\end{equation}

The transformation (\ref{sl2rchi}) acts non--linearly on the scalar fields
$u$ and $\chi$: 
\begin{eqnarray}
\label{newu}
e^{\bar{u}} = C^2 e^{-u} + (D +C \chi )^2 e^u \\
\label{newchi}
\bar{\chi} e^{\bar{u}} = AC e^{-u} + (B + A \chi )(D + C \chi ) e^u 
\end{eqnarray}
and the pseudo--scalar axion fields transform as
\begin{eqnarray}
\bar\sigma_1 = A\sigma_1 + B\sigma_2 \\
\label{endsl2rchi}
\bar\sigma_2 = C\sigma_1 + D\sigma_2  .
\end{eqnarray}

When $C=-B =-1$ and $A=D=0$, Eq. (\ref{sl2rchi}) interchanges the two
axion fields, $\sigma_i$, and inverts the 
ten--dimensional string coupling
$\bar{g}_s =\exp( {\bar{\Phi}} ) = g_s^{-1} = \exp ( -\Phi )$ that is
defined in terms of the 
ten--dimensional dilaton, $\Phi=2u$.  Thus, the
strongly--coupled regime of the theory is mapped onto the
weakly--coupled one, and vice--versa. The
effective four--dimensional dilaton field (\ref{shifteddilaton})
transforms as $\bar{\varphi}=-\varphi/2+\sqrt{3}v/4$. 

We refer to this as the ${\rm SL}(2,R)_{\chi}$ 
symmetry. It is the ${\rm SL}(2,R)$
symmetry of the ten--dimensional theory written in terms of the 
four--dimensional variables~\cite{Maharana97,Roy98}. 
The field $v$ determines the conformal factor 
\begin{equation}
\label{10Dfactor}
\Omega^2=e^{\Phi/4-\varphi}=e^{\sqrt{3}(\beta -\sqrt{3}\varphi)/4}  
\end{equation}
that relates the four--dimensional Einstein metric to the
corresponding part of the ten--dimensional Einstein metric (\ref{conften}). 
The
invariance of both $v$ and the four--dimensional metric $g_{\mu\nu}$ 
implies that the corresponding components,
$\hat{g}_{\mu\nu}=e^{\sqrt{3}v/2}g_{\mu\nu}$, of the ten--dimensional Einstein
metric are also invariant. The radius of the internal
space in the ten--dimensional Einstein frame is 
$e^{-v/\sqrt{3}}$ and the complete ten--dimensional Einstein metric
(\ref{conften}) is therefore invariant, as in Eq.~(\ref{schwarz}).

Finally, we note that although a general ${\rm SL}(2,R)$ matrix of the form
given in Eq.~(\ref{sigma}) has three independent real parameters,
there is a two-dimensional sub-group,
\begin{equation}
\label{sigma2d}
\Sigma_0 \equiv \left( \begin{array}{cc} A^{-1} & 0 \\ 
B  & A \end{array} \right) ,
\end{equation}
which leaves the Lagrangian (\ref{solitonicaction}) invariant term by
term. These transformations correspond either to a constant shift or
rescaling of the axion fields, such that the three
four-vectors, $e^{-\varphi}K_\mu$, 
$e^{(-\varphi+\sqrt{3} \beta )/2}(J_\mu+\chi K_\mu)$ and
$e^{(-\varphi-\sqrt{3} \beta )/2}(L_\mu-\sigma_2K_\mu)$ remain invariant.
Thus the only non-trivial transformation is the ``boost''
\begin{equation}
\label{sigma1d}
\Sigma_1 \equiv \left( \begin{array}{cc} \cosh\theta & \sinh\theta \\ 
\sinh\theta  & \cosh\theta \end{array} \right) ,
\end{equation}
which introduces at most one new parameter $\theta$.

We can employ the ${\rm SL}(2,R)_{\rm \chi}$ symmetry of the action to
generate new four-dimensional solutions of the field equations.  For
example, in Section \ref{Section8.2.1} we derive the homogeneous
cosmological solutions with non-trivial $\chi$ field by applying the
transformation given in Eq.~(\ref{sl2rchi}) to the homogeneous
dilaton-moduli-vacuum solutions discussed in Section \ref{Section7.1}.

\subsubsection{Z$_2$ and SL(2,R)$_{\sigma_2}$ symmetry}

The importance of the dual action 
(\ref{solitonicaction}) is that a further ${\rm SL}(2,R)$ symmetry
may be uncovered. The NS--NS sector of the reduced
action~(\ref{4daction}) is invariant under the `T--duality'
transformation $\bar{\beta}=-\beta$, corresponding to an inversion of the
internal space. This ${\rm Z}_2$ symmetry can be extended to the RR
sector of the theory and the dual action~(\ref{solitonicaction}) is
symmetric under the discrete transformation
\begin{equation} 
\label{td}
\bar{\beta} =-\beta , \qquad \bar{\chi} = \sigma_2 , \qquad 
\bar{\sigma}_2 = \chi , \qquad \bar{\sigma}_1 =-\sigma_1+\chi
\sigma_2 ,
\end{equation}
where the four--dimensional dilaton, $\varphi$, and 
four--dimensional Einstein frame metric
remain invariant. Note, however, that because the volume of the internal
space changes, the ten--dimensional Einstein frame metric 
(related to the four--dimensional Einstein
frame metric by the conformal factor given in Eq.~(\ref{10Dfactor})) is {\em
not} invariant under this transformation.
In terms of the conserved axion currents, defined in Eqs.~(\ref{Kcurrent}), 
(\ref{Jcurrent}) and~(\ref{Lcurrent}), the
reflection symmetry (\ref{td}) corresponds to
\begin{equation}
\bar{\beta} =-\beta , \qquad \bar{K}_\mu = -K_\mu, \qquad 
\bar{J}_\mu = L_\mu , \qquad \bar{L}_\mu = J_\mu .
\end{equation}

This reflection symmetry implies the existence of an
alternative ${\rm SL}(2,R)$ symmetry in the dual action which can be obtained
from a combination of the ${\rm SL}(2,R)_{\chi}$ transformation given in
Eq.~(\ref{sl2rchi}) plus the reflection symmetry in Eq.~(\ref{td}).
Analogously to Eqs.~(\ref{ufield}) and~(\ref{vfield}), but with
$\beta \to -\beta $, we introduce the new scalar fields:
\begin{eqnarray}
\label{wfield}
w \equiv \frac{1}{2} \varphi - {\sqrt{3}\over2} \beta \\
\label{xfield}
x \equiv \frac{\sqrt{3}}{2} \varphi + {1\over2} \beta  .
\end{eqnarray}
The dual effective action, 
Eq.~(\ref{solitonicaction}), then takes the form 
\begin{eqnarray}
\label{waction}
S_{4*} &=& 
 \int d^4 x \sqrt{-g} \left[ R - \frac{1}{2} \left( \nabla w \right)^2 
-\frac{1}{2} e^{2w} \left( \nabla \sigma_2 \right)^2 - 
\frac{1}{2} \left( \nabla x \right)^2
 \right. \nonumber \\
&& \left. \qquad
 -\frac{1}{2} e^{\sqrt{3}x} \left( 
e^{w} \left( \nabla \sigma_1 -\chi \nabla \sigma_2 \right)^2 
+e^{-w} \left( \nabla \chi \right)^2 \right) \right]  .
\end{eqnarray}
Defining the symmetric $2 \times 2$ matrix: 
\begin{equation}
\label{Nmatrix}
P \equiv   \left( \begin{array}{cc} e^{w} &  \sigma_2 
e^{w} \\ 
\sigma_2  e^{w} & e^{-w} +
\sigma_2^2 e^{w} \end{array} \right)
\end{equation}
and the vector:
\begin{equation}
\tau \equiv \left( \begin{array}{c} \sigma_1-\chi\sigma_2 \\ 
   \chi \end{array} \right) 
\end{equation}
allows us to express this action as 
\begin{equation}
\label{wacions}
S_{4*} = \int d^4 x \sqrt{-g} \left[ R +\frac{1}{4} {\rm Tr} 
\left[ \nabla P \nabla P^{-1} \right] -\frac{1}{2} \left( \nabla 
x \right)^2
 -\frac{1}{2} e^{\sqrt{3} x} \nabla\tau^T P \nabla\tau \right]  .
\end{equation}
This is manifestly invariant under the  ${\rm SL}(2,R)$ transformation
\begin{equation}
\label{sl2rsigma2}
\bar{P} = \Sigma' P \Sigma'^T, \qquad \bar{g}_{\mu\nu} = 
g_{\mu\nu} , \qquad \bar\tau = \left( \Sigma'^T \right)^{-1} \tau
\qquad \bar{x} = x 
\end{equation}
and this implies that
\begin{eqnarray}
\label{neww}
e^{\bar{w}} &=& C'^2 e^{-w} + (D' +C' \sigma_2 )^2 e^{w} \\
\label{newsigma2}
\bar{\sigma_2} e^{\bar{w}} &=& (B' + A' \sigma_2 )(D' + C' \sigma_2 ) e^{w} 
+A'C' e^{-w} \\ 
\label{chisigma2}
\bar\chi &=& -C'(\sigma_1-\sigma_2\chi) + D'\chi\\
\label{sigma1sigma2}
\bar\sigma_1-\bar\chi\bar\sigma_2 &=& A'(\sigma_1-\sigma_2\chi) - B'\chi  .
\end{eqnarray}

We refer to this as the  ${\rm SL}(2,R)_{\sigma_2}$ symmetry of the action. 
It should be emphasised that this is not the ten--dimensional ${\rm
SL}(2,R)_{\chi}$ symmetry  recast in terms 
of the four--dimensional action.  The
${\rm SL}(2,R)_{\sigma_2}$ transformation mixes the $\sigma_2$ axion
field with $w$. This latter field is not the ten--dimensional dilaton,
because it includes an 
additional  contribution from the modulus field, $\beta$. Thus, 
the radius of the internal dimensions transforms non-trivially and the
ten--dimensional Einstein metric is not invariant under~(\ref{sl2rsigma2}). 

Comparison of Eq.~(\ref{ufield}) with Eq.~(\ref{wfield}) and
Eq.~(\ref{vfield}) with Eq.~(\ref{xfield}) implies that the discrete
transformation $\beta \leftrightarrow -\beta$ is equivalent to $u
\leftrightarrow w$ and $v \leftrightarrow x$. Moreover,
Eqs.~(\ref{matrixM}) and~(\ref{Nmatrix}) imply that the
reflection symmetry (\ref{td}) is equivalent to $M \leftrightarrow
P$. Thus, the ${\rm SL}(2,R)_{\sigma_2}$
symmetry transformation is formally equivalent to the ${\rm Z}_2$
transformation (\ref{td}), followed by the ${\rm SL}(2,R)_{\chi}$
transformation (\ref{sl2rchi}), followed by another ${\rm Z}_2$
transformation (\ref{td}).

It should be emphasized that 
neither of the symmetries (\ref{sl2rchi}) or (\ref{sl2rsigma2}) 
coincide with the ${\rm SL}(2,R)$ symmetry (\ref{sphimodel})--(\ref{ssigmamodel}) 
of the NS--NS sector alone. 
The latter mixes the four--dimensional dilaton and the
NS-NS axion \cite{ShaTriWil91,Sen93a}. This symmetry 
is broken due to the interaction of the RR fields. 

\subsection{Cosmological Models}

\label{Section8.2}

\subsubsection{Cosmology with a Single RR Field}

\label{Section8.2.1}

We now discuss the cosmology of RR fields. The field equations
with a single NS--NS or RR form--field can be directly
integrated to yield simple analytic
expressions \cite{CopLahWan94,CopLahWan95,LukOvrWal97a,LukOvrWal97b,LuMukPop97,LuMahMuk98,CopLidWan97}. Solutions with more than one
form--field have also been found where the system reduces to an
integrable Toda model 
\cite{LukOvrWal97a,LukOvrWal97b,LuMukPop97,LuMahMuk98,Kaloper97,CopLidWan98a,CopLidWan98b}. Including the
interactions between the NS--NS and RR form--fields on the external
space leads in general to a more complicated system. However, 
it is still integrable due to the symmetries between the
fields. The solutions presented in this subsection 
generalize those previously obtained by a
single ${\rm SL}(2,R)$ transformation acting on the NS--NS sector
solutions discussed in Section \ref{Section7.1} \cite{PopSch97}.

The equations of motion for 
the five scalar fields in the action (\ref{solitonicaction}) 
are given by Eqs. (\ref{fieldsigma1})--(\ref{fieldsigma2}) 
and by Eqs. (\ref{fieldphi})--(\ref{fieldchi}). 
We consider the FRW models with arbitrary 
spatial curvature defined by the line element 
(\ref{FRW}). For consistency we 
require the pseudo--scalar axion 
fields to be spatially independent and the cosmologies 
must therefore be solitonic. The field equations then reduce to 
\begin{eqnarray}
\label{field1}
\varphi'' + 2{\tilde{a}'\over \tilde{a}} \varphi' =
  {1\over2} e^{\sqrt{3} \beta +\varphi} \chi'^2
  + {1\over2} e^{-\sqrt{3} \beta +\varphi} \sigma_2'^2
\nonumber \\
+ e^{2\varphi} (\sigma_1'-\chi\sigma_2')^2 \\
\label{field2}
\beta'' + 2{\tilde{a}'\over \tilde{a}} \beta' =
  {\sqrt{3}\over2} e^{\sqrt{3}\beta +\varphi} \chi'^2
  - {\sqrt{3}\over2} e^{-\sqrt{3}\beta +\varphi} \sigma_2'^2 \\
\label{field3}
\chi'' + \left( 2{\tilde{a}'\over \tilde{a}} 
+\sqrt{3}\beta' + \varphi' \right) \chi' =
\nonumber \\
- e^{-\sqrt{3}\beta +\varphi} \sigma_2' (\sigma_1'-\chi\sigma_2') \\ 
\label{field4}
\sigma_2'' + \left( 2{\tilde{a}'\over \tilde{a}} 
-\sqrt{3}\beta' + \varphi' \right) \sigma_2'
 = e^{\sqrt{3}\beta +\varphi} \chi' (\sigma_1'-\chi\sigma_2') \\ 
\label{field5}
(\sigma_1'-\chi\sigma_2')' + 2\left( {\tilde{a}'\over 
\tilde{a}} + \varphi' \right)
  (\sigma_1'-\chi\sigma_2') = 0
\end{eqnarray}
together with the Friedmann constraint
\bea
\label{friedmann}
12 \left[ \left({\tilde{a}'\over \tilde{a}}\right)^2 + 
\kappa  \right]
 = \varphi'^2 + \beta'^2 + e^{\sqrt{3}\beta +\varphi}\chi'^2 \nonumber \\
+ e^{-\sqrt{3} \beta +\varphi}\sigma_2'^2 + e^{2\varphi}
 (\sigma_1'-\chi\sigma_2')^2    ,
\eea
where $\tilde{a} \equiv a e^{-\varphi /2}$ and $a$ represent 
the cosmological scale factors in the Einstein 
and string frames, respectively, and a prime denotes 
differentiation with respect to $\eta$. 

The general FRW dilaton--moduli--vacuum solution (with vanishing RR
fields and NS-NS three--form field strength) was presented 
in Eqs. (\ref{dila}--\ref{dilbeta}). 
The general FRW solution containing a single excited RR form
field can be generated from this solution by applying the ${\rm SL}(2,R)$
transformations in Eq.~(\ref{sl2rchi}) or Eq.~(\ref{sl2rsigma2})
\cite{CopLahWan94,CopLahWan95,LukOvrWal97a,LukOvrWal97b,LuMukPop97,LuMahMuk98,CopLidWan97}. We obtain
\bea
\label{GENERALA}
a^{2n} &=& {a_*^{2n}\over2}
 \left[ { (\tau/\tau_*)^{n(1+\sqrt{3}\cos\xi_1)}  
+ (\tau/\tau_*)^{n(1-\sqrt{3}\cos\xi_2)} \over (1+\kappa\tau^2)^n}
 \right] \\
\label{GENERALPHI}
e^{\varphi} &=& {e^{\varphi_*} \over 2^{1/n}}
\left[ (\tau/\tau_*)^{n\sqrt{3}\cos\xi_1} 
+ (\tau/\tau_*)^{-n\sqrt{3}\cos\xi_2}
 \right]^{1/n} \\
\label{GENERALY}
e^{\beta} &=& {e^{\beta_*}\over2^{1/m}} \left[ (\tau/\tau_*)^{m\sqrt{3}\sin\xi_1}  
+ (\tau/\tau_*)^{-m\sqrt{3}\sin\xi_2} 
\right]^{1/m} \\
\label{GENERALFORM}
\psi &=&\psi_{*} + K^{-1} \left[
\frac{ (\tau/\tau_*)^{n\sqrt{3}\cos\xi_1}
 - (\tau/\tau_*)^{-n\sqrt{3}\cos\xi_2}}
 {(\tau/\tau_*)^{n\sqrt{3}\cos\xi_1}
 + (\tau/\tau_*)^{-n\sqrt{3}\cos\xi_2} } \right]   ,
\eea
where the time parameter $\tau$ 
is defined in Eq. (\ref{wanmimtime}), 
$K=\pm e^{(\varphi_*/n)+(\beta_*/m)}$ and the field $\psi$ 
represents the field $\chi$ or $\sigma_2$,
depending up on which of these fields is excited. 
These solutions interpolate between two asymptotic regimes
where the form--fields vanish and the trajectories 
in $(\varphi ,\beta )$ space become straight lines~\cite{CopLahWan94}.  
If the asymptotic trajectory comes in at an initial angle $\xi_1$ to the
$\varphi$ axis, it leaves at an angle $\xi_2$. 
The values of the parameters $n$, $m$, $\xi_1$ and $\xi_2$ for
different choices of form--field are given in Table (\ref{table1}). Note 
that for each form--field there is a characteristic angle $\theta$ such 
that $1/n = \cos \theta$, $1/m = \sin \theta$ and 
$\xi_2 = 2\theta - \xi_1$.    

\begin{table}
\begin{center}
\begin{tabular}{||c|c|c|c|c||}
\hline \hline
$\psi$     & $1/n$   & $1/m$          &  $\xi_1$   &  $\xi_2$             \\
\hline \hline
$\chi$     & $1/2$   & $\sqrt{3}/2$   &  $\xi_*$   &   $(2\pi /3) -\xi_*$  \\
\hline
$\sigma_2$ & $1/2$   & $-\sqrt{3}/2$  &  $\xi_*$   &  $-(2\pi /3)-\xi_*$  \\
\hline
$\sigma_1$ & $1$     & $0$            &  $\xi_*$   &  $-\xi_*$         \\
\hline \hline
\end{tabular}
\end{center}
\caption[shortname]{Parameters in the cosmological solutions of 
Eqs.~(\ref{GENERALA}--\ref{GENERALY}) for different
choices of the $\psi$ field in Eq.~(\ref{GENERALFORM}).}
\label{table1}
\end{table}

The general solution with non-trivial $\chi$ and constant $\sigma_i$  
(i.e., vanishing three--form field strengths $H^{(i)}$)
is obtained by applying the ${\rm SL}(2,R)$ transformation (\ref{sl2rchi}).
The transformed fields $\bar{u}$ and $\bar\chi$ have the form
\bea
\label{baru}
e^{\bar{u}} &=& \left| 2 C (D+C\chi) \right| \cosh \left( u + \Delta
\right) \ ,\\
\bar\chi &=& \chi_* \pm {1 \over |2C(D+C\chi)|} \tanh \left( u + \Delta
\right) \ ,
\eea
where $e^\Delta\equiv|(D+C\chi)/C|$. The introduction of a
non--constant $\bar\chi$ field places a lower bound on $\bar{u}$, and
hence the ten--dimensional dilaton field, $\Phi=2u$.  A typical solution with
$\chi'\neq0$ is shown in Fig.~\ref{fig1}.  
The RR field interpolates between two
asymptotic vacuum solutions where $\chi'\to0$. 
Trajectories that come in 
from infinity ($u\to\infty$) at an angle $\xi_1=\xi_*$, where
$-\pi/6\leq\xi_*\leq5\pi/6$, are then reflected in the line $u=u_*=
(\varphi_* + \sqrt{3}\beta_*)/2$ and emerge at an angle $\xi_2=
(2\pi/3)-\xi_*$. 

\begin{figure}
\begin{center}
\leavevmode\epsfysize=5.5cm \epsfbox{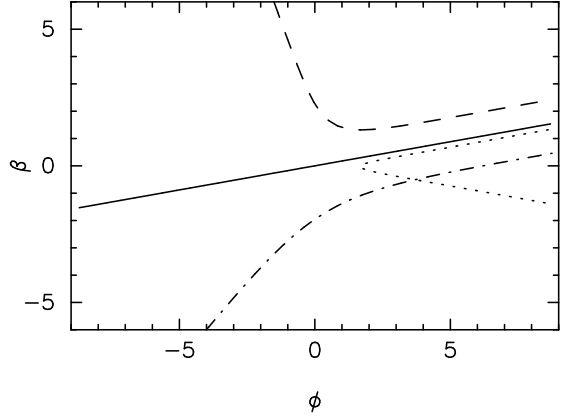}\\ 
\end{center}
\caption[Single form-field]
{Trajectories in $(\varphi,\beta )$ field--space for the 
dilaton--moduli--vacuum solution (solid line) with $\xi_*=\pi/9$. The
dashed, dot--dashed and dotted lines represent the three 
single form--field solution with $\psi=\chi$, $\sigma_2$ and
$\sigma_1$, respectively, obtained by the appropriate ${\rm SL}(2,R)$
transformation of the dilaton--moduli--vacuum solution.}
\label{fig1}
\end{figure}

The mirror image of the $\chi'\neq0$ solution under the reflection
symmetry Eq.~(\ref{td}) is a solution with $\sigma_2' \neq 0$ and
$\chi'=0$. Because the T-duality,
Eq.~(\ref{td}), leaves the four--dimensional dilaton, 
$\varphi$, as well as the four--dimensional
Einstein frame metric invariant, we find that the evolution of both
$\varphi$, and thus the original string frame metric, is the same for a
single excited RR field regardless of whether it is $\chi$ or
$\sigma_2$ that is excited. 

The solution with $\sigma_2' \neq 0$, can also be generated by applying the
transformation (\ref{sl2rsigma2}) to the dilaton--moduli--vacuum
solutions in Eqs. (\ref{dila}--\ref{dilbeta}). 
This transformation leaves 
$\chi'=0$  but leads to $\sigma_1' = \chi \sigma_2'$ and, consequently, to 
$H^{(1)}=0$ and $H^{(2)}\neq0$. 
Note that $\sigma_1$ is only constant when $\chi=0$. There is 
a lower bound on the field $w$, and the generic behaviour of
$\varphi$ and $\beta$ for this solution is plotted in Fig.~\ref{fig1}.  
The RR form--field again leads to a solution that
interpolates between two asymptotic vacuum solutions, where
$\sigma_2'\to0$. Trajectories that come from 
infinity ($w\to\infty$) at an angle $\xi_1=\xi_*$, where 
$-5\pi/6\leq\xi_*\leq\pi/6$, are reflected in the line $w=w_*=
(\varphi_* - \sqrt{3}\beta_*)/2$, 
and emerge at an angle $\xi_2= -(2\pi/3)-\xi_*$. 

For completeness we note that Eqs.~(\ref{GENERALA}--\ref{GENERALFORM})
also represent a solution with $\psi=\sigma_1$ (see
Table~\ref{table1}) which is the general `dilaton--moduli--axion'
solution presented in Eqs. (\ref{axiphi})-(\ref{axisigma}). 
For $\chi=0$ this corresponds
to vanishing RR field strengths and an excited NS--NS three-form field
strength, $H^{(1)}$.  Note that for $\chi\neq0$ (but constant) this
corresponds to a particular solution with non-vanishing RR three-form
field strength $H^{(2)}=-\chi H^{(1)}$ [see Eqs.~(\ref{defH1})
and~(\ref{defH2})]. The typical evolution of
the fields $\varphi$ and $\beta$ is shown in Fig.~\ref{fig1}.
The presence of a non-vanishing $\sigma_1'$ enforces a lower bound on
the value of the four--dimensional dilaton, $\varphi\geq\varphi_*$. 
Trajectories
that come from infinity ($\varphi\to\infty$) at an angle
$\xi_1=\xi_*$, where $-\pi/2\leq\xi_*\leq\pi/2$, are reflected in the
line $\varphi=\varphi_*$ back out at an angle $\xi_2=-\xi_*$. No
trajectories can reach $\varphi\to-\infty$ unless
$\sigma_1'=0$~\cite{CopLahWan94}.

\subsubsection{Cosmology with Two RR Fields}

We now consider solutions where two of the form--fields are non-vanishing 
but the third is zero. The field equations
(\ref{field1}--\ref{field5}) imply 
that the only consistent solution of this type arises when 
$H^{(1)}=0$. {}From Eq.~(\ref{defH1}), $\sigma_1' = \chi
\sigma_2'$ and this allows $\sigma_1$ to be eliminated. 
Eqs.~(\ref{field3}) and~(\ref{field4}) may be integrated directly 
to yield $\tilde{a}^2e^{\varphi+\sqrt{3}\beta }\chi'=L$ 
and $\tilde{a}^2e^{\varphi-\sqrt{3} \beta}\sigma_2'=J$,
where $J$ and $L$ are arbitrary constants. Defining 
a new time parameter $T\equiv \int^{\eta} d\eta'/ \tilde{a}^2
\propto \ln | \tau |$  
and new variables $q_{\pm} \equiv \varphi \pm(\beta /\sqrt{3})$
implies that the field
equations for the dilaton and moduli may be expressed as 
\bea
\label{fieldq1}
\ddot{q}_- =J^2 e^{q_+ -2q_-} \\
\label{fieldq2}
\ddot{q}_+ = L^2 e^{q_- -2q_+}
\eea
and the Friedmann constraint (\ref{friedmann}) gives
\be
\label{friedmannq}
{1\over8} \left( \dot{q}_+ + \dot{q}_- \right)^2
+ {3\over8} \left( \dot{q}_+ - \dot{q}_- \right)^2 + V
=\frac{3a_*^4}{2}  ,
\ee
where a dot denotes $d/dT$ and the potential 
\be
V \equiv  {1\over2}\left(J^2 e^{q_+ -2q_-} + L^2 e^{q_- -2q_+}\right)
\ee

Equations~(\ref{fieldq1}--\ref{friedmannq}) correspond to those of the
SU(3) Toda system~\cite{Kostant79}. This  has recently been studied in
similar models by a number of authors
\cite{LukOvrWal97a,LukOvrWal97b,LuMukPop97,LuMahMuk98,Kaloper97}.
The general solution is of the form~\cite{Kaloper97}
\be
e^{q_-} = \sum_{i=1}^3 A_i e^{-\lambda_iT} \ , \quad
e^{q_+} = \sum_{i=1}^3 B_i e^{\lambda_iT} \ ,
\ee
where $\sum_i\lambda_i=0$, so that $\lambda_{\rm
min}<0$ and $\lambda_{\rm max}>0$. 
This gives the asymptotic solution for $\varphi$ and $\beta$ as
$T\to-\infty$:
\be
e^\varphi \sim e^{-(\lambda_{\rm max}-\lambda_{\rm min})T/2}
\ ,\quad 
e^{\beta} \sim e^{\sqrt{3}(\lambda_{\rm max}+\lambda_{\rm min})T/2}  ,
\ee
while as $T\to+\infty$ we have
\be
e^\varphi \sim e^{(\lambda_{\rm max}-\lambda_{\rm min})T/2}
\ ,\quad
e^{\beta} \sim e^{\sqrt{3}(\lambda_{\rm max}+\lambda_{\rm min})T/2}  .
\ee

As in the single form-field solutions discussed above, the asymptotic
solutions correspond to straight lines in the $(\varphi,\beta )$ plane 
(see Fig. \ref{fig2}).  We
see that trajectories that come from infinity ($\varphi\to\infty$) at
an angle $\xi_*$ are reflected back out at an angle $-\xi_*$. This is
exactly the qualitative behaviour of the NS--NS dilaton--moduli--axion
solution (\ref{axiphi})--(\ref{axisigma}).  
However, the range of allowed asymptotic trajectories is
more restricted than in the pure NS--NS case. The potential in the
constraint Eq.~(\ref{friedmannq}) is bounded from above and we
therefore require that the value 
of the modulus field is  also bounded 
such that $|\beta |\leq\varphi/\sqrt{3}$ asymptotically. Thus,
the asymptotic solutions are restricted to the range
$-\pi/6\leq\xi_*\leq\pi/6$, where $V\leq3a_*^4/2$.

\subsubsection{Cosmology with Three RR Fields}

The general FRW solutions to the type
IIB string action presented in Eq.~(\ref{solitonicaction}) can 
in principle be generated from the dilaton--moduli--vacuum solutions 
(\ref{dila}--\ref{dilphi}) 
by a general ${\rm SL}(3,R)$ 
transformation. This is equivalent to the three (non--commuting)
${\rm SL}(2,R)$ transformations
${\rm SL}(2,R)_{\chi}\times {\rm SL}(2,R)_{\sigma_2}\times
{\rm SL}(2,R)_{\chi}$. 
Such a sequence of transformations yields a solution 
that is analytically very complicated. The generic feature 
of the solution is that it exhibits a sequence of bounces between asymptotic
vacuum states. A typical trajectory is shown in Figs.~\ref{fig2}
and~\ref{fig3}. 
We find that the time--dependence of the fields $\chi$ and
$\sigma_2$ induces lower bounds on the variables $u$ and $w$,
respectively, as seen in the single form--field solutions.  
In the general solution this results in a
lower bound on $\varphi = u+w$.

\begin{figure}
\begin{center}
\leavevmode\epsfysize=5.5cm \epsfbox{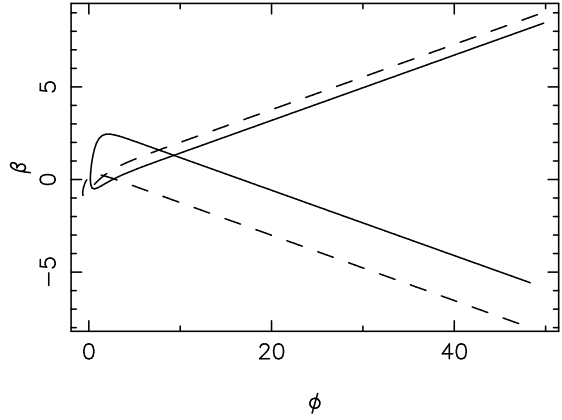}\\ 
\end{center}
\caption[Two and three form fields]
{Trajectories in $(\varphi, \beta )$ field--space for a typical
solution with two RR form--fields with $H^{(1)}=0$ (dashed line) and a
solution with all form--fields non--trivial (solid line). 
The two solutions are related by an ${\rm SL}(2,R)$ transformation.}
\label{fig2}
\end{figure}

\begin{figure}
\begin{center}
\leavevmode\epsfysize=5.5cm \epsfbox{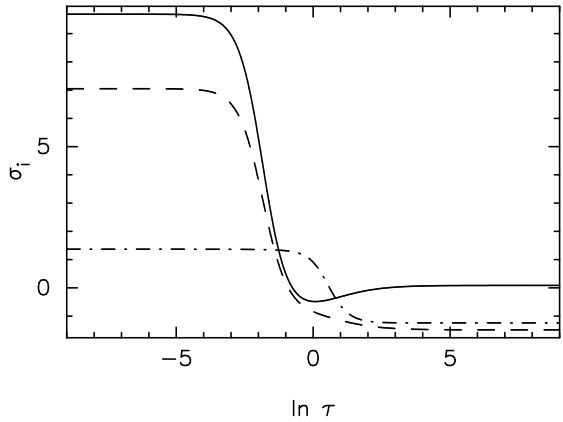}\\ 
\end{center}
\caption[Axion fields]
{The three axion fields $\sigma_1$ (solid line), $\sigma_2$
(dashed line) and $\chi$ (dot--dashed line) against $\ln\tau$ 
for the solution shown in Fig.~2 with all form--fields non--trivial.}
\label{fig3}
\end{figure}

The general type IIB cosmology contains a non-vanishing NS--NS
form--field, but can always be obtained from a Toda system with
$H^{(1)}=0$ by a single ${\rm SL}(2,R)$ transformation (\ref{sl2rchi}).  The
asymptotic behaviour of $\varphi$ and $\beta$ is invariant under this
transformation. This follows since  $u\to\infty$ asymptotically for all 
solutions in the Toda system\footnote{An exceptional case is when $u\to
u_*$ asymptotically, where $u_*$ is a  constant. In this case
$\bar{u}\to$constant, though not necessarily $u_*$, but the qualitative
behaviour is the same.}
and, from Eq.~(\ref{baru}), we obtain $\bar{u}\to u$ in the general
solution. We also have $\bar{v}=v$ and thus $\varphi$ and $\beta$ are
invariant in this limit. Thus, trajectories in $(\varphi,\beta )$ field-space
come in at an angle $\xi_*$ and leave at an angle $-\xi_*$, where
$-\pi/6\leq\xi_*\leq\pi/6$.

\subsubsection{Anisotropic RR Cosmology}

The symmetry transformations described in Section \ref{Section8.1}
may be employed to generate solutions with non--trivial 
NS--NS and RR form fields from {\em any} given solution 
to the dilaton--moduli--vacuum field equations 
\cite{CopLidWan98a,CopLidWan98b}. 
We now extend the analysis to the 
spatially homogeneous but anisotropic Bianchi universes discussed 
in Appendix \ref{appendixC}. 
A Lagrangian for the Bianchi class A models may be derived from
the dual action (\ref{solitonicaction}) by integrating over the
spatial variables. If we assume that all scalar fields 
are homogeneous on the spatial 
hypersurfaces (the solitonic ansatz), it follows that 
the reduced action is given in the Einstein frame by 
\begin{equation}
S_{4*} = \int dT \, \left[ L_g + L_m \right] \ , 
\end{equation}
where 
\begin{eqnarray}
\label{gravaction}
L_g &\equiv& -6\dot{\tilde{\alpha}}^2 
+ 6 \dot{\beta}_+^2 +6\dot{\beta}_-^2+ {^{(3)}}R (\tilde{\alpha}, 
\beta_{\pm} ) \ , \\
\label{matteraction}
L_m &\equiv& \frac{1}{2} \dot{\varphi}^2 
+\frac{1}{2} \dot{\beta}^2 +\frac{1}{2}
e^{\sqrt{3} \beta +\varphi} \dot{\chi}^2 
+\frac{1}{2} e^{-\sqrt{3} \beta +
\varphi} \dot{\sigma}_2^2 +
\frac{1}{2} e^{2\varphi} \left( \dot{\sigma}_1
-\chi \dot{\sigma}_2 \right)^2
\ ,
\end{eqnarray}
represent the gravitational and matter Lagrangians, 
respectively, a dot denotes differentiation with 
respect to the time coordinate \cite{Taub51} (cf. Eq. 
(\ref{neworthogtime})),
\begin{equation}
\label{taubtime}
T\equiv \int d\tilde{t} \, e^{-3\tilde{\alpha}}
 = \int dt \, e^{-3\alpha + \varphi} \ ,
\end{equation}
the scalar curvature of the three--surfaces, ${^{(3)}}R$, is 
given by Eq. (\ref{3bianchicurvature}) with $A_b=0$ and 
$\tilde{\alpha} \equiv \alpha -\varphi /2$, where $\alpha$ is defined in Eq. 
(\ref{3metricbianchi}). 

It follows, therefore, that the action for each specific Bianchi type 
is uniquely determined once the functional form 
of the three--curvature has been specified. 
The advantage of employing the time variable (\ref{taubtime}) 
is that the gravitational and matter sectors of the action 
(\ref{solitonicaction}) are effectively decoupled from one another. 
This implies that 
the vacuum solutions $( \dot\chi = \dot\sigma_i =0)$ for 
the dilaton and moduli fields are given linearly by 
\begin{eqnarray}
\label{dilatonvacuum2}
\varphi &=& \varphi_* + \left( \sqrt{2E} \cos\xi_* \right)  \, T \ ,\\
\label{dilatonvacuum3}
\beta &=& \beta_* + \left( \sqrt{2E} \sin\xi_* \right) \, T \ ,
\end{eqnarray}
for all Bianchi types, where $E$ is an arbitrary positive constant of
integration which represents the kinetic energy associated with
Lagrangian $L_m$. Consequently, these solutions  
correspond to straight-line trajectories in the 
($\varphi , \beta$) field space. This is important 
because it implies the field 
trajectories illustrated in Figs. (\ref{fig1})--(\ref{fig3}) for the 
isotropic FRW solutions also apply directly to the homogeneous Bianchi 
class A cosmologies with RR fields. 

\newpage
\section*{PART III}

\addcontentsline{toc}{section}{PART III}

\section{Pre-Big Bang Cosmology}
\setcounter{equation}{0}
\label{prebigbangSection}
\label{Section9}

In this and the following Section 
we address what is perhaps the first attempt to develop a new
cosmological scenario based on the underlying string symmetries that we
have discussed in this review. It was pioneered by Veneziano and
Gasperini \cite{Veneziano91,GasVen93a} and has led to a huge and
expanding wealth of literature (for a regularly maintained update see
\cite{Gweb}). A radically new scenario has emerged to
describe the very early universe in which we can effectively talk about a
period before the big bang. For this reason such a scenario has been 
called {\em pre-big bang} cosmology \cite{Veneziano91,GasVen93a}.  

The field equations derived from the string effective action admit 
inflationary solutions that are driven
by the kinetic energy associated with the massless fields rather than
any interaction potential. 
A crucial test of inflationary models of this sort is the spectra
of perturbations that they predict, as these can be measured today.
Conventional `slow-roll' inflationary models 
produce fluctuations in both the 
gravitational wave background and the quasi-massless inflaton
field (which drives the period of inflation). 
In general, it is difficult to
produce anything other than an almost scale-invariant spectrum of
adiabatic density perturbations since the inflaton field hardly
evolves during the inflationary era.  (For a review see, e.g., Refs. 
\cite{LidLyt93,LidLidKol97}). 

In contrast, there are many massless fields present in the pre--big bang
scenario, each producing their
own spectrum of perturbations. The dilaton, graviton and
moduli fields yield a blue spectrum that is strongly 
tilted towards higher frequencies \cite{BruGasGio95}. On the other 
hand, the axion fields that are present 
may have significantly different spectral slopes
due to their explicit couplings to the dilaton and moduli fields
\cite{CopEasWan97,CopLidWan97}. These 
can be consistent with the current constraints derived from observations of 
large--scale structure and microwave background anisotropies. Although 
this is a promising feature, a 
number of new problems also appear in this scenario, most notably the
graceful exit problem~\cite{BruVen94}. There are also concerns about
the specific initial conditions required for sufficient 
inflation to proceed \cite{TurWei97,ClaLidTav98,KalLinBou99,MahOnoVen98,BuoMeiUng98,Veneziano97,BuoDamVen99}. We address these problems 
in the remainder of this Section, after we have discussed the 
basics concepts behind the pre--big bang cosmology. 

\subsection{Dilaton--Vacuum solutions} 

\def\tbox{\stackrel{\sim}{\Box}}

\label{Section9.1}

In order to keep our discussion as straightforward as possible whilst
retaining the essential physics of the problem, we will concentrate on
the four--dimensional, NS-NS string effective action given in
Eq.~(\ref{reduced1}). This includes the four-dimensional 
dilaton field, $\varphi$, the pseudo-scalar axion field, $\sigma$, 
and a modulus field, $\beta$. The axion field 
represents the degrees of freedom of the antisymmetric three-form
field strength and the modulus field represents the
scale of the internal space. We assume that the universe is adequately 
described by a standard, spatially flat FRW metric with
scale factor $a(t)$. Integrating over the spatial variables in 
Eq. (\ref{reduced1}) then yields a reduced action:
\begin{equation}
S=\int dt e^{3\alpha -\varphi} \left[ 6 \dot{\alpha} \dot{\varphi} 
-6\dot{\alpha}^2 
-\dot{\varphi}^2 +{1\over2}\dot{\beta}^2 
+{1\over2}e^{2\varphi}\dot{\sigma}^2 \right]
\end{equation}
where $\alpha\equiv\ln a$.
Defining the shifted dilaton~\cite{Veneziano91,Tseytlin92,EasMaeWan96}
\begin{equation}
\label{sdd}
\bar{\phi} \equiv \varphi-3\alpha \, .
\end{equation}
implies that the reduced action simplifies further to
\begin{equation}
\label{actionreversal}
S = \int dt e^{-\bar\phi} \left[ 3\dot\alpha^2 - \dot{\bar\phi}^2 
+{1\over2}\dot{\beta}^2 
+{1\over2}e^{2\varphi}\dot{\sigma}^2 \right]
\end{equation}

The field equations derived from Eq. (\ref{actionreversal}) 
are symmetric under time reversal, $t \rightarrow -t$. However,  
Veneziano~\cite{Veneziano91} also noted that the action is 
invariant under the discrete ${\rm Z}_2$ transformation
$\alpha\to-\alpha$ and $\bar\phi\to\bar\phi$ when the axion 
field is trivial, $\dot\sigma=0$.
Together, these discrete symmetries correspond to the transformation
\begin{equation}
\label{sfdexample}
a(t)  \to  1/a(-t), \qquad 
\varphi(t)  \to  \varphi(-t) - 6 {\rm \ln}(a(-t)),
\end{equation}
and form the motivation behind the pre--big bang scenario. 

Equation~(\ref{sfdexample}) 
is an example of the scale factor duality discussed earlier in
the review in Section \ref{Section5.1}. 
Applying the duality transformation (\ref{sfdexample}) simultaneously 
with time reversal implies that 
the Hubble expansion parameter $H \equiv d(\ln a) /dt$ 
remains invariant, 
$H(-t) \to H(t)$, whilst its first derivative 
changes sign, $\dot{H} (-t) \to - \dot{H} (t)$. 
A decelerating, post--big bang solution -- characterized 
by $\dot{a}>0$, $\ddot{a}<0$ and $\dot{H}<0$ --  is therefore mapped onto 
a pre--big bang phase of inflationary expansion, since $\ddot{a} /a
= \dot{H}+H^2 >0$. The Hubble radius $H^{-1}$ 
decreases with increasing time and the expansion is therefore 
super-inflationary.

Thus, the pre-big bang cosmology is one that
has a period of super-inflation driven simply by the kinetic energy of the
dilaton and moduli fields. This is related by duality 
to the usual FRW post--big bang
phase. The two branches are separated by a curvature singularity, 
however, and it is not clear how the 
transition between the pre-- and post--big bang phases 
might proceed. This is the graceful exit problem of 
the pre--big bang scenario \cite{BruVen94}.

To be more precise, we now consider the dilaton-moduli-vacuum
solutions of Section \ref{Section7.1} within a pre--big bang
context. In these solutions, a single modulus field $\beta$ describes
the evolution of the internal space and the axion field is held
constant.  The solution for a flat ($\kappa=0$) FRW universe is
included in the general FRW solutions given in
Eqs.~(\ref{dila}--\ref{dilbeta}) and corresponds to the well--known
monotonic power-law, or `rolling radii', solutions\footnote {We do not
consider here the trivial flat spacetime solution
$\varphi'=\beta'=a'=0$.}  \cite{Mueller90}.
This class of models represents a particular case of the 
generalized Kasner backgrounds (\ref{DstringKasner})--(\ref{varphi})
discussed in Section {\ref{Section6}. The solution is 
given in terms of proper time in the string frame by
\begin{eqnarray}
\label{dilphit1}
e^\varphi & = & e^{\varphi_*} \left|{t\over
 t_*}\right|^{2\cos\xi_*/(\sqrt{3}+\cos\xi_*)} 
 \, ,\\
\label{dilat1}
a & = &
 a_* \left|{t\over t_*}\right|^{(1+\sqrt{3}\cos\xi_*)/(3+\sqrt{3}\cos\xi_*)}
 \, ,\\
\label{dilbetat1}
e^\beta & = &
 e^{\beta_*}
 \left|{t\over t_*}\right|^{2\sin\xi_*/(\sqrt{3}+\cos\xi_*)}
 \, ,
\end{eqnarray}
and is shown in figures \ref{figflatfrw1}--\ref{figflatfrw3} for
$\cos\xi_*=\pm1$. 
For $\cos\xi_*<-1/\sqrt{3}$ there is accelerated expansion, i.e.,
inflation, in the string frame for $\eta<0$ and $e^{\varphi} \to 0$ as
$t \to -\infty$, corresponding to the weak coupling regime.
The expansion is an example of `pole--law' inflation \cite{PolSah89}. 
We note that Levin and Freese
\cite{LevFre93,Levin95a,Levin95b,Levin95c} have also discussed
inflationary solutions based primarily on the kinetic energy of a
massless Brans--Dicke type field in a modified gravitational theory.

We see that these solutions have semi-infinite proper lifetimes. Those
starting from a singularity at $t=0$ for $t\ge 0$ are denoted as the
(--) branch in Ref.~\cite{BruVen94}, while those which approach a
singularity at $t=0$ for $t\leq0$ are referred to as the $(+)$ branch
(see figures \ref{figflatfrw1}--\ref{figflatfrw3}).
Our choice of origin for the time coordinate is arbitrary.  A more
fundamental definition of the $(+/-)$ branches may be given by
considering the evolution of the shifted
dilaton (\ref{sdd}):
\begin{equation}
\label{defshifted}
\bar{\phi} = {e^{\varphi_*} \over a_*^3} \left|
{t_*\over t} \right| \, .
\end{equation}
Its time derivative
is always positive on the $(+)$ branch (or $t<0$) and always
negative on the $(-)$ branch (or $t>0$).  These $(+/-)$ branches do
{\em not} refer to the choice of sign for $\cos\xi_*$. On either the
$(+)$ or $(-)$ branches of the dilaton-moduli-vacuum cosmologies we
have a one-parameter family of solutions corresponding to the choice of
$\xi_*$, which determines whether $e^\varphi$ goes to zero or infinity
as $t\to0$. 
These solutions become singular as the conformally invariant time
parameter $\eta \equiv \int dt /a(t) \to 0$ (as illustrated by the
divergence of the Hubble rate in figure~\ref{figflatfrw3}) and there is
no way of naively connecting the two branches based simply on these
solutions \cite{BruVen94}.

As we have seen a number of times in this review,
it often proves useful to consider the cosmic dynamics 
in the Einstein frame, where the dilaton field is minimally 
coupled to gravity. The Einstein frame 
is related to the string frame metric by the conformal transformation
given in Eq.~(\ref{DEinstein}), where $D=4$. Under such a
transformation, the scale factor in the Einstein frame, given in
Eq.~(\ref{einsteinscalefactor}), becomes
\begin{equation}
\label{Esf}
\tilde{a} = \tilde{a}_* 
\left|\frac{\tilde{t}}{\tilde{t}_*}\right|^{1/3}.
\end{equation}
In terms of conformal time, $\eta$,
we have $t\to\pm\infty$ and $\tilde{t}\to\pm\infty$ as
$\eta\to\pm\infty$.  In the Einstein frame, as $\eta \to 0$ on the (+)
branch, the universe is collapsing with $\tilde{a} \to 0$, and the
comoving Hubble length $|d(\ln\tilde{a})/d\eta|^{-1}=2|\eta|$
decreases with time. Thus, in both frames there is inflation
taking place~\cite{GasVen93b} in the sense that a given comoving
scale, which starts arbitrarily far within the Hubble radius in either
conformal frame as $\eta \to - \infty$, inevitably becomes larger than
the Hubble radius in that frame as $\eta \to 0$. The significance of
this is that it means that perturbations can be produced in the
dilaton, graviton and other matter fields on scales much larger than
the present Hubble radius from quantum fluctuations in flat spacetime
at earlier times -- this is a vital property of any inflationary scenario.

For completeness, it is worth reminding the reader that these
solutions can be extended to include a time-dependent axion field,
$\sigma(t)$, by exploiting the ${\rm SL}(2,R)$ S-duality invariance of
the four--dimensional, NS-NS action \cite{CopLahWan94}. This was
discussed in Section \ref{Section7.1}.

\subsection{Initial Conditions}

\label{Section9.2}

An intriguing (and to some disturbing) consequence emerges from the above 
discussion. Since both $\dot{H}$ and $\dot{\varphi}$ are positive in the 
pre--big bang phase, the initial values for these
parameters must be {\em very small}. This raises a number of important 
issues concerning fine--tuning in the pre--big bang 
scenario~\cite{TurWei97,ClaLidTav98,KalLinBou99,MahOnoVen98,BuoMeiUng98,Veneziano97,BuoDamVen99}. There needs to be enough
inflation in a homogeneous patch in order
to solve the horizon and flatness problems which means that the
dilaton driven inflation must survive for a 
sufficiently long period of
time. This is not as trivial as it may appear, however,  since the period of
inflation is limited by a number of factors. 
In this subsection, we review the criticisms 
that have been levelled at the initial conditions of the pre--big 
bang cosmology. We assume 
for simplicity that the moduli fields are trivial.

The fundamental postulate of the scenario  is that 
the initial data for inflation lies well within the 
perturbative regime of string theory, 
where the curvature and coupling are very 
small \cite{GasVen93a}. Inflation then proceeds 
for sufficiently homogeneous initial conditions 
\cite{Veneziano97,BuoMeiUng98}, 
where time derivatives are dominant with respect to
spatial gradients, and the universe evolves into a high curvature and 
strongly--coupled regime. 
Thus, the pre--big bang initial state should correspond to a
cold, empty and flat vacuum state.
The initial universe would have been huge relative
to the quantum scale and hence should have been well described by
classical solutions to the string effective action. This should be
compared to the initial state which describes the standard hot big
bang, namely a dense, hot, and highly curved region of spacetime.
This is quite a contrast and 
a primary goal of pre--big bang  cosmology must be to develop a 
mechanism for smoothly 
connecting these two regions, since we believe that the standard big
bang model provides a very good representation of the current evolution of
the universe.

At the present epoch, our observable universe 
appears very nearly homogeneous on sufficiently large scales. 
In the standard, hot big bang model, 
it corresponded to a region at the Planck time that was $10^{30}$ times larger 
than the horizon size, $l_{\rm Pl}$.  This 
may be viewed as an initial condition in the big bang model or as a final 
condition for inflation. 
It implies that the  comoving Hubble radius, $1/(aH)$, 
must decrease during 
inflation by a factor of at least $10^{30}$ if the horizon problem is to 
be solved. For 
a power law expansion, this implies that 
\begin{equation}
\label{bound}
\left| \frac{\eta_f}{\eta_i} \right| \le 10^{-30}
\end{equation}
where subscripts $i$ and $f$ 
denote values at the onset and end of inflation, respectively. 

In the pre--big bang scenario, Eq. (\ref{dilphi})
implies that the dilaton grows as $e^{\varphi} \propto |
\eta |^{-\sqrt{3}}$. At the start of the post--big bang epoch, the 
string coupling, $g_s =e^{\varphi /2}$, should be of order unity. Thus, 
the bound (\ref{bound}) implies that the initial value of the 
string coupling is strongly constrained, $g_{s,i} \le 10^{-26}$. 
Turner and Weinberg interpret this constraint as a severe fine--tuning 
problem in the scenario, because inflation in the string frame can 
be delayed by the effects of spatial curvature 
\cite{TurWei97}. It was shown by Clancy, Lidsey and Tavakol that the
bounds are further tightened when spatial anisotropy is introduced 
\cite{ClaLidTav98}. Indeed, the combined effects of anisotropy and 
spatial curvature can prevent pre--big bang inflation 
from occurring, in contrast to the negatively curved FRW model, where 
the late--time attractor corresponds to a dilaton--dominated expansion 
\cite{ClaLidTav98}.

Moreover, the dynamics of 
the NS--NS axion field also places a lower bound on the 
allowed range of values that the string coupling may take, as we 
saw in Section \ref{Section7.1} \cite{CopLahWan94,CopLahWan95}. 
A similar effect due to RR axion fields was 
discussed in Ref. \cite{FeiVaz98}. It should be noted, however, that 
this restriction may be relaxed if inflation also 
proceeds during the `string' phase that follows directly from  
the dilaton--driven era once higher--order and loop corrections 
have become important. This possibility was discussed by Maggiore 
and Sturani \cite{MagStu97}.

In the standard inflationary scenario,
where the expansion  is quasi--exponential, the Hubble radius is approximately 
constant and $a \propto (-\eta )^{-1}$. Thus, the homogeneous 
region grows by a 
factor of $|\eta_i /\eta_f |$ as inflation proceeds. During a pre--big bang 
epoch, however, $a \propto ( -\eta )^{-1/1+\sqrt{3}}$ and 
the increase in the size of a homogeneous region is 
reduced by a factor of at least $10^{30 \sqrt{3}/(1+\sqrt{3})} \approx 
10^{19}$ relative to that of the standard inflation 
scenario. This implies that the initial 
size of the homogeneous region should exceed $10^{19}$ in string units if 
pre--big bang inflation is to be successful in solving the problems 
of the big bang model 
\cite{Veneziano91,GasSanVen91,KalLinBou99}. The occurrence of such a large 
number was cited by 
Kaloper, Linde and Bousso as a serious problem of the pre--big 
bang scenario, because it implies that the universe must already 
have been large and smooth by the time inflation began \cite{KalLinBou99}. 

On the other hand, Gasperini has recently emphasized 
that the initial homogeneous region of the pre--big bang universe is 
not larger than the horizon even though it is large relative 
to the string/Planck scale \cite{Gasperini99}. 
Indeed, during pre--big bang inflation, the 
Hubble radius decreases linearly with respect to cosmic time and 
it follows that $H^{-1}_i /H^{-1}_f = 
| \eta_i /\eta_f |^{\sqrt{3} /(1+\sqrt{3} )}$. Consequently, when 
Eq. (\ref{bound}) is saturated one may 
verify that successful inflation 
is possible if the size of the homogeneous region at the onset of 
inflation is given by the horizon size at that time. 
The question that then arises when discussing the naturalness, or 
otherwise, of the above initial conditions is what 
is the basic unit of length that should be employed \cite{Gasperini99}. 
At present, this question has not been addressed in detail. 

One way of studying the question of naturalness 
is to ask whether there exists an
``attractor'' whose basin of attraction  
(i.e. the set of initial conditions which evolve 
to this state) has a large or full measure
in the space of all possible initial data.
In its fullness, this is an impossible question to address because 
we still lack a definitive non--perturbative formulation of string theory.
Furthermore, the nature of the generic attractor is not 
known even at the reduced level of the effective 
actions because the resulting field equations 
are non--linear partial differential equations. 
The crucial point here is to establish those attractors that 
have the largest basins and are therefore the most `natural' 
when further restrictions are imposed. 

The qualitative behaviour of 
pre--big bang cosmology in the asymptotic 
past before the onset of inflation has been addressed in a series of papers 
\cite{Veneziano97,BuoMeiUng98,ClaLidTav98,MahOnoVen98,ClaLidTav99}. 
Veneziano and collaborators conjectured that pre--big bang 
inflation generically evolves out of an initial state that 
approaches the Milne universe in the semi--infinite past,  
$t \rightarrow -\infty$ \cite{Veneziano97,BuoMeiUng98}. 
The Milne universe may be mapped onto the future (or past) light cone of the 
origin of Minkowski spacetime and therefore corresponds to a 
non--standard representation of the string 
perturbative vacuum. It is flat spacetime expressed in an expanding frame: 
\be
\label{m5}
ds^2 =-dt^2 +t^2 \left( dx^2 + e^{-2x} (dy^2 +dz^2 ) \right)  .
\ee
The proposal was that the Milne 
background represents an early time attractor, with a large
measure in the space of initial data. If so, this 
would provide strong justification for the postulate that 
inflation begins in the weak coupling and curvature regimes
and would render the pre-big bang assumptions regarding the 
initial states as `natural'.

Clancy {\em et al.} took a critical look at this conjecture and 
argued that the Milne universe is an unlikely 
past attractor for the pre--big bang scenario \cite{ClaLidTav99}. They 
suggested that plane wave backgrounds represent a more 
generic initial state for the universe \cite{ClaLidTav98}. 
In particular, they  
considered the class of orthogonal Bianchi B models. This has a non--zero 
measure in the space of homogeneous initial data (see Appendix 
\ref{appendixC}). 
The asymptotic states of the Bianchi type III and V universes are 
isomorphic to the Milne universe. 
However, these models represent a set of measure zero and  
a homogeneous plane wave background is the attractor for this 
class of models with a full measure of initial states \cite{HewWai93}. 
Indeed, we discussed one such plane wave in Section \ref{Section7.4} 
within the context of a tilted Bianchi type V cosmology 
\cite{ClaFeiLid99a}. Recently, Kunze has applied previous 
asymptotic analyses \cite{CarFei84,CarFei85} 
to investigate initial conditions of inhomogeneous 
$G_2$ cosmologies \cite{Kunze99}. 

Buonanno, Damour and Veneziano have subsequently proposed that 
the initial state of the pre--big bang universe should correspond to 
an ensemble of gravitational and dilatonic waves \cite{BuoDamVen99}. 
They refer to this as the state of `asymptotic past triviality'. 
When viewed in the Einstein 
frame these waves undergo collapse when certain conditions are 
satisfied. In the string frame, these gravitationally unstable 
areas expand into homogeneous regions on large scales. 

To conclude this Section, it is clear that the question 
of initial conditions in the pre--big bang scenario 
is currently unresolved. Nevertheless, 
in our view, it is premature to rule out the scenario 
purely on the subjective grounds of naturalness. Indeed, one of the 
important features of the scenario is that the pre-big bang 
era may have left behind unique observational signatures  that could be 
detectable today on large scales. As is the case for any cosmological 
model, it must be the observations  that decide whether 
it provides a viable description for the evolution of the early universe. 
We defer a detailed discussion of the 
observational consequences of the pre--big bang model until the next Section
and proceed by first addressing 
a further unresolved problem for the scenario.

\subsection{Graceful Exit in string cosmology}  

\label{Section9.3}

Throughout this article we have attempted to explain how duality
symmetries of string theory can lead to a new class of cosmological
scenarios. The pre-big bang solutions of
Eqs.~(\ref{dilphit1}--\ref{dilbetat1}) provide a particularly
interesting case \cite{Veneziano91,GasVen93a}, where the universe
starts very near the cold, empty and flat perturbative vacuum.
However, we also know that standard cosmology assumes that the early
universe was in a hot, dense, and highly curved state.  Thus, if we
want to make phenomenological sense of the initial conditions
associated with the pre--big bang, we have to explain how they would
evolve naturally into those of the standard scenario at some later
time, smoothing out the big bang singularity. It is to this key
question that we now turn our attention.  A number of authors have
addressed the thorny issue of the graceful exit transition from the
dilaton-driven kinetic inflation phase to the subsequent standard
radiation dominated evolution
\cite{BruMad97,BruMad98,GasMahVen96,LukPop97,Rey96,GasVen96,
DabKie97,MahMukPan97,BuoGasMag97,LalPop97,Lidsey97,GasMagVen97}. 

In \cite{BruVen94} it was argued, that such a transition cannot occur
while the curvature was below the string scale and the string coupling
was still weak, leading to the conclusion that an intermediate
``string phase" of high curvature or strong coupling is required
\cite{BruGasGio95b}. This no-go result was later extended to a wide
range of low energy actions
\cite{KalMadOli95,EasMaeWan96,KalMadOli96}. The situation is 
not promising when we first look at the tree level equations of
motion. The very low curvature and coupling that describes the early
evolution, means that the low-energy, tree-level string effective
action provides a good start for our understanding of the
dynamics. The field equations indicate that the string perturbative
vacuum, with vanishing coupling constant $g_s= e^{\varphi/2}=0$, is
unstable to small perturbations in the metric and dilaton. These
perturbations lead to an accelerated growth in the curvature and
coupling \cite{Veneziano91,GasVen93a}, which are unbounded, leading to
a singularity in the curvature and/or the coupling being reached in a
finite amount of cosmic time, for any realistic choice of the (local)
dilaton potential. To avoid such a singularity, we must turn our
attention to the impact that higher-order corrections can have on the
effective action \cite{BruVen94,KalMadOli95,KalMadOli96,EasMaeWan96}.

In string theory, such corrections come in two flavours, being
controlled by two independent expansion parameters. One is the
field-dependent (and thus in principle space-time-dependent) coupling
$g_s$, which controls the importance of string-loop corrections. They
represent the quantum corrections associated with the classical string
action.  The other parameter, $\alpha'$, controls the importance of
finite-string-size corrections, which are small if fields vary little
over a string-length distance $g_s = \sqrt{\alpha'}$.  When this
second expansion parameter is small, then the higher-derivative
corrections to the action can be neglected and string theory can be
described as an effective quantum field theory.  

We generally expect
that both string loop corrections and $\alpha'$ corrections will be
important as we approach the high curvature regime. This has been
investigated by a number of authors
\cite{AntRizTam94,RizTam94,KirKou95,Rey96,GasVen96,GhoMadVen99}, 
and arises because the
inclusion of loops is accompanied by the appearance of
higher-derivative terms in the effective action, which in turn  
requires the simultaneous inclusion of higher orders in
$\alpha'$ for consistency \cite{AntRizTam94}.  

One possible exception arises if the initial 
value of the string coupling is sufficiently small. It is then quite
possible that the universe reaches the high-curvature regime while 
the coupling is still weak. In this case, 
higher-derivative ($\alpha'$) corrections become important, but the
loop corrections remain negligible. We will
see, however, that in general the string corrections are a vital
component of any realistic attempt to modify the high curvature
singularity.  

In this Section we will review a number of attempts to
obtain a graceful exit in the context of string cosmology. One of
the most significant advances was made in \cite{GasMagVen97}, where
the authors considered the modification of the curvature singularity
due simply to ``stringy'' $\alpha'$ corrections, but at lowest order in
$g_s$.  They showed that there existed fixed point solutions to the
equations of motion, where a cosmological background evolving from the
perturbative vacuum could be attracted into a state of constant
curvature and linearly evolving dilaton, i.e., of constant $H$ and
$\dot\varphi$ (in the string frame). 
In particular, they considered a very
special class of Bianchi type I metrics given by 
\begin{eqnarray} ds^2 = - dt^2
+ \sum_{i=1}^{n} e^{2 H_i t} dx^i dx^i , ~~~~~~~ \varphi(t) = c t + \varphi_0
\label{31} 
\end{eqnarray} 
and parameterized by the $(n+1)$ constants $c$ and $H_i$.  

In this scenario, the shifted
dilaton $\bar{\phi}$ satisfies $\dot{\bar{\phi}} \equiv \dot \varphi -nH<0$
in $n$ isotropic spatial dimensions. This is a necessary
condition for the starting point of a standard cosmology, where
the background can be subsequently attracted by an appropriate
potential to a state with expanding metric ($H>0$) and frozen dilaton
($\dot\varphi=0$).  Of course, since the dilaton keeps growing after the
transition to the string phase, the effects of loops, of a
non-perturbative dilaton potential, and of the back-reaction from
particle production, must eventually become important, and as we shall
see, they will play an essential role in the second transition from
the string phase to the usual hot big bang scenario.  

Nevertheless, 
in \cite{GasMagVen97}, the authors showed that the case of constant
curvature and linear dilaton led to a system of $(n+1)$ algebraic
equations in $(n+1)$ unknowns ($n$ Hubble constants and $\dot{\varphi}$).
Providing an example to first order in $\alpha'$ (i.e. four
derivatives), they determined the fixed points, and showed, by
numerical integration, that any isotropic pre-big bang background
necessarily evolves smoothly towards the regular fixed points, thus
avoiding the singularity. Out of these solutions, one in
particular is of interest, namely the case where
$\dot{\bar{\phi}}\ne 0$. String-phase solutions with
$\dot{\bar{\phi}} < 0$ then play the role of late-time attractors for
solutions coming from pre-big bang initial conditions. 
An example of this behaviour was provided by Gasperini {\em et al.}
\cite{GasMagVen97}. To first--order in $\alpha'$,
the simplest effective action in the string frame that reproduces the
massless bosonic sector of the tree-level string $S$-matrix can be
written in the form \cite{MetTse87}: 
\begin{equation} S={1\over
16\pi \alpha'^{(n-1)/2}}\int d^{n+1}x \sqrt{|g|}e^{-\varphi} \left[ R+ (\nabla
\varphi)^2+{k\alpha' \over 4} R_{\mu\nu\alpha\beta}^2\right]\; ,
\label{41} 
\end{equation} where $k=1,1/2$ for the bosonic and
heterotic string, respectively. (We have assumed the
torsion background is trivial).  
A convenient field redefinition
that eliminates terms with higher than second derivatives from the
effective equations is achieved by replacing the square of the Riemann
tensor with the Gauss--Bonnet invariant $R^2_{GB} \equiv
R_{\mu\nu\alpha\beta}^2-4 R_{\mu\nu}^2+ R^2$ \cite{Zwi85}, at the price of
introducing dilaton-dependent $\alpha'$ corrections. The field
redefinition 
\begin{eqnarray} 
\tilde{g}_{\mu\nu}=g_{\mu\nu}+4k \alpha'
\left[ R_{\mu\nu}-\partial_\mu\varphi\partial_\nu \varphi+ g_{\mu\nu}(\nabla
\varphi)^2\right] \nonumber \\
\tilde{\varphi} =\varphi -k\alpha'
\left[R+(2n-3)(\nabla\varphi)^2\right], 
\label{42} 
\end{eqnarray}
truncated to first order in $\alpha'$, produces (dropping the tilde
over the redefined fields): 
\begin{equation} S={1\over
16\pi \alpha'^{(n-1)/2}}\int d^{n+1}x \sqrt{|g|}e^{-\varphi} \left[ R+ (\nabla
\varphi)^2+{k\alpha' \over 4} \left(R^2_{GB} - (\nabla
\varphi)^4\right)\right] . 
\label{43} 
\end{equation} 

Specializing to the
Bianchi type I background, and to the case in which the spatial sections are
the product of two isotropic, Ricci--flat manifolds, restricts 
the analysis to a metric of the form
\begin{equation} 
g_{00}=-1, ~~~~ g_{ij}=\delta_{ij} e^{\beta(t)}, ~~~~ g_{ab}=\delta_{ab}
e^{\gamma(t)}
\label{44}
\end{equation} 
where $(i,j) = (1, \ldots , n)$ and 
$(a,b)=(n+1, \ldots , n+N )$. 
The cosmological field equations then
yield constant curvature solutions in the isotropic case $N=0$ 
\cite{GasMagVen97}. By specifying 
$\dot\varphi=x={\rm const}$ and $\dot{\beta}=y={\rm const}$ 
Gasperini {\em et al.} found real solutions for all 
$1 \le n \le 9$.  In particular, for
$n=3,6,9$, the coordinates of the fixed points in the
plane $(\dot\varphi, \dot{\beta})$ are given by 
\begin{eqnarray} 
n&=&3 ~~~~~~~~~~x=\pm1.40...,
~y=\pm0.616...,\nonumber\\ n&=&6 ~~~~~~~~~~x=\pm1.37...,
~y=\pm0.253...,\nonumber\\ n&=&9~~~~~~~~~~x=\pm1.38...,
~y=\pm0.163..., ~~. \label{48} 
\end{eqnarray} 
respectively, in units  where $k\alpha'=1$. 

Significantly, they
showed by integrating numerically the field equations for $\beta$ and
$\varphi$, and imposing the constraint on the initial data, that for any
given initial condition corresponding to a state of pre-big bang
evolution from the vacuum (i.e. $0<\dot{\beta}<x$,
$\dot{\bar{\phi}}=\dot\varphi-n\dot{\beta} >0$), then the solution is
necessarily attracted to the expanding fixed points (\ref{48}). For
the string effective action (\ref{43}), the fixed points are
continuously joined to the perturbative vacuum
($\dot{\beta}=0=\dot\varphi$) by the smooth flow of the background in
cosmic time, so the action ({\ref{43}) exhibits a smooth transition
from the dilaton phase to the string phase of the pre-big bang
scenario, at least to first order in $\alpha'$, a result in agreement
with previous assumptions \cite{BruGasGio95b}\footnote{An intriguing
physical motivation for such a solution, namely asymptotic De Sitter
phase with linear dilaton, has been proposed in \cite{FofMagStu99},
where they point out that such a behaviour is driven in string theory
by the formation of a gravitino-dilatino condensate.}.  

Unfortunately,
in this model the dual of the expanding pre-big bang branch is not
regularized, which means that there is no smooth monotonic evolution
from growing to decreasing curvature.  However, the expanding fixed
point determined by the $\alpha'$ corrections does correspond to a
final configuration of the post-big bang type, with
$\dot{\bar{\phi}}<0$, $\dot{\beta}>0$, which offers the possibility
for the background to be attracted by an appropriate potential in the
expanding, frozen-dilaton state of the standard scenario. We now go on
to see how this was first achieved in the context of adding loop
corrections to the classical $\alpha'$ correction ({\ref{43}).  

The
first successful model of graceful exit transition from a
dilaton-driven inflationary phase to a decelerated FRW era was
proposed by Brustein and Madden \cite{BruMad97,BruMad98}. It made use
of the result just derived that classical corrections can stabilize a
high curvature string phase while the evolution is still in the weakly
coupled regime. The crucial new ingredient that they added was to show
that if additional terms of the type that may result from quantum
corrections to the string effective action are present and 
also induce violation
of the null energy condition (NEC)\footnote{The Null Energy Condition 
is satisfied if $\rho + p \ge 0$, where $\rho$ and $p$ 
represent the effective energy density and pressure of the additional 
sources.}, then evolution towards a
decelerated FRW phase is possible. Of course this violation of the
null energy condition can not continue indefinitely, and eventually it
needs to be turned off in order to stabilise the dilaton at a fixed
value, perhaps by capture in a potential minimum or by radiation
production.  

The analysis of \cite{BruMad97} resulted in a set of
necessary conditions on the evolution in terms of the Hubble
parameters $H_S$ in the string frame and $H_E$ in the Einstein frame
and the dilaton $\varphi$ (see Eq.~(\ref{HEHS})). They were 

\begin{itemize} 

\item Initial
conditions of a (+) branch and $H_S,\dot\varphi>0$ require $H_E<0$. 

\item
A branch change from (+) to $(-)$ has to occur while $H_E<0$. 

\item A
successful escape and exit completion requires NEC violation
accompanied by a bounce in the Einstein frame after the branch change
has occurred, ending up with $H_E>0$. 

\item Further evolution is
required to bring about a radiation dominated era in which the dilaton
effectively decouples from the ``matter" sources. 

\end{itemize} 

In the
work of \cite{BruMad98}, the authors employed both types of string
inspired corrections. The $\alpha'$ corrections \cite{GasMagVen97}
created an attractive fixed point for a wide range of initial
conditions which stabilized the evolution in a high
curvature regime with linearly growing dilaton.  This then caused the
evolution to undergo a branch change, all of this occurring for small
values of the dilaton (weak coupling), so the quantum corrections could
be ignored.  However, the linearly growing dilaton means that the
quantum corrections eventually become important. Brustein and Madden
employed these to induce NEC violation and allow the universe to
escape the fixed point and complete the transition to a decelerated
FRW evolution.  

For completeness, we introduce the notation
adopted in \cite{BruMad98}. The four--dimensional 
effective action in the string frame can be written as 
\begin{equation} 
S_{eff}=  {1 \over 16\pi \alpha'}   \int
d^4 x \sqrt{|g|} e^{-\varphi} \left[
R+(\nabla \varphi)^2 + \frac{1}{2} {\cal L}_c \right], 
\label{effacts} 
\end{equation} 
with the
Lagrangian ${\cal L}_c$ containing corrections to the lowest order 
four--dimensional action as specified below.  

Writing the metric as $ds^2=
-dt^2+a^2(t) dx_i dx^i$ and $\varphi=\varphi(t)$, we look for solutions
to the equations of motion. It proves useful to split ${\cal L}_c$ up
into components: 
\begin{equation} 
{\cal L}_c={\cal L}_{\alpha'}+{\cal
L}_{q}+{\cal L}_{m}
\label{actsplit} 
\end{equation} 
The Lagrangian
${\cal L}_{\alpha'}$ is taken to be of the specific form of the
$\alpha'$ classical corrections proposed in Eq.~(\ref{43}) by
\cite{GasMagVen97}, 
\begin{equation} 
\frac{1}{2} {\cal
L}_{\alpha'}= k \alpha' \left[\frac{R_{GB}^2}{4}- 
\frac{(\nabla \varphi)^4}{4} \right]. 
\label{gmvlag} 
\end{equation} 
The Lagrangian ${\cal L}_{q}$
represents quantum loop corrections parameterized by powers of
$e^{\varphi}$, and ${\cal L}_{m}$ represents radiation or a dilaton
potential, important for the final transition to radiation domination
and a fixed dilaton.  

Two useful quantities are the Hubble parameter in the Einstein frame 
and its time derivative:
\begin{eqnarray} 
\label{HEHS}
H_E &=& e^{\varphi/2}
(H_S-\frac{1}{2} \dot \varphi) \\ 
\dot H_E &=& e^{\varphi/2} (\dot H_S -
\frac{1}{2} \ddot \varphi + \frac{1}{2} \dot \varphi H_S -\frac{\dot
\varphi^2}{4}) 
\end{eqnarray} 
If we initially set ${\cal L}_{q}={\cal
L}_{m}=0$, we recover the fixed point solutions,
Eq.~(\ref{48}). However, as we have already noted, these do not lead to
a branch change and instead, the solution represents a singular collapse
in the Einstein frame. This follows because the dilaton is increasing linearly 
and  there 
is insufficient NEC violation. Moreover, the addition
of conventional sources to ${\cal L}_m$, such as radiation, does not
violate NEC. Consequently, 
these effects cannot help with the completion of the exit transition. 
and this suggests that sources for ${\cal L}_{q}$ are required.  

Unfortunately, the precise forms of the
quantum loop corrections have not been calculated, but it 
is anticipated that they should be 
parameterized by powers of $e^\varphi$.  In
\cite{BruMad98}, the authors introduced a series of trial terms which
could eventually emerge as the actual quantum corrections.  One of
these was, 
\begin{eqnarray} 
\frac{1}{2} {\cal L}_q^{\varphi}&=&- e^{\varphi} (\nabla
\varphi)^4. 
\label{gp4} 
\end{eqnarray} 
At the one loop level, the overall
coefficient of the corrections can be absorbed by a shift of $\varphi$,
and it therefore determines the value of $\varphi$ at which the quantum
corrections begin to be important, but does not lead to qualitatively
different behaviour.  

The relative factor of $e^\varphi$ in ${\cal
L}_q^{\varphi}$ compared to Eq.~(\ref{gmvlag}) leads to important
consequences, in particular ${\cal L}_q^{\varphi}$ can lead to
increasingly strong NEC violation as the dilaton increases. 
This produces 
the required bounce in the Einstein frame.  Unfortunately, the late
time solutions 
\begin{eqnarray} 
H_S&=&{2 t \over 3} \\ 
\varphi&=&\log \left( {
t^6 \over 39366} \right) 
\end{eqnarray} 
indicate unbounded growth in the
curvature and dilaton \cite{BruMad98}. The price that has been paid for
obtaining a bounce in the Einstein frame is to introduce a new
graceful exit problem.  In general, it appears that generic forms of
quantum corrections can complete the exit from the fixed point of
\cite{GasMagVen97} to the bounce region in the Einstein frame, showing
that NEC violation is not only necessary, but is in some sense
sufficient. However, they are accompanied with unbounded growth of the
dilaton at late times, continue to be dominated by corrections and
continue to violate NEC, leading to unstable $(-)$ branch
solutions. To overcome this problem we need to turn off the NEC
violating corrections and to stop the dilaton from growing indefinitely. 

A direct
approach to completing the exit transition is to assume that there
exists some mechanism that shuts off the correction terms, and hence,
NEC violation. Brustein and Madden proposed an explicit example to do
this \cite{BruMad98}, by replacing the quantum correction in the
action with a non-perturbative function $f(\varphi) {\cal L}_q^{\varphi}$,
where $f(\varphi)$ is a positive constant for $\varphi<\varphi_0$ for some
constant $\varphi_0$ and then smoothly goes to zero. Thus, $f(\varphi)$ has the
form of a smoothed step function. This successfully eliminates the
loop corrections at late times so the dilaton may be easily captured
by a potential or slowed by radiation.  

An alternative method of
suppressing the NEC violation is to add higher--order loop corrections
of the correct sign to counter the leading order corrections. 
For example, a two loop contribution of the form
\begin{eqnarray} 
\frac{1}{2} {\cal L}_q^{2 \varphi}&=&e^{2\varphi}(\nabla
\varphi)^4 
\end{eqnarray} 
can overwhelm the one loop NEC violation when
the dilaton becomes sufficiently large  \cite{BruMad98}. With this form of
correction the behaviour is mild enough
for the dilaton to be easily captured 
in a potential minimum. This can be accompanied with the onset 
of a decelerating phase of expansion. Similarly, these
solutions are stable enough that the growing dilaton can be halted by
introducing radiation. They can then pass into a radiation dominated
phase and be smoothly joined to standard cosmologies.  

Although
Brustein and Madden \cite{BruMad98} provided the first example of a
completed graceful exit based on classical evolution from an
effective action, the models were not derived from any particular particle
physics inspired examples and a number of unresolved questions remain. 
In particular, do specific string models
produce coefficients of the appropriate sign and magnitude? Do the actual
one-loop terms that are generated violate NEC? And finally, 
what is the physical mechanism that shuts off these effects at the correct 
epoch?

We now turn our attention to attempts that have been made 
in answering these interesting questions.  One of the most
promising directions has been proposed in \cite{FofMagStu99}. In this 
paper, the authors 
studied loop corrections derived from the heterotic string theory
compactified on a $Z_N$ orbifold. They included the effects of {\em
all-order} loop corrections to the K\"{a}hler potential together with 
corrections to the gravitational couplings, including both threshold
corrections and corrections due to the mixed K\"{a}hler-gravitational
anomaly. They demonstrated that in this
particular model, string loops lead to solutions that, in the
string frame, smoothly connect the pre-big bang evolution to a phase
where the curvature and the derivative of the dilaton are decreasing. 
In other words they achieve a successful graceful exit.  

The choice
of the order $\alpha'$ classical correction was the same as 
that proposed in Eq.~(\ref{43}) \cite{GasMagVen97}, 
but string loop corrections
emerged out of the details of the compactification on the $Z_3$
orbifold. The encouraging result is that the corrections include a
term which has some of the desired properties of the function
$f(\varphi)$ described above, including the correct sign. In fact, the loop
corrections due to the K\"{a}hler potential produce a bounce in
$H_E$, moving the solution into the required region
$H_E>0$. Unfortunately, the new scaling solution still leads to fixed
points $\dot{\varphi}$ and $H$ constant, as opposed to the required
$H,\dot{\varphi}$ decreasing with time. 

On the other hand, there is a remarkable turnaround
of fortunes when moduli-dependent threshold corrections are
included in the action. These turn the De Sitter phase with linearly
growing dilaton into a phase with $H,\dot{\varphi}$ decreasing as the
solution approaches the (-) branch. This 
is precisely the behaviour that we
require. However, without the K\"{a}hler potential corrections, there
would not have been a bounce in $H_E$. This emphasizes the importance of
the corrections to the overall success of the model. One worrying
aspect also emerges, namely that the string coupling $g_S^2 = e^\varphi$ is
of order one in these regions, i.e. we are beginning to enter the strong
coupling regime. This indicates the need to consider genuine
non-perturbative features as opposed to the perturbative corrections
that have so far been incorporated. Although 
this is beyond the scope of this
review, it should be noted that in \cite{FofMagStu99}, the authors
began to address this issue. In particular, they pointed out that at a
critical coupling, the loop corrections to the K\"{a}hler potential
induce a ghost-like instability, i.e., the kinetic term of the dilaton
vanishes. This is similar to what happens in the Seiberg--Witten theory and
signals the transition to a new regime, where the light modes in the
effective action are different and are related to the original ones by
S-duality \cite{SeiWit94a,SeiWit94b}. 
In a string context, this means that a D-brane dominated
phase is entered, and these should be taken into account
\cite{MagRio98}.  

We should ask why Eq.~(\ref{43}) is the
correct form of order $\alpha'$ classical corrections? Indeed, it was
proposed in \cite{GasMagVen97} because of its invariance under field
redefinitions and, as such, is a sensible first correction term. However,
there exist other terms which could be added at the same order and
which maintain invariance under field redefinition. These are
given in \cite{Maggiore97} and \cite{Meissner96,KalMei97} and 
lead to a generalization of Eqs.~(\ref{43}) and (\ref{gmvlag}), 
\begin{eqnarray} 
\frac{1}{2} {\cal L}_{\alpha'}= k \alpha' \left[ 
{1\over4}R_{GB}^2 + {A \over 2} (\nabla \varphi)^4 
+ {D \over 2} \Box \varphi (\partial \varphi )^{2} 
+ {C \over 2} \left(
R^{\mu\nu}-\frac{1}{2}g^{\mu\nu}R \right) \partial_{\mu} \varphi
\partial_{\nu} \varphi \right]
\label{Full} 
\end{eqnarray} 
where terms up to fourth order in the derivatives are kept.  
Recently, Brustein and Madden have analysed this
classical action in some depth \cite{BruMad99}. 
Requiring that the whole Lagrangian arises from a field redefinition
of the low-energy effective action then implies that the three remaining
coefficients satisfy \cite{BruFofStu99}, 
\begin{equation} 
C=-(2A+2D+1)
\label{constraint}
\end{equation} 

We can recover the original classical correction of Eq.~(\ref{43})
by setting $C=D=0$. However, it is clear that this corresponds to a particular
choice of parameters and there exist more general combinations
satisfying Eq.~(\ref{constraint}).  Indeed, 
a combination has been proposed with the interesting property that
the correction to the low energy action also possesses scale factor
duality \cite{Meissner96,KalMei97} . 
This can be achieved  by setting $C=-2,~D+1,~A=-\frac{1}{2}$. 
In some respects this form of the correction is better motivated than
the truncated combination of Eq. (\ref{43}), as it has introduced the
string symmetry of scale factor duality. As we have
mentioned earlier, the precise forms of the quantum corrections have
not been calculated, but choosing particular forms that 
are related to the
classical corrections enables the new solutions to violate the null
energy condition. This is a necessary condition for there to be a branch
change \cite{BruMad97,BruMad98}. We believe that the string coupling,
$g_s^2=e^\varphi$, controls the importance of the string loop corrections and 
therefore a reasonable first guess seems to be that we can model the
quantum corrections by multiplying each term of the classical
correction by a suitable power of the string coupling
\cite{BruMad97,BruMad98}. 

Thus, we can write down an expression for the
modified Lagrangian including classical and quantum corrections as
\begin{equation}
{\cal L} = {\cal L}^{(0)} + {\cal L}_{\alpha'} + 
p e^{\varphi}{\cal L}_{\alpha'} + q e^{2\varphi}{\cal L}_{\alpha'} + ...
\label{expansion}
\end{equation}
where ${\cal L}^{(0)}$ is the tree-level Lagrangian and ${\cal L}_{\alpha'}$ 
is the
classical correction of Eq. (\ref{Full}), with $p$ and $q$ representing 
arbitrary coefficients that multiply the quantum corrections. A detailed
analysis of the solutions to Eq.~(\ref{expansion}) has been performed
in \cite{CarCop99}. The general feature that emerges is that it is possible to
obtain a successful graceful exit when the quantum corrections are
included, but not if it is only the classical corrections of
Eq.~(\ref{Full}) that are present. (This result was first pointed out in
\cite{BruMad99}).  Of particular interest are the solutions which
correspond to the classical action preserving scale factor
duality. However, although the solutions are
non-singular, the problem of stabilising the dilaton field still
remains since these models have no method of providing the required potential
for the field.
     
Mukhanov and Brandenberger have proposed an alternative approach that
could also lead to non-singular dilaton cosmologies
\cite{MukBra92}. In the limited curvature hypothesis, one explicitly
constructs a model with curvature terms that remain non-singular for
all time. Recently, Brandenberger and Easson \cite{EasBra99} have
extended earlier work \cite{BraEasMai98} and applied this idea directly
to the pre--big bang scenario. 
By introducing specific higher derivative terms 
to the usual string frame dilaton--graviton 
action, they have obtained non--singular spatially
homogeneous solutions where all the curvature
invariants are bounded \cite{EasBra99}. 
In particular, they obtain
solutions to the graceful exit problem which at late time consist of an
expanding FRW universe with $\dot{H} <0$. However, as in all other
scenarios, a stabilizing potential for the dilaton has to be 
introduced by hand.  

Easther and Maeda \cite{EasMae96} have also proposed an
interesting mechanism for regulating the singularities present in the tree
level string effective action. By employing one-loop effects, 
they were able to numerically demonstrate the existence
of non-singular solutions which smoothly connect a contracting phase
to an expanding phase when spatial curvature is present in 
the model.

In conclusion, the question of how the universe exits from
the pre--big bang phase to the decelerating FRW phase has been a major 
problem for the pre--big bang scenario, and it still awaits a fully
satisfactory explanation. However, a great deal of progress has been
made in recent years.  At the perturbative level, it has become clear that both
classical $\alpha'$ as well as quantum loop corrections are required
if a successful resolution to the problem 
is to be found. In particular, quantum
corrections are necessary for achieving a bounce in the Einstein curvature,
$H_E$. Unfortunately, the form of these corrections is not yet
fully understood, although specific models that appear to 
work do now exist. 
The major issue of stabilizing the dilaton is still a thorny
one. Most scenarios lead to fixed point solutions where the dilaton
grows linearly with time and rapidly enters the strong coupling regime. 
This implies that non-perturbative features, such as those 
arising from a D-brane dominated phase, might have to 
be considered \cite{MagRio98}. 

\section{Primordial Perturbations from a Pre--Big Bang Era}
\setcounter{equation}{0} 
 
\def\tbox{\stackrel{\sim}{\Box}} 

\label{Section10}

In the previous Section, we considered only homogeneous classical
solutions to the lowest--order string equations of motion. We will now consider
inhomogeneous perturbations that may be generated due to vacuum
fluctuations. 
The formalism required to describe vacuum fluctuations about a
homogeneous background including first-order metric fluctuations was
pioneered by Mukhanov and collaborators
\cite{Mukhanov88,MukFelBra92}.
While the solutions for the homogeneous dilaton, axion and scale
factor in the different frames may lead to interesting behaviour in
the early universe, the success of the standard big bang model
suggests that the evolution should closely approach the conventional
general relativistic evolution at least by the time of
nucleosynthesis. If we are to see any trace of the earlier evolution
it will be in the primordial spectrum of inhomogeneities present on
large-scales that we observe today. Such large-scale structure can
only be generated by some unconventional physics, such as inflation
\cite{LidLyt93}, topological defects\cite{VilShe94} or a
pre-big bang epoch.  

The production of scalar and tensor metric perturbations in the
pre-big bang scenario has been studied by various authors (see for
example~\cite{BruGasGio95,BruGasGio95b,BruGasVen97,GalLitOcc97,Maggiore97b,Gasperini97,CopEasWan97,CopLidWan97,Hwang96,Hwang98,HwaNoh98}). 
During a period of accelerated expansion the comoving Hubble length,
$|d(\ln a)/d\eta|^{-1}$, decreases and vacuum fluctuations
which are assumed to start in the flat-spacetime vacuum state may be
stretched up to exponentially large scales. The precise form of the
spectrum depends on the expansion of the homogeneous background and
the couplings between the fields.
The comoving Hubble length,
$|d(\ln\tilde{a})/d\eta|^{-1}=2|\eta|$, does indeed decrease in the
Einstein frame during the contracting phase when $\eta<0$. Because the
dilaton, moduli fields and graviton are minimally coupled to this
metric, this ensures that small-scale vacuum fluctuations will
eventually be stretched beyond the comoving Hubble scale during this epoch.

As we remarked earlier, the axion field is taken to be a constant in
the classical pre-big bang solutions.  However, even when the
background axion field is set to a constant, there will inevitably be
quantum fluctuations in this field.  We will show that these 
fluctuations can not be neglected and, moreover, that 
they are vital if the pre-big bang scenario is to have
any chance of generating the observed density perturbations.

In this Section we will calculate the spectrum of semi-classical axion
perturbations as well as dilaton and moduli perturbation spectra
produced in the pre-big bang scenario. Then we will extend this
calculation to the more general dilaton--moduli--axion cosmologies
discussed in Section \ref{Section7.1}, by constructing S-duality
invariant combinations of the field perturbations. This enables us to
derive S-duality invariant solutions. It will turn out that the
late-time dilaton and axion spectra are {\em independent} of the
preceding evolution along different but S-duality related classical
solutions. Significantly, the tilt of the axion spectrum can be
appreciably different from the steep ``blue'' spectra of dilatons and
gravitons predicted by the pre-big bang scenario. This implies that
the pre--big bang scenario can indeed predict spectra consistent with
the observed scale-invariant form.

\subsection{Metric perturbations}

In this subsection we set out our formalism for describing
inhomogeneous linear perturbations about the spatially homogeneous and
isotropic background solutions.  
One can readily extend the four-dimensional formalism of
gauge-invariant metric perturbations~\cite{Bardeen80,MukFelBra92} to
higher-dimensional cosmologies~\cite{KodSas84,GasGio97}. 
Arbitrary linear perturbations about a $D$-dimensional cosmological
model, where the spatial hypersurfaces are maximally symmetric, can be
described by the perturbed line element
\begin{eqnarray}
\label{ds}
ds^2 = a^2(\eta) \left\{ -(1+2A) d\eta^2+2(B_{|i}-S_i)d\eta dx^i
\right. \nonumber \\
\left. + \left[(1-2\psi)\gamma_{ij}+2E_{|ij} +2F_{i|j}+h_{ij} \right] dx^idx^j
 \right\} \, ,
\end{eqnarray}
where $\gamma_{ij}$ is the metric on the spatial hypersurfaces, and
the subscript $|$ denotes covariant derivatives on these
hypersurfaces. In general we have four scalar perturbations ($A$, $B$,
$\psi$, and $E$), two vectors ($S_i$ and $F_i$) and one tensor
perturbation ($h_{ij}$), defined in terms of their transformation
properties on the $(D-1)$-dimensional spatial
hypersurfaces~\cite{Bardeen80,Stewart90}.  The advantage of splitting
the metric perturbations into scalar and tensor parts is that the
scalar and tensor modes evolve independently to first order with only
the scalar perturbations being coupled to scalar field
fluctuations~\cite{Bardeen80}.  In addition, linear perturbations
about the homogeneous background fields can be decomposed as a sum of
Fourier modes with comoving wavenumber $k$ which evolve independently
of other wavenumbers.

Two of the scalar functions and one vector function can be eliminated
by an appropriate gauge transformation~\cite{Bardeen80}. 
We will find it convenient to work in terms of the gauge-invariant
scalar quantities~\cite{KodSas84}
\begin{eqnarray}
\widetilde{A} &\equiv& A + \psi + \left( {\psi \over h} \right)^{\prime}
\ ,\\
\widetilde{B} &\equiv& B - E' - {\psi \over h} \ .
\end{eqnarray}
where a prime denotes derivatives with respect to conformal time, $\eta$.
These coincide with the scalar metric perturbations in the uniform
curvature gauge~\cite{Hwang96,HwaNoh98,MalWan98}\footnote{Called the
``off-diagonal gauge'' in Ref.~\cite{BruGasGio95}.} where the
metric perturbations on the constant-$\eta$ hypersurfaces
($\widetilde\psi$ and $\widetilde{E}$) have been eliminated by a gauge
transformation.

For completeness we also give the gauge-invariant vector perturbation
\begin{equation}
\widetilde{S}_i \equiv S_i + F_i' \ .
\end{equation}
The perturbed Einstein equations constrain the gauge-invariant vector
perturbation to be proportional to the vorticity of the velocity
field~\cite{KodSas84}. This necessarily vanishes (to first-order) in a
scalar field dominated universe and the vector metric perturbations
are therefore zero in a pre--big bang era.

In the Einstein frame, the first-order perturbed line element can
then be written as
\begin{equation}
\label{dds}
d\widetilde{s}^2  =  \widetilde{a}^2(\eta)
 \left\{ -(1+2\widetilde{A})d\eta^2
 + 2\widetilde{B}_{,i} d\eta dx^i
+ \left[\delta_{ij} + h_{ij}\right] dx^i dx^j \right\} \, .
\end{equation}
Note that the scalar metric perturbations are not invariant under a
conformal transformation. Even the spatially flat nature of the line
element in Eq.~(\ref{dds}) is not preserved under a conformal
transformation back to the string frame due to the first-order
perturbation in the conformal factor
$e^\varphi=e^{\varphi_0}(1+\delta\varphi)$.  However vector and tensor
perturbations do remain invariant under both conformal transformations
and time-coordinate transformations $\eta\to\eta+\delta\eta$.

We can relate the scalar metric perturbations $\widetilde{A}$ and
$\widetilde{B}$ to the more familiar ones introduced by
Bardeen\cite{Bardeen80} in terms of the gauge-invariant metric
potentials \cite{MukFelBra92},
\begin{eqnarray}
\label{APhiPsi}
\widetilde{A} & \equiv & \widetilde\Phi + \widetilde\Psi
 + \left( {\widetilde\Psi \over \widetilde{h}} \right)' \, ,\\
\label{BPhiPsi}
\widetilde{B} & \equiv & - \, {\widetilde\Psi \over \widetilde{h}} \, .
\end{eqnarray}
The gauge transformation
\begin{equation}
\label{gt}
\eta \to \eta-{\widetilde{\Psi}\over\widetilde{h}} \, ,
\end{equation}
brings the metric of Eq.~(\ref{dds}) into the more commonly used
longitudinal gauge~\cite{MukFelBra92}, where
\begin{eqnarray}
d\widetilde{s}^2 & \to & \widetilde{a}^2(\eta)
 \left\{ -(1+2\widetilde{\Phi})d\eta^2 
 + \left[(1-2\widetilde{\Psi})\delta_{ij} + h_{ij}\right]
 dx^i dx^j \right\} \, .
\end{eqnarray}


We will now consider the evolution of linear metric perturbations
about the four-dimensional dilaton-moduli-vacuum solutions given in
Eqs.~(\ref{dilphit1}--\ref{dilbetat1}).
Considering a single Fourier mode, with comoving wavenumber $k$,
the perturbed Einstein equations yield the evolution equation
\begin{equation}
\label{Aeom}
\widetilde{A}'' + 2\widetilde{h}\widetilde{A}' + k^2 \widetilde{A}
  =  0 \, ,
\end{equation}
plus the constraint
\begin{equation}
\label{AofB}
\widetilde{A} = - ( \widetilde{B}' +2\widetilde{h}\widetilde{B} ) \, .
\end{equation}
In the spatially flat gauge we have the simplification that the
evolution equation for the scalar metric perturbation, Eq.~(\ref{Aeom}), 
is independent of the evolution of the different
massless scalar fields (dilaton, axion and moduli), although they will
still be related by the constraint Eq.~(\ref{pBBAcon}).
The metric perturbation's evolution is dependent only up on the 
Einstein frame scale factor, $\widetilde{a}(\eta)$, given by
Eq.~(\ref{Esf}), which in turn is determined solely by the stiff fluid
equation of state for the homogeneous fields in the Einstein frame. 
It is independent of the time dependence of the individual scalar fields.

Eq. (\ref{Aeom}), using the background solution~(\ref{Esf}), can be
integrated to give the general solution
\begin{equation}
\label{genA}
\widetilde{A} = 
 \left[ A_+ H_0^{(1)}(-k\eta) + A_- H_0^{(2)}(-k\eta) \right] \, ,
\end{equation}
where $H_\nu^{(1)}(z)\equiv J_\nu(z)+iY_\nu(z)$ and $H_\nu^{(2)}\equiv
J_\nu(z)-iY_\nu(z)$ are  Hankel functions of the 
first and second kind. Using the recurrence relation between Bessel
functions, we obtain from Eqs.~(\ref{AofB}) and~(\ref{genA}),
\begin{equation}
\label{genB}
\widetilde{B} = {1\over k}
 \left[ A_+ H_1^{(1)}(-k\eta) + A_- H_1^{(2)}(-k\eta) \right] \, .
\end{equation}

One of the most useful quantities we can calculate 
is the curvature perturbation on uniform energy density hypersurfaces (as
$k\eta\to0$). It is commonly denoted by $\zeta$~\cite{BarSteTur83} and 
is given in terms of the gauge-invariant potential $\Phi$ by~\cite{MukFelBra92}
\begin{equation}
\zeta \equiv \widetilde{\Phi}
 - {\widetilde{h}^2 \over \widetilde{h}'-\widetilde{h}^2}
 \left( \widetilde\Phi + \widetilde{h}^{-1}\widetilde{\Phi}' \right) 
\, .
\end{equation}
Hence,  with $\widetilde{h}$ given by Eq.~(\ref{Esf}) for the scale
factor in the Einstein frame, we obtain
\begin{equation}
\zeta = {\widetilde{A}\over3} \, ,
\end{equation}
in any dilaton--moduli--vacuum or dilaton--moduli--axion cosmology.

The significance of $\zeta$ is that in an expanding universe it
becomes constant on scales much larger than the Hubble scale
($|k\eta|\ll1$) for purely adiabatic perturbations, even through
changes in the equation of state.  In single-field inflation models
this allows one to compute the density perturbation at late times,
during the matter or radiation dominated eras, by equating $\zeta$ at
``re-entry'' ($k=\widetilde{a}\widetilde{H}$) with that at horizon
crossing during inflation.  A number of authors have calculated the
spectrum of $\widetilde{A}$, and hence $\zeta$, in order to predict
the density perturbations induced in the pre-big bang
scenario~\cite{BruGasGio95,Hwang96,HwaNoh98}. One can either use the vacuum
fluctuations for the canonically normalised field at early times/small
scales (as $k\eta\to-\infty$) or use the amplitude of the scalar field
perturbation spectra (discussed in the next subsection) to normalise the
solution for $\widetilde{A}$ given in Eq.~(\ref{genA}). This yields
the curvature perturbation spectrum on large scales/late times (as
$k\eta\to0$):
\begin{equation}
\label{Pzeta}
\label{Aspectrum}
{\cal P}_{\zeta} = {8\over\pi^2}
 l_{\rm Pl}^2\widetilde{H}^2 (-k\eta)^3[\ln(-k\eta)]^2
 \, ,
\end{equation}
where $l_{\rm Pl}$ is the Planck length in the Einstein frame and remains 
fixed throughout.  
The scalar metric perturbations become large on superhorizon
scales ($|k\eta|<1$) only near the Planck era, $\widetilde{H}^2\sim
l_{\rm Pl}^{-2}$. Even though Bardeen's gauge invariant perturbations
$\widetilde{\Phi}$ and $\widetilde{\Psi}$, defined in
Eqs.~(\ref{APhiPsi}) and~(\ref{BPhiPsi}), actually become large much
earlier~\cite{BruGasGio95}, the fact that the perturbations remain
small in our choice of gauge implies that our linear calculation is in
fact valid up until the Planck epoch~\cite{BruGasGio95}.

The spectral index of the curvature perturbation spectrum is
conventionally given as~\cite{LidLyt93}
\begin{equation}
\label{zetaspecindex}
n \equiv 1+ {d\ln{\cal P}_{\zeta} \over d\ln k}
\end{equation}
where $n=1$ corresponds to the classic Harrison-Zel'dovich spectrum
for adiabatic density perturbations favoured by most models of
structure formation in our universe. By contrast the pre--big bang era
leads to a spectrum of curvature perturbations with $n=4$.
Such a steeply tilted spectrum of
metric perturbations implies that there would be effectively no primordial
metric perturbations on large (super-galactic) scales in our present
universe if the post-Big bang era began close to the Planck scale. The
metric fluctuations are of order unity on the Planck scale
($10^{-33}$cm) when $T\sim10^{32}$K in the standard post-big bang
model.  This corresponds to a comoving scale of about $0.1$cm today
(when $T=2.7$K), about $10^{-29}$ times the scale of perturbations
observed on the microwave background sky. Thus the microwave
background temperature anisotropies should be of order $10^{-87}$
rather than the observed $10^{-5}$.

Fortunately, as we shall see later, the presence of the axion field could
provide an alternative spectrum of perturbations more suitable as a
source of large-scale structure.
The pre-big bang scenario is not so straightforward as in the
single field inflation case, because the full low-energy string
effective action possesses many fields which can lead to non-adiabatic
perturbations. This implies that density perturbations at late times
may not be simply related to $\zeta$ alone, but may also be dependent
upon fluctuations in other fields. As we shall see, one of these
fields, the axion field, can lead to a markedly different spectrum. 


The gravitational wave perturbations, $h_{ij}$, are both gauge and
conformally invariant. They decouple from the scalar perturbations in
the Einstein frame to give a simple evolution equation for each
Fourier mode
\begin{equation}
\label{heom}
h_k'' + 2\widetilde{h}\, h_k' + k^2 h_k  = 0 \, .
\end{equation}
This is exactly the same as the equation of motion for the scalar
perturbation given in Eq.~(\ref{Aeom}) and has the same
growing mode in the long wavelength ($|k\eta|\to0$) limit given by
Eq.~(\ref{Pzeta}). We will not consider here gravitational waves
propagating in the $n$ internal dimensions~\cite{GasGio97,Giovannini97}.
The spectrum depends solely on the dynamics of the scale factor in the
Einstein frame given in Eq.~(\ref{Esf}), which remains the
same regardless of the time-dependence of the different dilaton,
moduli or axion fields. It leads to a spectrum of primordial
gravitational waves steeply growing on short scales, with a spectral
index\footnote
{Conventionally a scale-invariant spectrum is denoted 
by $n_T=0$ for tensor perturbations, in contrast to the choice $n=1$ for 
scalar perturbations~\cite{LidLidKol97}.} 
$n_T=3$~\cite{GasVen93a,GasVen93c,BruGasGio95,BruGasGio95b,GalLitOcc97},
in contrast to conventional inflation models which require
$n_T<0$~\cite{LidLyt93}. The graviton spectrum appears to be a robust
and distinctive prediction of any pre-big bang type evolution based on
the low-energy string effective action.

Such a spectrum could be observed by the next generation of
gravitational wave detectors such as the Laser Interferometric
Gravitational Wave Observatory (LIGO) if they are on the right scale
\cite{AllBru97,Maggiore99}. The current frequency of these waves
depends on the cosmological model, and in general we would require
either an intermediate epoch of stringy inflation~\cite{BruMad98}, or
a low re-heating temperature at the start of the post-big bang
era~\cite{CopLidLid98} to place the peak of the gravitational wave
spectrum at the right scale.  Nonetheless, the possible production of
high amplitude gravitational waves on detector scales in the pre--big
bang scenario is in marked contrast to conventional inflation models
in which the Hubble parameter decreases during inflation. This
produces a negative spectral tilt $n_T<0$ and the isotropy of the
cosmic microwave background on large scales then leads to an upper
limit on the amplitude of perturbations that is many orders of
magnitude below the LIGO sensitivity \cite{Liddle94}.

Because the scalar and tensor metric perturbations obey the same
evolution equation, their amplitude is directly related. The amplitude
of gravitational waves with a given wavelength is commonly 
described in terms of their energy density at the present epoch. For
the simplest pre--big bang models this is given in terms
of the amplitude of the scalar perturbations as
\begin{equation}
\Omega_{\rm gw} = {2\over z_{\rm eq}} {\cal P}_\zeta 
\end{equation}
where $z_{\rm eq}=24000\Omega_oh^2$ is the red-shift of
matter-radiation equality. The advanced LIGO configuration will be
sensitive to $\Omega_{\rm gw}\approx10^{-9}$ over a range of scales
around 100Hz. However, the maximum amplitude of gravitational waves on
these scales is constrained by limits on the amplitude of primordial
scalar metric perturbations on the same scale~\cite{CopLidLid98}. In 
particular, if
the fractional over-density when a scalar mode re-enters the horizon during
the radiation dominated era is greater than about $1/3$, then that
horizon volume is liable to collapse to form a black hole.

This is important, because a 
scale with comoving frequency $f_*\sim100$Hz re-enters
the Hubble radius during the radiation era when the temperature is
\begin{equation}
{T_* \over 10^9\,{\rm GeV}} \approx {f_* \over 100\, {\rm Hz}} \ ,
\end{equation}
The mass contained within the horizon at that time is
\begin{equation}
M_* \approx 10^{14} \left( {100\, {\rm Hz} \over f_*} \right)^2\ {\rm
g}\,.
\end{equation}
Thus, gravitational radiation at LIGO-observable
frequencies could be associated with primordial black holes with
masses of order $10^{14}$g. Such black holes 
have a lifetime of the order the Hubble time and 
would be evaporating today!

There are strong observational limits on the mass-fraction of
primordial black holes with masses greater than about
$1$g~\cite{CarGilLid94,GreLid97}. Only a
tiny fraction of the universe is allowed to form black holes, and this
in turn yields an upper limit on the allowed amplitude of
gravitational waves \cite{CopLidLid98}:
\begin{equation}
\Omega_{\rm gw} = {5\times10^{-6} \over \Omega_0h^2} \, .
\end{equation}
This is slightly below the upper limit on the total density in
gravitational waves coming from models of primordial
nucleosynthesis, $\Omega_{\rm gw}<5\times10^{-5}$,
but is well within the advanced LIGO sensitivity.
If we find PBH's and gravitational waves together then this would
indeed be an exciting result for string cosmology!

\subsection{Dilaton--Moduli Perturbation Spectra}

We will now consider inhomogeneous linear perturbations in the fields
about a homogeneous background given by
\begin{equation}
\varphi = \varphi(\eta) + \delta \varphi ({\bf x},\eta), ~~~\sigma = \sigma(\eta) 
+ \delta \sigma ({\bf x},\eta), ~~~\beta = \beta (\eta) + \delta \beta 
({\bf x},\eta) \ .
\end{equation}
The perturbations can be re-expressed as a Fourier series in terms of
Fourier modes with comoving wavenumber $k$.  
We will work in terms of the scalar field perturbations in the
spatially flat gauge. These can be defined in a gauge-invariant manner as
\begin{equation}
\widetilde{\delta{x}} = \delta x - x'{\psi \over \widetilde{h}} \,.
\end{equation}
In particular, we note that the scalar field perturbations in the 
longitudinal ($\delta x_l$) and spatially flat ($\widetilde{\delta{x}}$)
gauges are related by
\begin{equation}
\label{gaugephi}
\delta x_l
 = \widetilde{\delta x} + x' {\widetilde\Psi \over \widetilde{h}} \, .
\end{equation}
In what follows we will always work with field perturbations defined in the
spatially flat gauge and drop the tildes.

We will first consider the production of dilaton, moduli
and axion perturbations during a pre-big bang evolution where the
background axion field is constant, $\sigma'=0$, and 
the evolution of the homogeneous background fields is
given in Eqs.~(\ref{dilphit1}--\ref{dilbetat1}).
The dilaton and
moduli fields both evolve as minimally coupled massless fields
in the Einstein frame. In particular, the dilaton perturbations are
decoupled from the axion perturbations and the equations of motion in
the spatially flat gauge become
\begin{eqnarray}
\label{pBBdphieom}
\delta\varphi'' + 2\widetilde{h}\delta\varphi' + k^2\delta\varphi
 & = & 0 \, , \\
\label{pBBdbetaeom}
\delta\beta'' +2\widetilde{h}\delta\beta' +k^2\delta\beta
 & = & 0 \, ,\\
\label{pBBdsigmaeom}
\delta\sigma''+2\widetilde{h}\delta\sigma' + k^2\delta\sigma
 & = & - 2\varphi'\delta\sigma' \, ,
\end{eqnarray}
Note that these evolution equations for the scalar field perturbations
defined in the spatially flat gauge are automatically decoupled from
the metric perturbations, although they are still related to the
scalar metric perturbation, $\widetilde{A}$, by the constraint
\begin{equation}
\label{pBBAcon}
\widetilde{A} = {\varphi'\over4\widetilde{h}} \, \delta\varphi
  +{\beta'\over4\widetilde{h}} \, \delta\beta \, .
\end{equation}
We see that, to first-order, the metric
perturbation, $\widetilde{A}$, is determined solely by the dilaton and
moduli field perturbations. 


The canonically normalised dilaton and moduli field perturbations
are~\cite{Mukhanov88,BruGasGio95,GasGio97,Giovannini97} 
\begin{eqnarray}
\label{defu}
u & \equiv & {1\over\sqrt{16\pi}l_{\rm Pl}} \widetilde{a}\delta\varphi \, ,\\
w & \equiv & {1\over\sqrt{16\pi}l_{\rm Pl}} \widetilde{a}\delta\beta \, ,
\end{eqnarray}
which, from Eqs.~(\ref{pBBdphieom})
and~(\ref{pBBdbetaeom}), obey the same wave equation
\begin{equation}
\label{upp}
u'' + \left( k^2 - {\widetilde{a}''\over\widetilde{a}} \right) u = 0 
\, .
\end{equation}
After inserting the simple solution for the Einstein frame
scale factor given in Eqs.~(\ref{Esf}) we find that this equation
gives the general solution
\begin{equation}
\label{usol}
u  = 
 |k\eta|^{1/2} \left[ 
  u_+ H_0^{(1)}(|k\eta|) + u_- H_0^{(2)}(|k\eta|) 
 \right]
 \, .
\end{equation}

On the $(+)$ branch, i.e., when $\eta<0$, we can normalise modes at
early times, $\eta\to-\infty$, where all the modes are far inside the
Hubble scale, $k\gg|\eta|^{-1}$, and can be assumed to be in
the flat-spacetime vacuum. The reader may note that in conventional
inflation we have to assume that this result for a quantum field in a
classical background holds at the Planck scale. Here, however, the
normalisation is done in the zero-curvature limit in the infinite
past. Just as in conventional inflation, this produces perturbations
on scales far outside the horizon, $k\ll|\eta|^{-1}$, at late times,
$\eta\to0^-$.

Conversely, the solution for the $(-)$ branch with $\eta>0$ is dependent
upon the initial state of modes far outside the horizon,
$k\ll|\eta|^{-1}$, at early times where $\eta\to0$. The role of a period
of inflation, or of the pre-big bang $(+)$ branch, is precisely to set
up this initial state which otherwise appears as a mysterious initial
condition in the conventional (non-inflationary) big bang model.

Allowing only positive frequency modes in the flat-spacetime vacuum
state at early times for the pre-big bang $(+)$ branch
requires~\cite{BirDav82} that, as $k\eta\to-\infty$,
\begin{equation}
\label{shortwave}
u \to {e^{-ik\eta} \over \sqrt{2k}} \, ,
\end{equation}
and similarly for $w$, giving
\begin{equation}
\label{uplus}
u_+ = w_+ = e^{i\pi/4} {\sqrt\pi\over2\sqrt{k}} \, , \qquad u_-=w_-=0 \, .
\end{equation}

The power spectrum for perturbations is commonly denoted by
\begin{equation}
{\cal P}_{\delta x} \equiv {k^3\over2\pi^2} |\delta x|^2 \, ,
\end{equation}
and thus for modes far outside the horizon ($k\eta\to0$) we have 
\begin{eqnarray}
\label{pBBphi}
{\cal P}_{\delta\varphi} & = & {32\over\pi^2} l_{\rm Pl}^2\widetilde{H}^2
 (-k\eta)^3[\ln(-k\eta)]^2
 \,,\\
\label{pBBbeta}
{\cal P}_{\delta\beta} & = & {32\over \pi^2} l_{\rm Pl}^2\widetilde{H}^2
 (-k\eta)^3[\ln(-k\eta)]^2
 \,,
\end{eqnarray}
where
$\widetilde{H}\equiv\widetilde{a}'/\widetilde{a}^2=1/(2\widetilde{a}\eta)$
is the Hubble rate in the Einstein frame.
The amplitude of the perturbations grows towards small
scales, but only becomes large for modes outside the horizon
($|k\eta|<1$) when $\widetilde{H}^2\sim l_{\rm Pl}^{-2}$, i.e., the
Planck scale in the Einstein frame.
The spectral tilt of the perturbation spectra is given by
\begin{equation}
\label{specindex}
n-1 \equiv \Delta n_x =  {d\ln{\cal P}_{\delta x} \over d\ln k}
\end{equation}
which from Eqs.~(\ref{pBBphi}) and~(\ref{pBBbeta}) gives $\Delta
n_\varphi=\Delta n_\beta=3$ (where we neglect the logarithmic
dependence).  This of course is the same steep blue spectra we obtained
earlier for the metric perturbations. 

The constraint Eq.~(\ref{pBBAcon}), allows us to normalise the
amplitude of the metric perturbations $\widetilde{A}$ using
Eqs.~(\ref{dilat1}--\ref{dilbetat1}) for the background fields, and
Eqs. (\ref{pBBphi}) and~(\ref{pBBbeta}) for the scalar-field
perturbations. Together with setting $\zeta=\widetilde{A}/3$ this
yields the scalar curvature power spectrum given in Eq.~(\ref{Pzeta})
[remembering that in Eq.~(\ref{pBBAcon}) we are adding independent
random variables].  This spectrum of scalar metric perturbations is
entirely independent of the integration constants that parameterise
the dilaton-moduli-vacuum solutions given in
Eqs.~(\ref{dilat1}--\ref{dilbetat1}). As in the case of the
spectrum of tensor perturbations, 
this makes it a robust prediction of any pre-big
bang scenario where the universe collapses in the Einstein frame, and
becomes dominated by homogeneous scalar fields.

\subsection{Magnetic field perturbations in the Pre--Big Bang scenario}

One of the outstanding problems in modern cosmology is the origin of
the galactic and extra-galactic magnetic
fields~\cite{Kronberg94}. Conventional inflation models where the
photon field is minimally coupled to the scalar field driving
inflation cannot leave a magnetic field on large
scales~\cite{TurWid88,Ratra92a,Ratra92b}.  However in string theory
the dilaton is automatically coupled to the electromagnetic field
strength.  In the heterotic string effective action the photon field
Lagrangian is of the form (see Eq. (\ref{HET}))
\begin{equation}
\label{Lem}
{\cal L} = e^{-\varphi} F_{\mu\nu} F^{\mu\nu} \,,
\end{equation}
where the field strength is derived from the vector potential,
$F_{\mu\nu}=\nabla_{[\mu}A_{\nu]}$. Note that in an isotropic FRW
cosmology the magnetic field must vanish to zeroth-order, and thus
the vector field perturbations are gauge-invariant
and we can neglect the metric back-reaction to first-order. 
If we work in the radiation gauge ($A^0=0$, $A^i_{|i}=0$) then the
field perturbations can be treated as vector perturbations on the
spatial hypersurfaces. The field perturbation can then be written
as~\cite{LemLem95} 
\begin{equation}
A_i = e^{\varphi/2} \chi_k(\eta) Q_i({\bf x})
\end{equation}
where $\delta^{mn}\partial_m\partial_nQ_i=-k^2Q_i$.
The canonically normalised field, $\chi_k$, satisfies the
wave-equation~\cite{LemLem95,GasGioVen95a,GasGioVen95b} 
\begin{equation}
\chi_k'' + 
\left( k^2 - {(e^{-\varphi/2})''\over e^{-\varphi/2}} \right) \chi_k = 0
\,.
\end{equation}
Note that unlike scalar field perturbations these perturbations are
coupled only to the gauge coupling strength $e^\varphi$ and not to the
scale factor. 
The time dependence of the dilaton (rather than the scale factor)
leads to particle production during the pre--big bang from an initial
vacuum state~\cite{LemLem95,GasGioVen95a,GasGioVen95b}.

For the dilaton-moduli--vacuum pre--big bang solutions in
Eqs.~(\ref{dilphit1})--(\ref{dilbetat1}) the general solution can be
given in terms of Hankel Functions
\begin{equation}
\chi_k = 
 |k\eta|^{1/2} \left[ 
  \chi_+ H_\mu^{(1)}(|k\eta|) + \chi_- H_\mu^{(2)}(|k\eta|) 
 \right] \, ,
\end{equation}
where $\mu=|1+\sqrt{3}\cos\xi_*|/2$. Taking the flat spacetime vacuum
state at early times ($k\eta\to-\infty$) on the $(+)$ branch, we
obtain the power spectrum at late times/on large scales ($|k\eta|\ll1$)
\begin{equation}
{\cal P}_\chi = \left( {C(\mu)\over2\pi} \right)^2 k^2
(-k\eta)^{1-2\mu} \,.
\end{equation}
where the numerical coefficient, $C(\mu)$, is given by
Eq.~(\ref{Cofr}). 
This leads to an energy density in electromagnetic perturbations,
$\rho_{\rm em}\propto{\cal P}_\chi$, when these modes re-enter the horizon
($k=aH$) during a subsequent (post-big bang) radiation dominated
era, where the dilaton is fixed ($\phi'=0$). This yields the
dimensionless density relative to the critical density~\cite{LemLem95}
\begin{equation}
\Omega_{\rm em}
 \equiv {8\pi l_{\rm Pl}^2 \rho_{\rm em} \over 3\widetilde{H}^2} 
\sim \left( {k\over k_s} \right)^{4-|1-2\mu|} \ ,
\end{equation}
where we have neglected terms of order unity which can be calculated
by a careful matching of Bogoliubov coefficients between solutions in
succeeding cosmological
epochs~\cite{LemLem95,GasGioVen95a,GasGioVen95b}.

The minimum tilt possible for the pre--big bang solutions given in
Eqs.~(\ref{dilphit1})--(\ref{dilbetat1}) is obtained for $\xi_*=0$
when $\mu=(1+\sqrt{3})/2$ and the spectral tilt $\Delta n_{\rm
em}=4-\sqrt{3}\approx2.3$, which is still strongly tilted towards
smaller scales. It is impossible to obtain a less strongly tilted
spectrum for heterotic photons as that would require an even more
rapid evolution of the dilaton. However, $\xi_*=0$ already corresponds to 
the case where the evolution of all the moduli fields is frozen 
and this maximizes the rate
of change of the dilaton. Instead, one would have to consider a more
strongly coupled field than the one appearing in the Lagrangian 
Eq.~(\ref{Lem}).

\subsection{Axion perturbations in the Pre--Big Bang scenario}

While the dilaton and moduli fields evolve as massless minimally
coupled scalar fields in the Einstein frame, the axion field's kinetic term
still has a non-minimal coupling to the dilaton field. This is evident in
the equation of motion, Eq.~(\ref{pBBdsigmaeom}), for the axion field
perturbations $\delta\sigma$.
The canonically normalised field perturbation is 
\begin{equation}
\label{defv}
v \equiv {1\over\sqrt{16\pi}l_{\rm Pl}} e^{\varphi} \widetilde{a} \delta\sigma \, ,
\end{equation}
and since the background axion field is constant, the resulting
density perturbations are only second-order in the axion perturbation. This
allows us to neglect the back-reaction from the metric to linear order.
The field perturbation $\delta\sigma$ is gauge invariant when
$\sigma'=0$ [see Eq.~(\ref{gaugephi})]
and in any gauge, the axion perturbation obeys the
decoupled wave equation given in Eq.~(\ref{pBBdsigmaeom}). This can be
re-written in terms of $v$ as
\begin{equation}
\label{vpp}
v'' + 
 \left( k^2 - {(e^\varphi\tilde{a})''\over e^\varphi\tilde{a}} \right) v = 0
 \, . 
\end{equation}
The non-minimal coupling of the axion to the dilaton
leads to a significantly different evolution to
that of the dilaton and moduli perturbations. 
Substituting in the background power-law solutions from
Eqs.~(\ref{dilphit1}--\ref{dilbetat1}), we have 
\begin{equation}
\label{vsol}
v  = 
 |k\eta|^{1/2} \left[ 
  v_+ H_\mu^{(1)}(|k\eta|) + v_- H_\mu^{(2)}(|k\eta|) 
 \right] \, ,
\end{equation}
where we have used $\mu\equiv|\sqrt{3}\cos \xi_*|$.
Once again, we can normalise this by employing the flat spacetime vacuum
state at early times as $-k\eta\to\infty$ on the $(+)$ branch, as in
Eq.~(\ref{shortwave}). We obtain
\begin{equation}
\label{vplus}
v_+= e^{i(2\mu+1)\pi/4} {\sqrt{\pi} \over 2\sqrt{k}} \, , \qquad v_-=0 \, 
\end{equation}
and hence we have
\begin{equation}
\label{pBBdsigma}
\delta\sigma = {2\pi l_{\rm Pl} \over \sqrt{k}} e^{i(2\mu+1)\pi/4}
{\sqrt{-k\eta} \over e^{\varphi} \tilde{a}} H_{\mu}^{(1)}(-k\eta) \,
.
\end{equation}
At late times, as $-k\eta\to0$, we find\footnote
{When $\mu=0$ the dilaton remains constant and the 
axion perturbations evolve like those for the dilaton and moduli fields. 
The late time evolution in this case is logarithmic with respect
to $-k\eta$, as given in Eqs.~(\ref{pBBphi}) and~(\ref{pBBbeta}).}
(for $\mu\neq0$)
\begin{equation}
\label{pBBsigma}
{\cal P}_{\delta\sigma} = 64\pi l_{\rm Pl}^2 C^2(\mu) \left( {e^{-\varphi}\widetilde{H}
 \over 2\pi} \right)^2 (-k\eta)^{3-2\mu} \, ,
\end{equation}
where the numerical coefficient
\begin{equation}
\label{Cofr}
C(\mu) \equiv {2^\mu\Gamma(\mu) \over 2^{3/2}\Gamma(3/2)} \, ,
\end{equation}
approaches unity for $\mu\to3/2$.

The key result is that the spectral index can differ significantly
from the steep blue spectra obtained for the dilaton and moduli fields
that are minimally coupled in the Einstein frame. The spectral index for the
axion perturbations is given by
\begin{equation}
\label{Dnsigma}
\Delta n_\sigma = 3 - 2\sqrt{3}|\cos\xi_*|
\end{equation}
and depends crucially upon the evolution of the dilaton, parameterised
by the value of the integration constant $\xi_*$. The spectrum becomes
scale-invariant as 
$\sqrt{3}|\cos\xi_*|\to3/2$, which if we return to the
higher-dimensional underlying theory corresponds either to a fixed
dilaton field in ten-dimensions [$\phi\propto\varphi+\sqrt{3}\beta$
from Eq.~(\ref{4Dvarphi})] or its T-dual solution with isotropic
expansion, discussed in Section~\ref{Section6.4}.
The lowest possible value of the spectral
tilt $\Delta n_\sigma$ is $3-2\sqrt{3}\simeq-0.46$ which is obtained
when stable compactification has occurred and the moduli field $\beta$
is fixed. The more rapidly the internal dimensions evolve, the steeper
the resulting axion spectrum until for $\cos\xi_*=0$ we have $\Delta
n_\sigma=3$ just like the dilaton and moduli spectra. 

When the background axion field is constant these perturbations,
unlike the dilaton or moduli perturbations, do not affect the scalar
metric perturbations. Axion fluctuations correspond to isocurvature
perturbations to first-order).  However, if the axion field does
affect the energy density of the universe at later times (for
instance, by acquiring a mass) then the spectrum of density
perturbations need not have a steeply tilted blue spectrum such as
that exhibited by the dilaton or moduli perturbations. Rather, it
could have a nearly scale-invariant spectrum as required for
large-scale structure formation. Two possible scenarios are presented
in section~\ref{sect10.7}.

\subsection{SL(2,R) invariant perturbation spectra in
dilaton--moduli--axion cosmologies}

The general four--dimensional dilaton--moduli--axion solutions for the
NS-NS sector of the string effective action were presented in
Eqs.~(\ref{axiphi})--(\ref{axisigma}). They are related to the
dilaton--moduli--vacuum solutions presented in
Eqs.~(\ref{dila})--(\ref{dilbeta}) by an ${\rm SL}(2,R)$
transformation of the form given in Eqs.~(\ref{sphimodel})
and~(\ref{ssigmamodel}).

When we allow the background homogeneous axion fields to become
time-dependent, we must allow for the interaction between the
dilaton, moduli and axion fields and the metric to first-order.
But we have seen that in the spatially flat gauge the
evolution equations for both the scalar and tensor metric
perturbations [Eqs.~(\ref{Aeom}) and~(\ref{heom})] are independent of
the evolution of the different scalar fields and are determined solely
by the evolution of the Einstein frame scale factor given in
Eq.~(\ref{Esf}) which remains invariant under the ${\rm SL}(2,R)$
transformation.  The moduli field perturbations also remain decoupled
from both the axion and dilaton, and their evolution equation,
Eq.~(\ref{pBBdbetaeom}), is unaltered by the ${\rm SL}(2,R)$
transformation. Thus, the spectral tilts of the scalar and tensor
metric perturbations and the moduli spectrum, Eq.~(\ref{pBBbeta}),
remain the same as in the pre-big bang scenario.  

However, the dilaton and axion fields as well as their perturbations
will in general be affected by the S-duality transformations.


The dilaton and axion perturbation field equations
become coupled to first order when
$\sigma'\neq0$, and we have
\begin{eqnarray}
\label{dphieom}
\delta\varphi'' + 2\widetilde{h}\delta\varphi' + k^2\delta\varphi
 & = & 
2e^{2\varphi}\sigma'^2\delta\varphi + 2 e^{2\varphi}\sigma'\delta\sigma' \\
\label{dsigmaeom}
\delta\sigma''+2\widetilde{h}\delta\sigma' + k^2\delta\sigma
 & = & 
- 2(\sigma'\delta\varphi'+\varphi'\delta\sigma') 
\, ,
\end{eqnarray}
plus the constraint
\begin{equation}
\label{Aconstraint}
\widetilde{A}
 = {\varphi'\over4\widetilde{h}} \, \delta\varphi
 +{e^{2\varphi}\sigma'\over4\widetilde{h}} \, \delta\sigma
  +{\beta'\over4\widetilde{h}} \, \delta\beta \, .
\end{equation}
The chances of obtaining analytic solutions to Eqs.
(\ref{dphieom})--(\ref{Aconstraint}) might appear remote. However, the
presence of the ${\rm SL}(2,R)$ invariance in the underlying action
allows us to find linear combinations of the axion and dilaton
perturbations which remain straightforward to integrate even in the
more general case.

Remarkably, we can construct new variables
\begin{eqnarray}
\label{defx}
x &\equiv& e^\varphi \left( {\varphi'\over\widetilde{h}} \delta\sigma
 - {\sigma'\over\widetilde{h}} \delta\varphi \right) \,,\\
\label{defy}
y &\equiv& {\varphi'\over\widetilde{h}} \delta\varphi
 + {e^{2\varphi}\sigma'\over\widetilde{h}} \delta\sigma \, ,
\end{eqnarray}
such that the perturbation equations decouple and the field
equations~(\ref{dphieom}) and~(\ref{dsigmaeom}) become
\begin{eqnarray}
\label{xeom}
x'' + 2\widetilde{h} x'
 + \left[ k^2 - (\varphi'^2+e^{2\varphi}\sigma'^2) \right]
 x & = & 0 \\
\label{yeom}
y'' + 2\widetilde{h} y' + k^2 y & = & 0 \, .
\end{eqnarray}

These equations decouple, even in a general dilaton-axion background,
because these variables are invariant under an ${\rm SL}(2,R)$
transformation given in Eqs.~(\ref{sphi}) and~(\ref{schi}). This
follows from writing them in terms of the symmetric ${\rm SL}(2,R)$ matrix
${\cal M}$ defined in Eq.~(\ref{M}).  We have
\begin{eqnarray}
\label{9.90}
2 \widetilde{h} x & = & {\rm tr}(J{\cal M}J{\cal M}'J\delta {\cal M})
\,,\\
\label{9.91}
2 \widetilde{h} y & = & {\rm tr}(J{\cal M}'J\delta {\cal M}) \,,
\end{eqnarray}
These are the time-components of the ${\rm SL}(2,R)$ invariant four-vectors
defined in Eqs.~(\ref{vectoru}) and~(\ref{vectorv}) in
Section~\ref{Section4.4} and are the unique S-duality invariant
linear combinations of the axion and dilaton perturbations. 
They reduce to the
(decoupled) axion and dilaton perturbations in the pure dilaton-moduli-vacuum
background, as $\sigma'\to0$, where we have
\begin{eqnarray}
\label{xtosigma}
x & \to & {\varphi' \over \widetilde{h}} e^\varphi\delta\sigma
=2\sqrt{3}\cos\xi_* e^\varphi\delta\sigma \, ,\\
y & \to & {\varphi' \over \widetilde{h}} \delta\varphi
 =2\sqrt{3}\cos\xi_* \delta\varphi 
\label{ytophi}
\end{eqnarray}
and $\xi_*$ is the integration constant in
Eqs.~(\ref{axiphi})--(\ref{axisigma}).

Having found S-duality invariant variables, one can verify that the
evolution equations for these variables, Eqs.~(\ref{xeom})
and~(\ref{yeom}), are themselves invariant under S-duality. Remembering
that the general dilaton-moduli-axion cosmological solutions can always be
related to the dilaton-moduli-vacuum solutions by 
an ${\rm SL}(2,R)$ transformation, we
see that the evolution equations for $x$ and $y$ in an arbitrary
dilaton-moduli-axion cosmology are {\em exactly} the same as those 
for the axion and
dilaton perturbations in the dilaton-moduli-vacuum case. Just as in the
constant axion case, we can define canonically normalised variables
\begin{eqnarray}
u &\equiv& {1\over2\mu\sqrt{16\pi}l_{\rm Pl}} \widetilde{a} y \, ,\\
v &\equiv& {1\over2\mu\sqrt{16\pi}l_{\rm Pl}} \widetilde{a} x \, ,
\end{eqnarray}
where $\mu=\sqrt{3}\cos\xi_*$, 
which reduce to the definitions given in Eqs.~(\ref{defu})
and~(\ref{defv}) in the dilaton-moduli-vacuum case.
In general, $u$ obeys the S-duality invariant equation of motion given in
Eq.~(\ref{upp}) and whose general solution is given by
Eq.~(\ref{usol}). The equation of motion for $v$ given in Eq.~(\ref{vpp}),
however, is not invariant under an S-duality transformation. Instead
the S-duality invariant version of the equation of motion becomes
\begin{equation}
v'' + \left( k^2 - {\mu^2-1/4 \over \eta^2} \right) v = 0 \, .
\end{equation}
which reduces to Eq.~(\ref{vpp}) when $\sigma'=0$. The general
solution for $v$ is therefore still given by Eq.~(\ref{vsol}).

We can still normalise cosmological vacuum perturbations at early
times on the $(+)$ branch as $\eta\to-\infty$ because we have seen
that in this limit the general dilaton-moduli-axion solution given in
Eqs.~(\ref{axiphi}--\ref{axisigma}) approaches the constant axion
solutions with $\sqrt{3}\cos\xi_*=+\mu\geq0$. This in turn implies that 
the constants $u_\pm$ and $v_\pm$ are
given by Eqs.~(\ref{uplus}) and~(\ref{vplus}). 
By picking S-duality invariant field perturbations we have been able to
calculate the general dilaton-moduli-axion cosmological perturbation spectra
using the pure dilaton-moduli-vacuum cosmological vacuum states.
We can then write
\begin{equation}
\label{Py}
{\cal P}_y = {128\mu^2\over\pi^2} l_{\rm Pl}^2\widetilde{H}^2
 (-k\eta)^3[\ln(-k\eta)]^2
 \,.
\end{equation}
and the generalised axion perturbation spectrum is given by
\begin{equation}
\label{Px}
{\cal P}_x = {16\mu^2C^2(\mu)\over\pi}
 l_{\rm Pl}^2\widetilde{H}^2 (-k\eta)^{3-2\mu} \, .
\end{equation}

We have already seen that at late times on the $(+)$ branch, as
$\eta\to0$, the general dilaton-moduli-axion solutions approach
dilaton-moduli-vacuum solutions with $\sqrt{3}\cos\xi_*=-\mu\leq0$.
It follows from Eqs.~(\ref{xtosigma}) and~(\ref{ytophi}) that as
$\eta\to0_-$ the final power spectrum for vacuum fluctuations in the
general dilaton-moduli-axion cosmologies given by Eqs.~(\ref{Py})
and~(\ref{Px}) reduce to those given for the dilaton and axion
fluctuations in Eqs.~(\ref{pBBphi}) and~(\ref{pBBsigma}) for the ${\rm
SL}(2,R)$-duality related dilaton-moduli-vacuum case.
The tilt and amplitude of the spectra are determined solely by
the parameter $\mu=|\sqrt{3}\cos\xi_*|$ and are insensitive to the
specific time dependence of the axion field in different, but
S-duality related, solutions~\cite{CopEasWan97}. 
This invariance of the perturbation spectra produced along different
duality related solutions extends to perturbation spectra in arbitrary
cosmological solutions related by an ${\rm SL}(2,R)$ transformation,
not just those derived from string theory~\cite{Wands99}.

The constraint equation for the metric perturbations $\widetilde{A}$,
Eq.~(\ref{Aconstraint}), is only dependent on the ${\rm
SL}(2,R)$-invariant perturbation $y$ [defined in Eq. (\ref{yeom})] and
the moduli perturbations $\delta\beta$:
\begin{equation}
 \widetilde{A} = {1\over4}y +
 {\beta'\over4\widetilde{h}}\delta\beta \, .
\end{equation}
Thus, the scalar metric perturbations are invariant under ${\rm SL}(2,R)$ 
transformations that leave the Einstein frame metric unchanged. 
The metric is unaffected by the specific time--dependence of the axion
field and the perturbation spectrum is the same as that obtained in
the constant axion case, given in Eq.~(\ref{Aspectrum}).

\subsection{Perturbation spectra with more degrees of freedom}

Thus far we have only considered a single axion field in addition to
the dilaton and moduli fields. However there are many massless degrees
of freedom in the low energy string action which will all be excited
during a pre--big bang era. 
In this context we will refer to all additional fields which have canonical 
kinetic terms in the Einstein frame as moduli fields whereas axion fields, 
$\sigma_i$, retain a non-minimal
coupling to the dilaton or moduli fields in the Einstein frame.
The Lagrange density for the axion field has the form
\begin{equation}
\label{defaxionL}
\widetilde{\cal L}_{\sigma_i} = - {1\over2} e^{2\omega_i} (
\widetilde\nabla \sigma_i )^2 \ ,
\end{equation}
where $\omega_i$ is a linear function of the dilaton and
moduli fields.
For the specific example of the NS-NS axion discussed earlier we have
$e^{\omega_i}=e^\varphi$ (see Eq. (\ref{reduced2})).

All dilaton and moduli perturbations, minimally coupled in the
Einstein frame, will yield steep blue spectra, as given in
Eqs.~(\ref{pBBphi}) and~(\ref{pBBbeta}), in a pre--big bang scenario.
As a consequence the spectrum of scalar metric perturbations given in
Eq.~(\ref{Aspectrum}) is independent of the number of additional
fields.
However the various (pseudo--) scalar axion fields present in low
energy effective actions can have different perturbation spectra due
to their different couplings to the dilaton and moduli fields. In
general, however, these numerous fields are coupled to the {\em same}
dilaton and moduli which will lead to distinctive relations between
the corresponding perturbation spectra.  This non-minimal coupling for
each field can
be eliminated by a conformal transformation to a specific
conformally related metric, which we will refer to as the
corresponding axion frame.  In order to understand the perturbation
spectra produced in different fields it is revealing to look at the
evolution of the conformally related metrics,
$g_{\mu\nu}\to e^{2\omega_i}g_{\mu\nu}$.
Quite generally we can define the rescaled scale factor in the axion
frame~\cite{CopLahWan94,CopEasWan97,CopLidWan98c}
\begin{equation}
\label{defaxionsf}
\bar{a}_i = e^{\omega_i} \widetilde{a} \,.
\end{equation}
The time-dependence of these axionic scale factors follows from the
dilaton-moduli-vacuum solutions given by Eqs.~(\ref{dila})
and (\ref{dilphi}) and can be written in terms of conformal time as
\begin{equation} 
\label{axionsf}
\bar{a}_i = \bar{a}_{*i} |\eta|^{r_i+(1/2)} \,.
\end{equation} 
In terms of the proper time in the axion frame we have
\begin{equation}
\bar{a}_i = \bar{a}_{*i} \left( {\bar{t}_i\over\bar{t}_{*i}}
 \right)^{(1+2r_i)/(3+2r_i)}
\end{equation}
For $r_i<-3/2$ we have conventional power-law inflation (not
pole-inflation) during the pre-big bang era ($\eta<0$) with $\bar{a}_i
\sim \bar{t}_i^{\bar{p}_i}$, where
$\bar{p}_i=1+[2/(-2r_i-3)]>1$. This has important
consequences for the tilt of the power spectrum of semi-classical
perturbations in the axion field produced on large scales.

The canonically normalised axion field perturbations are given 
by~\cite{Mukhanov88,CopEasWan97,CopLidWan98c} 
\begin{equation}
\label{defvi}
v_i \equiv {1\over\sqrt{16\pi}\l_{\rm Pl}} \bar{a}_i\delta\sigma_i \, 
\end{equation}
and the equation of motion can be written in terms of $v_i$ as
\begin{equation}
\label{vipp}
v_i'' + \left( k^2 - {\bar{a}_i''\over\bar{a}_i} \right) v_i = 0 \, .
\end{equation}
In the terminology of Ref.~\cite{BruGasVen98}, the pump field $S$ for the
perturbations in each axion field is given by the square of the scale
factor in the corresponding conformal frame, $S_i=\bar{a}_i^2$.
For pre--big bang solutions, i.e., $\eta<0$, we can normalise modes 
on small scales at early times by requiring that $v_i \to
{e^{-ik\eta}/\sqrt{2k}}$ as $k\eta\to-\infty$ \cite{BirDav82},
and this gives
\begin{equation}
\label{pBBdsigmai}
\delta\sigma_i = {2\pi l_{\rm Pl} \over \sqrt{k}} e^{i(2\mu_i+1)\pi/4}
{\sqrt{-k\eta} \over \bar{a}} H_{\mu_i}^{(1)}(-k\eta) \, .
\end{equation}
where $\mu_i=|r_i|$. Thus for modes far outside the horizon
($-k\eta\to0$) we have
\begin{equation}
\label{pBBsigmai}
{\cal P}_{\delta\sigma_i} = 16\pi l_{\rm Pl}^2
 \left( {C(\mu_i) \over 2\pi} \right)^2
 {k^2\over\bar{a}^2} (-k\eta)^{1-2\mu_i} \, ,
\end{equation}
where the numerical coefficient $C(\mu_i)$ is defined in
Eq.~(\ref{Cofr}).

The expression for the axion power spectrum can be written in terms of
the field perturbation when each mode crosses outside the horizon
\begin{equation}
{\cal P}_{\delta\sigma_c}
 = 16\pi l_{\rm Pl}^2 
 \left[{C(\mu_i)\over r_i+(1/2)}\right]^2
 \left( {\bar{H}_{ic} \over 2\pi} \right)^2 \, ,
\end{equation}
where $\bar{H}_{ic}$ is the Hubble rate in the axion frame when $|k\eta|=1$. 
This is the power spectrum for a
massless scalar field during power-law inflation which approaches the
famous result
${\cal P}_{\delta\sigma}/16\pi l_{\rm Pl}^2=(\bar{H}_{ic}/2\pi)^2$ as
$r_i \to -3/2$, and
the expansion in the axion frame becomes exponential\footnote
{The factor $16\pi l_{\rm Pl}^2$ arises due to our dimensionless definition of
$\sigma$.}. 

The amplitude of the power spectra at the end of the pre--big bang
phase can be written as (see Eq. (\ref{pBBsigma}))
\begin{equation}
\label{Pdsend}
\left. {\cal P}_{\delta\sigma_i} \right|_s
 = 64\pi l_{\rm Pl}^2 C^2(\mu_i) e^{-2\omega_i}
 \left( {\widetilde{H} \over 2\pi} \right)_s^2
 \left( {k\over k_s} \right)^{3-2\mu_i}\, ,
\end{equation}
where $k_s$ is the comoving wavenumber of the scale just leaving the
Hubble radius at the end of the pre--big bang phase, $k_s\eta_s=-1$. 
The spectral tilts for the axion perturbation spectra are thus given by
\begin{equation}
\Delta n_i = 3 - 2\mu_i \,,
\end{equation}
which generalises the result given for the single NS-NS axion in
Eq.~(\ref{Dnsigma}). 
The tilts depend crucially upon the value of $\mu_i$. The spectrum
becomes scale-invariant in the limit $\mu_i\to3/2$. The
lowest possible value of the spectral index for any of the axion
fields is $3-2\sqrt{3}\simeq-0.46$.  Requiring conventional power-law
inflation, rather than pole inflation, in the axion frame, guarantees
a negatively tilted spectrum ($\Delta n_i<0$)\footnote {Note that
although the power spectrum for axion perturbations diverges on large
scales for $\Delta n_i<0$, the energy density is proportional to
$k^2{\cal P}_{\delta\sigma_i}$ and this remains finite.}.

The axion perturbation spectra can have different spectral indices,
but in a given string model there is a {\em specific} relationship
between them. This follows as a direct consequence of the symmetries
of the effective action. These symmetries relate the coupling
parameters between the various fields and are manifested in the
spectra.  Such perturbation spectra could provide distinctive  
signatures of the early evolution of our universe.  The analysis
presented above is applicable to a wide class of non-linear
sigma models coupled to gravity.  In such models, the couplings
between the massless scalar fields are specified by the functional
form of the target space metric. These couplings determine the
appropriate conformal factors analogous to those in Eq.~(\ref{Omega})
that leave the fields minimally coupled and it is the evolution of
these couplings that directly determines the scale dependence of the
perturbation spectra.

As an example, we consider the perturbation spectra produced within
the context of a triple axion system derived from the type IIB
superstring reduced to four-dimensions~\cite{CopLidWan98c}. The dual
effective action for the type IIB superstring in four dimensions was
presented in Eq.~(\ref{solitonicaction}).  In four dimensions the
three-form field strengths from the NS-NS and RR sectors are dual to
the gradients of two pseudo-scalar axion fields, $\sigma_1$ and
$\sigma_2$. The third axion field, $\sigma_3\equiv\chi$, is the RR
axion already present in the ten--dimensional theory.  The scalar fields
parametrise an ${\rm SL}(3,R)/{\rm SO}(3)$ coset of a non-linear sigma model in
Einstein gravity~\cite{CopLidWan98b},
where the conformal factors appearing in
Eq.~(\ref{defaxionL}) are given by
\begin{equation}
\label{Omega}
e^{2\omega_i} = \left\{
\begin{array}{ll}
e^{2\varphi} & {\rm for}\ \sigma_1 \\
e^{\varphi-\sqrt{3}\beta} & {\rm for}\ \sigma_2 \\
e^{\varphi+\sqrt{3}\beta} & {\rm for}\ \sigma_3
\end{array}
\right. \,,
\end{equation}
These factors 
reflect the different couplings that each of the axion fields 
has to the dilaton
and moduli fields in the effective action.
The exponents in the power-law solutions, Eq.~(\ref{axionsf}), for the
axion scale factors are then given by
\begin{equation}
r_i = \left\{
\begin{array}{ll}
\sqrt{3} \cos\xi & {\rm for}\ \sigma_1 \\
\sqrt{3} \cos(\xi+\pi/3) & {\rm for}\ \sigma_2 \\
\sqrt{3} \cos(\xi-\pi/3) & {\rm for}\ \sigma_3
\end{array}
\right. \,.
\end{equation}
All three spectral indices for the axion fields in the truncated type
IIB model where the fields parameterise an ${\rm SL}(3,R)/{\rm SO}(3)$ 
coset are
determined by the single integration constant $\xi$. 
The spectral tilts are shown in Figure \ref{figuretilt}. They take the values
\begin{equation}
\Delta n_i = 3 - 2\sqrt{3}|\cos(\xi-\xi_i)|
\end{equation}
where
\begin{equation}
\xi_i = \left\{
\begin{array}{ll}
0 & {\rm for}\ \sigma_1 \\
 -\pi/3 & {\rm for}\ \sigma_2\\
\pi/3 & {\rm for}\ \sigma_3
\end{array}
\right. \,.
\end{equation}
One of the axion fields {\em always} has a red spectrum ($\Delta
n_i<0$) while the other two spectra are blue ($\Delta n_i>0$), except
in the critical case $|\cos\xi|=\sqrt{3}/2$, where two of the spectra
are scale-invariant and only one is blue.  This provides an example of
the important phenomenological role that the RR sector of string
theory can play in cosmological 
scenarios~\cite{LukOvrWal97a,LukOvrWal97b,Kaloper97,LuMukPop96a,PopSch97}.

\begin{figure}[t]
\centering 
\leavevmode\epsfysize=5cm \epsfbox{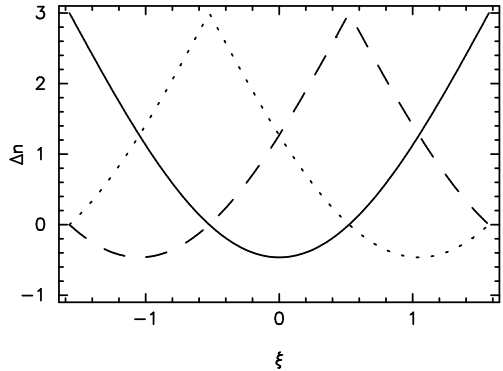}\\ 
\caption[Spectral indices]{\label{ni} Spectral tilts $\Delta n_i$ for three
axion fields' perturbation spectra in the truncated type IIB action as a
function of integration constant $\xi$ in the pre--big bang solutions. 
The solid line corresponds to $\Delta n_1$, the dotted line to $\Delta
n_2$ and the dashed line to $\Delta n_3$.}
\label{figuretilt}
\end{figure}

The requirement that at least one axion field has a red or
scale-invariant perturbation spectra in the ${\rm SL}(3,R)$ model could have
damaging observational consequences, as we will discuss shortly. 
This raises the possibility that realistic models with additional
degrees of freedom might also inevitably produce red perturbation
spectra. This would then put in doubt 
the ability of the pre--big bang scenario to
produce a sufficiently homogeneous universe on large
scales~\cite{CopLidWan98c}. However, 
such a conclusion depends on whether the effect of introducing more 
moduli fields is more or less significant than the introduction 
of further axion fields. 

For example, we can extend the analysis to 
the ${\rm SL}(4,R)$ non-linear sigma model. 
This introduces a further modulus field
and three more axion fields. A study of the perturbation spectra
produced during a pre--big bang era in such a model has
shown~\cite{BriWan99} that in fact the constraint on the upper bound
of the minimal spectral tilt is {\em relaxed} by increasing the size
of the group. 
In such an ${\rm SL}(4,R)$ model we require only that
$n_i\leq3-\sqrt{6}\simeq0.55$ for at least one axion
field~\cite{BriWan99}. As any ${\rm SL}(n,R)$ group with $n>4$ includes
${\rm SL}(4,R)$ sub-groups, this demonstrates that increasing the group size in
this way evades the requirement of a red spectrum threatened in
Ref. \cite{CopLidWan98c}.

Alternatively one can simply increase the number of effective moduli
fields by dropping the requirement of isotropy in the 
three--dimensional external
space~\cite{Giovannini99}. This is reasonable in the pre-big bang era
as isotropy is only required by observations in the post-big bang
phase. In the simplest case of the Bianchi I cosmology discussed in
Section~\ref{Section7.2}, the shear evolves like an
additional massless field and introduces a new free parameter in the
cosmological solution. The anisotropy complicates the perturbation
analysis (here based on a maximally symmetric external space) and is
beyond the scope of the present review, but it is possible to
show~\cite{GasGio97,Giovannini97} that 
introducing shear allows all the other
dilaton and moduli background fields to remain constant in the pre--big
bang era. This results in a maximally steep blue spectrum for all the axion
fields.

\subsection{Large-scale structure from a Pre-Big Bang era}
\label{sect10.7}

We have seen how vacuum fluctuations in massless fields present in the
string effective action predict different spectra of inhomogeneities
on super-horizon scales at the end of a pre-big bang era, but we have
not yet discussed how these perturbations evolve through to the
present post-big bang era. Indeed, in the absence of a definite model
for the graceful exit from one phase to the other, it is impossible to
make definite predictions. The minimal assumption usually made is that
large-scale perturbations should be ``frozen-in'', at least during a
sufficiently rapid transition. This is indeed the standard assumption
made in most models of conventional inflation, where the 
detailed dynamics of reheating at the end of the inflationary
era is usually neglected~\cite{LidLidKol97}. 
This constancy of both the scalar curvature
perturbation, $\zeta$, and the tensor perturbations, $h_{ij}$, on
super-horizon scales was implicit in our earlier discussion of both
gravitational waves and primordial black holes.

\paragraph{Isocurvature axion perturbations.}

In simple toy models of the transition from pre- to post-big bang
phases, where the fields remain effectively massless, the large-scale
perturbation spectra do indeed seem to be
frozen-in~\cite{BuoMeiUng98b}. The curvature perturbations, and all
the dilaton-moduli fields which contribute to the energy density at
first order, have steep blue spectra which are completely
inconsistent with the observed large-scale structure in our
universe. In this case the axion field perturbations,
$\delta\sigma_i$, only contribute to the energy density at
second-order, but as the first-order perturbations are so small, these
second-order effects could dominate on large scales. We can estimate
the energy density associated with the massless axion field in the Einstein
frame using Eq.~(\ref{Pdsend}) to
give~\cite{BuoMeiUng98b,DurGasSak98,DurGasSak99} 
\begin{equation}
\widetilde\rho_i \sim {k^2\over \widetilde{a}^2} \, {e^{2\omega_i}\, {\cal
P}_{\delta\sigma_i}|_s \over 16\pi l_{\rm Pl}^2}
= C^2(\mu_i)\, {k^2\over \widetilde{a}^2}\, \left( {\widetilde{H} \over
2\pi} \right)^2_s \left( {k \over k_s} \right)^{3-2\mu_i} \ ,
\end{equation}
where the subscript $s$ denotes quantities to be evaluated at the end of the
dilaton-moduli-driven pre-big bang phase.
Note that although the amplitude of the field perturbations depends
upon the conformal factor $e^{2\omega_i}$, the effective energy
density of perturbations with $k\sim k_s$ is the same for all the
fields calculated in the Einstein frame, and depends only up on the
Hubble rate $\widetilde{H}_s$ which is naively expected to be given by 
\begin{equation}
\label{Hs}
l_{\rm Pl}^2 \widetilde{H}_s^2
 \sim e^{\varphi_s} \sim 10^{-2} \ .
\end{equation}
The different conformal factors affect only the tilt of the spectrum
and for $\mu_i=3/2$ we have a scale-invariant spectrum of density
perturbations.

Temperature anisotropies on the cosmic microwave background sky due to
these second-order density perturbations have been estimated to
be~\cite{DurGasSak98,DurGasSak99}
\begin{equation}
{\Delta T \over T} \sim \left. {\widetilde\rho_i \over \rho_{\rm
crit}} \right|_{k=\widetilde{a}\widetilde{H}}
 \sim l_{\rm Pl}^2 \widetilde{H}_s^2 \left( {k \over k_s}
\right)^{3-2\mu_i} \,.
\end{equation}
To be compatible with the observed level of anisotropies this requires
either a lower than expected normalization compared with that given in
Eq.~(\ref{Hs}), or a slightly blue-tilted spectrum, $\Delta
n_i=3-2\mu_i\sim+0.1$.

\begin{figure}[t]
\centering 
\leavevmode\epsfysize=7cm \epsfbox{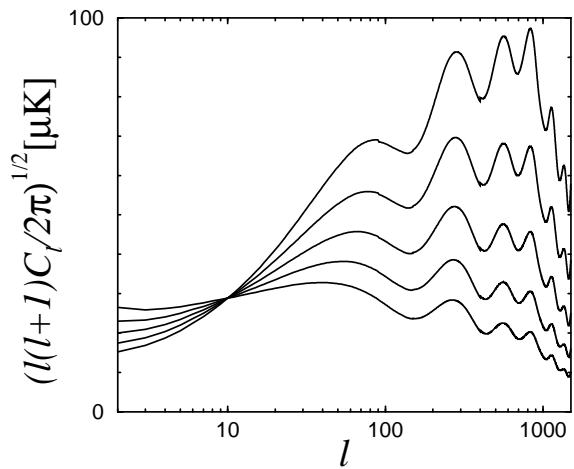}\\ 
\caption[Cosmic microwave background angular power spectrum]
{\label{cmb} Cosmic microwave background angular power
spectra induced by axionic seeds for five different spectral tilts in
a universe with critical matter density. The spectral tilts, increasing
from bottom to top, are $\Delta n_i= 0.1$, $0.2$, $0.3$, $0.4$
and $0.5$. (Figure kindly supplied by Filippo Vernizzi based on results
presented in Ref.~\cite{MelVerDur99}.)}
\label{cmbplot}
\end{figure}

Recently, Melchiorri {\em et al.} 
\cite{MelVerDur99} have determined the perturbation 
spectra for the massless 
axionic seeds and extended the analysis into the acoustic-peak region
by employing a full Boltzmann code. 
They obtain encouraging results. The evolution of the internal 
dimensions (see Eq.~(\ref{dilbetat1})) 
during the pre-big bang phase is vital in determining the overall 
normalisation of the peaks as well as the value of the spectral tilt, 
$\Delta n_i$, given in 
Eq.~(\ref{Dnsigma}). We do not summarize the details of 
their calculation here, but for completeness we 
include some of their key results. 
Figure~\ref{cmbplot} shows the angular 
power spectra for five different tilts in the range 
$0.1 \le \Delta n_i \le 0.5$. The generic features 
exhibited by these power spectra are the 
significant acoustic peaks in the multipole 
range $250 < l < 300$, 
with a corresponding blue power spectrum for the 
scalar component of the perturbations. 
Ref.~\cite{MelVerDur99} compares predictions with the published
microwave background data and shows that those with $0.3 \le \Delta
n_i \le 0.5$.  appear to be consistent with the present data.
Of most significance is the fact that the isocurvature nature of the
perturbations results in a `hump' in the spectrum at $l < 100$. This
differs from the spectra that arise from adiabatic perturbations in
standard, potential--driven inflationary models.  The location of the
first acoustic peak in these isocurvature spectra is also different to
the standard scenario. Thus, these models are predictive and can in
principle be distinguished from the more conventional inflationary
scenarios.

It is quite clear that this type of approach should prove to be a
strong test for the models. Indeed, it may be that the simplest models
are already ruled out. Of course, these models have yet to fully
incorporate the behaviour of the perturbations through the graceful
exit phase, and it is important to understand the impact that this
epoch may have on the perturbed fields.  A first attempt to understand
this process suggests that the impact is potentially significant
\cite{KawSod99,KawSod99a}.

\paragraph{Primordial curvature perturbations.}

The result that the curvature perturbation on uniform density
hypersurfaces, $\zeta$, remains fixed on super-horizon scales is
strictly true only for adiabatic perturbations
\cite{BarSteTur83,WanMalLyt00}. The large number of fields present in
the low energy action can support entropic as well as adiabatic
perturbations. We define adiabatic perturbations as those
perturbations for which
\begin{equation}
{\delta\sigma_i \over \sigma_i'} = {\delta\sigma_j \over \sigma_j'}
\ \ \forall\  i,j \,. 
\end{equation}
By contrast, for entropy perturbations we have 
\begin{equation}
\Gamma_{ij} \equiv {\delta\sigma_i \over \sigma_i'} - {\delta\sigma_j
\over \sigma_j'} \neq 0 \,.
\end{equation}
These entropy perturbations can alter the value of the curvature
perturbation, $\zeta$, even on super-horizon scales if they give rise
to a significant non-adiabatic pressure
perturbation~\cite{WanMalLyt00}, $\delta
p_{\rm nad}=\delta p - c_s^2\delta\rho$ where
$c_s^2\equiv\dot{p}/\dot\rho$ is the adiabatic sound speed.

Massless axion fields only contribute to the pressure or energy
density at second order and therefore cannot lead to a first-order
change in the curvature perturbation. However if they interact with
other fields and/or acquire a mass~\cite{BruHad98} 
they can lead to a first-order effect.
Consider a simple toy model with a second-order phase-transition,
where the density is continuous but there is an abrupt change in the
pressure on a hypersurface, $\Sigma$, triggered by the axion field
reaching a critical value $\sigma_c$.  The curvature perturbation,
$\psi$, on the hypersurface $\Sigma$ can be given in terms of the
field fluctuations on spatially flat ($\psi=0$) hypersurfaces as
\begin{equation}
\label{defpsiSigma}
\psi_\Sigma = {\widetilde{a}' \over \widetilde{a}}
 {\delta\sigma_i \over \sigma_i'} \,.
\end{equation}
At the same time the initial curvature perturbation on
uniform density hypersurfaces, $\zeta_1$, can be written as
\begin{equation}
\zeta_1 = \psi_\Sigma + {\widetilde{a}' \over \widetilde{a}}
 {\delta\rho_\Sigma \over \rho'} = 
\psi_\Sigma - {1\over 3\gamma_1} {\delta\rho_\Sigma \over \rho} \,.
\end{equation}
where the pressure $p_1=(\gamma_1-1)\rho$ and
$\delta\rho_\Sigma$ is the density perturbation on $\Sigma$.
An analogous expression can be written for the subsequent curvature
perturbation, $\zeta_2$, when the pressure $p_2=(\gamma_2-1)\rho$.
Eliminating $\delta\rho_\Sigma$ then yields 
\begin{equation}
\zeta_2 = {\gamma_1\over\gamma_2} \zeta_1
 + \left( 1 - {\gamma_1\over\gamma_2} \right) \psi_\Sigma \,,
\end{equation}
where $\psi_\Sigma$ is given by Eq.~(\ref{defpsiSigma}).
Thus the final value of $\zeta$ depends not only up on the initial
value, but also up on the fractional change in the equation of state,
$\gamma_1/\gamma_2$, and the curvature of $\Sigma$ given by
$\psi_\Sigma$.
The curvature perturbation remains unaltered ($\zeta_2=\zeta_1$) only
for $\gamma_1=\gamma_2$, or adiabatic perturbations where the uniform
density and uniform field hypersurfaces necessarily coincide (i.e.,
$\psi_\Sigma=\zeta_1$). 
If the initial curvature perturbation, $\zeta_1$, is negligible on
large scales, as seems likely in the pre-big bang scenario, then the
amplitude of the final perturbation $\zeta_2$ depends up on the
equation of state and the time-evolution of the background field
$\sigma_i'$, both of which are unspecified here. But because 
$\zeta$ is linearly dependent on $\delta\sigma_i$, the
spectral index of the curvature perturbation,
Eq.~(\ref{zetaspecindex}), is simply given by
\begin{equation}
n = 1 + \Delta n_i = 4 - 3\mu_i\,.
\end{equation}
and we can recover a Harrison-Zel'dovich spectrum of curvature
perturbations on super-horizon scales for $\mu_i=3/2$.

\newpage
\section{Outlook: Ho\v{r}ava--Witten Cosmology}
 
\label{Section11}

\setcounter{equation}{0}

\def\theequation{\thesection.\arabic{equation}}

In recent years two of the most important 
ideas to have been developed in particle physics and cosmology 
are those of superstring theory and cosmological inflation. 
Significant advances have been made in both fields. 
{}From an astrophysical point of view, 
models of inflation can now be strongly constrained by a host 
of cosmological  observations, most notably those arising 
from microwave background
anisotropies  and high redshift surveys of galaxy clusters and 
superclusters \cite{LidLyt93,LidLidKol97,KamKos99,LytRio99}. In
the near future, a flood of forthcoming data will only 
serve to improve the current situation. In particular, 
gravity wave interferometers such as LIGO 
will constrain any primordial gravitational wave background 
that may have been 
generated during an inflationary epoch \cite{AllBru97}. 

On the theoretical side, 
a major change of emphasis has occurred with the appearance 
of eleven--dimensional M--theory as the (proposed) fundamental 
quantum theory of gravity \cite{Witten95}. 
The moduli space of M-theory encompasses all 
five, anomaly free, ten--dimensional superstring theories. In this 
sense, no particular string theory is necessarily 
more fundamental than any of the others. Rather, they are 
all related by 
the S--, T-- and U--duality symmetries. Moreover, 
another region of the M--theory 
moduli space corresponds to eleven--dimensional 
supergravity \cite{Witten95,Townsend95}, 
implying that M--theory is more than a theory of superstrings. 
This re--establishes the importance of 
eleven--dimensional supergravity in the study 
of particle physics and cosmology.
In view of the excitement that has been generated by this discovery,
there is a pressing need to study the cosmology associated with M-theory
and superstring theory, especially within the context of the duality 
symmetries of the theories. Such a programme is still in the 
early stages of its development, but already 
progress has been made in understanding the nature of 
cosmologies that may arise. 

A key question to be addressed is 
whether a definitive inflationary model can be developed 
from string/M--theory and ultimately 
constrained by astrophysical observation. The pre--big bang scenario 
represents the first step towards such a goal \cite{Veneziano91,GasVen93a}. 
Whilst there 
are certainly a number of unresolved problems with this scenario, 
the duality symmetries that arise within 
string theories can have important cosmological consequences, 
both at the theoretical and observational levels. 
It is quite possible, therefore, that some manifestation of string duality 
could be detectable within the foreseeable future. 

The precise region of the M--theory moduli space that describes  
our present--day universe is uncertain, but from 
a particle physicists' point of view, the most favoured 
location is that of ${\rm E}_8 \times {\rm E}_8$ 
heterotic string theory . Ho\v{r}ava and Witten 
have shown that the strongly--coupled limit of this theory 
is M--theory on an eleven--dimensional orbifold $R^{10} \times S^1/{\rm Z}_2$
\cite{HorWit96a,HorWit96b}. The weakly--coupled 
heterotic string theory is then recovered in the limit where the radius 
of $S^1$ tends to zero. The orbifold $S^1/{\rm Z}_2$ 
may be viewed as a segment of the real line that is bounded by 
two fixed points on the circle. The effect of the 
${\rm Z}_2$ transformation is to reverse the orientation of the circle, 
$y \rightarrow -y$, where $y$ is the coordinate of the eleventh dimension, 
and to
change the sign of the three--form potential. This latter change of sign is 
necessary for the eleven--dimensional 
supergravity theory to remain invariant. The two sets of  
${\em E}_8$ gauge supermultiplets are located on 
each of the ten-dimensional orbifold fixed planes \cite{HorWit96a,HorWit96b}.
In other words, they propagate only at the ${\rm Z}_2$ fixed points and 
restricting the fields in this way cancels the gravitational anomalies. 

In view of these developments, 
we conclude the review with a brief discussion on the 
cosmological solutions admitted by the  Ho\v{r}ava--Witten 
theory. Despite its potential importance, 
relatively little work has been done thus far in 
deriving and analysing tine--dependent solutions in this theory. 
An early study was made by 
Benakli, who  found a class of cosmological solutions 
from certain $p$--brane configurations \cite{Benakli98a,Benakli98b}.
Our aim is to present 
a flavour of the type of solutions that can be found and we 
will therefore be brief on the technical subtleties. An excellent review of 
the detailed solutions can be found in \cite{LukOvrWal98b}. 

Witten subsequently considered a 
further compactification to four dimensions on a 
deformed Calabi-Yau manifold and showed that the resulting 
theory has $N=1$ supersymmetry \cite{Witten96}. 
{}From this a remarkable conclusion is deduced; comparison of the 
gravitational and GUT couplings implies that the orbifold must be {\em larger} 
than the radius of the Calabi--Yau space \cite{Witten96,BanDin96}. 
Indeed, the relative sizes may differ by more than an order 
of magnitude. This immediately implies that the early universe 
may have undergone a phase where it was five--dimensional. 

Motivated by these considerations, Lukas {\em et al.} 
derived an effective five--dimensional theory by a direct 
compactification of the Ho\v{r}ava--Witten theory on a Calabi--Yau space
\cite{LukOvrSte98a,LukOvrSte98b}. The form of the eleven--dimensional 
metric they assumed is given by
\begin{equation} 
 ds^2_{11} =V^{-2/3}g_{\mu\nu}dx^{\mu}dx^{\nu}+
 V^{1/3}\Omega_{mn}dy^mdy^n, 
\end{equation} 
where $x^{\mu}$ ($0\le\mu , \nu \le 4$) and $y^m$ $(5 \le m , n \le 
10)$ are the coordinates on the five--dimensional spacetime and 
Calabi--Yau space, respectively, and the metrics are denoted by 
$g_{\mu\nu}(x)$ and $\Omega_{mn}(y)$. The deformation 
of the Calabi--Yau manifold is parametrized by the scalar degree 
of freedom  $V(x)$. The five--dimensional metric is comprised of the 
four--dimensional spacetime and the orbifold dimension. 

As emphasized by Lukas {\em et al.} \cite{LukOvrSte98a,LukOvrSte98b}, 
a necessary 
condition that must be satisfied for a consistent compactification 
of Ho\v{r}ava--Witten theory 
to five dimensions is that a non--zero mode of the four-form 
field strength on the Calabi--Yau three--fold 
has to be included. For this reason, such a compactification 
differs from that of the standard Kaluza--Klein reduction 
of eleven--dimensional supergravity. 

In what follows we adopt the normalization of Reall \cite{Reall98}. 
A consistent truncation of the effective, five--dimensional theory is given by
\cite{LukOvrSte98a,LukOvrSte98b,Reall98}
\begin{eqnarray} 
\label{HWaction}
S=\int_{M_5} d^5 x \sqrt{-g}\left(\frac{1}{2}R_5 -
\frac{1}{2}(\nabla\phi)^2
- e^{-\sqrt{2}\phi}\nabla\xi\cdot\nabla\bar{\xi}-
\frac{1}{6}\alpha^2e^{-2\sqrt{2}\phi}\right) \nonumber\\
+ \sqrt{2}\int_{M_4^{(1)}}d^4 x\sqrt{-\tilde{g}}\alpha e^{-\sqrt{2}\phi}-
\sqrt{2}\int_{M_4^{(2)}} d^4 x\sqrt{-\tilde{g}}\alpha e^{-\sqrt{2}\phi}, 
\end{eqnarray}
where $\alpha$ is a constant, 
the orbifold fixed planes and five--dimensional spacetime are denoted 
by $M_4^{(1,2)}$ and $M_5$, respectively, and the pull--back of the 
five--dimensional metric onto $M^{(1,2)}_4$ is denoted by 
$\tilde{g}_{ij}$ ($0\le i,j \le 3$). The complex scalar mode, $\xi$, 
determines the non--trivial components of the three--form potential 
on the Calabi--Yau manifold such that  $A_{mnp} =\xi \omega_{mnp}/6$, 
where $\omega_{mnp}$ is the harmonic $(3,0)$ form. However, 
the consistent truncation of the five-dimensional action 
that we are considering here is only 
possible if the imaginary component of $\xi$ is constant. 
The scalar field $\phi$ is related to the deformation parameter 
by $\phi \equiv (\ln V)/\sqrt{2}$. 

The equations of motion derived from the action (\ref{HWaction}) are
given by \cite{Reall98}
\begin{eqnarray} 
R_{\mu\nu} -\frac{1}{2} Rg_{\mu\nu} 
=\nabla_{\mu}\phi\nabla_{\nu}\phi+2e^{2\rho-\sqrt{2}\phi}
 \nabla_{\mu}\rho\nabla_{\nu}\rho {}\nonumber\\
 - g_{\mu\nu}\left(\frac{1}{2}
 (\nabla\phi)^2+e^{2\rho-\sqrt{2}\phi}(\nabla\rho)^2+\frac{1}{6}
 \alpha^2 e^{-2\sqrt{2}\phi}\right) \nonumber\\
+ \sqrt{2}\alpha
 \sqrt{\frac{\tilde{g}}{g}}\tilde{g}^{ij}g_{i\mu}g_{j\nu}e^{-\sqrt{2}\phi}
 (\delta(y)-\delta(y-\pi\lambda)), 
\end{eqnarray} 
\begin{eqnarray} 
\Box\phi = -\sqrt{2}
e^{2\rho-\sqrt{2}\phi}(\nabla\rho)^2-\frac{\sqrt{2}}{3}\alpha^2 
e^{-2\sqrt{2}\phi} \nonumber \\
+
2\alpha\sqrt{\frac{\tilde{g}}{g}}e^{-\sqrt{2}\phi}
(\delta(y)-\delta(y-\pi\lambda)) 
\end{eqnarray} 
\begin{equation} 
\partial_{\mu}(\sqrt{-g}e^{\rho-\sqrt{2}\phi}
 \partial^{\mu}\rho)=0, 
\end{equation} 
where $\xi \equiv e^{\rho +i\theta}$ $(\theta ={\rm constant})$ and 
$y\in  [-\pi\lambda,\pi\lambda]$ denotes the the coordinate along the 
orbifold. The ${\rm Z}_2$ fixed points are located at $y=0$ 
and $y=\pi\lambda$.

The constant $\alpha$ in Eq. (\ref{HWaction}) 
arises due to the non--zero mode of the 
four--form field strength on the Calabi--Yau space. It results 
in a self--interaction potential for the scalar field, 
$\phi$, that does not exhibit a global minimum. 
This is highly significant, because it implies that 
flat space is {\em not} a solution to this theory. However, 
there does exist a static solution to the field equations
that may be interpreted as a pair of parallel three-branes
\cite{LukOvrSte98a,LukOvrSte98b}.  
The existence of a three--brane solution is suggested by the 
potential term in the five--dimensional effective action. This 
may be interpreted as a $0$--form field strength and, in general, 
a solitonic $p$--brane is supported by the magnetic charge of a 
$(D-p-2)$--form field strength in $D$ spacetime dimensions 
\cite{DufKhuLu95}. 

When $\rho$ is constant, the 
analytical form of this three--brane
solution is given by \cite{LukOvrSte98a,Reall98} 
\begin{equation}
\label{threebrane}
ds^2_5= -e^{2U_2} d\eta^2 +e^{2A_2} d\Omega^2_{3,\kappa} + e^{2B_2} dy^2 ,
\end{equation}
where
\begin{eqnarray}
\label{specificbrane}
e^{U_2(y)} = e^{A_2 (y)} = a_0 H^{1/2} \nonumber \\
e^{B_2(y)} =b_0 H^2 \nonumber \\
V(y) =b_0 H^3 \nonumber \\
H(y) = \frac{\sqrt{2}}{3} \alpha |y| +c_0 ,
\end{eqnarray}
the three--metric, 
$d\Omega_{3,\kappa}^2$, is the line element on the 
three--dimensional hypersurfaces defined in Eq. (\ref{constantline})
with positive $(\kappa > 0)$, negative 
$(\kappa < 0)$ and zero $(\kappa =0)$ spatial curvature, respectively, and 
$\{ a_0 , b_0 , c_0 \} $ are constants. 
The solution (\ref{threebrane}) preserves half of the 
$D=5$ supersymmetries. After a further reduction to four dimensions, 
the four--dimensional spacetime may be interpreted as the world--volume 
of the three--brane \cite{LukOvrSte98a,LukOvrSte98b}. This 
solution also applies when $\xi$ has the non--trivial form
\cite{LukOvrSte98a,LukOvrWal98a,Reall98}
\begin{equation} 
 \xi=e^{i\theta}(d_0H^4+\xi_0) ,
\end{equation} 
where $\{ \theta$, $d_0 , \xi_0 \}$ are real constants.

The solution (\ref{threebrane}) has been 
generalized to allow for a cosmological time--dependence 
\cite{LukOvrWal98a,Reall98}. Thus far, 
all solutions of this type have been found by employing the 
technique of separation of variables. They have the generic form 
\begin{eqnarray}  
\label{reallsolution}
ds^2_5=-e^{2U(\eta ,y)}d\eta^2+e^{2A(\eta 
,y)}d\Omega_{3,\kappa}^2+e^{2B(\eta ,y)}dy^2 \nonumber \\
\phi= \phi (\eta ,y),\qquad\rho=\rho(\eta ,y), 
\end{eqnarray} 
where $U(\eta ,y)=U_1(\eta )+U_2(y)$, etc. This separable 
ansatz makes it possible to search independently 
for $y$-- and $\eta$--dependent solutions. The spatially 
flat models $(\kappa =0)$ were considered in 
\cite{LukOvrWal98a}, and generalized to the open and closed models 
$(\kappa \neq 0)$ \cite{Reall98}. In particular, 
the $y$--dependence of the solutions found by 
Reall \cite{Reall98} when $\xi$ is trivial is given by Eq. 
(\ref{specificbrane})
and the time--dependence is 
\begin{eqnarray} 
\label{realltimesolution}
e^{A_1}= e^{U_1} &=& 
\frac{\tau^{(1-\delta)/2}}{\sqrt{1+\kappa \tau^2}} \nonumber \\
e^{B_1}&=&\tau^{\delta} \nonumber \\
e^{\phi_1} &\propto& \tau^{\epsilon\delta}, 
\end{eqnarray}
where $\delta=\pm\sqrt{3} /{2}$, $\epsilon =1/\sqrt{2}$ and 
the new time variable, $\tau$, is defined in Eq. (\ref{wanmimtime}) 
to be 
\begin{equation} 
\label{realltime} 
\tau \equiv \left\{
\begin{array}{ll}
\kappa^{-1/2} |\tan(\kappa^{1/2}\eta)| & {\rm for}\ \kappa>0 \\
|\eta|  & {\rm for}\ \kappa=0 \\
|\kappa|^{-1/2} |\tanh(|\kappa|^{1/2}\eta)| & {\rm for}\ \kappa<0
\end{array}
\right.
\ .
\end{equation}

These solutions incorporate the $\kappa =0$ 
cosmological solutions first derived in \cite{LukOvrWal98a}. 
All models evolve out of a five--dimensional curvature singularity and the 
spatially closed models also develop into a singularity 
after a finite time. On the other hand, 
the negatively--curved model has interesting 
asymptotic behaviour in the late--time limit. The metric for 
the $\kappa=-1$  solution has components of the form
\begin{eqnarray}
e^{2A} = e^{2U} = a_0^2 |\sinh \eta |^{1\mp\frac{\sqrt{3}}{2}} (\cosh 
\eta )^{1\pm\frac{\sqrt{3}}{2}}H(y) \nonumber \\
e^{2B} = b_0^2 |\tanh 
\eta |^{\pm\sqrt{3}}H^4(y) 
\end{eqnarray} 
and  the scalar field is given by
\begin{equation} 
e^{\sqrt{2}\phi}=b_0 |\tanh \eta |^{\pm\sqrt{3}/2}H^3(y)   .
\end{equation} 
It follows that at late times, 
$e^{A_1}\propto e^{\eta}$ and $B_1$ asymptotically becomes constant. 
This corresponds to the static 
domain wall solution (\ref{threebrane}) and 
implies that this solution evolves  
from an initial curvature singularity towards the supersymmetric 
vacuum solution of the theory. 
 
The time-dependence of the solutions (\ref{reallsolution}) reduces to
a particular case of the dilaton--moduli--vacuum solutions
(\ref{dila})--(\ref{dilbeta}) derived in Section \ref{Section7}
\cite{MimWan95a,CopLahWan94}.  The reason for this similarity is due
to the fact that in both cases, the time--dependence of the solutions
originates from the dynamics of massless fields. This follows because
the potential energy terms in the Ho\v{r}ava--Witten equations of
motion arise only in the separated equations containing spatial $(y)$
derivatives. In effect, any dynamical effect that these terms may have
is eliminated by the static, domain wall sector of the separable
solution. 
Recently it has been shown \cite{Lidsey00} that any anisotropic and
inhomogeneous cosmological solution to the lowest-order
dilaton-graviton string equations of motion (including anisotropic and
inhomogeneous solutions discussed in section~\ref{Section7}) may be
employed as a seed to derive a curved, three-brane cosmological
solution to five-dimensional heterotic M-theory compactified on a
Calabi-Yau three-fold.

In the Ho\v{r}ava-Witten context, the size of the orbifold, i.e., the
physical separation of the two domain walls is parametrized by $B_1$
and this plays the role of the dilaton field in four--dimensions.
Thus, the solution (\ref{realltimesolution}) has the same analytical
form as the dilaton--moduli--vacuum solution
(\ref{dila})--(\ref{dilbeta}), where $\cos \xi_* =\pm 1/2$. These
specific values on the exponents of the solution arise because the
$\delta$--function terms in the Ho\v{r}ava-Witten equations of motion
must be time--independent if separable solutions are to be found.
This condition implies that the orbifold dimension and Calabi--Yau
space must expand in a well--defined way $(B_1 = \sqrt{2} \phi_1)$ and
this leads to the restriction on the allowed values of $\xi_*$. For
these values, the string--frame scale factor for the spatially flat
model is given [see Eq.~(\ref{dilat1})] by $a \propto t^{(9\pm
4\sqrt{3})/33}$ and this solution does not represent an inflationary
cosmology.  It is not presently clear whether there exist other
cosmological solutions that lead to realistic inflationary behaviour
within the context of the Ho\v{r}ava-Witten theory. The open question
that remains, therefore, is whether this theory is compatible with the
inflationary scenario. Perhaps this question can be addressed by
searching for non--separable solutions to the field equations 
\cite{ChaRea99,KimKim99} or by
including further non--trivial form--fields in the effective action.

\vspace{1in}

\section*{Acknowledgements}

\addcontentsline{toc}{section}{Acknowledgements}

We thank J. Barrow, A. Billyard, 
H. Bridgman, R. Brustein, C. Cartier, D. Clancy, A. Coley, R. Durrer, 
R. Easther, A. Feinstein, M. Gasperini, D. Holden, N. Kaloper, I. Kogan, 
K. Kunze, 
A. Lahiri, A. Liddle, R. Madden, K. Maeda, J. Maharana, H. Reall, 
K. Stelle, R. Tavakol, M. V\'azquez-Mozo, G. Veneziano and F. Vernizzi
for many helpful discussions and communications. 
JEL and DW are supported by the Royal Society. EJC 
is supported by the Particle Physics and Astronomy Research 
Council, PPARC (UK). 


\newpage
\appendix

\section{Conformal transformations}

\setcounter{equation}{0}

\def\theequation{A.\arabic{equation}}

\def\tbox{\stackrel{\sim}{\Box}}

\label{appendixA}
\label{confsec}

In this Appendix we will show how various geometric and physical
quantities transform under a conformal rescaling of the metric in $D$
spacetime dimensions. Quantities in the conformal metric will be denoted
by a tilde.

We shall denote the conformal factor by $\Omega^2$ which must be positive
(to leave the signature of the metric unaltered) and is some
function of the spacetime coordinates, $x^{\mu}$.  The conformally
transformed metric is then
\begin{equation}
\label{CTg}
\tilde{g}_{\mu\nu} = \Omega^2 g_{\mu\nu}
\end{equation}
and the infinitesimal line element is scaled:
\begin{equation}
d\tilde{s}^2 = \Omega^2 ds^2 .
\end{equation}
Notice that the space-/time-like or null properties of vectors remain
unaltered. The determinant of the metric scales as
\begin{equation}
\label{CTrootg}
\sqrt{-\tilde{g}} = \Omega^D \sqrt{-g}
\end{equation}

\subsection{Intrinsic curvature}

Geometric quantities can then be defined relative to the conformally
rescaled metric~(\ref{CTg}). The Christoffel connection for instance is
\begin{eqnarray}
\tilde{\Gamma}^{\lambda}_{\mu\nu} & = & \frac{1}{2} \tilde{g}^{\lambda\kappa}
 \left( \tilde{g}_{\mu\kappa ,\nu} +
        \tilde{g}_{\nu\kappa ,\mu} - \tilde{g}_{\mu\nu ,\kappa} \right) \\
& = & \Gamma^{\lambda}_{\mu\nu} + \frac{1}{\Omega} 
 \left( g^{\lambda}_{\mu} \Omega_{,\nu} +
   g^{\lambda}_{\nu} \Omega_{,\mu} -
   g_{\mu\nu}g^{\lambda\kappa}\Omega_{,\kappa} \right)
\end{eqnarray}
The Riemann and Ricci tensors can similarly be defined, yielding a Ricci
scalar which can be given terms of the old metric as
\begin{equation}
\tilde{R} = \Omega^{-2} \left( R - 2(D-1)\Box \ln\Omega 
	- (D-2)(D-1) g^{\mu\nu} 
		\frac{\Omega_{,\mu}\Omega_{,\nu}}{\Omega^2} \right)
\end{equation}
The d'Alembertian operator itself transforms as
\begin{equation}
\tbox\sigma = \Omega^{-2} \left( \Box\sigma + 
 (D-2) g^{\mu\nu} \frac{\Omega_{,\mu}}{\Omega} \sigma_{,\nu} \right)
\end{equation}
and we can write the original Ricci scalar in terms of the conformally
transformed metric:
\begin{equation}
\label{CTR}
R = \Omega^2 \left( \tilde{R} + 2(D-1) \tbox ( \ln \Omega ) 
 - (D-2)(D-1) \tilde{g}^{\mu\nu} \frac{\Omega_{,\mu}\Omega_{,\nu}}{\Omega^2}
\right) 
\end{equation}

\subsection{Extrinsic curvature}

The extrinsic curvature tensor on a hypersurface orthogonal to a unit
vector field, $n^A$, is given by
\begin{equation}
K_{AB} = h_A^C h_B^D \nabla_C n_D \ , 
\end{equation}
where the induced metric orthogonal to $n^A$ is
\begin{equation}
\label{hAB}
h_{AB} = g_{AB} \pm n_A n_B
\end{equation}
and upper or lower signs correspond to time-like or space-like
vector fields respectively.
The extrinsic curvature tensor can be decomposed into three parts:
\begin{equation}
K_{AB} = \frac{K}{3} h_{AB} + \sigma_{AB} + \omega_{AB}
\end{equation}
The expansion is given by
\begin{equation}
K = h^{AB} K_{AB}  , 
\end{equation}
the shear $\sigma_{AB}$ is the symmetric traceless part and the
vorticity $\omega_{AB}$ is the anti-symmetric part, which is
necessarily zero for a hypersurface-orthogonal vector field.

Under the conformal transformation~(\ref{CTg}), the unit vector field
must be rescaled to give
\begin{equation}
\tilde{n}^A = \Omega^{-1} n^A \ .
\end{equation}
This leads to the new extrinsic curvature tensor in the conformally
rescaled metric
\begin{equation}
\tilde{K}_{AB} = \Omega K_{AB} + h_{AB}t^C\Omega_{,C} \ .
\end{equation}
The shear and vorticity are simply rescaled with
$\tilde{\sigma}_{AB}=\Omega\sigma_{AB}$ and 
$\tilde{\omega}_{AB}=\Omega\omega_{AB}$, but the expansion becomes
\begin{equation}
\label{CTK}
\tilde{K} = \Omega^{-1} K + (D-1)\Omega^{-1} \tilde{t}^A \Omega_{,A} \ .
\end{equation}

\subsection{Dilaton--gravity action}

\label{CTdilgravaction}

We now consider the dilaton-gravity action in $D$ dimensions:
\begin{equation}
\label{dilgravaction}
S = \int_M d^Dx \sqrt{|g|} e^{-\phi} \left[ R - \omega (\nabla\phi)^2
\right] + 2 \int_{\partial M} d^{D-1}x \sqrt{|h|} e^{-\phi} K 
\end{equation}
where $R$ is the intrinsic curvature of the metric $g_{AB}$ on the
manifold $M$ and $K$ is the extrinsic curvature of the boundary
hypersurface $\partial M$ orthogonal to the vector field $n^A$, with
the induced metric $h_{AB}$. [See Eq.(\ref{hAB})]\footnote{The 
boundary term is necessary if the field equations are to be
derived from requiring that the action is extremised with respect to
any field variations that vanish on the boundary. Otherwise one requires in
addition that the first derivatives of the variations also vanish on
the boundary~\cite{York72,GibHaw77,Wands94}.}

Under the conformal transformation (\ref{CTg}) the
action becomes
\begin{eqnarray}
S = \int_M d^Dx \Omega^{-D} \sqrt{|\tilde{g}|} e^{-\phi} 
\left[ \Omega^2 \tilde{R}
+2(D-1)\Omega^2\tbox(\ln\Omega) \right. \nonumber \\
\left. 
-(D-2)(D-1)(\tilde\nabla\Omega)^2 - \omega \Omega^2(\tilde\nabla\phi)^2
\right] \nonumber\\
+ 2 \int_{\partial M} d^{D-1}x \Omega^{1-D} \sqrt{|h|} e^{-\phi}
\left[ \Omega K - (D-1)(\tilde\nabla\Omega)_A\tilde{n}^A \right]
\ ,
\end{eqnarray}
where the Ricci scalar is given by (\ref{CTR}) and the extrinsic
curvature by~(\ref{CTK}).
This reduces to
\begin{eqnarray}
S = \int_M d^Dx \sqrt{|\tilde{g}|} e^{-\phi} \Omega^{2-D} \left[
 \tilde{R} - \omega (\tilde\nabla\phi)^2
+(D-2)(D-1)\Omega^{-2} (\tilde\nabla\Omega)^2 \right. \nonumber \\
\left. +2(D-1)\Omega^{-1}(\tilde\nabla\Omega)(\tilde\nabla\phi)
\right] + 2 \int_{\partial M} d^{D-1}x \sqrt{|h|} e^{-\phi} \Omega^{2-D}
\tilde{K} \ . 
\end{eqnarray}

There are two important cases:
\begin{enumerate}
\item
$\Omega=e^{f(\phi)}$, where $f(\phi)\neq\phi/(2-D)$\\
In this case the form of the action given in Eq.~(\ref{dilgravaction})
remains conformally invariant if we define
\begin{equation}
\tilde\phi = \phi + (D-2)f(\phi)
\end{equation}
and 
\begin{equation}
\tilde\omega = \frac{\omega - (D-1) f'(\phi) \left[ 2 + (D-2) f'(\phi)
\right]}{[1+(D-2)f'(\phi)]^2} \ .
\end{equation}
In the particular case $f(\phi)=-2\phi/(D-2)$, the dimensionless
Brans-Dicke parameter $\omega$ remains invariant, but we have 
$\tilde\phi=-\phi$. This invariance of the gravitational Lagrangian
plays a key role in string-string duality discussed in
Section~\ref{Section5}.
\item
$\Omega^{2-D}=e^\phi$\\
In this case we have
\begin{equation}
\label{ste}
S = \int_M d^Dx \sqrt{|\tilde{g}|} \left[ \tilde{R}
 - \left( \omega + {D-1 \over D-2} \right) (\tilde\nabla\phi)^2
\right]
+ 2 \int_{\partial M} d^{D-1}x \sqrt{|h|} \tilde{K} \ . 
\end{equation}
This is usually referred to as the Einstein frame, as the
gravitational action has the standard Einstein-Hilbert form, and the
dilaton is minimally coupled to the metric in this frame.
\end{enumerate}

\subsection{Energy--momentum tensor}

We now consider the part of the action contributed by the integral of the
matter Lagrangian:
\begin{equation}
S_{\rm matter} = \int_{M} \sqrt{-g} d^Dx {\cal L}_{\rm matter}
\end{equation}
The energy-momentum tensor is defined as
\begin{equation}
T^{\mu\nu} = \frac{2}{\sqrt{-g}} \frac{\partial}{\partial g_{\mu\nu}}
 \left( \sqrt{-g} {\cal L}_{\rm matter} \right)
\end{equation}
so that a first-order variation of the matter action
with respect to the metric is given by
\begin{equation}
\frac{\delta S_{\rm matter}}{\delta g_{\mu\nu}} = 
 \frac{1}{16\pi} \int_M \sqrt{-g} d^Dx \left[ 8\pi T^{\mu\nu} \right]
\end{equation}

In terms of the conformally transformed metric
\begin{equation}
S_{\rm matter} = \int_{M} \sqrt{-\tilde{g}} d^Dx \Omega^{-D} 
 {\cal L}_{\rm matter}
\end{equation}
so the corresponding energy-momentum tensor must be
\begin{eqnarray}
\tilde{T}^{\mu\nu} & = & \frac{2}{\sqrt{-\tilde{g}}} 
 \frac{\partial}{\partial \tilde{g}_{\mu\nu}}
 \left( \sqrt{-\tilde{g}} \Omega^{-D} {\cal L}_{\rm matter} \right) 
 \\ \nonumber
& = & \Omega^{-D} \frac{2}{\sqrt{-g}} 
 \frac{\partial g_{\lambda\kappa}}{\partial \tilde{g}_{\mu\nu}}
 \frac{\partial}{\partial g_{\lambda\kappa}}
 \left( \sqrt{-g} {\cal L}_{\rm matter} \right) \\ \nonumber
& = & \Omega^{-D-2} T^{\mu\nu}
\end{eqnarray}
Therefore one can, mathematically speaking, describe the
energy-momentum tensor $T^{\mu\nu}$ as conformally invariant with
weight $-(D+2)$. (Its trace $T$ has weight $-D$.) However the common
statement that only traceless ($T=0$) matter (i.e. radiation) is
``conformally invariant'' follows from the further, physical,
requirement that if the original energy-momentum tensor is conserved
($\nabla^\mu T_{\mu\nu} = 0$) then the conformally transformed
energy-momentum tensor is also conserved. In general we find
\begin{equation}
\tilde{\nabla}^{\mu} \tilde{T}_{\mu\nu} = - \frac{\Omega_{,\nu}}{\Omega}
 \tilde{T} \ .
\end{equation}

If the matter in the original metric evolves as a perfect fluid its
energy-momentum tensor can be given in terms of its density $\rho$,
pressure $p$ and (time-like) velocity field $u^{\mu}$, normalised such
that $u_{\mu}u^{\mu}=-1$:
\begin{equation}
T^{\mu\nu} = (\rho +p) u^{\mu}u^{\nu} + p g^{\mu\nu}
\end{equation}
Under a conformal transform the unit time-like velocity field transforms
as $\tilde{u}^{\mu} = \Omega^{-1} u^{\mu}$ and the energy-momentum
tensor can be rewritten as
\begin{equation}
\tilde{T}^{\mu\nu} = (\tilde{\rho} +\tilde{p}) \tilde{u}^{\mu}\tilde{u}^{\nu}
 + \tilde{p} \tilde{g}^{\mu\nu}
\end{equation}
where
\begin{eqnarray}
\tilde{\rho} & = & \Omega^{-D} \rho \\
\tilde{p} & = & \Omega^{-D} p
\end{eqnarray}
Notice that a barotropic fluid, $p=(\gamma -1)\rho$, retains the same
barotropic index under the conformal transformation. However it will not
remain a perfect fluid conserving energy and momentum unless
$T=\{(D-1)\gamma-D\}\rho=0$.

\subsection{Form Fields}

A specific type of massless field that arises many times in this review is 
the antisymmetric $(n-1)$--form potential. This has a field strength 
defined by 
$H_{A_1A_2\ldots} \equiv n \partial_{[A_1}B_{A_2\ldots]}$. Since this 
definition is independent of the metric, it 
is conformally invariant. However, the
scalar quantity
\begin{equation}
H^2\equiv g^{A_1B_1}g^{A_2B_2}\ldots
H_{A_1A_2\ldots}H_{B_1B_2\ldots}
\end{equation}
transforms as
\begin{equation}
\label{CTH}
\tilde{H}^2 = \Omega^{-2n} H^2 \ .
\end{equation}


\section{Modular Group of the Torus}

\setcounter{equation}{0}

\def\theequation{B.\arabic{equation}}

\label{appendixB}

In this appendix, we discuss the relationship between  
a 2--torus and the group ${\rm SL}(2,Z)$ 
\cite{AlvMee99,GreSchWit87,GivMalRab89,ShaWil89,SeiWit86,LusThe89}. 
Let the coordinates spanning the 2--plane $\Re^2$ be 
represented by $\sigma_1$ and $\sigma_2$. 
A two--dimensional torus $T^2$ can then be constructed 
by imposing a suitable equivalence relation in the 2--plane 
\cite{GreSchWit87}: 
\begin{equation}
\label{equivalent}
(\sigma_1 ,\sigma_2 ) \approx (\sigma_1 +p , \sigma_2 +q)
\end{equation}
where $p$ and $q$ are arbitrary integers. Hence, a 2--torus 
may be viewed as $\Re^2/\Lambda$, where $\Lambda $ is a two--dimensional 
lattice. 

There exists a group of transformations where 
points that are equivalent under the identification
(\ref{equivalent}) are mapped onto equivalent points. 
These transformations correspond to the diffeomorphisms 
\begin{equation}
\label{diff}
(\sigma_1 ,\sigma_2) \rightarrow (d\sigma_1 +c\sigma_2 , 
b\sigma_1 +a\sigma_2)
\end{equation}
where $\{ a,b,c,d \} $ are integers. Eq. (\ref{diff}) represents an 
invertible and one--to--one map of the 
2--torus onto itself if these constants also satisfy the 
additional constraint $ad-bc =1$. 
Eq. (\ref{diff}) gives the action of the matrices 
\begin{equation}
\label{sl2z}
U=\left( \begin{array}{cc}
d & c \\ 
b & a \end{array} \right) , \quad ad-bc =1 , \quad
\{ a,b,c,d, \} \in Z
\end{equation}
by fractional linear transformations. These 
matrices form a group ${\rm SL}(2,Z)$ corresponding to 
a subgroup of ${\rm SL}(2,R)$. As we discussed in Section \ref{Section4.2}, 
the action of ${\rm SL}(2,Z)$ on the complex 
plane is determined by $\tau \rightarrow (a\tau +b)/(c\tau +d)$ and the 
group is generated by the transformations 
\begin{eqnarray}
\label{tauz1}
\tau \rightarrow -1/\tau \\
\label{tauz2}
{\rm Re} \, \tau \rightarrow {\rm Re} \, 
\tau +1
\end{eqnarray} 
If we consider transformations away 
from the imaginary axis, ${\rm Re} \, \tau \ne 0$, Eq. (\ref{tauz1}) 
interchanges 
the interior region of the unit circle with the exterior region 
\cite{ShaWil89} and is equivalent to the 
simultaneous interchange $\sigma_1 \rightarrow \sigma_2$
and $\sigma_2 \rightarrow - \sigma_1$ \cite{GreSchWit87}. The 
transformation (\ref{tauz2}) leaves $\sigma_1$ invariant and 
induces the shift $\sigma_2 \rightarrow \sigma_2+\sigma_1$. 

A 2-torus has complex dimension of one and may also 
be parametrized in terms of a complex variable $z$ defined 
by \cite{GreSchWit87}
\begin{equation}
z \equiv \sigma_1 +\sigma_2 \tau
\end{equation}
where $\tau$ is an arbitrary complex number with positive 
definite imaginary part. Then, from Eq. (\ref{equivalent}), 
a 2--torus may be described  
(up to a conformal equivalence) in terms of the 
complex plane, where the identifications $z 
\approx z+1$ and $z \approx  z+\tau$ 
are made \cite{Schwarz98a,SeiWit86}. In other words, a 2--torus 
is constructed by identifying the opposite ends of the parallelogram, where  
one length of the 
parallelogram is identified with the point 1 and the 
other with the complex parameter $\tau$ \cite{Vafa97}. 
It then follows that $\tau$ and $\tau' =(a \tau +b )/(c \tau +d)$ 
describe equivalent tori. 

To summarize, the group ${\rm SL}(2,Z)$ may be realised 
as the group 
of reparametrizations of the 2--torus, corresponding to 
a change in basis vectors of the lattice in $\Re^2$. 
For these reasons it 
is referred to as the modular group of the torus and the parameter 
$\tau$ then represents the modular parameter. This provides a 
geometrical interpretation of the S--duality of the type IIB superstring
discussed in Section \ref{Section5.3}. 

We also remark that the appearance 
of the ${\rm SL}(d,R)/{\rm SO}(d)$ non--linear 
sigma--model is a generic feature of 
(super)--gravity theories compactified 
on $T^d$. The group ${\rm SL}(d,R)$
is the remnant of the higher--dimensional 
diffeomorphism (Lorentz) invariance
that is unbroken by the toroidal compactification 
For example, the toroidal compactification of Einstein gravity 
leads to the action (\ref{slaction}). Since 
$f_{ab}$ is an arbitrary, symmetric, $d \times d$ 
matrix with unit determinant,
the moduli fields arising from the higher--dimensional metric 
parametrize the ${\rm SL}(d,R)/{\rm SO}(d)$ coset. 
With this connection in mind, 
the S--duality of the ten--dimensional type IIB superstring 
has been cited as evidence for the existence 
of a twelve--dimensional theory that is referred 
to as F--theory \cite{Vafa96}.

\section{Bianchi Classification of Homogeneous Spacetimes}

\setcounter{equation}{0}

\def\theequation{C.\arabic{equation}}

\label{appendixC}

The Bianchi models are four--dimensional spacetimes 
that admit three--dimensional, 
space-like hypersurfaces, $\Sigma_t$, on which a three--parameter 
group of isometries acts simply transitively 
\cite{RyaShe75,Wald84,MacCallum79}. 
The group of isometries is a Lie group, $G_3$.  
By simply transitive, we mean that 
for all points $p$ and $q$ on $\Sigma_t$ there 
exists a unique element of the Lie group that maps $p$ onto $q$. 
This implies there is a one--to--one 
correspondence between the points on $\Sigma_t$ and the elements 
of $G_3$ and the structure of the 
spacetime is therefore $M_4=R\times G_3$. Coordinates can then 
be chosen such that the four--dimensional 
line element is given by 
\begin{equation}
\label{bianchimetric}
ds^2 =-dt^2  +h_{ab} (t) \omega^a\omega^b , \qquad a,b =1,2,3
\end{equation}
The surfaces of homogeneity, 
$\Sigma_t$, then represent surfaces of constant proper 
time, $t$.  
The one--forms $\omega^a$ determine the isometry 
of the three--surfaces and satisfy the 
Maurer--Cartan equation
\begin{equation}
\label{omegadefinition}
d\omega^a =\frac{1}{2} {C^a}_{bc}\omega^b \wedge \omega^c
\end{equation}
where ${C^a}_{bc}$ are the structure constants of the Lie algebra 
of $G_3$. These are antisymmetric, 
\begin{equation}
\label{antisymmetric}
{C^a}_{(bc)} =0
\end{equation}
and satisfy the Jacobi identity
\begin{equation}
\label{jacobi}
{C^e}_{d[a}{C^d}_{bc]} =0
\end{equation}

Bianchi was the first to determine 
and classify all three--dimensional 
Lie algebras into nine types, I, II, $\ldots$, IX \cite{Bianchi98}.
Here we consider the classification due to Ellis and MacCallum 
\cite{EllMac69}. 
The antisymmetric condition (\ref{antisymmetric}) 
implies that ${C^a}_{bc}$ has at most nine 
independent components. These can be separated into the six 
components of a symmetric $3 \times 3$ matrix $M^{ab}$ 
and the three components of a $3\times 1$ vector $A_b$. Defining 
the latter as the trace: 
\begin{equation}
A_b \equiv {C^a}_{ab}
\end{equation}
implies that the former may be defined by 
\begin{equation}
\label{Mab}
{C^c}_{ab} \equiv M^{cd}\epsilon_{dab} + {\delta^c}_{[a}
A_{b]}
\end{equation}
where $\epsilon^{abc}$ is the totally antisymmetric tensor 
and $\epsilon^{123} =1$. 
Substitution of Eq. (\ref{Mab}) into Eq. (\ref{jacobi}) 
implies that $A_b$ is transverse to $M^{ab}$ \cite{EllMac69}: 
\begin{equation}
\label{jacobi1}
M^{ab}A_b =0
\end{equation}
Thus, only six components of 
${C^a}_{bc}$ can be independent. 
If $A_b$ is non--trivial, it may be viewed as an eigenvector 
of $M^{ab}$ with a zero eigenvalue. 
Without loss of generality, $M^{ab}$ may 
be diagonalized, $M^{ab}={\rm diag}[m_1, m_2 , m_3]$, 
and $A_b$ may be written as $A_b =(A,0,0)$. By a 
suitable rescaling, the eigenvalues of $m^{ab}$ 
can then be made equal to $0$, $\pm1$ and this implies that 
Eq. (\ref{jacobi1}) simplifies to 
\begin{equation}
\label{jacobi2}
m_1A =0
\end{equation}
Models where $A=0$ are referred to as the Bianchi class A \cite{EllMac69}. 
Models where $A \ne 0$ $(m_1 =0)$ belong to the Bianchi class B. 
In the former case, the Lie algebras are classified by the rank (the 
number of non--zero elements) of $M^{ab}$ and the modulus of 
its signature. Thus, there are 
six types in this class. 
The different Bianchi types  are labelled in Table \ref{bianchitypes} 
\cite{MacCallum79,Siklos84}.

\begin{table}
\begin{center}
\begin{tabular}{||c|c|c|c|c||}
\hline \hline
Class       &     Type               &     $m_1$ & $m_2$ & $m_3$   \\
\hline \hline
A           &       I                &       0   &    0  &   0     \\
            &       II               &       1   &    0  &   0     \\
            & ${\rm VI}_0$           &       0   &    1  &  -1     \\
            & ${\rm VII}_0$          &       0   &    1  &   1     \\
            &      VIII              &       1   &    1  &  -1     \\
            &       IX               &       1   &    1  &   1     \\
\hline 
B           &       V                &       0   &    0  &   0     \\
            &       IV               &       0   &    0  &   1     \\
            & III $({\rm VI}_{-1})$  &       0   &    1  &  -1     \\
	    & ${\rm VI}_h(h < 0)$    &       0   &    1  &  -1     \\
            &  ${\rm VII}_h(h >  0)$ &       0   &    1  &   1     \\
\hline \hline
\end{tabular}
\end{center}
\caption[shortname]{The Bianchi types and eigenvalues of $M_{ab}$. 
See the text for details.}
\label{bianchitypes}
\end{table}

In the class B there are four possibilities for the rank
of $M^{ab}$ and the modulus of its signature, since $m_1=0$. 
When ${\rm rank}[M^{ab}] =2$, one may define a scalar, $h$: 
\begin{equation}
h \equiv \frac{A^2}{m_2m_3}
\end{equation}
Thus, there are two one--parameter families of Lie algebras 
in the class B. These are labelled by the parameter $h$
and $h<0$ for the type ${\rm VI}_h$ and $h>0$ for 
the type ${\rm VII}_h$. We remark that the Bianchi type III 
is the same as the type ${\rm VI}_{-1}$.  

In some cases the $G_3$ group of isometries may only represent a subgroup 
of the full symmetry. There may also exist a group of transformations 
such that all points $p$ on $\Sigma_t$ are mapped onto 
themselves. This group of transformations is called the 
isometry group. Spacetimes admitting an isometry group are 
called locally rotationally symmetric (LRS) spacetimes \cite{Ellis67}. 
LRS spacetimes arise for Bianchi types I, II, 
${\rm VII}_0$, VIII, IX, III and 
${\rm VII}_h$ \cite{EllMac69}. The FRW metrics admit a three--dimensional 
isotropy group. They are the isotropic limits of the 
type IX $(k=1)$, types I and ${\rm VII}_0$ $(k=0)$,  and 
types V and ${\rm VII}_h$ $(k=-1)$. 

\begin{table}
\begin{center}
\begin{tabular}{||c|c||}
\hline \hline
Type    & $\omega^a$ \\
\hline \hline
I       & $dx$, $dy$, $dz$ \\
II      & $dx-zdy$, $dy$, $dz$ \\
IV      & $dx$, $e^xdy$, $e^x(xdy+dz)$ \\
V       & $dx$, $e^xdy$, $e^xdz$     \\
VI(III) & $dx$, $e^{Ax}(\cosh xdy -\sinh xdz)$, $e^{Ax}(-\sinh x dy +\cosh x dz)$ \\
VII     & $dx$, $e^{Ax}( \cos x dy -\sin x dz)$, $e^{Ax}(\sin x dy +\cos x dz)$ \\
VIII    & $\cosh y \cos z dx -\sin z dy$, $\cosh y \sin z dx+\cos z dy$, $\sinh ydx +dz$ \\
IX      & $\cos y\cos z dx-\sin zdx$, $\cos y\sin zdx +\cos zdy$, $-\sin ydx+dz$ \\
\hline \hline
\end{tabular}
\end{center}
\caption[shortname]{The standard form of the one--forms 
$\omega^a$ for the Bianchi types 
that arise in the spacetime metric (\ref{bianchimetric}).} 
\label{bianchiforms}
\end{table}

The one--forms that arise in the spacetime metric 
(\ref{bianchimetric}) are tabulated in Table \ref{bianchiforms} for 
all Bianchi types. (For further details see, e.g., Ref. \cite{MacCallum79}). 
The three--metric may be parametrized by 
\begin{equation}
\label{3metricbianchi}
h_{ab}(t) =e^{2\alpha (t)} \left( e^{2\beta (t)} \right)_{ab}
\end{equation}
where 
\begin{equation}
\label{betametric}
\beta_{ab} \equiv {\rm diag} \left[ 
\beta_++\sqrt{3}\beta_-, \beta_+ -\sqrt{3}\beta_- ,-2\beta_+ \right]
\end{equation}
is a traceless matrix. Thus, $[{\rm det} h_{ab}]^{1/2} \equiv 
h^{1/2}  =e^{3\alpha}$ and the volume parameter and shear are 
entirely determined by $\alpha$ and $\beta$, respectively.

The scalar curvature, ${^{(3)}}R$, of the homogeneous hypersurfaces, 
$\Sigma_t$, is uniquely determined by the structure constants 
${C^a}_{bc}$ of the corresponding Lie algebra for each Bianchi type 
\cite{Wald83,Wald84}. It is given by 
\begin{equation}
\label{3bianchicurvature}
{^{(3)}}R =- \frac{3}{2} A_bA^b -h^{-1}\left( M_{ab}M^{ab} 
-\frac{1}{2} {M^a}_a{M^b}_b \right)
\end{equation}
where indices are raised and lowered with $h^{ab}$ 
and $h_{ab}$, respectively. Thus, the Bianchi 
type I is spatially flat. In this sense, it represents 
the simplest anisotropic cosmology and the 
$G_3$ is the abelian translation 
group $T^3$. For closed spatial sections, $\Sigma_t$ has 
the topology of a 3--torus, $T^3 =S^1 \times S^1 \times S^1$. 
It follows from Table \ref{bianchitypes} 
that all Bianchi models, except 
type IX, have non--positive spatial curvature, ${^{(3)}}R \le 0$ 
\cite{Wald83}. 
For type IX, $G_3={\rm SO}(3)$ and the topology of $\Sigma_t$ 
is $S^3$. 

In the class B models $(A \ne 0)$, a divergence term may arise 
when integrating over the spatial variables in the effective action due to 
the term proportional to $A_bA^b$ 
in Eq. (\ref{3bianchicurvature}). 
This renders the Lagrangian 
and Hamiltonian formulations of the field equations 
ambiguous for these models \cite{Sneddon75}. 
For the class A, on the other hand, 
the reduced action for the four--dimensional dilaton--graviton 
sector of the string effective action (\ref{NSNSactionreduced}) 
is derived by substituting in the {\em ansatz}
(\ref{bianchimetric}) and 
integrating over the spatial variables. 
This leads to 
\begin{equation}
S=\int dt e^{3\alpha -\varphi} \left[ 6 \dot{\alpha} \dot{\varphi} 
-6\dot{\alpha}^2 +6\dot{\beta}_+^2 +6\dot{\beta}^2_- -
\dot{\varphi}^2 +{^{(3)}}R \right]
\end{equation}

Finally, we note that when the dilaton field 
is constant on the surfaces of homogeneity, $\varphi =\varphi (t)$, 
the Einstein and string frame metrics correspond to 
the same Bianchi metric. In other words, the conformal transformation 
(\ref{ste}) relating the two metrics does not 
alter the one--forms $\omega^a$. Indeed, when the three--metric is given by 
(\ref{3metricbianchi}), the conformal transformation is formally equivalent 
to a rescaling of the time parameter and volume of the spatial 
sections.

\end{document}